\documentclass[twoside,a4paper,10pt]{report} 
\usepackage{graphicx,psfrag,epsfig}
\usepackage{amsmath,amssymb,latexsym} 
\usepackage{fancyhdr,cite}

\newcommand{\beq}{\begin{equation}}
\newcommand{\eeq}{\end{equation}}
\newcommand{\ordo}{\mathcal{O}}

\DeclareMathOperator{\Tr}{Tr}
\DeclareMathOperator{\tr}{tr}

\begin{document}
\pagestyle{empty}

%%%%% Titlepages %%%%%
\pagestyle{empty}
  \vspace{2cm}
  \sffamily\upshape\mdseries
  \noindent
  {\Huge \textbf{Light Hadron Spectroscopy in \vspace*{5mm}\\
    Quenched Lattice QCD with \vspace*{5mm}\\
    Chiral Fixed-Point Fermions }}
  
  \vspace{10.5cm}

  \noindent
  Inauguraldissertation \\
  der Philosophisch-naturwissenschaftlichen Fakult\"at \\
  der Universit\"at Bern

  \vspace{1.5cm}

  \noindent
  vorgelegt von

  \vspace{4mm}

  \noindent  {\large \textbf{Simon Hauswirth}}

  \vspace{3mm}

  \noindent
  von Gsteig (BE)

  \vspace{1.5cm}

  \noindent
  Leiter der Arbeit: \parbox[t]{7cm}{Prof.~Dr.~P.~Hasenfratz \\
    Institut f\"ur theoretische Physik \\
    Universit\"at Bern}
  
  \normalfont
\newpage
\cleardoublepage

%%% Local Variables: 
%%% mode: latex
%%% TeX-master: "LGTwFPA"
%%% TeX-master: t
%%% End: 

\pagestyle{fancy}
\pagenumbering{roman}
\tableofcontents
\newpage
\pagestyle{empty}
\cleardoublepage

\pagestyle{fancy}
\bibliographystyle{h-physrev4}
\pagenumbering{arabic}
\setcounter{page}{1}

\fancyhead[RE]{\nouppercase{\small\it Abstract and Summary}}
\chapter*{Abstract and Summary}
\addcontentsline{toc}{chapter}{Abstract and Summary}

Quantum Chromodynamics (QCD), the theory of the strong interaction, is one
of the most prominent examples for a beautiful and successful physical 
theory. At large distance, or equivalently at low energy, perturbative
expansions in the coupling constant---the standard tool to treat quantum field
theories analytically---break down, and a non-perturbative
formulation is required to calculate physical quantities. 
In this thesis, we construct the Fixed-Point fermion action
for lattice QCD, which is a highly improved discretization of the
continuum theory that preserves the chiral symmetry inherent in the
original formulation. We perform studies in quenched light hadron spectroscopy
to examine the properties of this action and investigate in detail the chiral
limit of pseudoscalar mesons, which is inaccessible to non-chiral lattice
formulations.

To start with, Chapter~\ref{ch:intro} provides a brief
introduction to the field of elementary particle physics, to Quantum
Chromodynamics and the lattice as a tool to 
probe the non-perturbative regime of the strong interaction, and
motivates the construction of improved transcriptions of the theory to
discrete space-time. A long standing problem, namely the formulation
of chiral symmetric lattice fermions, is addressed in 
Chapter~\ref{ch:chiral_fermions}. An elegant solution has been found
using
Renormalization Group methods, leading to the
classically perfect Fixed-Point actions. Chapter~\ref{ch:d_fp}
describes the parametrization and construction of the Fixed-Point fermion 
action for lattice QCD and presents some elementary properties of the
resulting Dirac operator. A different
possibility to obtain chiral lattice fermions is the overlap construction. We
combine the Fixed-Point and the overlap approach in Chapter~\ref{ch:d_ov} to
remove the 
residual chiral symmetry breaking of our parametrized Dirac
operator, getting a fermion action which inherits the
advantages of both formulations at a higher computational cost. The
chirality and locality properties of this overlap-improved 
Dirac operator are then tested in the artificial framework of smooth
instanton gauge configurations. 

Next, we turn to one of the most
fundamental applications of lattice 
QCD, namely the calculation of hadron masses.
Chapter~\ref{chapter:had_spect} gives an 
introduction to the technical details of how the light
hadron mass 
spectrum is extracted from lattice simulations. With chiral symmetric
fermion actions, it is 
possible to perform lattice simulations at quark masses very close to
or even at the physical mass of up and down quarks, thus allowing to study the
chiral limit, which is complicated by non-analytic
terms in the quenched approximation to QCD. At such small quark
masses, additional quenching effects appear in a finite lattice volume
which contaminate in particular 
the pseudoscalar meson channel and are related to the zero modes of
the Dirac operator. We devote Chapter~\ref{ch:zeromodes} to the study of
these topological finite-volume effects and examine possible solutions for the
problem of extracting reliable pseudoscalar meson masses at small
volumes and quark masses.

In Chapter~\ref{ch:results_fp}, we present the results of a
spectroscopy simulation with the Fixed-Point fermion and gluon lattice
actions. This study is the one of the first hadron spectroscopy
calculations with a chiral symmetric action including checks for
cut-off and finite-volume 
effects. After estimating the magnitude of the topological 
quenching effects, we closely examine the chiral limit of the
pseudoscalar meson and extract the coefficient of the quenched chiral
logarithm in two different ways. We also consider the chiral
extrapolations for vector mesons and baryons and present part of the light
hadron spectrum at finite lattice spacing. Then we study the
dependence of the hadron masses on 
the physical volume and the lattice spacing for the parametrized
Fixed-Point Dirac operator. The scaling properties
of the vector meson mass is compared to other formulations of lattice
fermions. Finally, we investigate how well the continuum
energy-momentum hadron dispersion relation is preserved by our lattice
action, and examine the effect of overlap-improvement on the spectrum
and dispersion relation.
The final chapter contains our conclusions and prospects for the future.

The work covered in this thesis is part of an ongoing project of
parametrizing, 
testing and applying Fixed-Point fermions in lattice QCD, carried out
in collaboration with Thomas   
J\"org, Peter Hasenfratz, Ferenc Niedermayer and Kieran Holland. The
simulations in the last chapter were performed in the framework of the
BGR collaboration.
Part of the results presented here have already been 
published in papers \cite{Hasenfratz:2000xz,Hasenfratz:2002}
and conference proceedings
\cite{Hasenfratz:2000qb,Hasenfratz:2001hr,Hasenfratz:2001qp}. 
While the focus of this thesis is on simulations of the light hadron
spectrum, we will recapitulate some of the basic issues discussed in the
PhD thesis of Thomas J\"org \cite{thomas:diss} which are relevant for
understanding the 
applications and results in the later chapters in
order to keep this work as self-contained as possible.

\fancyhead[RE]{\nouppercase{\small\it Introduction}}
\chapter{Introduction}
\label{ch:intro}

This introductory chapter provides some background information
for the work covered in the body of the thesis. We start at the very
beginning and give a 
short overview of the history and evolution of the field of elementary particle physics. Then we briefly present in Section
\ref{sect:qcd} the foundations of Quantum
Chromodynamics, the theory of the strong nuclear force, and introduce
the important concepts of symmetries and asymptotic freedom. In order 
to calculate physical quantities in a quantum field theory, it is
necessary to introduce a regularization. The lattice, described in 
Section \ref{sect:lattice}, provides a regularization that allows to probe the
non-perturbative regime of strong coupling, where 
phenomena related to the hadronic world can be examined. We define
the most simple lattice actions and the basic tools needed to carry
out lattice computations. Finally, in Section
\ref{sect:motivation} we present
arguments why it is worthwhile to search for improved formulations of
lattice QCD. This motivates the construction and application of the
Fixed-Point Dirac operator that we perform in this thesis.

\section{The Search for the Fundamental Properties of Nature}
\label{sect:history}
Understanding nature is the ultimate goal of every physicist. The
basic questions lying at the foundations of a work like this are: How
does nature work? Can we explain the phenomena we see? Can we make predictions
about what can be seen? 
From the beginnings of history people have witnessed the phenomena
of nature and tried to explain
them. Starting at observations accessible to everyday life experience,
the interest 
has moved to objects beyond human perception. At
the end of this journey towards finding the fundamental laws of nature,
there are two areas: the very small and the very large.
The world of the very large is studied in cosmology, where one tries
to understand the origin, evolution and fate of the universe as a whole.
At the other end of the spectrum one
asks what the basic building blocks of the universe are and how they 
interact. These questions are addressed by the field that is today
called elementary particle 
physics, and it is there where this work tries to add an almost infinitely
small fraction to scientific knowledge.

\subsubsection{The World beneath the Atom}
For most people, including those working in sciences like biology and classical
chemistry, the smallest structures of interest are atoms or even molecules,
and the subatomic world is not considered relevant. This is justified
if one is dealing with objects large compared to the atom, but if our
interest lies in how
nature works at the fundamental level, the fact that the
atom is not undividable, as its Greek name implies, can no longer
be ignored and the subatomic structure of matter needs to be
examined. Thanks to Rutherford's experiments it has been known for
more than 100 years that atoms
are built from a tiny nucleus and a surrounding cloud of electrons. Rutherford
concluded that the nucleus is made of positively charged particles
which he called protons, and for a certain time in the early 20th
century, it seemed like with protons and electrons and Einstein's photon the
basic constituents of matter were found. Paul Dirac's formulation of Quantum
Electrodynamics (QED) in 1926 explained beautifully how electrons interact
by exchange of photons. However, Dirac's equation implied the existence of
an electron with exactly the same properties, but
opposite charge. This looked first as if the theory would be
wrong, since such a particle had never been seen before. As a theoretical
physicist however, Dirac trusted the beauty of his theory more than
the experimental possibilities at that time and
drew the conclusion that this antiparticle---the 
so-called positron---had to exist. Dirac's prediction turned true when
in 1932 the existence of the positron was confirmed in experiments.
The observation that our universe is mainly made of matter, and not of
antimatter like positrons and antiprotons, is related to a small
asymmetry known as 
CP-violation and is a subject of present research.

There were also a number of other problems which implied that protons,
electrons, photons and the electromagnetic force alone were not sufficient
to explain the structure of matter. Among them was the unsolved question why
the atomic nucleus is stable: Protons are 
positively charged, so there should be a strong electromagnetic
repulsion between the protons in the 
nucleus, which drives them apart. The newly discovered neutron could not
help in solving this problem, as it 
is not electrically charged and therefore not able to hold the nucleus
together. Obviously there had to be some other
force which would explain why atomic nuclei didn't fall into pieces. 
Another problem was the anomalous magnetic moment of the proton. While for the
electron the measurements for this quantity were in perfect agreement
with the theoretical prediction of QED, there was almost a factor of 3
difference for the proton, which was a sign that the proton has some
non-trivial internal structure and is not an elementary
particle. Again, Quantum Electrodynamics alone was not able to explain
this phenomenon.
Yet another problem was found in the nuclear beta decay, where in
an unstable atomic nucleus a
proton decays into a neutron and a positron. Here the energy
of the positron leaving the nucleus was found to be considerably smaller than
the energy difference 
between the proton and the neutron, and  it was not clear where the
missing energy was lost. To solve this problem,
Wolfgang Pauli postulated in 1931 the existence of the neutrino, an
uncharged particle which carries the remaining energy in the beta
decay.
This particle would be very difficult to observe, as its interactions
with other matter are very limited, and in fact the neutrino was
experimentally found only in 1956.
Altogether, it became clear that while for some time it seemed as if
the world of elementary particles was almost fully explained, the
theory was obviously not complete and 
there had to be other, yet unknown mechanisms responsible for these
phenomena. 
The situation changed dramatically with the discovery of a wealth of
new particles 
in cosmic ray observations and in experiments with the newly invented
particle accelerators.

\subsubsection{Handling Elementary Particles}

The way to get experimental information on subatomic particles is to collide
two particles with as much energy as possible and then to observe what
happens. In 
general new particles are created, and one just needs to detect them and check
their properties. In the early 20th century, the only way to observe
such high-energy
collisions was to wait for cosmic particles to
crash into the atmosphere. These particles are emitted
in cosmic events like supernovae and therefore carry a lot more
energy than what was possible to reach on earth at that time. When such a
fast-moving particle 
hits a nitrogen or oxygen atom of
the earth's atmosphere, the collision products can be
examined in suitable detectors. It was in cosmic ray experiments where
in 1937 the muon and ten years later the pion and kaon particles were found.
Unfortunately almost all cosmic radiation is absorbed in the
outmost layers of the 
atmosphere. Hence for many interesting experiments with cosmic
radiation it is necessary to equip a balloon  
or an airplane with the appropriate instruments and send them into the
stratosphere.
Furthermore, it is not possible
to design a cosmic ray experiment at own
will, as the properties of the incoming and the target particles can
not be set up freely.
 
These drawbacks were overcome by the development of particle accelerators. With
such a device one takes a particle, accelerates it to very high
energies and lets it collide with a target. It is then possible 
to measure all interesting quantities of the collision products. The
accelerated particles, which can be charged particles like electrons
or protons, move in a ring-like structure, where they are kept by strong
magnetic fields. The larger the diameter of the ring and the stronger
the magnetic field, 
the faster the particles can move and the more energy is set free in the
collision. As an example, the LEP collider at CERN which was running
until 2001 has a diameter
of 27 km and reaches a total energy of 100 GeV in electron-proton collisions.
The Large Hadron Collider (LHC) which is under construction at CERN
will collide protons and antiprotons at energies of 14~TeV.

Reaching high energies in a collider experiment is crucial because the
total energy 
provides a threshold for the mass of the created particle. If the rest energy
of a particle is larger than the total energy of the collided particles, it
can not be created in the collision process. Thus for example to
create a $\rho$~meson, a total 
energy of 770 MeV, which corresponds to its mass, is
required. The problem is that often the particles predicted by
theorists have masses too large to be created in current colliders,
and therefore larger and larger colliders have to be constructed in
order to confirm or falsify the theoretical predictions.

\subsubsection{Bringing Order into the Chaos}
The availability of particle accelerators lead to an enormous growth in the
number of newly found particles in the 1950s and 1960s,  and
there was a definite need for 
a theory which explained why all these particles were there. All one could do
at that time was to bring some order into the wealth of particles and
to classify
them according to their properties. While most of the particles were very
short-lived and had life-times on the order of $10^{-24}$ seconds, a
few of them 
decayed only after a much longer time of about $10^{-10}$ s. These
particles were called 
``strange'' due to this unexplained property by Murray Gell-Mann in
1953. Gell-Mann found that this whole wealth of particles could be
explained in a systematic way when assuming an underlying structure,
namely a small number of constituents which, when grouped in different
combinations, form the experimentally found particles.
These constituents, introduced by Gell-Mann and Zweig,
were called quarks, a name taken from James Joyce's novel
``Finnegans Wake''.  The quark model could not only explain the known
particles, but also predict new ones, which were needed in order to
fill the gaps in the tables of possible combinations of quarks. 
The problem with the quark model was just that
no one had ever seen a quark as a separate object in an
experiment. All the detected collision products 
were made out of two or three quarks. In 1973, work of t'Hooft,
Politzer, Gross  
and Wilczek explained this puzzle with the concept of asymptotic freedom,
implying that the strong force between two quarks increases when the
quarks are pulled apart. In particular, a state with a single quark is not
allowed, as it would need infinite energy to separate it from the
others. Moreover, when the force between quarks pulled apart
reaches a certain threshold, new quarks can be created out of the
vacuum, and what 
remains are again bound states of two or three quarks.  Taking the quarks as
fundamental building blocks and  
the color force introduced by Gell-Mann, Fritzsch and Leutwyler as
an interaction between the quarks, the quantum theory of the strong
nuclear force, Quantum Chromodynamics (QCD), was born.
Finally there was a tool to describe the strong
interaction, and all the different particles that were found could be
explained from common grounds with only a few basic elements and from
underlying symmetry principles.

At about the same time, Glashow, Weinberg and Salam developed a
quantum theory for 
the weak interaction, which is responsible for the
nuclear beta decay mentioned before. They postulated the existence of
the $W$ and $Z$ 
bosons as mediators of the weak interaction, and these particles
were indeed found at CERN in 1983.
Furthermore, the theory of Glashow, Weinberg and Salam allowed unifying the 
electromagnetic and weak interactions into the so-called electroweak theory.
Today, QCD as a theory of the strong interactions and Glashow, Weinberg
and Salam's electroweak theory form the Standard Model (SM)
of elementary particle physics, which has been very successful up to date in
explaining what nature does at a very small scale. The Standard Model
does not include gravitation, which
is the last of the four fundamental forces listed in Table
\ref{tab:fund_interactions}. At the subatomic level 
however, the gravitational force is
negligibly small, and thus it is ignored in SM particle
physics. The constituents of matter appearing in the Standard Model
are on one hand the six quarks listed in Table \ref{tab:quarks} and
the six leptons 
$e$, $\nu_e$, $\mu$, $\nu_\mu$, $\tau$, $\nu_\tau$, which all are fermions
and thus follow the Pauli exclusion principle, and on the other hand
the photon, gluon and the $W^\pm$ and $Z$ particles which are bosons and
carry the electromagnetic, strong and weak interactions between the
fermions. Finally, the SM 
predicts the existence of a Higgs particle, which gives a
non-vanishing mass to the
weak bosons. The existence of the Higgs boson is not yet confirmed by
experiment, but it is expected that the particle will be
found as soon as the next generation of colliders start operation.

It is obvious that the Standard Model is not yet the ultimate theory
of nature, not only because it does not contain gravity, but also because
quite a large number of unknown input parameters are needed. Therefore
many theoretical physicists work on finding candidates
for an even more fundamental theory that unifies all the four
interactions. These attempts lead to 
exciting discoveries like superstring theories living in
10-dimensional space-time, and more recently
11-dimensional $M$-Theory. While from the theoretical point of
view these theories are very attractive, from what we know today it is
extremely difficult to 
connect them to phenomenological information and thus to test their
predictions, as 
the typical energy scales involved are far beyond reach of any
foreseeable experiment. 

In the following, we will stay within the bounds of the Standard
Model. We concentrate on the strong interaction and
the particles participating therein, the quarks and gluons. Many
fundamental questions in particle physics are related to the strong
force, hence the study of Quantum Chromodynamics is a highly rewarding
task, both from 
the phenomenological and the theoretical point of view.

\begin{table}
\begin{center}
\begin{tabular}{l|l|l|l|l} \hline\hline
interaction & mediator & gauge group & acts on & rel. strength  \\ \hline
electromagnetic & photon & U(1) & e.m. charged & 1 \\
weak & $W^\pm, Z$ & SU(2) & quarks, leptons & $10^{-4}$\\
strong & gluon & SU(3) & quarks, gluons & 60 \\
gravitational &  &  & all & $10^{-41}$\\ \hline
\end{tabular}
\end{center}
\caption{The four fundamental forces of nature, with the particle
mediating the interaction and the corresponding gauge group characterizing
the underlying symmetry. The relative strength 
is given by the force between two up-quarks at distance $3\cdot
10^{-17}$ m. The Standard Model describes the first three of these
forces, while gravitation is treated in General Relativity.} 
\label{tab:fund_interactions}
\end{table}

\section{Quantum Chromodynamics}
\label{sect:qcd}
The strong interactions between elementary particles are described by
Quantum Chromodynamics (QCD), the 
quantum theory of the color force. The basic degrees of freedom of the theory
are the quark and gluon fields. Like all quantum field
theories in the Standard Model, QCD is a local gauge theory. The
gluons, which are the gauge 
fields of the theory, are introduced to ensure local gauge invariance
and thus generate the interaction among the particles.
The gauge group has to be chosen as an external input when
constructing the theory. From particle phenomenology follows that the
quarks appear in three different colors, and 
that in nature the gauge group of the color force is the special unitary group
$SU(3)$.\footnote{ As a theoretical generalization, the theory can also be set
up with the gauge group $SU(N_c)$ for an arbitrary number of colors $N_c$.}
The beauty and strong predictive power of QCD lies in the fact that
only a small number of parameters need to be fixed to define the
theory and to get physical predictions.

\subsection{The QCD Lagrangian}
The fermions from which QCD is constructed are the $n_f$ flavors of
quark fields  
$q^k(x) \in \{u,d,s,c,t,b\}$, $k=1,\dots, n_f$, which are
Grassmann-valued Dirac spinors and $SU(3)$ triplets in color
space. Thus, under a local gauge 
transformation $U(x)\in SU(3)$ the quark and antiquark fields $\bar q^k
= (q^k)^\dag\gamma_0$ transform like
\begin{eqnarray}
 q^k(x)    & \longrightarrow& U(x)q^k(x), \\
 \bar q^k(x) &\longrightarrow& \bar q^k(x) U^\dag(x),
\end{eqnarray}
The gauge bosons are the $N_c^2-1$ gluon fields $A_\mu^a(x)\in SU(N_c)$.

\begin{table}
\begin{center}
\begin{tabular}{l|c|l|c|l|c} \hline\hline
 quark & $m$ [GeV]  & quark & $m$ [GeV] & quark & $m$ [GeV] \\ \hline
 $u$ (up) & 0.003(2)  &  $s$ (strange) & 0.120(50)	& $t$ (top) & 175(5) \\
$d$ (down) & 0.006(3) & $c$ (charm) & 1.25(10) & $b$ (bottom) & 4.2(2)\\ \hline
\end{tabular}
\end{center}
\caption{The three generations of quarks flavors with their respective
masses in natural units taken from the 2000 Review of Particle Properties
\cite{Groom:2000in}. The values in brackets estimate the uncertainty in
the mass value. The $u$-type quarks in the first row have an
electromagnetic charge of  
$2/3$, while for the $d$-type quarks in the second row the charge is
$-1/3$. The $u$, $d$, and $s$ masses are current-quark masses at the
scale $\mu=2$ GeV.} 
\label{tab:quarks}
\end{table}

A field theory is defined by its Lagrangian density ${\mathcal L}$, from
which the equations of motion  
and thus the dynamics of the theory can be derived. The QCD Lagrangian
\begin{equation} \label{eq:cont_qcd_l}
{\mathcal L}_{QCD}(x) = {\mathcal L}_F(x) + {\mathcal L}_G(x),
\end{equation}
can be split into the fermionic (quark) part
\begin{equation}\label{eq:cont_quark_l}
{\mathcal L}_F(x) = \sum_{k=1}^{n_f}\bar q^k(x) (i\gamma^\mu D_\mu - m
) q^k(x), 
\end{equation}
and the purely gluonic part
\begin{equation}\label{eq:cont_gluon_l}
{\mathcal L}_G(x) = -\frac{1}{4} F_{\mu\nu}^a(x) F^{\mu\nu a}(x), 
\end{equation}
which in itself defines a non-trivial Yang-Mills theory and describes the
kinematics of the gluons. The sum over the repeated color index
runs from $a=1,\dots,N_c^2-1$. The gluon field strength tensor
appearing in the Lagrangian ${\mathcal L}_G$ is defined by
\begin{equation}
 F_{\mu\nu}^a(x) = \partial_\mu A_\nu^a(x) - \partial_\nu A_\mu^a(x) -
 g_s f_{abc} A_\mu^b(x) A_\nu^c(x),
\end{equation}
where $g_s$ is the strong coupling constant and
$f_{abc}$ are the structure constants of the gauge group $SU(N_c)$.
To ensure local gauge invariance, in the fermionic Langrangian
${\mathcal L}_q(x)$ the covariant derivative 
\begin{equation}
 D_\mu(x) = \partial_\mu - i g_s A_\mu(x),
\end{equation}
has to be taken, with the gauge field $A_\mu(x)$ being an element of the
gauge group,
\begin{equation}
 A_\mu(x) = A_\mu^a(x) \frac{\lambda^a}{2},
\end{equation}
where the group generators $\lambda^a/2$ follow the commutation
relation
\begin{equation}
 \left[ \frac{\lambda^a}{2}, \frac{\lambda^b}{2} \right] = i
 f_{abc}\frac{\lambda^c}{2}.
\end{equation}
Requiring the Lagrangian \eqref{eq:cont_qcd_l} to be gauge invariant,
the transformation 
rules for the gauge field and the field strength tensor are found to be
\begin{eqnarray}
 A_\mu(x) &\longrightarrow& U(x)A_\mu(x) U^\dag(x) -
 \frac{1}{g_s}\partial_\mu U(x)U^\dag(x), \\
 F_{\mu\nu}(x) &\longrightarrow& U(x) F_{\mu\nu}(x) U^\dag(x).
\end{eqnarray}

Having specified the QCD Lagrangian, the theory is defined,
and it remains to prescribe how to extract 
physical quantities. This is done most elegantly in the Feynman path
integral formalism, 
thus promoting the classical field theory to a quantum theory.
Let us now switch to Euclidean space (see Appendix \ref{app:euclidean_space}),
which will be natural for setting up a lattice
formulation. Expectation values for physical observables 
${\mathcal O}$, that can be arbitrary operators built from quark and
gluon fields, are defined by the path integral 
\begin{equation} \label{eq:cont_pathintegral}
 \langle {\mathcal O} \rangle = \frac{1}{Z} \int D \bar q
 Dq DA\ {\mathcal O}\ e^{-S^E[\bar q,q,A] },
\end{equation}
where the normalization in the denominator is given by the partition function
\begin{equation} \label{eq:cont_partfunc}
 Z = \int D \bar q Dq DA\ e^{-S^E[\bar q,q,A] }.
\end{equation}
Eq.~\eqref{eq:cont_partfunc} shows that a quantum field theory in imaginary
time formally resembles a system in classical 
statistical mechanics, where the probability of a state is
proportional to the
Boltzmann factor $\exp(-E/kT)$. In QCD, the Euclidean action 
\begin{equation} \label{eq:cont_action}
 S^E [\bar q,q,A] = \int d^4x\ {\mathcal L}^E_{QCD}(x),
\end{equation}
where ${\mathcal L}^E_{QCD}$ is the QCD Lagrangian
transformed to Euclidean space,
appears in the exponent of the Boltzmann factor.

\subsection{Global Vector and Axial Symmetries} \label{sect:cont_symm}
In the limit of $n_f$ massless quarks, the QCD Lagrangian 
\eqref{eq:cont_qcd_l}--\eqref{eq:cont_gluon_l} exhibits a global symmetry
\begin{equation} \label{eq:chiral_symmetry}
U_V(1) \times SU_V(n_f) \times U_A(1) \times SU_A(n_f),
\end{equation}
acting on the flavor and spin degrees of freedom. Writing
the $n_f$ quark fields as a vector, the corresponding symmetry
transformations are 
\begin{equation}
 q(x) \longrightarrow e^{-i\phi (T_D \otimes T_F)} q(x),
\end{equation}
where $T_D \in \{{\bf 1},\gamma_5\}$ acts on the Dirac structure and
generates the vector ($V$) or axial vector ($A$) transformations, and
$T_F$ works in 
flavor space to create the $U(1)$ (for $T_F={\bf 1}$) or $SU(n_f)$
transformations. The conserved currents related to these global
symmetries through the Noether theorem are
\begin{equation}
 j^\mu(x) = \bar q(x) \gamma^\mu (T_D \otimes T_F) q(x).
\end{equation}
The vector $U_V(1)$ symmetry is unbroken even for finite 
quark mass and gives rise to baryon number 
conservation. The $SU_V(n_f)$ leads to the multiplet structure of the
hadrons.  The axial $U_A(1)$ is explicitly broken on the quantum level
by instanton contributions, leading to
the Adler-Bell-Jackiw (ABJ) anomaly \cite{Adler:1969er,Bell:1969ts} of the
flavor-singlet axial current and the
massiveness of the $\eta^\prime$ meson. The $SU_A(n_f)$ is believed to be
spontaneously broken by a non-zero vacuum expectation value of the
quark condensate $\langle\bar qq\rangle$, and the associated
$(n_f^2-1)$ massless Goldstone bosons for $n_f=2$ are the pions.
In the real world, the global symmetry \eqref{eq:chiral_symmetry}
arises from the smallness of the light quark masses (see Table
\ref{tab:quarks}), where setting 
$m_u = m_d = 0$ and in some cases even $m_s=0$ is a good
approximation. For non-zero, but small quark masses, the pions are no
longer real Goldstone bosons, but quasi-Goldstone particles that
acquire a small mass. This would explain why the experimentally observed
pion masses $m_{\pi^0}=135$ MeV 
and $m_{\pi^\pm}=140$ MeV are so small compared to the masses of other
hadrons.

\subsection{Asymptotic Freedom}
The coupling constant $g_s$ of the strong interaction is actually not
a constant, but depends on the momentum transfer $Q$ of a given
process through quantum 
corrections, leading to the emergence of a generic scale $\Lambda$.  Often
not the coupling constant $g_s$ 
itself is used, but the fine-structure constant $\alpha_s =
g_s^2/4\pi$, which is to leading order given by
\begin{equation} \label{eq:running_coupling}
\alpha_s(Q^2) = \frac{\alpha_s(\Lambda)} {1+\alpha_s(\Lambda)
\frac{33 - 2n_f}{12\pi} \ln (\frac{Q^2}{\Lambda^2})} \,.
\end{equation}
In the running coupling \eqref{eq:running_coupling}, asymptotic
freedom of QCD shows up in the fact that $\alpha_s$
gets small at large momenta $Q$. On the other hand, the coupling 
increases with larger momenta or equivalently smaller distance,
leading to confinement of quarks\footnote{Quark confinement is a
non-perturbative phenomenon which does not follow from perturbation
theory.}. At the 
mass of the $Z$-boson $m_Z = 91$ GeV, measurements of the coupling
constant give a 
value of $\alpha_s(m_Z) = 0.118$, which is a reasonably small value
that a perturbative expansion in $\alpha_s$  around the free
theory makes sense. For deep inelastic scattering
processes studied in collider experiments, the momentum transfer is
of this order, so in this region QCD can be treated 
perturbatively. At scales around 1 GeV however that are typical for
the hadronic world, 
$\alpha_s$ is on the order of 1 and thus no longer a small parameter in
which an expansion is possible. Perturbation theory therefore breaks
down when small momenta or large distances are involved. In this
non-perturbative region of Quantum Chromodynamics, where one would
like to investigate issues like the hadron spectrum, hadronic matrix
elements of operators, spontaneous chiral symmetry breaking,
confinement or the topological structure of the vacuum, it is
necessary to use another approach to 
perform calculations. This is where lattice QCD comes into play.

\section{QCD on the Lattice}
\label{sect:lattice}
The lattice formulation of Quantum Chromodynamics in Euclidean space,
originally proposed 
in 1974 by Wilson \cite{wilson:74}, was designed as a tool to calculate
observables in the non-perturbative region of QCD from first
principles. Lattice QCD is at present the only method which allows to compute
low-energy hadronic
quantities in terms of the fundamental quark and gluon degrees of
freedom without having to tune additional parameters. The only input
parameters are the bare quark masses and the 
bare coupling constant, and from these all other quantities
like the masses of the hundreds of experimentally observed hadrons can
be calculated. Hence the lattice is a very 
powerful tool in checking that QCD is the correct theory for the
strong interactions and in making predictions for the dependence of
hadronic quantities on the input parameters. Formulating QCD on a
discrete space-time lattice opens the possibility to treat these
problems on computers, using methods analogous to those in
Statistical Mechanics. However, due to the large number of degrees of
freedom involved, lattice QCD simulations are computationally
very demanding, and it is still necessary to use a number of
tricks and approximations in order to cope with these demands. It is
then also
important to examine whether the effects introduced by these approximations
are under control.
Since the first numerical measurements in a
lattice gauge theory by
Creutz, Jacobs and Rebbi in 1979 \cite{Creutz:1979zg}, the progress in computer
technology and the theoretical developments in the field have allowed
to get closer to examining in a systematical manner the deep questions
which Lattice QCD is able to answer.
We present in the following a brief introduction to the basics of
lattice QCD that is necessary to follow the rest of the work. For a
more detailed discussion, we refer to the standard textbooks
\cite{Creutz:1992,rothe:97,montvay_muenster} or recent introductory articles
\cite{Gupta:1997nd,DeGrand:1996sc,DiPierro:2000nt,Munster:2000ez,Richards:1999px}. 
An extensive overview of the status of current research in lattice QCD can be
found in the proceedings of the annual lattice conference
\cite{Bhattacharya:2001xg}. 

\subsection{The Lattice Regularization}
Quantum field theories have to be regularized in order to give the path
integrals in Eq.~\eqref{eq:cont_pathintegral} that define physical
observables a meaning. In perturbation theory, a 
convenient way to do this is by dimensional
regularization, where the space-time dimension $d$ is
modified by a small parameter $\epsilon$ to $d=4-\epsilon$, or by
introducing a momentum cut-off $\Lambda$. At the end of a calculation,
the regularization has to be removed by taking the limit
$\epsilon\rightarrow 0$ or $\Lambda\rightarrow\infty$. The
lattice is nothing else than such a regulator for the theory.
In the lattice regularization, the continuum Euclidean space-time variable
$x_\mu$ is replaced by a discrete hypercubic space-time lattice,
\begin{equation}
 x_\mu \longrightarrow n_\mu a, \quad n_\mu\in{\bf Z},
\end{equation}
with lattice spacing $a$. This introduces an ultraviolet cut-off by
restricting the momenta to lie within the Brioullin zone $|p_\mu| < 
\pi/a$, removing the ultraviolet divergent behaviour of
integrals. Restricting the space-time extent to a finite lattice $n_\mu
< N_\mu$, $n_\mu \in {\bf N_0}$, the momenta take
discrete values $p_\mu = k_\mu\pi / N_\mu a$, with $|k_\mu|<N_\mu$ and
$k_\mu \in {\bf Z}$. Every quantity that is calculated in the lattice
regularized 
theory is finite, since the integrals are transformed into finite sums.

The
continuum quark and gluon fields are replaced by lattice fields living
on the sites and links of the lattice, respectively, 
and also the derivatives in the QCD Lagrangian
\eqref{eq:cont_qcd_l}--\eqref{eq:cont_gluon_l} have 
to be discretized in some way. It is
obvious that in the process of discretization some of the original
symmetries of the Lagrangian are partially or fully lost. As an
example, the Poincar\'{e} symmetry in continuum space-time is replaced
by a cubic symmetry on the lattice.
As mentioned before, it is necessary to remove the
regularization at the end of the calculation to get physical results,
and for the lattice regularization this means 
that the continuum limit $a \rightarrow 0$ has to be taken. In this
process one expects the lost symmetries to be restored. However, one
requires that the most important symmetries like gauge invariance,
which lies at the foundations of QCD, are
also present at finite lattice spacing. The lattice formulation of the
QCD Lagrangian should
therefore respect these symmetries.

The lattice quark and antiquark field are Grassmann variables
$\Psi(n)$, $\bar\Psi(n)$ defined at every lattice site $n$.
In natural units physical quantities can be expressed in units of
powers of length or inverse mass. For numerical applications, it is
convenient to work with dimensionless quantities. This can be done by
absorbing the dimension through appropriate powers of the lattice spacing
$a$. For the quark fields, the transcription on the
lattice is then given by 
\begin{equation} \label{eq:lattice_field}
 q(x)  \longrightarrow  a^{-3/2} \Psi(n). 
\end{equation}
The lattice gauge fields $U_\mu(n)$ are defined by the path-ordered
Schwinger line integral 
\begin{equation} \label{eq:schwinger_line}
 U_\mu(n) = {\mathcal P} \exp \left( ig \int_{na}^{(n+\hat\mu) a}\ dx A_\mu(x) \right),
\end{equation}
acting as parallel transporters of color between neighbouring
lattice points. They are thus elements of the gauge group $SU(3)$
defined on the links between lattice sites.
To lowest order in the lattice spacing, \eqref{eq:schwinger_line} reduces to
\begin{equation}
 U_\mu(n) \simeq e^{ iag A_\mu(x) }.
\end{equation}
Under a gauge transformation $G(n)$, the lattice quark and gluon
fields transform like
\begin{eqnarray}
 \Psi(n) &\longrightarrow& G(n) \Psi(n), \label{eq:gtr_psi} \\ 
 \bar\Psi(n) &\longrightarrow & \bar \Psi(n) G^\dag(n), \\
 U_\mu(n) &\longrightarrow& G(n) U_\mu(n) G^\dag(n+\hat\mu). \label{eq:gtr_u}
\end{eqnarray}
 The construction
of the lattice gauge fields is done such in order to ensure gauge invariance of
non-local quark operators. With these definitions, there are two
different types of gauge invariant objects: color traces of closed
loops of gauge links like the Wilson plaquette
\begin{equation} \label{eq:plaquette}
U_{\mu\nu}(n) = U_\mu(n) U_\nu(n+\hat\mu) U_\mu^\dag(n+\hat\nu)
U_\nu^\dag(n),
\end{equation}
and quark bilinears like $\bar\Psi(n) U_\mu(n) \Psi(n+\mu)$, where
gauge links connect the quark and the 
antiquark field along an arbitrary path in space-time.

Integrals over continuum space-time variables, as they appear in the action
\eqref{eq:cont_action}, are replaced by sums over the lattice sites,
\begin{equation}
 \int d^4x\ f[\bar q(x), q(x), A_\mu(x) ] \longrightarrow a^4 \sum_n \hat f
 [\bar \Psi(n), \Psi(n), U_\mu(n) ],
\end{equation}
where $\hat f$ is a discretized version of the function $f$.
The path integral over the quark and gluon fields in the expression
for the expectation value of observables
\eqref{eq:cont_pathintegral} and in the partition function
\eqref{eq:cont_partfunc} is  
transformed into a product of ordinary integrals over the fields at
all lattice sites,
\begin{equation}
 \int D\bar q Dq DA \longrightarrow \prod_{\alpha,l} \int d\bar\Psi_\alpha(l)
 \prod_{\beta,m} \int d\Psi_\beta(m) 
 \prod_{\rho,n} \int dU_\rho(n),
\end{equation}
which yields finite expressions on a finite lattice and can be evaluated
numerically.\footnote{ The integration over the gluon
fields is an integration over the gauge group SU(3).}

\subsection{Simple Lattice Actions}
Although the discretization of the continuum QCD Lagrangian
\eqref{eq:cont_qcd_l}--\eqref{eq:cont_gluon_l} might appear trivial at
first sight, there are some complications. The
lattice action given by Wilson \cite{wilson:74} is the most simple
working version and
is still widely used in simulations, although it is not free of
problems, as we will see later.
The action 
\begin{equation}
 S[\bar\Psi,\Psi,U] = S_G[U] + S_F[\bar\Psi,\Psi,U],
\end{equation}
can again be split in separate gauge and fermion parts.
The Wilson gauge action $S_G^{(\rm W)}$ is
constructed from the plaquette $U_{\mu\nu}$ in \eqref{eq:plaquette} by
\begin{equation}
 S_G^{(\rm W)}[U] =  \beta \sum_n \sum_{\mu <\nu} \big( 1 - \frac{1}{N_c}
 {\rm Re} \Tr\ 
 U_{\mu\nu}(n) \big) ,
\end{equation}
which in the limit $a\rightarrow 0$ goes over to the continuum form up
to ${\mathcal O}(a^2)$ errors. The parameter $\beta = 2N_c/g_s^2$
takes over the role of the bare coupling constant. The  fermionic lattice
action
\begin{equation} \label{eq:wilson_quarks}
 S_F[\bar\Psi,\Psi,U] = \sum_{n,n^\prime} \bar\Psi(n)
 D(n,n^\prime) 
 \Psi(n^\prime),
\end{equation}
 is bilinear in the quark fields.
The Wilson fermion action $S_F^{(\rm W)}$ is defined by setting $D=D^{\rm W}$,
with the Wilson Dirac operator
\begin{multline} \label{eq:d_wilson}
 D^{\rm W}(n,n^\prime) =  \frac{1}{2a}  \sum_\mu \left[
(\gamma_\mu - r) \delta_{n^\prime, n +\hat\mu} U_\mu(n)  -
 (\gamma_\mu + r) \delta_{n,n^\prime +\hat\mu} U_\mu^\dag(n-\hat\mu)
\right] \\
  + \left( m+\frac{4r}{a} \right) \delta_{nn^\prime},
\end{multline}
where the bare quark mass $m$ is another parameter of the theory.
The Wilson Dirac operator differs from the naive discretization of the
Euclidean continuum Dirac 
operator $\gamma_\mu D_\mu + m$ by a dimension $d=5$ term
proportional to the unphysical parameter $r$,
\begin{equation} \label{eq:wilson_term}
 S_F^{(\rm W)} = S_F^{(\rm naive)} - a\frac{r}{2} \sum_n
 \bar\Psi(n)\Box\Psi(n),
\end{equation}
which is called the Wilson or doubler term and is needed to remove
unphysical particles that appear through poles at the corners of the
Brioullin zone from the spectrum. In the continuum limit
$a\rightarrow 0$, the Wilson term vanishes as required. However, as
we show later, the cost for introducing the Wilson term
is the explicit breaking of chiral symmetry, leading to many
theoretical and practical problems and limitations.

\subsection{Monte Carlo Integration}
The integral over the fermion fields, which are anticommuting Grassmann
variables, in the lattice version of the
partition function \eqref{eq:cont_partfunc} can be performed
analytically. For a bilinear fermion action \eqref{eq:wilson_quarks},
the integration over quark and antiquark fields gives the
determinant of the fermion matrix, and the partition function  on the lattice 
then reads
\begin{equation} \label{eq:int_partfunc}
 Z = \prod_{\mu,n}\int dU_\mu(n) \det D\ e^{-S_G[U]},
\end{equation}
where $D$ is the lattice Dirac operator and $S_G[U]$ is a lattice version of
the gluon action. Expectation values for observables ${\mathcal O}$ are calculated from
\begin{equation} \label{eq:lat_expval}
 \langle {\mathcal O}\rangle = \frac{1}{Z}\prod_{\mu,n}\int dU_\mu(n)
 \ {\mathcal O}\det D\ e^{-S_G[U]}.
\end{equation}
At this point, it is obvious that the theory is
ready to be put on a computer, since in Eq.~\eqref{eq:lat_expval} only an
integration over the $SU(3)$ gauge fields is left. However, for standard
numerical integration, the number of degrees of freedom is still far too
large, therefore one has to resort to statistical methods. The way the
gauge field integral is usually handled is by Monte Carlo
integration:
A finite number $N$ of gauge configurations $U^{(i)}$,
$(i=1,\dots,N)$, are statistically sampled with the probability
distribution given by the 
fermion determinant $\det D$ times the
Boltzmann factor $\exp(-S_G[U^{(i)}])$. Observables are
then estimated from the sample mean
\begin{equation}
 \langle {\mathcal O} \rangle \simeq \bar{\mathcal O} = \frac{1}{N}
 \sum_{i=1}^N  {\mathcal O}[U^{(i)}].
\end{equation}
In practice this amounts to generating a set of $N$ independent gauge
configurations with a
Markov chain algorithm that respects the
required probability distribution and measuring  
the observable on the resulting set of gauge configurations. As we have seen,
after integrating out the fermions in the
partition function \eqref{eq:int_partfunc}, their influence on the
weighting is given 
by the determinant. It turns out that the calculation of the fermion
determinant is by far the most time-consuming part in a lattice
simulation. This is why most lattice QCD calculations up to date have
been done in the quenched approximation, which is explained in the
next section.

\subsection{The Quenched Approximation}
The determinant of the Dirac operator is a non-local quantity.
Even for moderate lattice sizes, its exact calculation is not feasible
on today's computers. Various algorithms have
 been developed to tackle this problem, but keeping dynamical fermion
 loops in the simulation is still a very demanding task. The easiest
 way out is to consider quenched QCD, as done in this work, where the
 fermion determinant in Eq.~\eqref{eq:lat_expval} is set to  
\begin{equation}
\det D = 1.
\end{equation}
 This approximation is equivalent to making the virtual quarks infinitely
heavy, leading to the complete suppression of internal quark loops.
Neglecting the fermion determinant simplifies the technical treatment
enormously, as then in the generation of the gauge configurations the
probability distribution is given by the Boltzmann factor alone,
whereas in unquenched QCD the determinant has to be calculated in
every Monte Carlo update step.
It is however obvious that quenched QCD is not the correct theory to describe
nature, as for example two 
quarks can be pulled apart to an arbitrary distance\footnote{The
energy needed increases linearly with distance.} in quenched QCD,
while in 
nature at a certain point two additional quarks are created 
and string breaking occurs. Quenched QCD is not even
mathematically clean, as it is not unitary. The
only reason for using the quenched approximation is that the computation of
the fermion determinant is extremely demanding, and by setting $\det D
= 1$ a factor of several orders of magnitude in time is gained.

The reason why the investigation of quenched lattice QCD is nevertheless
interesting 
is that since neglecting of the determinant amounts just to a different
weighting of 
the gauge configurations, the quenched theory still shows the crucial
properties of QCD like 
asymptotic freedom and spontaneous symmetry breaking. Therefore it is
possible to examine many
non-trivial questions first in the quenched theory, giving qualitative
hints what in full, unquenched QCD might occur.
A phenomenological argument why ignoring virtual quark loops is not a
completely useless approximation is given by the Zweig rule, which states
that processes where the constituent quarks do not survive are suppressed.
Many years of
lattice simulations have shown that the errors due to quenching in
physical observables are in most cases only on
the order of 10\%, allowing to make also quantitative
predictions. However, it is very important that quenching 
effects are well investigated. 

\subsection{Continuum Limit, Renormalization and Scaling}
The parameters  $g_s$ (or $\beta$)
and $m$ which are put into a lattice simulation are bare quantities,
and when taking the continuum limit $a\rightarrow 0$, physical 
quantities and not the bare parameters have to be kept fixed. 
Consider the
physical observable ${\mathcal O}_{\rm phys}$ with mass dimension
$d_{\mathcal O}$ and its dimensionless lattice counterpart ${\mathcal
O}_{\rm lat}$, which depends on the lattice spacing through the
coupling $g_s(a)$ and the quark mass $m(a)$. The continuum limit
\begin{equation} \label{eq:contlim}
 {\mathcal O}_{\rm phys} = \lim_{a\rightarrow 0} a^{-d_{\mathcal O}} {\mathcal
 O}_{\rm lat}(g_s(a),m(a)),
\end{equation}
is taken by measuring ${\mathcal O}_{\rm lat}$ at different values of
$g_s$. At sufficiently small $a$, the
dependence of the coupling constant on the lattice spacing $g_s(a)$
should be a universal function, independent of the
observable under consideration. The same holds for the function
$m(a)$ for the quark mass in unquenched QCD. This property is called
scaling, and the 
range of lattice spacings or gauge couplings the hypothesis is valid is
called the scaling window. Lattice simulations have to be performed
within this scaling window in order to extract reasonable continuum results.

The scale dependence can be removed by forming ratios of particle
masses
\begin{equation}
\frac{am_1(a)}{am_2(a)} = \frac{m_1(0)}{m_2(0)} + {\mathcal O}(m_1a) +
{\mathcal O}((m_1a)^2) + \dots\, .
\end{equation}
The $a$-dependent terms on the right hand side are artifacts
from discretization errors and depend on the choice of lattice
action. If these terms are small, 
a controlled continuum extrapolation is possible, and we speak of scaling
of the quantity under consideration. It is certainly desirable that
a lattice action shows good scaling, that is small scaling violations.

When results in physical units are wanted, the lattice results, which
are always dimensionless, have to be converted 
to physical units by matching the result of one observable with the
experimental data. This observable might be the 
mass of a hadron like $m_\rho \simeq 770$ MeV or $m_N \simeq 940$ MeV
or a decay constant like $f_\pi \simeq 93$ MeV.
In the quenched approximation, the
lattice spacing does not depend on the quark mass, as there are only
external quarks. It is then also possible to fix the scale from a
purely gluonic quantity  
like the string tension $\sqrt{\sigma} \approx 420$ MeV. More reliable
than the string tension are Sommer-type scales \cite{Sommer:1994ce},
which are also related to the quark-antiquark potential.

\section{Why Improved Formulations of Lattice QCD?}
\label{sect:motivation}

In principle the choice of how to discretize the continuum QCD Lagrangian
\eqref{eq:cont_qcd_l}--\eqref{eq:cont_gluon_l} is free, as long as the
correct continuum limit is 
reached. However, there are a number of reasons why it is worthwhile
to search for improved lattice formulations of the Lagrangian. One reason is
the reduction of discretization errors:
Working at finite lattice spacing  $a$ introduces
discretization errors which affect simulation results and have to be
identified and removed. Simulations are normally carried out at
lattice spacings between 0.05 fm and 0.2 fm, and the typical size of
hadrons is on the order of 1 fm. As explained above, the continuum
limit \eqref{eq:contlim}
is taken by measuring observables at several lattice spacings and
extrapolating the results to $a\rightarrow 0$. 
While the discretization errors should disappear in the continuum limit, it is
advantageous to have a lattice formulation of the theory with small
discretization errors. First of all it is often not clear how
controlled and safe the continuum extrapolation 
is, therefore having results which show smaller $a$-dependence leads
to a more reliable extrapolation. On the other hand, working at small lattice
spacings is numerically very demanding, as the computational effort grows
roughly like $a^{-6}$ for quenched simulations and like $a^{-10}$ for
the unquenched case \cite{Gupta:1997nd}, thus it would obviously be
helpful to have a 
formulation of the theory which gives results of the same quality at
larger lattice spacing. Hence, a lattice QCD action with small
lattice artifacts allows either for a more reliable extrapolation
or to work at larger lattice spacings and to save computation time.
The usual way to reduce discretization errors is to improve the 
lattice action and operators systematically in orders of the lattice
spacing by adding irrelevant terms which remove the artifacts order by 
order. This improvement program, proposed by Symanzik
\cite{symanzik:83}, has been applied to various cases, the best known
of which is the ${\mathcal O}(a)$-improved Wilson clover fermion action
\cite{Sheikholeslami:1985ij}.  

Improved actions are also expected to show better behaviour in
restoring rotational and internal symmetries. Most prominent is the
$U_A(1)\otimes SU_A(n_f)$ chiral symmetry in \eqref{eq:chiral_symmetry}, which
is explicitly broken for Wilson-type fermions by the Wilson
term in Eq.~\eqref{eq:wilson_term}. Even in the continuum limit, this
explicit breaking of chiral 
symmetry leads to unwanted effects like an additive renormalization of
the quark mass, which means that in lattice simulations the bare quark
mass is a parameter which needs to be tuned. Another consequence of
explicit chiral symmetry breaking in the quenched theory is the appearance of
exceptional configurations, for which the quark propagator diverges
although the bare quark mass is still far from the critical
value. This makes it impossible to simulate
quarks much lighter than the strange quark, and therefore a long and
unreliable chiral 
extrapolation from the simulated quark masses to the physical
masses of the up and down quarks is needed.
The partially conserved
axial vector current also needs to be renormalized. Furthermore,
mixing between operators of different chiral representations occurs,
leading to technical difficulties in calculations of weak matrix
elements. There also exists a close connection between chiral zero
modes of the Dirac operator and
the topological structure of the gauge fields, and with standard
formulations of lattice QCD neither topology nor chiral fermion zero
modes are well-defined notions.
For a long time, it has been believed that chiral symmetry
can not be preserved on the lattice. Only after the resurrection of
the Ginsparg-Wilson 
relation \cite{Ginsparg:1982bj,Hasenfratz:1998ft}, it has been
realized that it is possible to retain an
exact, slightly modified chiral symmetry \cite{Luscher:1998pq} on the
lattice.

A radical approach to improvement is the classically perfect
Fixed-Point action \cite{Hasenfratz:1994sp}, 
which is defined at the fixed point of Renormalization Group
transformations. The Fixed-Point action gives exact continuum results
for classical predictions even at non-zero lattice spacing. Thus it allows
for scale invariant instanton solutions, satisfies the fermionic index
theorem and preserves chiral symmetry \cite{Hasenfratz:1998jp}. Even
for quantum results, discretization errors are expected to be
considerably reduced.
In this work we will construct and apply a parametrization of the
Fixed-Point fermion action. Since we will also make use of a recently
constructed 
Fixed-Point action \cite{Niedermayer:2000ts} for the gluons, the
results in this thesis serve as a first extensive test for Fixed-Point
actions in QCD.

%%% Local Variables: 
%%% mode: latex
%%% TeX-master: t
%%% End:

\fancyhead[RE]{\nouppercase{\small\it Chiral Fermions and Perfect Actions}}
\chapter{Chiral Fermions and Perfect Actions}
\label{ch:chiral_fermions}
Only very recently, it has become possible to simulate chiral fermions
on the lattice. This exciting discovery lead to growing activity in
applying and testing chiral lattice actions. In this chapter we
recapitulate the problems with formulating chiral lattice fermions and
different solutions, which all obey the ubiquitous
Ginsparg-Wilson relation. Fixed-Point (FP) fermions are not only
chiral, but also 
classically perfect. We present the conceptual basics of perfect actions
and the application to free fermions.

\section{Chiral Symmetry on the Lattice}

In Section \ref{sect:cont_symm} we have presented the global flavour
symmetries inherent in
the continuum QCD Lagrangian. In this section we describe the
problems arising when the theory is transcribed onto the lattice, and
how it is possible to retain chiral symmetry in lattice QCD.
Consider the global flavour-singlet $U_V(1)$ vector transformation
\begin{subequations}
\begin{eqnarray}
 \Psi(n) &\longrightarrow& e^{i\phi} \Psi(n), \\
 \bar \Psi(n) &\longrightarrow& \bar \Psi(n) e^{-i\phi} ,
\end{eqnarray}
\end{subequations}
and the $U_A(1)$ axial vector transformation
\begin{subequations} \label{eq:u_a}
\begin{eqnarray}
 \Psi(n) &\longrightarrow& e^{i\phi\gamma_5} \Psi(n), \\
 \bar \Psi(n) &\longrightarrow& \bar \Psi(n) e^{i\phi\gamma_5}, 
\end{eqnarray}
\end{subequations}
acting on the lattice fermion fields.
It is obvious that the fermion lattice action $\sum_{n,n^\prime} \bar \Psi(n)
(D(n,n^\prime) + m) \Psi(n^\prime)$ 
satisfies the $U_V(1)$ symmetry for all quark masses $m$. Setting $m=0$, the
chiral $U_A(1)$ symmetry is present only if the Dirac operator anticommutes with
$\gamma_5$:
\begin{equation} \label{eq:cont_chir_symm}
 \{D,\gamma_5 \} = 0.
\end{equation}
The major obstacle for formulating lattice fermions respecting
chiral symmetry is the Nielsen-Ninomiya no-go theorem
\cite{Nielsen:1981rz,Nielsen:1981xu}, which states that it is not
possible to have a lattice Dirac operator which is local, has the correct
continuum limit, is free of doublers and satisfies
Eq.~\eqref{eq:cont_chir_symm}. If the continuum fermion action is
discretized naively, chiral symmetry is preserved, but instead of one
fermion there appear 16 massless particles. To remove these doublers,
in the Wilson action \eqref{eq:wilson_term}
a term is added to the action which gives the doublers a mass, but
breaks chiral symmetry explicitly by violating the anticommutation relation
\eqref{eq:cont_chir_symm}. It is clear that all the other
properties in the Nielsen-Ninomiya theorem have to be conserved in
order to obtain a reasonable lattice Dirac operator, and therefore breaking
chiral symmetry seems to be the only way to get around the
theorem. However, instead of the 
hard breaking by the Wilson term, a better approach is to slightly modify
Eq.~\eqref{eq:cont_chir_symm} to the so-called Ginsparg-Wilson
relation \cite{Ginsparg:1982bj} 
\begin{equation} \label{eq:gw_d}
 \{D,\gamma_5 \} = a D\gamma_5 2R D,
\end{equation}
where the newly introduced term on the right hand side vanishes in the
continuum limit. It is useful to express Eq.~\eqref{eq:gw_d} in terms of the
quark propagator:
\begin{equation} \label{eq:gw_prop}
 \{D^{-1},\gamma_5 \} = a\gamma_5 2R.
\end{equation}
The term $2R$, which is denoted like this for historical reasons, is a
local operator. From this requirement follows that Eq.~\eqref{eq:gw_prop} is a
highly non-trivial condition, since the quark propagator $D^{-1}$ on the
left-hand side is a non-local object.

It has been shown by L\"uscher \cite{Luscher:1998pq} that for a Dirac
operator fulfilling the 
Ginsparg-Wilson relation \eqref{eq:gw_d}, it is possible to define an
exact lattice 
chiral symmetry, which is a modified version of the continuum
$U_A(1)$ symmetry. When the transformation \eqref{eq:u_a} is
replaced by 
\begin{subequations}
\label{eq:uam}
\begin{eqnarray}
 \Psi(n) &\longrightarrow& e^{i\phi\gamma_5(1-aD/2)} \Psi(n), \\
 \bar \Psi(n) &\longrightarrow& \bar \Psi(n) 
 e^{i\phi(1-aD/2)\gamma_5}, 
\end{eqnarray}
\end{subequations}
the fermion action is invariant. An analogous statement holds for the
flavour non-singlet axial transformation. At this point, it might seem
that there is more symmetry than expected, because due to the
ABJ~anomaly the $U_A(1)$ symmetry should be broken at the quantum level. The
solution comes from the observation that the fermionic integration
measure is not invariant under the modified transformation
\eqref{eq:uam}, but transforms like
\begin{equation} \label{eq:measure_transform}
 D\bar\Psi D\Psi \longrightarrow \exp{(2N_f\times {\rm index}(D))} D\bar\Psi D\Psi,
\end{equation}
thus creating the expected anomaly for topologically non-trivial gauge
configurations. The fermionic index in \eqref{eq:measure_transform},
\begin{equation}
{\rm index}(D) \equiv n_- - n_+,
\end{equation}
is the difference between the number of zero eigenmodes of the Dirac operator
with positive and negative chirality, and is related to the
topological charge $Q_{\rm top}$ through the Atiyah-Singer index
theorem \cite{Atiyah:1971rm}
\begin{equation} \label{eq:index_theorem}
 {\rm index}(D) = Q_{\rm top} = \frac{1}{32\pi^2} \int\!
 d^4x\ \epsilon_{\mu\nu\rho\sigma} \tr(F_{\mu\nu}F_{\rho\sigma}).
\end{equation}

\section{Fermions with Exact or Approximate Chiral Symmetry}
While the Ginsparg-Wilson relation \eqref{eq:gw_d} has been known for
a long time, no solution was found until recently, when three
different formulations were independently discovered which all fulfill
the Ginsparg-Wilson relation and thus retain exact chiral symmetry
on the lattice. These solutions are the  
domain wall \cite{Kaplan:1992bt,Shamir:1993zy,Furman:1995ky} and
overlap fermions
\cite{Narayanan:1993wx,Narayanan:1993ss,Narayanan:1995gw}, which were
originally proposed to formulate chiral gauge theories, and the
Fixed-Point fermions \cite{Hasenfratz:1994sp}. 

Domain wall fermions are defined by extending Wilson fermions into a
non-physical fifth dimension 
with lattice spacing $a_s$, lattice size $N_s$ and a negative
mass. The different chiralities are then located on the two opposite
domain walls, with 
the mixing exponentially suppressed by the size of the fifth dimension
$N_s$. In the limit $N_s\rightarrow\infty$, exact chiral symmetry is
obtained. 

For overlap fermions, there exists an explicit construction
with exact chiral symmetry: Defining the kernel
\begin{equation} \label{eq:ov_kernel}
 A = 1 - a D^{\rm W},
\end{equation}
with the Wilson operator $D^{\rm W}$, the overlap Dirac operator
$D^{(\rm ov)}$ is given by \cite{Neuberger:1998fp}
\begin{equation} \label{eq:overlap}
 D^{(\rm ov)} = \frac{1}{a} \left( 1 - \frac{A}{\sqrt{A^\dag A}}
 \right) . 
\end{equation}
Domain wall fermions with infinite
fifth dimension $N_s$ are in fact equivalent to overlap fermions, when
a different kernel $A$
\cite{Neuberger:1998bg,Kikukawa:1999sy} is put into
\eqref{eq:overlap}. It is not obvious that the overlap 
construction generates a local operator, which requires that the
couplings decrease exponentially with distance. Losing
locality would render the whole formulation useless. However, it has
been shown that both the overlap operator with Wilson kernel
\eqref{eq:overlap} and the 
4-d effective formulation for the domain wall operator are local
\cite{Hernandez:1998et, Kikukawa:1999dk}.

Fixed-Point fermions are defined through Renormalization Group
transformations, as discussed in detail in Section
\ref{sect:perfact}. It has been first shown for FP fermions that the
index theorem on the lattice remains valid \cite{Hasenfratz:1998ri}. 
For FP fermions, there is no explicit expression,
except for the non-interacting case. They have to be
constructed in an iterative procedure, which we present in Section
\ref{sect:parametrization}. An important difference to domain wall and
overlap fermions is that FP fermions do not only respect chiral
symmetry, but are classically perfect and therefore are expected to have
small cut-off effects. 

Having presented lattice fermion formulations with exact chiral
symmetry, it is important 
to show where approximations have to be taken which introduce again some
residual chiral symmetry breaking. For domain wall fermions, it is
obvious that the extension of the fifth dimension $N_s$ has to be
finite in actual simulations. Mixing of the two chiralities is then
still possible,
and in recent simulations by the RBC \cite{Blum:2000kn,Orginos:2001xa}
and CP-PACS 
\cite{AliKhan:2000iv,Aoki:2001de} collaborations and in
\cite{Jung:2000fh} the effects of this residual chiral symmetry 
breaking have been investigated closely. The
exponential decay was found to be surprisingly slow, and
although rather large extensions $N_s$ of ${\mathcal O}$(32--64) have
been used 
in the simulations,
getting close to the limit of exact chiral symmetry does not seem to
be easily possible. This might be an effect arising from the small
eigenvalues of the hermitean Wilson operator
\cite{Aoki:2001ai}. Although there are several proposals how to cope
with this problem  \cite{Hernandez:2000iw,Edwards:2000qv}, it 
is not obvious why one should work with 
domain wall fermions, given the equivalence to
overlap fermions, where the chiral symmetry breaking effects are much
better under control and can even be eliminated completely.

Also for FP fermions, approximations have to be taken. First of all,
the Dirac operator has to be restricted to finite 
extension, in our case to the hypercube. All couplings outside the
hypercube are truncated. Second, only a limited set of gauge paths are
considered for the couplings in the hypercube. It is therefore clear
that only approximate chiral symmetry is present for the parametrized
FP operator. The question then is whether the residual chiral symmetry
breaking is negligibly small for the task under consideration. We took
advantage of the freedom in the choice of Dirac operator for the
overlap kernel \eqref{eq:ov_kernel} and used the overlap 
construction \eqref{eq:overlap} with the parametrized FP operator as
an input kernel to remove the residual chiral symmetry breaking of our 
parametrization for some applications.  

The difficulty in simulations with overlap fermions arises 
from the inverse square root in Eq.~\eqref{eq:overlap}, which is hard to
calculate numerically and requires again an iterative procedure.
Using tricks like the exact treatment of small 
eigenvalues of $A^\dag A$, it is however possible to make the calculation of
the inverse square root up to machine precision feasible within a few
hundred iterative steps, rendering the chiral symmetry exact.

From the above considerations, one can quantify the computational
demand of the different fermion formulations compared to the standard
case of Wilson fermions: The simulation of domain wall fermions requires
a factor of $N_s$ more computer time due to the additional fifth
dimension. For 
overlap fermions, the factor is given by the number of iterative steps
that has to be taken in order to compute the inverse square root. For
typical applications, this factor is of order ${\mathcal O}(200)$. The
computational cost of FP fermions depends on the
parametrization that is chosen. We will come back to this point in
Section \ref{sect:parametrization}.

Another way to get an approximately chiral symmetric
fermion action is to optimize a parametrization of a general Dirac
operator for solving the Ginsparg-Wilson relation
\cite{Gattringer:2000js,Gattringer:2000qu}. By truncating the
expansion of the general operator in terms of the number of gauge
paths and couplings and putting 
the truncated operator into the Ginsparg-Wilson relation, the free
parameters can be fixed. The resulting operator approximates the
Ginsparg-Wilson relation to a precision which depends on the
truncation.

\section{Perfect Actions from Renormalization Group Transformations}
\label{sect:perfact}

\begin{figure}[tb]
\begin{center}
\includegraphics[width=5cm]{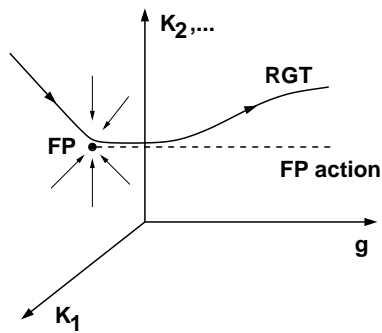}
\end{center}
\caption{Renormalization Group trajectory of asymptotically free theories.}
\label{fig:rgt}
\end{figure}

The perfect action approach to improving lattice actions 
followed to construct the Fixed-Point Dirac operator is inspired by the
Renormalization Group flow of asymptotically-free theories
\cite{Hasenfratz:1994sp}.
A Renormalization Group (RG) transformation
\cite{wilson_kogut:74,bell_wilson:74,bell_wilson:75} reduces the number of 
degrees of freedom by integrating some of them out in the path integral,
taking into account their effect on the remaining variables
exactly. This allows to get rid of short-distance fluctuations without
changing the physical content of the theory.
 Consider a lattice 
action which contains all possible interactions. The RG transformation is
defined by some blocking function which averages over the fields to produce a
new action on a coarser lattice with fewer fields. The new action generally
has different couplings from the original action, thus we can imagine the
blocking step as a flow in the coupling parameter space. Repeated
RG steps generate a trajectory in this space. In
Figure \ref{fig:rgt}, we show the RG trajectory for QCD with massless
quarks. The fixed point has the property that the couplings are
reproduced after a blocking step. For asymptotically free theories, the
fixed point is on the surface of vanishing coupling $g = 0$. 

Starting on this surface, the RG trajectory flows to the
fixed point. If one starts close to this surface at some small
coupling $g$, or equivalently small lattice spacing $a$, the RG
trajectory flows quickly towards the  
fixed point and then flows away from it. Let us assume we have an
action with couplings lying on the RG trajectory at an 
arbitrarily small $a$, thus having arbitrarily small lattice
artifacts. From there one can
reach any point on the RG trajectory by making sufficiently many blocking
steps, and all actions on the RG trajectory describe the same physics. The
physical observables of the continuum quantum theory are thus
identical to those of 
any lattice quantum theory on the RG trajectory, independently of the lattice
spacing. Such lattice actions are called quantum perfect. The
Fixed-Point action is an approximation to the RG trajectory for small
couplings $g$ and is classically perfect, which means it completely
describes the continuum classical theory without discretization errors
\cite{Hasenfratz:1994sp}. 

Fixed-Point actions have many desirable features. By closely approximating the
RG trajectory, they are expected to have largely reduced quantum lattice
artifacts. They can be optimized for locality. The FP Dirac operator
satisfies the Ginsparg-Wilson relation and so has good chiral behavior. The
FP QCD action has well-defined topology and satisfies the index
theorem on the lattice.
The properties of FP actions have first been tested in
models like the two-dimensional non-linear $\sigma$-model
\cite{Hasenfratz:1994sp,Blatter:1996ik} and the
$CP^3$-model \cite{Burkhalter:1996dr}. The approach has then been
extended to $SU(2)$ and $SU(3)$ Yang-Mills theories and fermions in 2
and 4 dimensions
\cite{Wiese:1993cb,Farchioni:1998xg,Bietenholz:1998rj,Lang:1998ib,Bietenholz:1997qc,Feurstein:1998cf,DeGrand:1996ih,DeGrand:1995jk,DeGrand:1995ji,D'Elia:1995ez,Bietenholz:1996cy,Chernodub:2000ax},
and first steps towards applications in partial differential equations have
been taken \cite{katz_wiese:97,Hauswirth:2000ri}.
Recently, a new parametrization for $SU(3)$ Yang-Mills was
constructed, showing reduced scaling violations in glueball masses and
finite temperature measurements \cite{Niedermayer:2000ts}. We use this
gluon action 
together with our parametrization of the FP Dirac operator for the
simulations in Chapters~\ref{ch:zeromodes} and \ref{ch:results_fp}. An
extension of this FP gluon 
action to anisotropic lattices was constructed and tested in
\cite{Rufenacht:2001pi}. For a pedagogical introduction to perfect
actions, consult \cite{Hasenfratz:1998bb}.

\section{Free Fixed-Point Fermions}

For the case of free fermions without mass, the Renormalization Group
construction is relatively easy. 
Because the fermionic action is quadratic in the fermion fields, the
Renormalization Group step for the fermion fields amounts to Gaussian
integration, which can be done exactly. On the lattice, a RG
transformation relates an action on a fine lattice with spacing $a$ to
a different action on a coarser lattice with spacing $2a$.
The blocking step thus connects the 
Dirac operators $D_{\rm f}$ on the fine and $D_{\rm c}$ on the
coarse lattice by \cite{Bietenholz:1996cy,DeGrand:1997nc}
\begin{equation}
  \label{RGeq}
  D^{-1}_{\rm c} = \frac{\bf 1}{\kappa} + \omega  D^{-1}_{\rm f}  \omega^{\dagger},
\end{equation}
provided $D_{\rm f}$ has no zero modes, where $\kappa$ is an optimizable free
parameter of the blocking and $\omega$
is the blocking function that relates the fine fields to the 
coarse fields. The Fixed-Point Dirac operator $D^{\rm FP}$ is
reproduced under the blocking step,
\begin{equation}
  (D^{\rm FP})^{-1} = \frac{\bf 1}{\kappa} + \omega  (D^{\rm FP})^{-1}
  \omega^{\dagger}, 
\end{equation}
and depends on the choice of the blocking function $\omega$.
For free fermions, that is in the absence of gauge fields, this
equation can be solved analytically. The FP Dirac 
operator is local, and the rate of fall off for the couplings can be
maximized by varying the parameter $\kappa$. However, $D^{\rm FP}$
contains infinitely 
many couplings. For practicality, the FP Dirac operator is
approximated with an ultralocal  
operator, for which each point is only coupled to its neighbors on the
hypercube. The effect of this truncation can be examined for the
energy-momentum dispersion relation, which is equivalent to the
continuum for the exact FP Dirac operator. The truncated operator
deviates from the exact result, but shows still considerably smaller
discretization errors than for example the Wilson operator
\cite{Hasenfratz:2000qb}. 

In fact, the Renormalization Group procedure does not only generate a
FP Dirac operator, but also a FP $R$ operator appearing on the right
hand side of the Ginsparg-Wilson relation \eqref{eq:gw_d}. Combining
\eqref{eq:gw_d} with the
blocking transformation \eqref{RGeq} that connects the propagators 
$D^{-1}$ on the coarse and fine lattices, we get the Renormalization
Group relation
\begin{equation}
  \label{RGeqR}
  R_{\rm c} = \frac{\bf 1}{\kappa} + \omega  R_{\rm f}  \omega^{\dagger},
\end{equation}
for the $R$ operator, and at the fixed point, $R_{\rm c} = R_{\rm f} = R^{\rm
FP}$. For free fermions, this equation 
can also be solved analytically. Choosing a symmetric overlapping
block transformation $\omega$ with a scale factor 2 that averages over 
hypercubes \cite{Kunszt:1998db}, the exact $R^{\rm FP}$ has only 
hypercubic couplings, and therefore a truncation like for the Dirac
operator is not required. With other methods to build a 
Dirac operator satisfying the Ginsparg-Wilson relation, for example the
overlap construction \eqref{eq:overlap}, $R$ is unconstrained and
typically $R = 1/2$ is 
taken for simplicity.

\fancyhead[RE]{\nouppercase{\small\it The Parametrized Fixed-Point
Dirac Operator}}
\chapter{The Parametrized Fixed-Point Dirac Operator}
\label{ch:d_fp}

The concept of perfect actions is theoretically very attractive. The
main obstacle for its application to a theory of general interest like
QCD is to find a 
parametrization that is rich enough to capture all the beautiful
properties of perfect actions, but is still feasible to
use in numerical simulations. While the Fixed-Point Dirac operator is local,
which means its 
couplings decrease exponentially, an ultralocal
parametrization will introduce a truncation. This truncation does of
course disturb the FP
properties, and it is a non-trivial task to find a parametrized form
whose properties do not deviate strongly from those of the
FP operator. A parametrization of the FP Dirac operator will
be more costly 
to simulate in terms of computer time than comparably simple Dirac operators
like the Wilson or clover 
operators, as there are more couplings between lattice sites than just
those to the nearest neighbor involved, and also the Clifford
structure can be richer.
However, one expects that the rewards compensate the
additional cost of a more complicated action. In the case of the Dirac
operator, a strong argument is certainly that chiral symmetry is
preserved, in contrast to the standard actions. Additionally, the
scaling violations are expected to be reduced for a parametrized FP
operator.
Smaller scaling violations allow
to simulate at larger lattice spacings, while the results are still of
unchanged quality. Since the computer time for a quenched lattice QCD
simulation increases like $a^{-6}$--$a^{-7}$, being able to simulate
at lattice spacing $2a$ instead of $a$ brings a factor of
$\mathcal{O}(100)$ in computational savings. Even more pronounced is
the situation in the unquenched case, where the cost increases like
$a^{-8}$--$a^{-10}$ with the lattice spacing, so that the expected
gain can even be of $\mathcal{O}(1000)$. 

In this chapter we first derive the structure of general lattice Dirac
operators respecting the appropriate discrete symmetries, and show how 
complicated operators with a large number of couplings and gauge paths can be
calculated efficiently. This has been
examined in detail in \cite{Hasenfratz:2000xz}, and we present here the
key concepts of the paper. Then we explain our procedure of fitting
the parameters of the general Dirac operator to the FP operator, using the
 Renormalization Group recursion relations. Finally we show
as a test for the properties of the resulting parametrization the 
eigenvalue spectrum, which measures how well the Ginsparg-Wilson
relation is obeyed.

%\section{The Parametrization}
%\label{sect:parametrization}
\section{General Lattice Dirac Operators}
The starting point for constructing any lattice Dirac operator is the question
what general structure is allowed if the basic lattice symmetries
have to be respected. These symmetries are discrete translation invariance,
gauge invariance, $\gamma_5$-hermiticity, charge conjugation,
permutation and reflection symmetry. Let us summarize the constraints
which the discrete symmetries impose on any lattice Dirac operator
$D(n,n';U)$.

\subsection{Discrete Symmetries and Gauge Invariance}
From translation invariance follows that $D(n,n+r)$ depends on the
lattice variable $n$ only through the $n$-dependence of the gauge
fields. In particular, the coefficients in front of the different
gauge paths which enter the Dirac operator do not depend on $n$.
The hermiticity properties of the lattice operator are required to be
the same as in the continuum,
\begin{equation} 
D(n,n') = \gamma_5 \, D(n',n)^\dagger \,\gamma_5,
\label{eq:herm}
\end{equation}
where the hermitean conjugation acts in color and Dirac space.
Permutations of the coordinate axes are defined in a straightforward
way, as just the Dirac indices appearing in $D$ can be
permuted.\footnote{Note 
that cubic rotations on the lattice can be replaced
by reflections and permutations of the coordinate axes.}
In Appendix \ref{app:axis_refl} and
\ref{app:charge_conjg}, we derive the conditions from charge conjugation,
\begin{equation}
  D(U_\mu) = C^{-1} D(U_\mu^*)^T C,
\end{equation}
where $C \gamma_\mu^T C^{-1} = - \gamma_\mu$, and reflections of a
coordinate axis $\eta$,
\begin{equation} 
D(n,n';U_\mu(n)) = P^{-1}_\eta 
D(\tilde{n},\tilde{n}';U_\mu^{{\cal P}_\eta}(\tilde n)) 
 P_\eta ,
\label{eq:permsym}
\end{equation}
where in our convention $P_\eta=\gamma_\eta \gamma_5$, and $\tilde n$ is
the reflected space-time variable defined in \eqref{eq:reflect_n}.

It remains to ensure gauge invariance, which is the most crucial ingredient. 
If the Dirac operator transforms under the gauge transformation
$G(n)\in SU(N)$ in \eqref{eq:gtr_psi}--\eqref{eq:gtr_u} like
\begin{equation}
D(n,n';U) \longrightarrow  G(n)^\dagger \,D(n,n';U^g) \,G(n'), 
\label{gau}
\end{equation}
where $U^g$ is the gauge transformed background field,
the fermion action stays gauge invariant. This can be achieved
by connecting the lattice sites $n$ and $n^\prime$ along an arbitrary
path
\begin{equation}
l=[l_1,l_2,\dots,l_k] ,
\end{equation}
of length $k$, where $|l_i|= 1,\dots,4$ is the direction of the path
at step $i$, 
with a parallel transporter
\begin{equation} \label{eq:gaugepath}
U^l(n) = U_{l_1}(n)U_{l_2}(n+\hat l_1)\cdot \dots \cdot U_{l_k}(n+\hat
l_1+ \dots +\hat l_{k-1}),
\end{equation}
made from products of gauge links.
For every offset $r=n^\prime-n$ appearing in the Dirac operator, one
or several paths $l$ connecting both end points have to be chosen to
make $D$ gauge covariant.

\subsection{General Construction}
 The symmetry  
conditions \eqref{eq:herm}--\eqref{eq:permsym} prescribe in which
combinations the 
different permutations and reflections of the gauge paths
\eqref{eq:gaugepath} have to enter the Dirac operator. 
\begin{figure}
\begin{center}
\setlength{\unitlength}{3mm}
\begin{picture}(26,12)
\put(1,5){\line(2,1){4}}
\multiput(1,5)(4,2){2}{\circle*{0.3}}
\put(0,6){\line(3,0){6}}
\multiput(0,6)(6,0){2}{\circle*{0.3}}
\put(3,3){\line(0,3){6}}
\multiput(3,3)(0,6){2}{\circle*{0.3}}
\put(3.5,5){\small $n$}
\put(3,6){\circle*{0.5}}
\put(-1,0){\small nearest neighbour}

\multiput(14,2)(3,0){3}{\multiput(0,0)(0,3){3}{\line(2,1){4}}}
\multiput(14,2)(2,1){3}{
\multiput(0,0)(0,3){3}{\line(3,0){6}}
\multiput(0,0)(0,3){3}{\multiput(0,0)(3,0){3}{\circle*{0.3}}}
\multiput(0,0)(3,0){3}{\line(0,3){6}}
}
\put(19.2,5.2){\small $n$}
\put(19,6){\circle*{0.5}}
\put(16,0){\small hypercubic}
\end{picture} 
\end{center}
\caption{Offsets reached from a given lattice site $n$ for a
nearest-neighbor (Wilson-type) and a hypercubic (FP) Dirac
operator. For obvious 
reasons, the figure is limited to the $d=3$ case.}
\label{fig:nn_vs_hyp}
\end{figure}
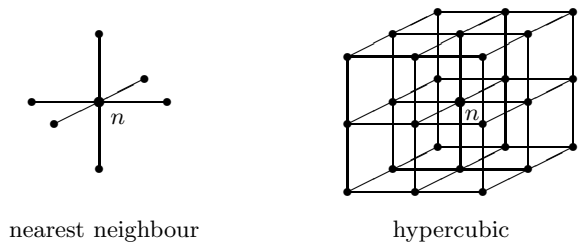
To be more specific, a general gauge covariant lattice operator with
color, space and Dirac indices can be written as 
\begin{equation}
  \label{Dgen}
  D(n,n') = \sum_{i=1}^{16} \Gamma_i \sum_l c(\Gamma_i,l) U^l(n) \, ,
\end{equation}
where $\Gamma_i$ are elements of the Clifford algebra and
$c(\Gamma_i,l)$ is the coupling
for the given path $U^l(n)$ and Clifford algebra element. 
The basic symmetries of the Dirac operator impose the following restrictions on Eq.~(\ref{Dgen}):

Translation invariance requires that the couplings $ c(\Gamma_i,l)$ 
  are constants in space-time or gauge invariant functions of gauge
  fields, respecting locality and invariance under the symmetry
  transformations. 
Charge conjugation and $\gamma_5$-hermiticity  together imply that the
  couplings $ c(\Gamma_i,l)$   are real
  for our choice of the Clifford algebra basis.
  From hermiticity it follows that the path $l$ and the
  opposite path $\bar{l} = [-l_k,\dots,-l_1]$,
  or equivalently  $U^l(n)$ and  $U^l(n)^\dagger$,
  should enter in the combination
  \begin{equation}
    \label{GUUd}
    \Gamma \left( U^l(n) + \epsilon_\Gamma U^l(n)^\dagger \right) \,,
  \end{equation}
  where the sign $\epsilon_\Gamma$ is defined by
  $\gamma_5 \Gamma^\dagger \gamma_5 =\epsilon_\Gamma \Gamma$.
Permutations and reflections of the coordinate axes (hypercubic
  rotations) imply  
  that for a given reference path $l_0$ a whole class of paths belongs
  to the Dirac 
  operator. These paths are related to $l_0$ by all the $16 \times 24
  = 384$ reflections and permutations 
  of the coordinate axes. Under such a symmetry transformation
  $\alpha=1,\ldots,384$ the Clifford algebra  
  element $\Gamma_0 \in \{{\bf 1}, \gamma_\mu,\sigma_{\mu\nu}, \gamma_5,
\gamma_\mu\gamma_5 \}$ associated with $l_0$ generally transforms\footnote{The transformed Clifford algebra
  element $ \Gamma^{(\alpha)}$ is of the same type (S, V, P, T,
  A) as $\Gamma_0$. } into 
  $\Gamma^{(\alpha)}$ and the parallel transporter $U^l(n)$ transforms 
  to $U^{l^{(\alpha)}}(n)$. Furthermore the sign   
  of the couplings may change, whereas their absolute value remains unchanged.

A Dirac operator satisfying all the basic symmetries can be written as
\begin{equation}
\label{eq:general_d}
D(n,n^\prime) = \sum_{\Gamma_0,l_0} c(\Gamma_0,l_0) \sum_\alpha O^\alpha(n),
\end{equation}
where the sum runs over a set of reference paths defined by $\Gamma_0$
and $l_0$ as well as over all 
the symmetry transformations $\alpha$ defined by the group of
permutations and reflections of the coordinate axes. The operators
$O^\alpha$ are defined by
\begin{equation}
O^\alpha(n) = \Gamma^{(\alpha)} \left( U^{l^{(\alpha)}}(n) +
\epsilon_\Gamma U^{l^{(\alpha)}}(n)^\dagger \right).
\end{equation}
To make 
the above construction clear, consider the example of the Wilson Dirac
operator \eqref{eq:d_wilson}, which contains only the elements ${\bf 1}$ and
$\gamma_\mu$ of the Clifford algebra and extends to nearest neighbors
as sketched in Fig.~\ref{fig:nn_vs_hyp}. For the scalar element
$\Gamma_0={\bf 1}$, the 
reference paths are $l_0 = []$, which amounts to the contact term, and the
nearest neighbor coupling  $l_0=[1]$, while for the vector element
$\Gamma_0=\gamma_1$ there is only the nearest neighbor $l_0=[1]$.
The sum over these reference paths and Clifford algebra elements in
Eq.~\eqref{eq:general_d} gives then the Wilson Dirac operator
\eqref{eq:d_wilson}, when the coefficients
\begin{eqnarray}
c({\bf 1},[]) &=& m+4r,  \nonumber\\
c({\bf 1},[1]) &=& -r/2,  \nonumber\\
c(\Gamma_1,[1]) &=& 1/2, 
\end{eqnarray}
 are taken. 
It is quite natural to include the full Clifford algebra to
parametrize the FP Dirac operator.
The scalar and vector elements are already present in
the continuum. The tensor element $\sigma_{\mu\nu}$ appears in 
the ${\mathcal O}(a)$-improved Sheikoleslami-Wohlert clover
operator and thus already in lowest order of the Symanzik improvement
program. Without the pseudoscalar element $\gamma_5$, the
Ginsparg-Wilson relation \eqref{eq:gw_d} could not be fulfilled, so it
is crucial for 
the chiral properties of the Dirac operator. Moreover, the topological
charge is proportional to $\Tr (\gamma_5D)$, which would be zero if
$D$ does not contain the pseudoscalar element. Finally, as the
Renormalization Group procedure which leads to the Fixed-Point
operator generates all the elements of the Clifford algebra, 
also the axial vector element $\gamma_\mu\gamma_5$ should be included.

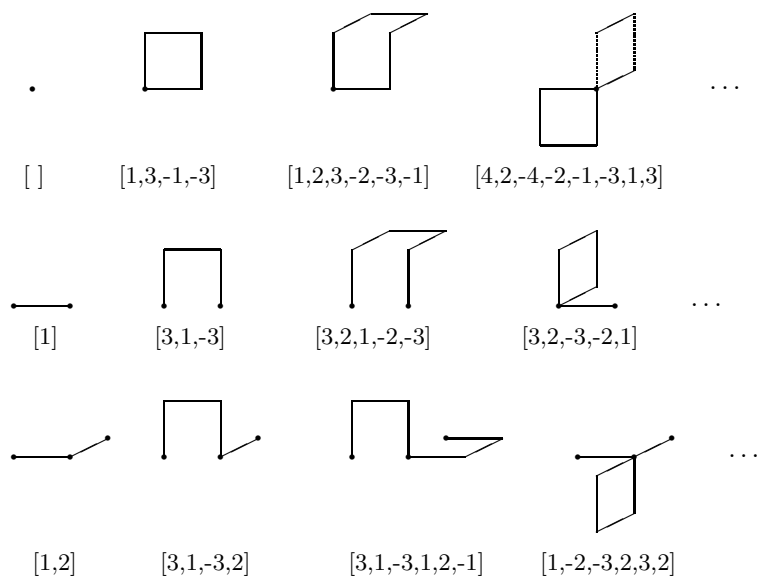
\begin{figure}
\begin{center}
\hspace{1cm}
\setlength{\unitlength}{2.5mm}
\begin{picture}(42,12.5)

\put(0,8)
{
\put(0,0){\circle*{0.3}}
\put(-0.5,-5){\small  [ ]}
}

\put(6,8)
{
\put(0,0){\circle*{0.3}}
\multiput(0,0)(0,3){2}{\line(3,0){3}}
\multiput(0,0)(3,0){2}{\line(0,3){3}}
\put(-1.4,-5){\small  [1,3,-1,-3]}
}

\put(16,8)
{
\put(0,0){\circle*{0.3}}
\multiput(0,0)(3,0){2}{\line(0,3){3}}
\multiput(0,3)(3,0){2}{\line(2,1){2}}
\multiput(0,0)(2,4){2}{\line(3,0){3}}
\put(-2.5,-5){\small  [1,2,3,-2,-3,-1]}
}

\put(30,8)
{
\multiput(0,0)(0,3){2}{\line(2,1){2}}
%\multiput(0,0)(2,1){2}{\line(0,3){3}}
\multiput(0,0)(2,1){2}{\dashbox{0.1}(0,3)}
\multiput(-3,0)(0,-3){2}{\line(3,0){3}}
\multiput(-3,0)(3,0){2}{\line(0,-3){3}}
\put(0,0){\circle*{0.3}}
\put(-6.5,-5){\small  [4,2,-4,-2,-1,-3,1,3]}
}

\put(36,8){$\dots$}

\end{picture} 
\end{center}

\begin{center}
\hspace{5mm}
\setlength{\unitlength}{2.5mm}
\begin{picture}(42,6)

\put(0,4)
{
\multiput(0,0)(3,0){2}{\circle*{0.3}}
\put(0,0){\line(3,0){3}}
\put(1,-2){\small  [1]}
}

\put(8,4)
{
\multiput(0,0)(3,0){2}{\circle*{0.3}}
\put(0,3){\line(3,0){3}}
\multiput(0,0)(3,0){2}{\line(0,3){3}}
\put(-0.5,-2){\small  [3,1,-3]}
}

\put(18,4)
{
\multiput(0,0)(3,0){2}{\circle*{0.3}}
\multiput(0,0)(3,0){2}{\line(0,3){3}}
\multiput(0,3)(3,0){2}{\line(2,1){2}}
\put(2,4){\line(3,0){3}}
\put(-2,-2){\small  [3,2,1,-2,-3]}
}

\put(29,4)
{
\multiput(0,0)(0,3){2}{\line(2,1){2}}
\multiput(0,0)(2,1){2}{\line(0,3){3}}
\put(0,0){\line(3,0){3}}
\multiput(0,0)(3,0){2}{\circle*{0.3}}
\put(-2,-2){\small  [3,2,-3,-2,1]}
}

\put(36,4){$\dots$}

\end{picture} 
\end{center}

\begin{center}
\hspace{5mm}
\setlength{\unitlength}{2.5mm}
\begin{picture}(42,10.5)

\put(0,8)
{
\put(3,0){\line(2,1){2}}
\put(0,0){\line(3,0){3}}
\multiput(0,0)(3,0){2}{\circle*{0.3}}
\put(5,1){\circle*{0.3}}
\put(1,-6){\small  [1,2]}
}

\put(8,8)
{
\put(3,0){\line(2,1){2}}
\put(0,3){\line(3,0){3}}
\multiput(0,0)(3,0){2}{\line(0,3){3}}
\multiput(0,0)(3,0){2}{\circle*{0.3}}
\put(5,1){\circle*{0.3}}
\put(-0.2,-6){\small  [3,1,-3,2]}
}

\put(18,8){
\put(6,0){\line(2,1){2}}
\multiput(3,0)(2,1){2}{\line(3,0){3}}
\put(0,3){\line(3,0){3}}
\multiput(0,0)(3,0){2}{\line(0,3){3}}
\multiput(0,0)(3,0){2}{\circle*{0.3}}
\put(5,1){\circle*{0.3}}
\put(-0.2,-6){\small  [3,1,-3,1,2,-1]}
}

\put(30,8)
{
\put(3,0){\line(2,1){2}}
\multiput(3,0)(0,-3){2}{\line(-2,-1){2}}
\multiput(3,0)(-2,-1){2}{\line(0,-3){3}}
\put(0,0){\line(3,0){3}}
\multiput(0,0)(3,0){2}{\circle*{0.3}}
\put(5,1){\circle*{0.3}}
\put(-2,-6){\small  [1,-2,-3,2,3,2]}
}

\put(38,8){$\dots$}

\end{picture} 
\end{center}
\vspace{-5mm}
\caption{Example gauge paths $l$ appearing in the parametrized FP Dirac
operator. For each pictorial representation of products of link
matrices, the corresponding $l$ is 
given. The paths in the first row appear in the contact term, the
second row shows nearest neighbor couplings to $r=(1,0,0,0)$ and the
third row paths leading to offset $r=(1,1,0,0)$.}
\end{figure}

\section{Efficient Implementation of General Dirac Operators}

At first sight one might think that it is not feasible to calculate 
such a general Dirac operator
with many different couplings, where every coupling might contain as 
many as $768$ paths. But one has to
keep in mind that in applications like hadron spectroscopy, the
calculation of propagators for small quark masses needs  
several hundreds of conjugate gradient steps and therefore one can
afford to spend some time to precalculate and store the whole Dirac
operator before starting to calculate the propagator. The preparation
of the Dirac operator then only needs a small fraction of the overall
time for a calculation. On top of this 
there are two reasons why the calculation of the gauge paths for
general operators can be done in a very efficient way:
First, there are a lot of paths which are invariant under certain
  subgroups of the reflections and  
  permutations. This reduces the number of terms significantly and in
  some case even 
  leads to a cancellation of certain terms because they have opposite signs.  
 A less trivial fact is that the sum of paths for many couplings 
  can be factorized in an efficient way, which means that large sums of many 
  paths can be written as a product of smaller sums of fewer paths.

As an example, consider the nearest neighbor coupling with $\Gamma_0 =
\gamma_5$ and reference path $l_0 = [2,1,-2,3,4,-3,-4]$. All the paths of 
this coupling can be written in the following compact way,
\begin{equation}
  \label{g5_1}
  \gamma_5 \sum_{\mu\nu\rho\sigma} \epsilon_{\mu\nu\rho\sigma}
  \left( S_{\mu \nu} P_{\rho\sigma}
    + P_{\rho\sigma} S_{\mu \nu} + \mbox{h.c.} \right) ,
\end{equation}
where the color matrices $S_{\mu \nu}$ and $P_{\rho\sigma}$ are
certain combinations of staples or plaquettes, respectively. 
When all the plaquettes and staples and the most frequent
combinations like $P_{\rho\sigma}$ 
are precalculated, operators like the one in Eq.~(\ref{g5_1}) can be
calculated very quickly. As an illustration of this we  
consider the parametrization of the Fixed-Point Dirac
operator used in our spectroscopy simulations, which has $2\cdot
41$ couplings in total,  
at least one per offset on the hypercube and per type of Clifford algebra
element. Building this operator on a workstation takes only ${\mathcal
O}(20)$ times as long as the multiplication of the operator with a vector 
and therefore it is a very small fraction of the time used to perform
a calculation of one propagator. On a supercomputer, this relation
gets worse, because the construction of the gauge paths is not as
well-suited for optimization or vectorization like the matrix-vector 
multiplication used in the inversion of the Dirac operator. The
measurements from our spectroscopy runs in Appendix
\ref{app:perf} show that on the Hitachi SR8000, the time to construct $D$ is
on the order of 10--20\% of the calculation time for 12 quark
propagators that are needed to construct hadron correlators,
which is not negligible anymore. However, when using the parametrized operator
in an overlap construction, the build-up time is negligible again,
because in that case the matrix-vector multiplication gets more
expensive by a factor which is given by the expansion order of the inverse
square root in the overlap operator.

Another question is how fast manipulations with the Dirac operator can
be executed after it has been constructed.
The basic operation required to calculate propagators or eigenvalues is the
multiplication of the Dirac operator on a vector. For a Dirac
operator with 81 hypercubic fermion offsets which contains the
complete Clifford algebra, the matrix-vector multiplication requires
$9\times 4 = 36$ times
more $c$-number multiplications than for the Wilson Dirac operator,
which only connects 9 offsets and whose Dirac structure can be treated
trivially. The actual performance on a specific computer
however depends a lot on the architectural and implementational
details, and as we do not have an optimized code for the Wilson
operator, we can not confirm this number from performance
measurements. There are 
however additional arguments why in actual simulations with the
parametrized FP operator the factor is considerably smaller than 36:
The most striking one is that for small quark masses, the Wilson operator
runs into problems with exceptional 
configurations, where the inversion converges very slowly or not at
all. We did not encounter such problems for the parametrized FP Dirac
operator at the comparably small quark masses covered in our
spectroscopy simulations.

\section{Parametrization of the Fixed-Point Dirac Operator in QCD}
\label{sect:parametrization}

Finding a good parametrization of a FP action is a non-trivial task.
 In the last years, some
attempts were taken to parametrize the FP Dirac operator
\cite{Bietenholz:2000iy,DeGrand:1999gp}, but
these were limited to moderate generalizations of the Wilson
operator, including only a few additional couplings and part of the
Clifford structure. These parametrizations were rather thought to be
taken as a 
starting point for the overlap construction than to be used on their
own. We took a different approach: Our goal was to get a 
parametrization as close as possible to the massless FP Dirac
operator, that can be directly used in QCD simulations near the
chiral limit. We therefore use a general Dirac operator as defined in
Eq.~(\ref{eq:general_d}) with all the 
couplings of the hypercube (see Fig.~\ref{fig:nn_vs_hyp}) and all
elements of the Clifford algebra.  In order to get an operator which
is close to the fixed point for a range of values of gauge couplings
$\beta$, the 
coefficients $c(\Gamma_0,l_0)$ in Eq.~\eqref{eq:general_d} were chosen
 not to be constants, but gauge invariant polynomials in local
fluctuations of the 
gauge fields. Furthermore we apply a RG-inspired smearing procedure
\cite{thomas:diss} to the gauge configurations and
project them back to SU(3), that is we are 
using so-called fat links. For the gluon sector, we use the recent
parametrization of the FP gauge 
action \cite{Niedermayer:2000ts}, which also makes use of fat gauge links.

Like in the case of free fermions, the QCD FP action is quadratic in the
fermion fields and the Renormalization Group 
step can again be done analytically. The QCD FP equation is the generalization
of the free case given by including gauge fields. In case
$D_f$ has zero modes, it is most conveniently written as  
\begin{equation}
\label{RGO}
D_c(V) = \kappa \, {\bf 1}  - \kappa^2 b^2 \omega(U) [ D_f(U)  + \kappa b^2
 \omega^\dagger(U)\omega(U) ]^{-1} \omega^\dagger(U) ,
\end{equation}
where $\kappa$ is an optimizable free parameter and $b$ is a scale
factor of the blocking transformation,
while $U$ and $V$ are the gauge fields on the fine and 
coarse lattice, respectively. They are related through the FP equation of 
the pure $SU(3)$ gauge theory
\begin{equation}
  S_G^{\rm FP}(V)=\min_{ \{U\} } \left( S_G^{\rm FP}(U) +T(U,V)\right),
  \label{RGG}
\end{equation}
where $S_G^{\rm FP}$ is the FP action of the pure $SU(3)$ gauge theory and
$T(U,V)$ is the kernel of the blocking
transformation. 
%As the fluctuations of the gauge fields on the fine
%lattice are much smaller than those of the coarse gauge 
%fields, the Dirac operator used on the fine lattice has to have good
%chiral properties on gauge fields with small  
%fluctuations. This however is by far easier than to have a Dirac
%operator with good chiral properties on gauge  
%fields with large fluctuations. 
An important fact for the
parametrization of the FP Dirac operator is that 
Eq.~(\ref{RGO}) can also be given in terms of the propagators,
\begin{equation}
  D_c^{-1}(V) = \frac{\bf 1}{\kappa} + b^2 \omega(U) D_f^{-1}(U) \omega^\dagger(U), 
  \label{RGP}
\end{equation}
as long as $D_f$ has no zero modes. In contrast to
Eq.~\eqref{RGO} the equation for the propagator gives much more  
weight to the small physical modes of the Dirac operator and can therefore be used to improve the properties of
the small modes of the parametrized FP Dirac operator.

\begin{figure}
\begin{center}
\includegraphics[width=11cm]{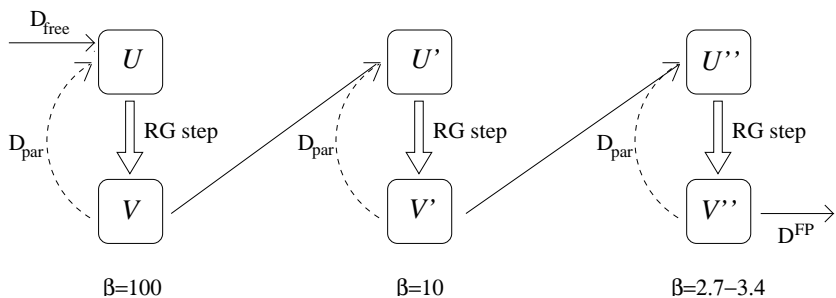}
\end{center}
\caption{Iterative procedure for parametrizing the FP Dirac
operator. The starting point at the top left is the FP Dirac operator
for free fermions $D_{\rm free}$. The RG blocking is then done on
three different 
sets of coarse and fine gauge configurations $U$ and $V$ at 
decreasing values of the gauge coupling 
$\beta$. As long as the free field limit of the parametrization is
fixed, the blocking can optionally be repeated at the same value of $\beta$ one
or several times (dashed lines). At the end, the parametrized FP Dirac operator
$D^{\rm FP}$ for
$\beta\approx 3$ is obtained.}
\label{fig:rg_iter}
\end{figure}

\subsection{Fitting the Parameters}
The parametrization is an iterative procedure, as sketched in
Fig.~\ref{fig:rg_iter}. 
We start at a large value of $\beta=100$, generate a few thermal
coarse gauge 
configurations $V^{(i)}$, $i=1,\dots,n_{\rm conf}$, with the FP gauge
action and determine the corresponding 
fine configurations $U^{(i)}$ through the minimization in Eq.~(\ref{RGG}).
As a Dirac operator on the right-hand side of the FP equations
\eqref{RGO} and \eqref{RGP}, we choose the free FP Dirac
operator. This is justified from the fact that the minimizing 
configurations $U^{(i)}$ have very small fluctuations at such a large
value of $\beta$ and are therefore very close to the free field
case. For each configuration, we  
take two sets of vectors $\{x^{(i)}_k\}$ and $\{y^{(i)}_k\}$,
$k=1,\dots,n_{\rm vec}$, of dimension $12N_{\rm site}$ on the coarse
lattice. The $x^{(i)}_k$ can either be 
random vectors or approximate small eigenmodes of $D_c$, and the
$y^{(i)}_k$ are just random vectors. From the right-hand sides of
Eqs.~\eqref{RGO} and \eqref{RGP}, the vectors
\begin{eqnarray}
\xi^{(i)}_k &=& D_c x^{(i)}_k, \\ 
\eta^{(i)}_k &=& D_c^{-1}y^{(i)}_k,
\end{eqnarray}
 are calculated. The couplings of the parametrized Dirac operator
$D_{\rm par}$ are then determined by minimizing
\begin{equation}
  \label{eq:chi_function}
  \chi^2 = \sum_{i,k} \big\| D_{\rm par}(V^{(i)}) x^{(i)}_k -
  \xi^{(i)}_k \big\|^2 +  
 \lambda \sum_{i,k} \big\| D^{-1}_{\rm par}(V^{(i)}) y^{(i)}_k -
  \eta^{(i)}_k \big\| ^2 , 
\end{equation}
where the sum runs over the different configurations and the set of
vectors per configuration and $\lambda$ is   
a weighting factor that has to be appropriately chosen.
During this procedure we keep leading terms in the naive continuum
limit fixed  such that the tree
level mass is zero,  
the $O(a)$ Symanzik condition is fulfilled, the dispersion relation
starts with slope $1$ and the normalization of 
the topological charge is correct \cite{Hasenfratz:2000xz,thomas:diss}
\footnote{In Eq.~(31) of \cite{Hasenfratz:2000xz}, 
the sign for the condition on the Dirac operator which fixes the
topological charge is incorrect. This influenced the first steps in
our parametrization 
procedure. In the last steps this condition was not used. The forcing
of the topological charge to the 
wrong sign in the earlier part of the iterative procedure
does not affect the end result significantly, because the
overlap reparametrization in the very last step straightened out this
error. In fact, 
even during the phase when the wrong condition was applied, the
linear terms in the fluctuation polynomials, which are not
affected by this condition, helped keeping the FP properties 
in the pseudoscalar sector present (see also \cite{thomas:diss}). }.
 Furthermore we fix
the free field limit such that the truncated free  
FP operator is recovered on the trivial gauge configuration $U=1$.
The minimization of the $\chi^2$-function \eqref{eq:chi_function}
yields a parametrized FP 
Dirac operator $D_{\rm par}$ which has good chiral properties over a 
larger range of gauge couplings than the initial truncated free FP Dirac
operator. 
The fitted parametrized operator $D_{\rm par}$ is now used on fine
configurations $U'$, determined via 
minimization from coarse configurations $V'$ generated thermally at a smaller
value of $\beta$. Minimizing the $\chi^2$-function
\eqref{eq:chi_function} again gives  
$D_{\rm par}(V')$, which performs well on an even larger range of gauge
couplings. The 
whole procedure is repeated until we reach $\beta \approx 3.0$,
corresponding to $a\approx 0.16$ fm. 
In the  final phase of this
iterative procedure, some of the naive continuum limit
constraints on the parameters of the Dirac operator are
released. Furthermore, in the last blocking step a low-order overlap
expansion is applied to the parametrization which is put into the
right hand side in order to reduce the remaining fluctuations of the
small eigenvalues even further \cite{thomas:diss}. 

Let us make a few comments on this procedure of iteratively finding
the parameters for the FP Dirac operator at lattice spacings typically used in
simulations. First, the use of vectors for the calculation of a
$\chi^2$-function for $D_{\rm par}$ is mandatory because the  
definition of $D_c$ requires a matrix inversion which we can only afford
for a limited number of vectors. Even then, on our workstations
we were restricted to lattices of maximum size $5^4$ for the coarse
and $10^4$ for the fine configurations. We usually worked with sets of
$5\leq n_{\rm conf} \leq 15$ different configurations and $n_{\rm vec}=5$ vectors per configuration.
 The use of large enough lattices is
important because when the lattice is too small, there are essentially no
small eigenvalues of $D$, which are however crucial for fitting the
chiral properties of the operator and become particularly
important when going to smaller values of $\beta$. If these small
eigenvalues are missing, the fit only captures the FP properties in
the region of large eigenvalues of $D$ well, and the resulting
parametrization then suffers for example from large additive mass
renormalization.  For the same reason it is
important to include the RG relation \eqref{RGP} for the propagator in
the fit.  At the largest value of $\beta$ however, the procedure was not so
sensitive for the presence of small eigenvalues, and we worked on
smaller lattices of size $3^4$ and $6^4$, respectively.

Second, we checked that the whole procedure does not strongly depend on
the choice 
of input Dirac operator for the right-hand side of the RG relation at the
largest value of $\beta$. When instead of the free-field FP operator
the Wilson operator is used in the first step, the set of
parameters after a few iterative steps agrees well with the one
resulting from the free FP operator as an input. This observation
confirms that at $\beta=100$, the 
minimized gauge configurations $U$ have such small fluctuations that
essentially any Dirac operator can be taken as an input for the
blocking without changing the result. The fluctuations can be
quantified by measuring the average 
value of the plaquette
\begin{equation}
 \langle u \rangle = \frac{1}{6N_{\rm site}}\sum_n \sum_{\mu<\nu}
 {\rm Re} \Tr\ 
 U_{\mu\nu}(n), 
\end{equation}
normalized to $\langle u \rangle = 3$ in the free field limit, which
is listed for the fine and coarse gauge 
configurations at different values of the coupling in Table
\ref{tab:avplaq}. The measurements show that the minimizing fine
configurations $U$ have very small fluctuations already at $\beta=3$,
and are very close to the free field limit at the largest
$\beta$. We have to remark that the average value of the plaquette hides the
somewhat different distribution for thermal and minimizing
configurations, which one also has to take into account.

\begin{table}
\begin{center}
\begin{tabular}{c|c|c|c} \hline\hline
$\beta$ & $\langle v\rangle $ &  $\langle w\rangle $ & $\langle u \rangle$ \\ \hline
100	& 2.92 & 2.989 & 2.998	\\
10	& 2.33 & 2.90 & 2.987	\\
5	& 1.68 & 2.76 & 2.97  \\
3.4	& 1.24 & 2.62 & 2.95  \\
3.0	& 1.16 & 2.56 & 2.94	\\
2.7	& 1.05 & 2.49 & 2.94  \\ \hline 
\end{tabular}
\end{center}
\caption{Average plaquette values $\langle v\rangle$ for the unsmeared
coarse configurations $V$, $\langle w \rangle$ for the smeared coarse
configurations 
and $\langle u \rangle$ for the minimizing fine configurations $U$ 
at different values of the gauge coupling which were used in the
parametrization procedure. The configurations at $\beta=5$ were only
used for checking whether an additional intermediate step improves the final
result.} 
\label{tab:avplaq}
\end{table}

\begin{figure}[htb]
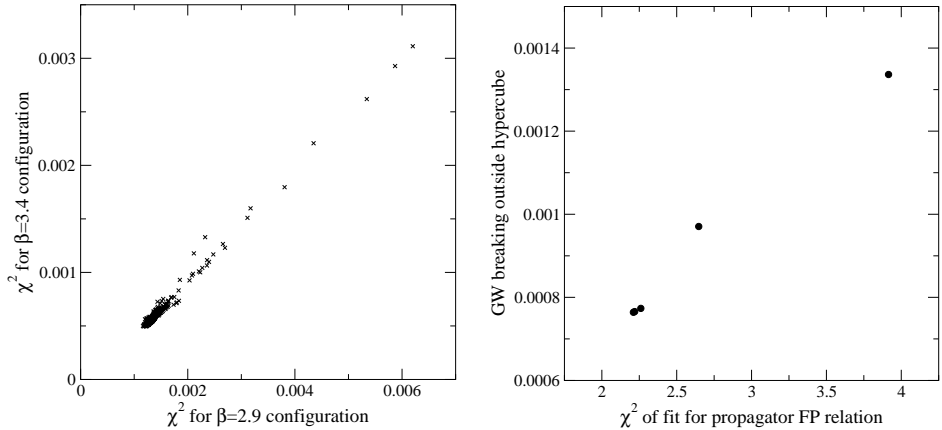

\begin{center}
\begin{tabular}{cc}
\hspace{-6mm}
\includegraphics[width=6cm]{epsf/chi_correlation.eps}
& 
\includegraphics[width=6cm]{epsf/chi_gw_correlation.eps}
\end{tabular}
\end{center}
\caption{Correlation plots from a parameter fit for the Dirac
operator. The figure on the left shows 
the correlation between the $\chi^2$ values of the propagator FP
relation (the second term on the right hand side of
Eq.~\eqref{eq:chi_function}) between two vectors on 
gauge configurations with $\beta=2.9$ and $\beta=3.4$ respectively, which
were the smallest and largest values of $\beta$ used in the fit. The right
figure shows the correlation between the total  $\chi^2$ of the
propagator FP relation and the breaking of the Ginsparg-Wilson
relation outside the hypercube as defined in Eq.~\eqref{eq:gw_breaking}. The
different points are taken from the Dirac operator parameter sets
created in the process of minimizing the $\chi^2$-function
\eqref{eq:chi_function} at a given set of coarse and fine configurations.}  
\label{fig:fit_coorelations}
\end{figure}

The last set of coarse and fine gauge configurations was chosen to
cover a range of values of the gauge coupling in the interval
$\beta\in[2.7,3.4]$, corresponding to lattice spacings in the range
$0.1$ fm $<a<0.22$ fm. These values were chosen such that the final 
parametrization can be used at somewhat
larger lattice spacings than typically used for simulations of
unimproved actions. That the fit 
captures the FP properties equally well on both ends of this range can 
be seen in the plot on the left of Fig.~\ref{fig:fit_coorelations}, where we
show the correlation between the values in the 
propagator part of the $\chi^2$-function \eqref{eq:chi_function} from two
gauge configurations at $\beta=2.9$ and $\beta=3.4$,
respectively. Configurations with 
$\beta<2.9$ were only used in the fitting of the FP relation \eqref{RGO} for
$D$, that is for the first summand in the $\chi^2$-function. 

We also tested whether adding
another intermediate step with fine and coarse configurations at
$\beta=5$ changes the outcome of the parametrization significantly,
but this was not the case.
Furthermore we performed a simple check whether it makes sense to explicitly
minimize the breaking of the Ginsparg-Wilson relation \eqref{eq:gw_prop} by
including it in the fit. This is a redundant constraint, as the FP
relation itself makes sure that the operator fulfills the
Ginsparg-Wilson relation. The right-hand side of \eqref{eq:gw_prop}
should be zero outside the hypercube, therefore we measured the
Ginsparg-Wilson breaking by calculating
\begin{equation} \label{eq:gw_breaking}
 \chi^2_{\rm GW} = \sum_x \big\| x \big[ \{ D_{\rm par}^{-1},\gamma_5
 \} - 2 \gamma_5 
 \big] y \big\|, 
\end{equation}
where the vector $y$ only has non-zero entries at the lattice origin
(0,0,0,0) and the sum runs over vectors $x$ with non-zero entries at
one single lattice site outside the hypercube around the origin.
This is computationally quite expensive due to the inversion of the
Dirac operator which is needed and therefore increases the time
for an iteration step in the minimization of $\chi^2$ considerably. The
plot on the right-hand side of 
Fig.~\ref{fig:fit_coorelations} shows that $\chi^2_{\rm GW}$ from the
Ginsparg-Wilson breaking seems to
correlate highly with the  $\chi^2$ from the propagator FP relation,
so we did not pursue this path further.

In order to parametrize the operator $R$ in Eq.~(\ref{RGeqR}) we
proceed in a similar way as for the Dirac operator. We also  
use a general operator with fat links and fluctuation polynomials. The
parametrization of $R$ is however simpler as  
it is trivial in Dirac space and therefore contains a smaller number
of operators. In contrast to the equation for the 
blocking transformation of $D^{\rm FP}$  the corresponding equation for
$R$ \eqref{RGeqR} contains no inversion and therefore 
the $\chi^2$-function which we minimize in order to find the optimal
parametrization of $R_{\rm par}$ can be defined as
\begin{equation}
  \label{eq:chi_function_r}
  \chi^2 = \| R_{\rm par} - R_{\rm C} \|^2 \, ,
\end{equation}
where the norm here is the matrix norm $\| A \|^2 = \sum_{i,j} \| a_{i
j} \|^2$.

A final remark on the hypercubic truncation: The Fixed-Point $R$ operator is
hypercubic, hence no truncation is needed. For the free FP Dirac operator,
which is known analytically, the couplings outside the hypercube
are very small, and thus the truncation does not distort the FP
properties too much. As our results will show, also in the interacting
case the FP operator can be well described by a hypercubic
parametrization, although some couplings which lie outside the
hypercube and are therefore left out tend to
grow larger than for the free case. We did however not consider
extending the parametrization beyond the hypercube, as then the
computational demand both 
for the construction of $D$ and for the matrix-vector multiplication
would grow very rapidly.

We have to mention another approximation that was taken in this work
when parametrizing the FP Dirac operator:
We only constructed 
the FP operator for zero quark mass, whereas in principle for every mass value
a different parametrization would be necessary. At larger masses, our
parametrization is therefore expected to deviate from the fixed point.

\section{Eigenvalue Spectrum}

\begin{figure}[htb]
\begin{center}
\includegraphics[width=80mm]{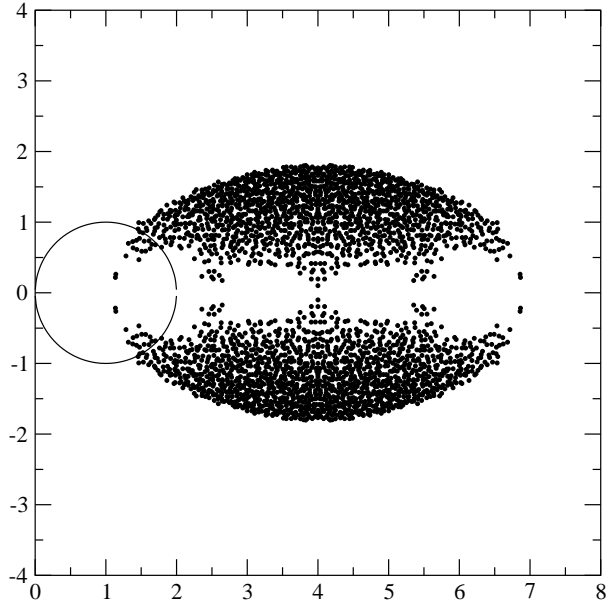}
\end{center}
\caption{Complex eigenvalue spectrum of the Wilson Dirac operator $D^{\rm
W}$ on a lattice of size $4^4$ at gauge coupling
$\beta=3.0$. For exactly chiral Dirac operators, the eigenvalues lie
on the circle 
with center (1,0) and radius 1 represented by the solid line. }  
\label{fig:wilson_ev}
\end{figure}

\begin{figure}[htb]
\begin{center}
\includegraphics[width=80mm]{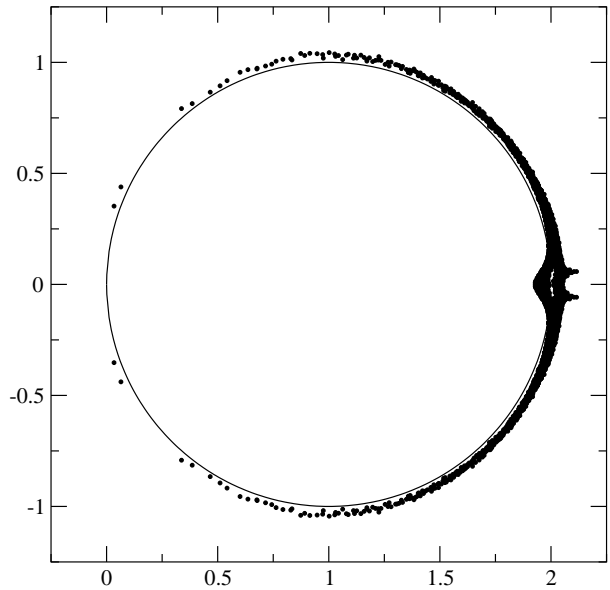} 
\end{center}
\caption{Complex eigenvalue spectrum of the rescaled parametrized FP  
Dirac operator $\tilde D^{\rm FP}$ on a lattice of size $4^4$
at gauge coupling $\beta=3.0$. For exactly chiral Dirac operators, the
eigenvalues lie on the circle 
with center (1,0) and radius 1 represented by the solid line. The
rescaling of $D^{\rm FP}$ is performed 
to account for the fact that due to the non-trivial $R$ in the
Ginsparg-Wilson relation, the eigenvalues would not be restricted to
the circle without rescaling, and then the chirality properties could
not be depicted in such a clear manner.}  
\label{fig:fp_ev}
\end{figure}

An easily accessible observable that quantifies the quality of a given
Dirac operator in terms of fulfilling the Ginsparg-Wilson relation is
the eigenvalue spectrum. Consider the case of a non-trivial $R$, and
define a rescaled Dirac operator $\tilde D = 
\sqrt{2 R}  D  \sqrt{2 R}$. Setting the lattice spacing $a=1$, the 
Ginsparg-Wilson relation \eqref{eq:gw_d} can be written as
\begin{equation}
  \label{eq:circle}
  \tilde D + \tilde D^{\dagger} = \tilde D^{\dagger}  \tilde D,
\end{equation}
implying that the eigenvalues of $\tilde D$ lie on a circle of radius
1 and center $(1,0)$. Using $R^{\rm FP}$ and the hypercubic approximation of
$D^{\rm FP}$, we compare
the eigenvalue spectrum of $\tilde D$ on a gauge configuration at 
$\beta=3.0$ with the spectrum of the Wilson
operator in Figs.~\ref{fig:wilson_ev} and \ref{fig:fp_ev}. While the
eigenvalues of the 
Wilson operator spread over a large region in the complex plane, they
lie almost exactly on the circle for the parametrized FP
Dirac operator, indicating that the hypercubic truncation and
the restriction to a finite set of gauge paths has only 
slightly affected the Ginsparg-Wilson property. Another very striking
observation is that the additive mass renormalization, which is given
by the point where the smallest eigenvalues intersect with the real
axis, is of ${\mathcal O}(1)$ and thus very large for the Wilson
operator, while for the 
parametrized FP Dirac operator no additive mass renormalization is seen in the
eigenvalue spectrum at this value of the gauge coupling. It has to
be stressed that this property is by no means enforced in the
parametrization procedure by shifting or constraining the
parameters. It originates only from the fact that the parametrization
describes the FP properties well. 

Checking the eigenvalue spectrum on
a set of different gauge 
configurations at given $\beta$, the fluctuation of the smallest or near-zero
eigenvalues can be measured, which is an important quantity for
simulating light quarks. Large fluctuations lead to the appearance of
exceptional configurations, making it impossible to invert the
Dirac operator at small quark mass. While the fluctuations are very
large for the Wilson 
operator, they are on the order of $10^{-2}$ for the parametrized FP
operator at $\beta=3.0$. Due to this crucial
property it is possible to perform lattice 
calculations at pseudoscalar to vector meson mass ratios as small as
$m_{\rm PS}/m_{\rm V}\approx 0.27$ for lattice spacings of 
$a=0.16$, as we show in the results of our spectroscopy simulations in
Chapter \ref{ch:results_fp}.

\fancyhead[RE]{\nouppercase{\small\it The Overlap-Improved Fixed-Point
Dirac Operator}}
\chapter{The Overlap-Improved Fixed-Point Dirac Operator}
\label{ch:d_ov}

The overlap construction proposed by Neuberger
\cite{Neuberger:1998fp} allows to formulate a lattice Dirac
operator with exact chiral symmetry. In the last few years, a lot of
activity has been going on testing and applying overlap fermions,
but while extensive
calculations of quenched QCD spectroscopy in the chiral limit have
been done with the 
approximately chiral domain-wall fermions \cite{Blum:2000kn,AliKhan:2000iv}, 
studies of lattice spectroscopy at small quark mass with overlap fermions
are only very recently emerging
\cite{Dong:2001yf,Dong:2001kv,Giusti:2001yw}.

Most groups working with overlap fermions use the Wilson Dirac  
operator as a starting point for the overlap, which might not be an
optimal choice. Wilson fermions show large scaling violations, and in the
overlap construction only the ${\mathcal O}(a)$ artifacts are
removed. The ${\mathcal O}(a^2)$ effects however will remain
present in the resulting Dirac operator and might even get
enhanced. Furthermore, the 
ultralocality of the Wilson operator is lost in the overlap, and while
the Wilson overlap operator is still local, the couplings 
decrease only with a rather small exponent
\cite{Hernandez:1998et}. Due to the strong chiral symmetry breaking of
the Wilson  
operator, also the numerical calculation of $1/\sqrt{A^\dag A}$ is not
easy because the condition number of the matrix $A^\dag A$ is large.
Better kernels for the overlap have been considered only by a few
groups \cite{Bietenholz:2000cc,DeGrand:2000tf} up to now. The clover
action as a kernel seems to perform even worse than the Wilson operator
\cite{Dong:2001fm}. 

The FP Dirac operator, being already an exact Ginsparg-Wilson
 operator, remains unchanged under the overlap
construction. Because our parametrization of the FP Dirac operator is
approximating the Ginsparg-Wilson relation very well, it will be
 changed only to a small extent by the overlap. The good scaling
 properties of the FP operator are then expected to be preserved in the end
 result. Also only relatively few iterative steps in the overlap
 expansion are necessary to ensure exact chirality. This property has been
used to calculate the finite-volume scaling of the chiral condensate
\cite{Hasenfratz:2001qp}, 
which is a task where chiral symmetry is required to be present
to a very high level. For the
 spectroscopy calculations in this work, we follow a slightly
 different strategy. We use an expansion to very low order in the
 inverse square root of the overlap, thus removing the already
 small chiral
 symmetry breaking effects introduced by the parametrization to a
 large extent, but not to machine precision. This strategy is from a
computational point of view more than competitive to standard 
 Wilson overlap simulations, and we expect to have the additional
 advantages of better localization of the resulting operator and
 improved scaling.

\section{Implementation of the Overlap}
\begin{figure}[tb]
\begin{center}
\includegraphics[width=95mm]{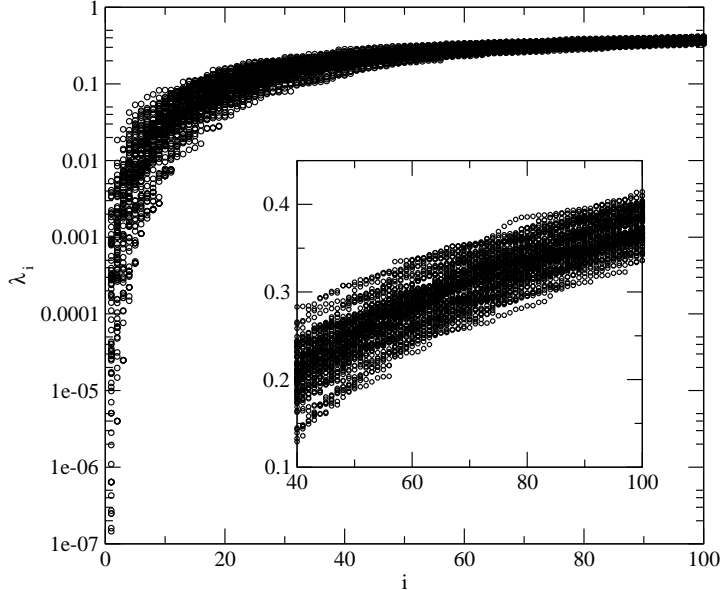}
\end{center}
\vspace{-2mm}
\caption{The 100 smallest eigenvalues of $A^\dag A$ with the FP kernel
for 80 different 
$12^3\times 24$ gauge configurations at $\beta=3.0$. The $x$-axis labels the
$i^{\rm th}$ smallest eigenvalue on a given gauge configuration, and
the $y$-axis shows its value $\lambda_i$.} 
\label{fig:adaga}
\end{figure}

Since we have parametrized the FP Dirac operator with a non-trivial $R$
operator in the Ginsparg-Wilson relation \eqref{eq:gw_d}, the massive
overlap-improved FP Dirac operator $D(m)$ which we use in our
simulations has to be specified with the corresponding covariant scalar
density \cite{Hasenfratz:2002} by
\begin{equation} \label{eq:d_ov_m}
D(m) \equiv \left( 1-\frac{m}{2} \right) D + \frac{m}{2R},
\end{equation}
where the massless overlap Dirac operator $D$ for non-trivial
Ginsparg-Wilson $R$ is
\begin{equation} \label{eq:dov}
D \equiv \frac{1}{\sqrt{2R}} \left( 1 - \frac{A}{\sqrt{A^\dag A}} \right) \frac{1}{\sqrt{2R}}.
\end{equation}
In the kernel $A$ of the overlap expansion we use the parametrized FP
Dirac operator $D^{\rm FP}$ from Chapter \ref{ch:d_fp}:
\begin{equation} \label{eq:ov_fp_kernel}
A \equiv 1- \sqrt{2R} D^{\rm FP} \sqrt{2R}.
\end{equation}
The inverse square root in Eq.~\eqref{eq:dov} is approximated using a
Legendre expansion. The convergence of this expansion can be
tremendously improved when treating the smallest eigenmodes of $A^\dag
A$ exactly, as then the condition number of the matrix becomes much
smaller. We show in Fig.~\ref{fig:adaga} the smallest 100 eigenvalues
on 80 gauge configurations from our spectroscopy simulation. There are
some isolated eigenvalues very close to zero and a 
rapidly increasing density of eigenvalues closer to 1. The threshold
where the eigenvalues get dense decreases with increasing lattice
spacing and volume. We showed in \cite{Hasenfratz:2001qp} that this
threshold lies much higher for the FP kernel
\eqref{eq:ov_fp_kernel} than for the standard kernel with the Wilson operator,
implying that the overlap expansion for the FP kernel needs orders
of magnitudes fewer iterations. In the simulations in
Chapter~\ref{ch:results_fp},  
we treat on all gauge configurations the smallest 100 eigenmodes
exactly. The eigenvalue of the largest exactly treated mode ($i=100$ in
Fig.~\ref{fig:adaga}) lies in this case between
$0.33 < \lambda < 0.42$ for all considered gauge configurations. The
largest eigenvalue of 
$A^\dag A$ is typically close to $\lambda=1.7$, and therefore the
subspace where the iterative solver works is very well-behaved.

\begin{figure}[tb]
\begin{center}
\includegraphics[width=8cm]{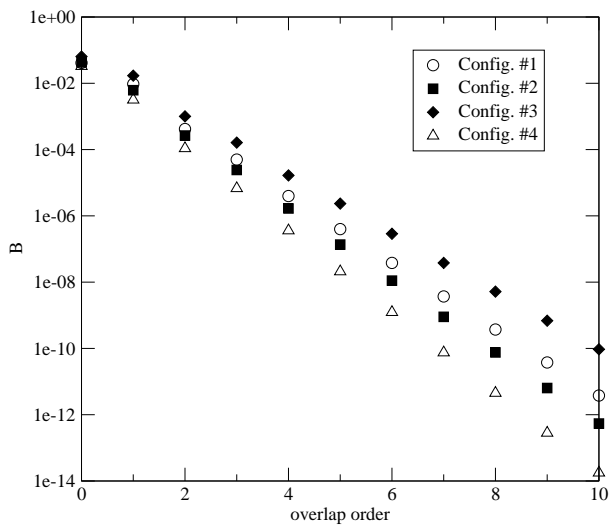}
\end{center}
\vspace{-2mm}
\caption{Breaking of the Ginsparg-Wilson relation for the overlap
Dirac operator with FP kernel on different $10^4$ configurations at
$\beta=3.2$. Only the smallest 10 eigenvalues are treated exactly.} 
\label{fig:gw_breaking}
\end{figure}

Due to the good chiral properties of the starting FP operator, we can
restrict the Legendre expansion to rather low order $n \leq
10$. In Fig.~\ref{fig:gw_breaking} we plot the residual breaking of the
Ginsparg-Wilson relation determined by 
$ B = |(\tilde D + \tilde D^\dag - \tilde D^\dag \tilde D)v|^2$
where $v$ is a random vector normalized to 1 and $\tilde D =
\sqrt{2R}D\sqrt{2R}$ is the rescaled Dirac operator, as a function
of the overlap order. We see that the chiral 
symmetry breaking decreases very quickly.
While $n \approx 10 $ has been used for measurements
of the chiral condensate \cite{Hasenfratz:2001qp}, for 
the spectroscopy calculations in this work we take $n=3$,
which gives a Dirac operator with improved, 
but not exactly chiral properties relative to the parametrized FP Dirac
operator. We will therefore call this the overlap-improved FP Dirac
operator. As an order $n$ Legendre 
expansion of the inverse square root of $A^\dag A$ requires $2n$
multiplications of the Dirac operator on a vector, 
the computational requirements of
the overlap-improved Dirac operator from calculating $A/\sqrt{A^\dag
A}$ are by a factor of 7 larger than for the parametrized FP Dirac
operator.\footnote{The time for multiplications with
$R$ is neglected here.} 

In a multi-mass inverter, the massive overlap Dirac
operator can not be used in the form \eqref{eq:d_ov_m} due to the
presence of the $R$ operator. Instead we actually invert the operator
$D2R$ by writing 
\begin{equation}
D(m)^{-1}=\frac{1}{1-m/2} 2R \left( D 2R + \frac{m}{1-m/2}\right)^{-1},
\end{equation}
where the term in brackets defines a shifted linear system that can be
solved by the inverter. As one can
see, every matrix-vector multiplication in the algorithm requires 
a multiplication of both the $D$ and the $R$ operator on the
vector. Because the $R$ operator is trivial in Dirac space, this leads however
only to a comparably small computational overhead.

\begin{figure}[tb]
\begin{center}
\includegraphics[width=9cm]{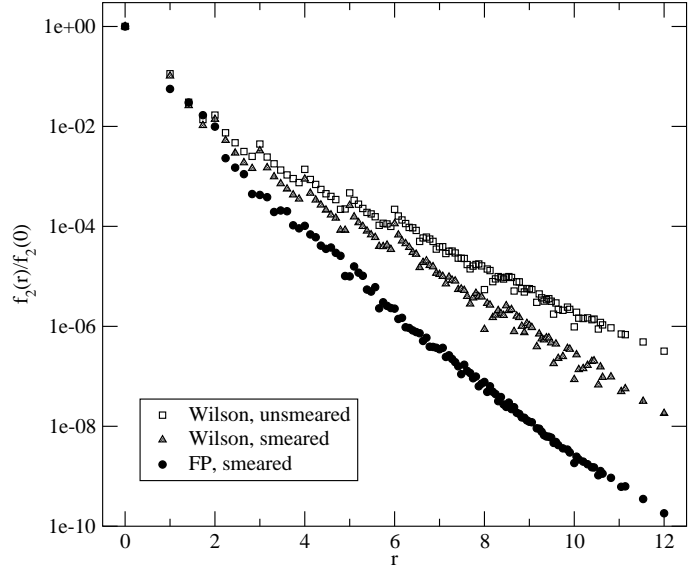}
\end{center}
\vspace{-3mm}
\caption{Locality of overlap Dirac operators with Wilson and FP
kernels given by the exponential decrease of couplings with 
distance in Eq.~\eqref{eq:locality}. The Wilson operator is examined
both on smeared and unsmeared gauge configurations with
$\beta=3.0$. The plot is taken from 
\cite{thomas:diss}.}   
\label{fig:ov_locality}
\end{figure}

\section{Locality of Couplings}
Every reasonable lattice Dirac operator has to be local, with
exponentially decreasing couplings. If the Dirac
operator is restricted to a finite number of lattice sites, like the
Wilson or the parametrized FP Dirac operator, it is called 
ultralocal. The overlap construction takes an ultralocal operator as
an input, but the result is no longer ultralocal. For the Wilson
kernel, locality of the resulting overlap operator has been shown,
but the exponential decay rate of the couplings is quite small. In
Fig.~\ref{fig:ov_locality} we compare
the locality, measured by
\begin{equation} \label{eq:locality}
f(r) = \underset{y}{\rm max}\{ |D\cdot v|; ||y-x||=r \},
\end{equation}
where $v$ is vector with a point source at $x$, for the  overlap Dirac
operator with the Wilson and the FP kernel. Together with other tests
for the locality of the Wilson overlap operator
\cite{Hernandez:1998et,Bietenholz:2001nu}, it follows that the locality is  
significantly improved when the FP kernel is used.

\section{Locality of Instanton Zero Modes}
The properties of approximately or exactly chiral Dirac
operators can be tested on smooth instanton gauge
configurations. Phenomenological
models suggest that instantons might be responsible  for
dynamical chiral symmetry breaking \cite{Schafer:1998wv}. Single
instantons, which are gauge configurations with 
topological charge $Q=1$, produce a zero mode of the Dirac 
operator through the index theorem \eqref{eq:index_theorem}. A pair of
an instanton and an anti-instanton produces two 
complex-conjugate near-zero eigenmodes. The chiral condensate, which
is the order parameter for 
chiral symmetry breaking, is related to the density of eigenvalues of
the Dirac operator near the origin through the Banks-Casher relation.
In the infinite volume limit, the density of exact zero modes becomes
negligible compared to the density of near-zero modes.
Only with the contribution of near-zero modes, the eigenvalue density
at zero therefore does not vanish at infinite volume. Initiated by
\cite{Horvath:2001ir}, many groups have recently
studied whether the local chirality properties of these near-zero
modes are consistent with the instanton model of chiral symmetry
breaking
\cite{DeGrand:2001pj,Hip:2001hc,Gattringer:2001mn,Hasenfratz:2001qp}. 

In this section, we analyze
eigenvalues and eigenvectors of the overlap-improved 
FP Dirac operator on gauge configurations describing a discretized
exact instanton, as 
done in \cite{Gattringer:2001cf} for the chirally improved Dirac operator. For
the above mentioned studies on local chirality, it is helpful to 
know how well a given Dirac operator reproduces the continuum zero mode
of an instanton. We show that the Wilson overlap Dirac operator is in
this respect 
not optimal, which might be due to its comparably bad locality properties.
The gauge configurations are constructed from discretized
$SU(2)$-in\-stan\-tons trivially embedded as
$2\times2$-submatrices in $SU(3)$ \cite{Gattringer:2001cf}. We work on
lattices of size $12^4$ 
and apply antiperiodic boundary conditions in time for the
fermions. Eigenvalues and eigenvectors are calculated with the
implicitly restarted Arnoldi algorithm \cite{sorensen:1992}.

\begin{figure}[tb]
\begin{center}
\includegraphics[width=90mm]{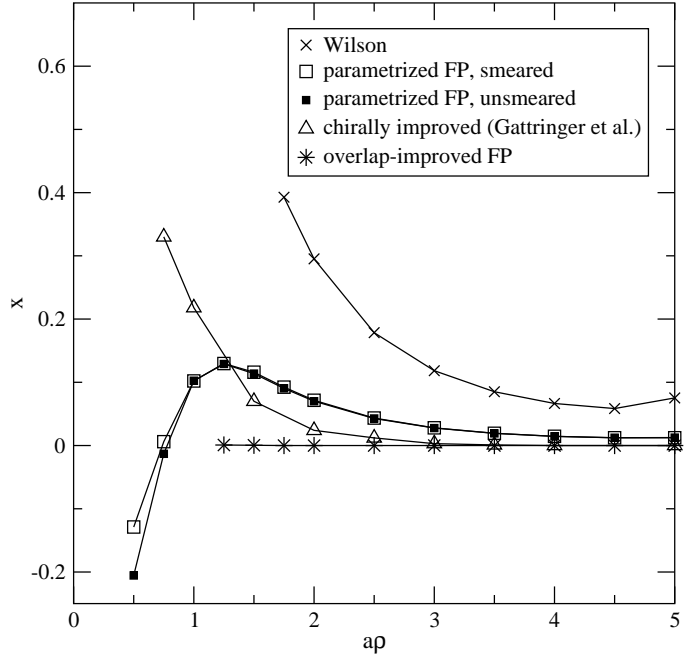}
\end{center}
\vspace{-3mm}
\caption{Position $x(a\rho)$ of the zero eigenvalue on the real axis in
dependence of the instanton radius for different Dirac operators.}
\label{fig:inst_ev}
\end{figure}

First, we compare the flow of the zero eigenvalue with instanton size 
for various Dirac operators in Fig.~\ref{fig:inst_ev}. For an exactly
chiral operator (and also 
for the overlap-improved FP operator on the scale of this plot), the
eigenvalue is exactly zero. Due to the residual chiral
symmetry breaking, it can move away from zero for
approximately chiral Dirac operators. The eigenvalue is however
restricted to the real axis, as long as the Dirac operator respects
$\gamma_5$-hermiticity. For varying instanton radius $\rho$, the
position $x$ of the zero eigenvalue on the real axis can be monitored,
providing a measure for the chiral properties of the Dirac operator.

\begin{figure}[tb]
\begin{center}
\includegraphics[width=8cm]{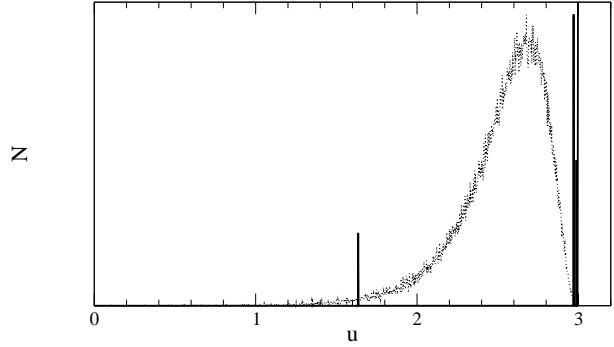}
\end{center}
\vspace{-3mm}
\caption{Histogram of plaquette values $u= {\rm Re\ Tr\ } U_{\mu\nu}$ on a
RG smeared Monte 
Carlo configuration at $\beta=3.0$ (smooth curve) and on a small, RG smeared
instanton with $a\rho=0.5$ (peaks).}
\label{fig:inst_plaq}
\end{figure}

For the Wilson operator, the eigenvalue quickly flows away from
zero towards the center of the
circle with decreasing instanton radius $\rho$. Calculating only the
few smallest eigenmodes, we lost track of it already at $a\rho<2$.
 The parametrized FP Dirac operator shows a much better 
behavior, with the eigenvalue staying close to zero. For
$a\rho<1.25$, it even moves back and takes a negative value at
$a\rho=0.5$. This can be interpreted as an 
'overimprovement' caused by the fairly large coefficients for the
fluctuation polynomials \cite{thomas:diss} in our parametrization: If
at some locations the gauge fields fluctuate very 
strongly, as it is the case for such an artificial
small instanton, the fluctuation polynomials shift the couplings
in the Dirac operator away from reasonable values. The histogram of
plaquette values in Fig.~\ref{fig:inst_plaq} illuminates the
qualitative difference of the fluctuations in these instanton and in
thermal Monte Carlo gauge configurations. While on a Monte Carlo
configuration the plaquette distribution is smooth, most of the
plaquettes of an instanton configuration are very close to $u=3$, but there is
also a peak at $u\approx 1.6$ from the 
plaquettes at the center of the instanton. As a consequence, the 
Dirac operator is strongly affected there through the fluctuation
polynomials which are proportional to terms like $3-u$,  causing
this unusual behavior of the real eigenvalue. 
 We never observed such real eigenvalues shifted towards negative values in
Monte Carlo gauge configurations used in actual 
lattice simulations, which were also used to parametrize the FP
operator. Only in such an artificial
environment as given by the discretized instantons the parametrized FP
Dirac operator shows this effect. 

We also investigate the effect of smearing for the FP operator. Obviously
the RG inspired smearing, which we use together with the FP Dirac
operator, does not change the flow of the real eigenvalue much. Only
for the smallest instanton $a\rho=0.5$ the 
eigenvalue is pushed back slightly towards zero when the gauge
configurations are smeared.
Another approximately chiral operator that behaves very well on 
these instanton configurations is the 
chirally improved Dirac operator by Gattringer et
al.~\cite{Gattringer:2001cf}, which 
is optimized for fulfilling the Ginsparg-Wilson relation. 

\begin{figure}[tb]
\begin{center}
\includegraphics[width=90mm]{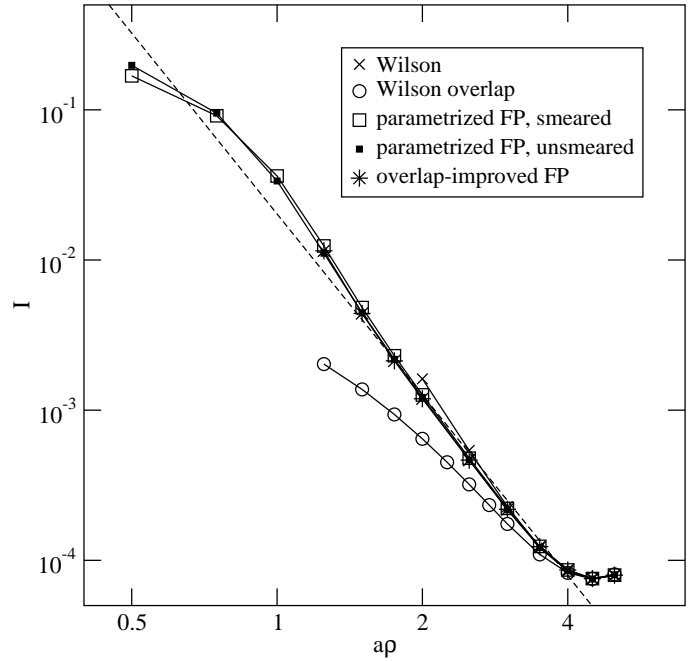}
\end{center}
\caption{Inverse participation ratio $I(a\rho)$ of the zero eigenmode in
dependence of the instanton radius for different Dirac operators. The
values for the Wilson overlap are taken from
\cite{Gattringer:2001cf}. The dashed line denotes the continuum value
as given in Eq.~\eqref{eq:ipr}.}
\label{fig:inst_ipr}
\end{figure}

As a second check we investigate the locality properties of the
corresponding zero eigenmode $\phi_0$, which is centered at the place
where the instanton sits. With the gauge invariant density
\begin{equation}
 p(x) \equiv \sum_{\alpha, a} \phi_0^{\alpha a}(x)^* \phi_0^{\alpha a}(x) 
 \stackrel{\rm cont.}{=} \frac{2\rho^2}{\pi^2(\rho^2+x^2)^3},
\end{equation}
where $\int d^4x\ p(x) = 1 $ for normalized eigenvectors,
we can define as a measure for the localization of the zero mode
$\phi_0$ the inverse participation ratio
\begin{equation} \label{eq:ipr}
 I \equiv \int\! d^4x\ p(x)^2 \stackrel{\rm cont.}{=} \frac{1}{5\pi^2\rho^4},
\end{equation}
which is plotted in Fig.~\ref{fig:inst_ipr} and compared to the
continuum value. All Dirac operators agree
for the largest instanton radii $\rho\geq 4$, where the lattice
results deviate from the continuum due to finite-volume
effects. The Wilson overlap 
operator however deviates strongly from the continuum values also for smaller
instantons, seemingly
having problems to reproduce the continuum zero mode. This can be
interpreted as a consequence of the relatively 
bad localization properties of the overlap with Wilson kernel.  Varying
the optimizing parameter $s$ in the Wilson overlap does not lead to a
significant 
change in these result \cite{Gattringer:2001cf}.
The situation is different for the ultralocal Dirac operators like the
parametrized FP and the Wilson operator. The parametrized FP operator
reproduces the continuum value well even for very small instantons,
while the Wilson operator performs somewhat worse, but still much
better than the Wilson overlap. 

The main result from this study is that the third order
overlap-improved FP Dirac 
operator almost shows no change compared to the parametrized FP
operator and captures the localization of the zero mode very well. The
same observation has been made for an overlap operator with a chirally
improved kernel, giving evidence that the
Wilson kernel for the overlap construction, which misses the localization
properties of the instanton zero mode, is not the best choice.
Another observation we make when comparing
the parametrized FP operator on both smeared
and unsmeared gauge configurations is that the RG-inspired smearing of the
gauge fields does not
change the results for the inverse participation ratio. The smearing
therefore does not 
seem to lead to a significant modification of the locality properties
of the Dirac operator.

\fancyhead[RE]{\nouppercase{\small\it Hadron Spectroscopy in Lattice QCD}}
\chapter{Hadron Spectroscopy in Lattice QCD}
\label{chapter:had_spect}

Among the most basic quantities that are calculable in lattice
QCD are the masses and decay constants of the various bound states of quarks
and gluons. Reproducing the experimentally observed spectrum of
hadronic particles is one of the strongest tests that QCD is the
correct theory to describe nature at the corresponding energy
scale. In this chapter we present how 
masses of light hadrons, which are made from up, down and strange
quarks\footnote{Because the masses of the charm and bottom quarks are on
the order of the lattice cutoff, lattice systems with heavy quarks
have to be treated either in a non-relativistic approach  
\cite{Lepage:1992tx,El-Khadra:1997mp}, in the static approximation
\cite{Eichten:1987xu} or on anisotropic lattices.}, are 
extracted from quantities accessible on the lattice, and we
discuss some refinements to 
improve the quality of these measurements. 

First we derive in Section
\ref{sect:corrfuncs} the basic observation that the 
exponential decay of hadronic correlation functions is related to the
hadron mass. The correlators can be calculated by inverting the lattice Dirac
operator and contracting the resulting quark propagators together with the
Dirac matrices corresponding to a particular hadronic state. Then we
show in Section \ref{sect:extended_sources} how the overlap of the creation and
annihilation operators with the ground state 
of the hadron can be increased by using extended wave functions at the
source and sink locations. In the last section, we concentrate on the
technical issues related to the fitting of lattice data to the
predicted functional forms and show how the statistical error of the
resulting hadron masses is estimated.

\section{Fermionic Observables from Correlation Functions}
\label{sect:corrfuncs}

The typical quantity to measure in lattice hadron spectroscopy
is the two-point function
\begin{equation} \label{eq:2point_corrfunc}
C(x)  \equiv  \langle 0| T\{{\mathcal
O}_f(x) {\mathcal O}^\dag_i(0)\} | 0 \rangle,
\end{equation}
describing the space-time propagation of a particle created at
the origin by the operator $ {\mathcal O}^\dag_i$ and 
annihilated at space-time coordinate $x=(\vec x, t)$ by the operator
$ {\mathcal O}_f$. To create a meson, a quark bilinear operator of the form
\begin{equation} \label{eq:general_meson_op}
{\mathcal O}^{\rm M}(x) \equiv \bar\psi^{a, f_1}_\mu(x)
\Gamma_{\mu\nu} \psi^{a, f_2}_\nu(x), 
\end{equation}
is used. The Clifford algebra element $\Gamma$ determines the quantum
numbers of the desired quark-antiquark state, and $f_1$, $f_2$ denote the
flavors $u$,$d$,$s$ of the quark constituents.
A baryon is created by the three-quark operator 
\begin{equation} \label{eq:general_baryon_op}
{\mathcal O}^{\rm B}(x) \equiv \epsilon_{abc} \chi_{\mu\nu\rho}
\psi_\mu^{a, f_1}(x) \psi_\nu^{b, f_2}(x) \psi_\rho^{c, f_3}(x),
\end{equation}
with some appropriate spin function $\chi_{\mu\nu\rho} $. In the
simulations in Chapter \ref{ch:results_fp}, we will work with
the baryon operators used by the MILC collaboration
\cite{milccode}, which create an equal mixture of a forward propagating
baryon and a backward propagating antibaryon on a periodic lattice. In
Tables \ref{tab:meson_operators} and \ref{tab:baryon_operators} we
list the spin content of our meson and baryon operators.

\begin{table}[tb]
\begin{center}
\begin{tabular}{c|c|c|c} \hline\hline
 Label & State  & $I^{G}(J^{PC})$ & Operator \\ \hline
 S & scalar    & $1^-(0^{++})$ & $\bar u(x)  d(x)$ \\
 P & pseudoscalar  & $1^-(0^{-+})$ & $\bar u(x) \gamma_5 d(x)$ \\
 A &    & $1^-(0^{-+})$ & $\bar u(x) \gamma_4\gamma_5 d(x)$ \\
 V & vector    & $1^+(1^{--})$ & $\bar u(x) \gamma_i d(x)$ \\ \hline
% N & Nucleon octet & $J=1/2$ & $[d^a
% {\mathcal C} \gamma_5 u^b] u^c_1 \epsilon_{abc}$ \\
% N0 &  & $J=1/2$ & $[d^a
% {\mathcal C} \gamma_4\gamma_5 u^b] u^c_1 \epsilon_{abc}$ \\
% D & Delta decuplet & $J=3/2$ & $(2[d^a
% {\mathcal C} \gamma^- u^b] u^c_1  + [u^a
% {\mathcal C} \gamma^- u^b] d^c_1 )\epsilon_{abc}$ \\
% D0 &  & $J=3/2$ & $(2[d^a
% {\mathcal C} \gamma_4\gamma^- u^b] u^c_1  + [u^a
% {\mathcal C} \gamma_4\gamma^- u^b] d^c_1 )\epsilon_{abc}$ \\ \hline
\end{tabular}
\end{center}
\caption{Local interpolating field operators for mesons.  In the
 vector meson, the sum over all polarizations $i=1,\dots,3$ is taken.}
\label{tab:meson_operators}
\end{table}

\begin{table}[tb]
\begin{center}
\begin{tabular}{c|c|c} \hline\hline
 Label & State & Operator \\ \hline
 N & octet, $|{\rm n}, s_3=\frac{1}{2}\rangle$ & $(d^a
 {\mathcal C} \gamma_5 u^b) u^c_1 \epsilon_{abc}$ \\
 N0 &   & $(d^a
 {\mathcal C} \gamma_4\gamma_5 u^b) u^c_1 \epsilon_{abc}$ \\ \hline
 D & decuplet, $|\Delta^+, s_3=\frac{3}{2}\rangle$ & $  2(d^a
 {\mathcal C} \gamma^- u^b) u^c_1 \epsilon_{abc}  + (u^a
 {\mathcal C} \gamma^- u^b) d^c_1 \epsilon_{abc}$ \\
 D0 &   & $2(d^a
 {\mathcal C} \gamma_4\gamma^- u^b) u^c_1 \epsilon_{abc} + (u^a
 {\mathcal C} \gamma_4\gamma^- u^b) d^c_1 \epsilon_{abc}$ \\ \hline
\end{tabular}
\end{center}
\caption{Local interpolating field operators for baryons.  The charge
 conjugation matrix is  $ {\mathcal C} =
 \gamma_2\gamma_4$.}
\label{tab:baryon_operators}
\end{table}

Let us demonstrate how to extract physical quantities from the correlation
function \eqref{eq:2point_corrfunc} on a lattice of infinite
volume. In order to single out particles
with defined momenta, consider the spatial Fourier transform
\begin{equation} \label{eq:2point_corrfunc_ft}
C(\vec p,t)  =  \sum_{\vec x} e^{i\vec p \vec x} \langle 0| T\{{\mathcal
O}_f(\vec x,t) {\mathcal O}^\dag_i(0,0)\} | 0 \rangle.
\end{equation}
Inserting a complete set of eigenstates $|n,\vec q\rangle$ with 
spatial momentum $\vec q$, we get 
\begin{equation} \label{eq:2point_corrfunc_inserted}
C(\vec p,t)  =  \sum_{\vec x} e^{i\vec p \vec x} \sum_n \int \frac{d^3\vec q}{(2\pi)^3 2 E_n(\vec q)}
\langle 0| {\mathcal O}_f(\vec x, t) |n,\vec q\rangle \langle n,\vec q| {\mathcal O}^\dag_i(0,0) | 0 \rangle ,
\end{equation}
where $E_n(\vec q)$ is the energy of the intermediate state $|n,\vec q\rangle$.
Applying the space-time translation ${\mathcal O}_f(x) = e^{i{\mathcal P 
x}}{\mathcal O}_f(0)e^{-i{\mathcal P x}}$  with four-momentum ${\mathcal
P} = ({\mathcal H}, \vec{\mathcal P})$ to the annihilation operator,
the correlation function \eqref{eq:2point_corrfunc_inserted} 
can be written as an exponentially weighted sum over all intermediate states,
\begin{align} \label{eq:2point_corrfunc_series}
C(\vec p,t)  & =  \sum_n \int \frac{d^3\vec
q}{(2\pi)^3 2 E_n(\vec q)} e^{-iE_n(\vec q) t}
\langle 0| {\mathcal O}_f |n,\vec q\rangle \langle n,\vec q| {\mathcal
O}^\dag_i | 0 \rangle \sum_{\vec x} e^{i(\vec p-\vec q) \vec x} \nonumber
\\
 & = \sum_n \frac{\langle 0| {\mathcal O}_f |n,\vec p \rangle
\langle n,\vec p| {\mathcal O}^\dag_i | 0 \rangle}{2 E_n(\vec p)}
\ e^{-iE_n(\vec p) t} ,
\end{align}
where $\sum_{\vec x} e^{-i\vec p \vec x} = (2\pi)^3 \delta(\vec p)$
has been used to
get rid of one momentum variable, and $ {\mathcal O}_{i,f} \equiv
{\mathcal O}_{i,f}(0,0)$. For large Euclidean time $\tau = 
it$, only the state with lowest energy  
contributes, therefore the asymptotic form of the correlation function becomes
\begin{equation}
C(\vec p,\tau) \xrightarrow{\tau\rightarrow\infty} \frac{\langle 0|
{\mathcal O}_f |1,\vec p\rangle 
\langle 1,\vec p| {\mathcal O}^\dag_i | 0 \rangle}{2 E_1(\vec p)}
\ e^{-E_1(\vec p)\tau} .
\end{equation}
Considering only intermediate states with zero momentum by setting 
$\vec p = 0$, the mass of the lightest state can be extracted from the
exponential decay of the Euclidean time correlation function
\begin{equation} \label{eq:zero_mom_correlator}
C(\tau) \xrightarrow{\tau\rightarrow\infty} \frac{A_f A_i^*}{2 m_1}
\ e^{-m_1\tau} ,
\end{equation}
where the amplitudes $A_{i,f} = \langle 0| {\mathcal O}_{i,f} |1\rangle$ 
represent the overlap of the operators ${\mathcal O}_{i,f}$ with the
lightest particle state having the quantum numbers of ${\mathcal O}_{i,f}$.

\begin{figure}[tb]
\begin{center}
\includegraphics[width=80mm]{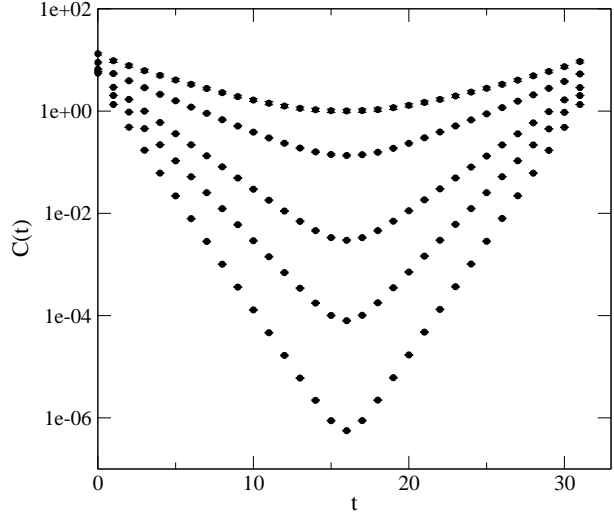}
\end{center}
\vspace{-3mm}
\caption{Pseudoscalar meson propagators $C(\tau)$ on a $16^3\times 32$
lattice for 
various quark masses. The data comes from our simulation at
$\beta=3.0$ presented in Chapter \ref{ch:results_fp}. Due to the
periodic boundary conditions, the 
correlators are symmetric around $\tau=16$.}
\label{fig:pion_corrs}
\end{figure}

In Fig.~\ref{fig:pion_corrs}, we plot typical examples of meson
correlation functions on the lattice at varying quark masses, showing clearly
the exponential decay with Euclidean time.

\subsection{Lattice Quark Propagators}
\label{sect:qprop}
In order to measure the correlation function
\eqref{eq:2point_corrfunc} in lattice QCD, it is expressed
in terms of the Euclidean quark propagator 
\begin{equation}
G_{\mu\nu}^{ab}(x,y) \equiv \langle 0| T\{\psi_\mu^a(x)
\bar\psi_\nu^b(y) \} |0 \rangle,
\end{equation}
using Wick contractions of the quark fields. Consider for example a flavor
non-singlet meson operator \eqref{eq:general_meson_op}, where $f_1
\neq f_2$. The two-point function is then
\begin{align} \label{eq:meson_corr_deriv}
\langle {\mathcal
O}^{\rm M}(\vec x,t) {\mathcal O}^{\rm M \dag} (0,0) \rangle & \equiv
\langle 0| 
T\{{\mathcal O}^{\rm M}(\vec x,t) {\mathcal O}^{\rm M\dag}(0,0)\} | 0
\rangle \nonumber\\ 
 & = \langle 0| T\{ \bar\psi_\sigma^{a, f_1}(x)\Gamma_{\sigma\mu} \psi_\mu^{a,
f_2}(x) \bar\psi_\nu^{b, f_2}(0)\Gamma_{\nu\rho} \psi_\rho^{b,
f_1}(0) \} |0\rangle \nonumber \\
 & = \langle G^{ab, f_2}_{\mu\nu}(x,0) \Gamma_{\nu\rho}
G^{ba, f_1}_{\rho\sigma}(0,x) \Gamma_{\sigma\mu} \rangle \nonumber \\
 & = \langle \Tr G^{f_2}(x,0) \Gamma G^{f_1}(0,x)\Gamma  \rangle, 
\end{align}
where the trace is taken over spin and color indices. Making use of the
$\gamma^5$-hermiticity of the quark 
propagator $G(x,y) = \gamma^5 G(y,x)^\dag \gamma^5$, where the
hermitean conjugation also acts in spin and color space, 
Eq.~\eqref{eq:meson_corr_deriv} becomes
\begin{equation} \label{eq:meson_corr_with_qprop}
\langle {\mathcal
O}^{\rm M}(\vec x,t) {\mathcal O}^{\rm M\dag}(0,0) \rangle  = \langle
\Tr G^{f_2}(x,0) \Gamma \gamma_5 G^{f_1}(x,0)^\dag \gamma_5  \Gamma  \rangle.
\end{equation}
The importance of this last step comes from the fact that in lattice QCD
calculations, the quark propagator $G(x,y) =
D^{-1}(x,y)$ is determined by a matrix inversion of the lattice Dirac
operator, which is by far the most expensive part of quenched simulations. A
complete inversion amounts to solving the linear system of $12V$ equations
\begin{equation}
D_{\mu\nu}^{ab}(x,y) G_{\nu\rho}^{bc}(y,z) =
\delta_{ac}\delta_{\mu\rho}\delta(x-z) ,
\end{equation} 
for $G$, which is in most cases not feasible. For the calculation of the 
right hand side of \eqref{eq:meson_corr_with_qprop}, it
is however only necessary to know the quark propagator $G(x,0)$ from a
fixed source 
point at the origin, where the particle is created, to all points on the
lattice. Hence it is sufficient to evaluate 12 columns of the inverted
matrix (one per spin and color) by solving
\begin{equation}
D_{\mu\nu}^{ab}(x,y) G_{\nu\rho}^{bc}(y,0) =
\delta_{ac}\delta_{\mu\rho}\delta(x), 
\end{equation} 
reducing the numerical size of the problem by a factor of $V$. This
trick does not work 
in the case of flavor singlet mesons, where the disconnected
contribution in the quark line graph requires the knowledge of the
full quark propagator, which is the reason why their treatment on the
lattice is much more demanding.

Finally, to get the zero momentum
correlator \eqref{eq:zero_mom_correlator}, it is sufficient to sum
Eq.~\eqref{eq:meson_corr_with_qprop} over all sink locations on a
given timeslice,
\begin{equation} \label{eq:pp_meson_correlator}
C(\tau) =  \Big\langle
\sum_{\vec x} \Tr G^{f_2}(x,0) \Gamma \gamma_5 G^{f_1}(x,0)^\dag \gamma_5  \Gamma  \Big\rangle,
\end{equation}
which projects out the $\vec p=0$ contribution in
\eqref{eq:2point_corrfunc_ft}.

\section{Extended Source and Sink Operators}
\label{sect:extended_sources}
In order to get a good signal-to-noise ratio for the measured hadron
correlators, the operators ${\mathcal O}_i$ and ${\mathcal O}_f$ in
Eq.~\eqref{eq:2point_corrfunc} should have a 
large overlap with the  
desired state. Local operators like \eqref{eq:general_meson_op} and
\eqref{eq:general_baryon_op} are not 
expected to fulfill this criterion well, as they do not take 
into account the spatial extension of the hadron. Especially
for small quark masses or small lattice spacings, neglecting the
hadron extension becomes a
problem, as light hadrons typically have a size of {$\mathcal O$}(1
fm), and lattice spacings in current simulations are mostly between
0.05 fm and 0.2 fm. Maximum  
overlap for a meson would be reached for a non-local operator
\begin{equation} \label{eq:general_meson_op_smeared}
{\mathcal O}^{\rm M}_{\rm ext}(t) = \sum_{\vec x, \vec y} \varphi(\vec x, \vec
y)\bar\psi^{a, f_1}_\mu(\vec x, t)
\Gamma_{\mu\nu} \psi^{a, f_2}_\nu(\vec y, t), 
\end{equation}
where $\varphi(\vec x, \vec y)$ is the wave function of the
meson. For a delta function
$\varphi(\vec x, \vec y) =  \delta(\vec x - \vec y)$, the local
operator \eqref{eq:general_meson_op} is reproduced. 

Spatially extended operators like
\eqref{eq:general_meson_op_smeared} are often 
referred to as smeared operators. 
They are in this form not gauge 
invariant quantities, and therefore their average over gauge
configurations would 
vanish due to Elitzur's theorem \cite{Elitzur:1975im}. To
prevent that, one either has to include the parallel
transporters in the operator or to work in a fixed gauge
background. While gauge fixing is technically easy and imposes no
restrictions on the wave function $\varphi(\vec x,\vec y)$, it
introduces a possible 
source of errors due to the Gribov copy problem (see Appendix
\ref{app:gaugefix}). 
To avoid this problem, various kinds of gauge invariant operators like 
Jacobi-smeared \cite{Allton:1993wc} or Wuppertal sources
\cite{Gusken:1990qx,Gusken:1989ad,Gupta:1991sn},  have been
constructed. In the following however, we will
concentrate on gauge non-invariant operators and measure them on gauge-fixed
configurations, which is a common procedure adopted in many
large-scale simulations
\cite{Iwasaki:1994gr,Aoki:1997xe,Bernard:1997ib,Aoki:1998bz,McNeile:1998cp,Bernard:2001av,AliKhan:2001tx}.

The wave
function $\varphi(\vec x, \vec y)$ of the simulated particle is a
priori not known, so one has 
to make a more or less reasonable guess.
A convenient, but not very physical choice is
the totally factorized
shell-model wave function \cite{Bacilieri:1988fh,DeGrand:1991yb,DeGrand:1991dz}
\begin{equation} \label{eq:shell_model_wf}
\varphi(\vec x,\vec y) = \phi(\vec x) \phi(\vec y),
\end{equation}
where the quark and antiquark move independently inside a region given
by the function $\phi(\vec x)$, which might for example chosen to be a wall
\cite{Bitar:1990cb}, a hard sphere or cube \cite{Bacilieri:1988fh}, a
Gaussian \cite{DeGrand:1998pr,DeGrand:1998jq}, or a  
radial exponential \cite{Aoki:1998bz,AliKhan:2001tx}. General
experience from lattice simulations has shown 
that using these kinds of extended operators, it is significantly easier 
to extract a reliable estimate for hadron masses from a fit to the
correlation function \eqref{eq:2point_corrfunc}, because due to the larger
overlap with the ground state the contributions from higher states in
\eqref{eq:2point_corrfunc_series} vanish at much smaller time 
separation $\tau$ than for local hadron operators
\cite{Marinari:1988aa,Aoki:1998bz,AliKhan:2001tx,Bitar:1994rk,Daniel:1992ek,Iwasaki:1994gr}\footnote{To
account for the higher state contributions, refined
strategies are to 
perform a double exponential fit to both the lowest and the first
excited state or to make Bayesian fits \cite{Lepage:2001ym}.}.

The advantage of the   
shell-model wave function \eqref{eq:shell_model_wf} over more
physical functions which depend on the relative 
coordinates between the quark and antiquark is that due to the
factorization into separate quark and antiquark parts, the quark
propagator can be calculated like for a local operator with only one
inversion per spin and color.
In the following, we suppose that $\phi(\vec x)$ is a real,
radial symmetric function,
\begin{equation} \label{eq:radial_wave_function}
\phi_{\vec r}(\vec x) = \phi ( ||\vec r - \vec x||) 
\end{equation}
around a center $\vec r$ of the source or sink.
Consider first a meson correlation function with a shell-model source operator
\eqref{eq:general_meson_op_smeared}--\eqref{eq:radial_wave_function}
\begin{equation}
{\mathcal O}^{\rm M}_{\rm sm}(\vec r, t) = \sum_{\vec x, \vec y}
\phi_{\vec r}(\vec x) \bar\psi^{a, f_1}_\mu(\vec x, t)
\Gamma_{\mu\nu} \phi_{\vec r}(\vec y) \psi^{a, f_2}_\nu(\vec y, t), 
\end{equation}
at timeslice $t=0$ and centered at  $\vec r=0$, and a local sink
operator at time $t$: 
\begin{align} \label{eq:meson_prop_local-sm}
\big\langle {\mathcal
O}^{\rm M}  (\vec x,t)  {\mathcal O}^{\rm M \dag}_{\rm sm} &
(0,0) \big\rangle 
   = \Big\langle 0 \,\Big|\, \bar\psi^{f_1}(\vec x, t)\Gamma \psi^{f_2}(\vec
x, t) \nonumber \\
 &  \hspace{18mm}\times \Bigl(\sum_{\vec x_0, \vec y_0} \phi_0(\vec x_0
)\bar\psi^{f_2}(\vec x_0, 0)\Gamma 
\phi_0(\vec y_0 )\psi^{ f_1}(\vec y_0,0)\Bigr)^\dag \,\Big|\, 0 \Big\rangle 
 \nonumber \\
  & = \Big\langle \Tr \sum_{\vec x_0}
G^{f_2}(\vec x,t;\vec x_0,0)\phi_0 (\vec x_0) \Gamma \sum_{\vec y_0}
G^{f_1}(\vec y_0,0; \vec x, t) \phi_0(\vec y_0) \Gamma \Big\rangle
\nonumber \\
  & = \Big\langle \Tr 
G^{f_2}_{\rm sm}(\vec x,t;0,0) \Gamma \gamma_5
G^{f_1}_{\rm sm}(\vec x, t; 0,0)^\dag \gamma_5 \Gamma
\Big\rangle.
\end{align}
The spatial distribution of the source is taken into account when
inverting the Dirac operator on the vector $\phi_0(\vec x)\delta(t)$
instead of $\delta(x)$,
defining the smeared source quark propagator $G_{\rm sm}$ in
Eq.~\eqref{eq:meson_prop_local-sm} as the solution of  
\begin{equation} \label{eq:smeared_quark_prop}
D(x,y) G_{\rm sm}(y,0) = \phi_0(\vec x) \delta(t).
\end{equation}
While introducing an extended source amounts to inverting the  
Dirac operator on a different source vector, using an
extended sink leads to 
a weighting of the quark propagator at different lattice sites on a
given timeslice, as can be seen from the
smeared-source, smeared-sink meson correlator
\begin{multline} \label{eq:ss_meson_correlator}
\left\langle {\mathcal
O}^{\rm M}_{\rm sm}  (\vec r, t) {\mathcal O}^{\rm M \dag}_{\rm sm}
(0,0) \right\rangle
  = \Big\langle \Tr \sum_{\vec x_t}
\phi_{\vec r}(\vec x_t) G^{f_2}_{\rm sm}(\vec x_t,t;0,0) \Gamma
\gamma_5 \\
\times \sum_{\vec y_t} \phi_{\vec r}(\vec y_t)  G^{f_1}_{\rm sm}(\vec
y_t, t;0,0)^\dag \gamma_5 \Gamma 
\Big\rangle,
\end{multline}
where the smearing of the source again is absorbed in the quark propagator.

As shown above, the hadron mass is extracted from
the Fourier transform of the correlation function at zero momentum,
which implies summing Eq.~\eqref{eq:ss_meson_correlator} over all sink
locations $\vec r$. For a smeared sink, 
the numerical effort can get quite large, as there are then three sums
over all lattice points on a given time slice. A technical trick to
accelerate the calculation of smeared-sink meson correlators is
to rewrite Eq.~\eqref{eq:ss_meson_correlator} in Fourier 
space, which allows to make use of efficient Fast Fourier
Transform (FFT) algorithms to speed up the calculation. With the discrete
Fourier representations of both the sink wave function 
\begin{equation}
\phi_{\vec r}(\vec x) = \frac{1}{N_s^{3/2}}\sum_{\vec k} e^{-i\frac{2\pi}{N}\vec k (\vec r - \vec x)} \phi(\vec k) ,
\end{equation}
where $k_i=0,\dots,N_s-1$ for $i=1,\dots,3$, and the quark propagator
\begin{equation}
G(\vec x,t;0,0) =  \frac{1}{N_s^{3/2}} \sum_{\vec k} e^{-i\frac{2\pi}{N}\vec k \vec x}  G(\vec k,t;0,0),
\end{equation}
the convolutions in the meson correlator
\eqref{eq:ss_meson_correlator} can be expressed as
\begin{equation}
\sum_{\vec x} \phi_{\vec r}(\vec x_t)  G(\vec x_t, t;0,0) = \sum_{\vec
k} e^{-i\frac{2\pi}{N}\vec k \vec r} \phi(\vec k) G(\vec k,t;0,0),
\end{equation}
leading to the zero-momentum smeared-source, smeared-sink meson correlator
\begin{multline}
\sum_{\vec r} \left\langle {\mathcal O}^{\rm M}_{\rm sm} (\vec r, t)
{\mathcal O}^{\rm M \dag}_{\rm sm} (0,0) \right\rangle = N_s^3\, \Big\langle
\sum_{\vec k} \Tr \phi(\vec k) 
G^{f_2}_{\rm sm}(\vec k,t; 0,0) \Gamma\gamma_5 \\
\times \phi(\vec k) G^{f_1}_{\rm sm}(\vec k,t; 0,0)^\dag
\gamma_5\Gamma \Big\rangle .
\end{multline}
With this trick, the smeared-sink meson correlator at zero momentum is
calculated in the same manner as the point-sink case
\eqref{eq:pp_meson_correlator} after replacing the 3-space
fields by their Fourier transforms.

\begin{figure}[tb]
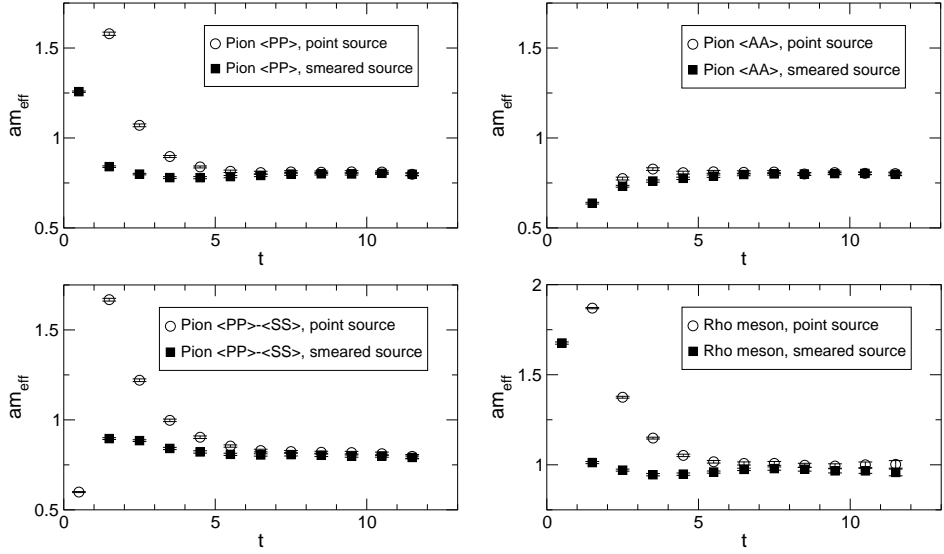

%\begin{center}
\hspace{-6mm}
\begin{tabular}{cc}
\includegraphics[width=60mm]{epsf/smeared_sources_P.eps} &
\includegraphics[width=60mm]{epsf/smeared_sources_A.eps} \\
\includegraphics[width=60mm]{epsf/smeared_sources_PS.eps} &
\includegraphics[width=60mm]{epsf/smeared_sources_V.eps}
\\
\end{tabular}
%\end{center}
\caption{Effective mass plots for meson
correlators with point and Gaussian smeared sources at bare quark mass
$am=0.23$. The lattice size
is $9^3\times 24$, the lattice spacing $a=0.16$ fm, and the data comes from
70 configurations evaluated with the FP Dirac operator. Although the
lattice spacing is rather large, a clear enlargement of the plateau
can be seen for all but the zero component axial vector $\langle AA\rangle$
correlator for the pion.}
\label{fig:smeared_source_1}
\end{figure}

\begin{figure}[tb]
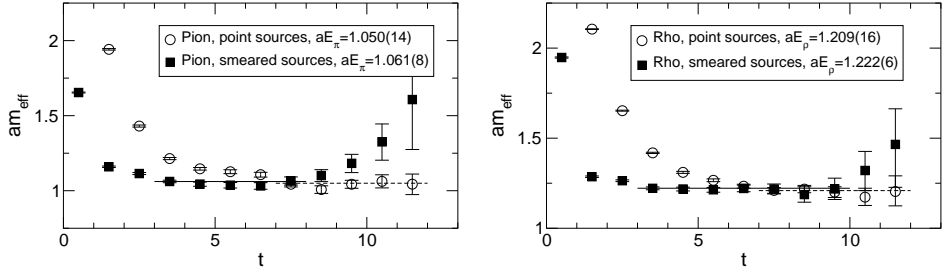

%\begin{center}
\hspace{-6mm}
\begin{tabular}{cc}
\includegraphics[width=60mm]{epsf/smeared_sources_P_p1.eps} &
\includegraphics[width=60mm]{epsf/smeared_sources_V_p1.eps} \\
\end{tabular}
%\end{center}
\caption{Effective energy plots at momentum $|ap|=2\pi/9$ for meson 
correlators with point and smeared sources. The simulation parameters
are the same as in Fig.~\ref{fig:smeared_source_1}. The straight
and dashed lines denote the time interval where the fit is done for
the smeared and point sources. The statistical error for the fit
results $aE_\pi$ and $aE_\rho$ is considerably 
smaller using smeared sources.} 
\label{fig:smeared_source_2}
\end{figure}

The effect of improving the overlap of the interpolating operators
with the desired hadron state can be seen 
in the effective mass plots\footnote{Effective masses are explained in the
next section.} in
Fig.~\ref{fig:smeared_source_1}. Compared are effective masses of
pseudoscalar and vector mesons at large quark mass $am=0.23$ and
lattice spacing $a=0.16$ fm for 
meson correlators made with point and Gaussian smeared
sources, respectively. In both cases point sinks were chosen, and 70 gauge 
configurations of lattice size $9^3\times 24$ were used. The Gaussian smearing
is defined by the shell-model wave function
\begin{equation} \label{eq:gaussian_source}
\phi_{\vec r}(\vec x) = e^ {-\gamma(\vec x-\vec r)^2},
\end{equation}
and the source center $r_i= N_i/2$ with $i=1,\dots,3$ is located at the center of the
time slice. The extension of the source used in the plots
 is given by $\gamma=0.15$. The plateau in the effective mass starts
much earlier in most cases for smeared sources, 
 providing a larger time interval over which the correlators can be
 fitted. The lattice spacing in the plots is rather large, and the
effect will be enhanced at smaller $a$, where  
 the point source correlators might not even reach a plateau within
 the given number of temporal lattice
points. Fig.~\ref{fig:smeared_source_2} shows 
plots for the effective energy of the pseudoscalar and vector meson at the
lowest non-zero momentum $|ap|=2\pi/9$. There the signal is worse, and
the length of the plateau for
smeared correlators is only slightly increased, as the signal starts
to deviate at large 
$t$. However, again the plateau region starts much earlier when using
smeared sources, thus raising the confidence in that 
really the asymptotic behavior of the correlator is reached. Also, the
statistical error of the energy resulting from the fit is significantly smaller
for the smeared source correlators.

\section{Fitting Hadron Propagators}
As we have shown, hadron masses are extracted from the
exponential fall-off of Euclidean time correlation functions at zero
momentum \eqref{eq:zero_mom_correlator}, which are expressed in terms
of the quark propagator and can thus be evaluated on the
lattice. Consider the case of flavor non-singlet mesons
Eq.~\eqref{eq:pp_meson_correlator}, of which the correlation
function $C(t)$ is
measured  on a
lattice of temporal size $T$ for all $0\leq  
t<T$.\footnote{Staying from now on in Euclidean
space, we will use the notation $t$ also for Euclidean time.} Using periodic boundary conditions, the data points can be
fitted against the
asymptotic form 
\begin{equation} \label{eq:asympt_meson}
\begin{split}
 C_\infty(t) & =  Z \big( e^{-mt} +
 e^{-m(T-t)} \big),  \\
  & =  2Z  e^{-mT/2}
 \cosh\big(m(T/2-t)\big), 
\end{split}
\end{equation}
of the meson correlator in Euclidean time~$t$ in order to determine
the mass $m$ and the coefficient 
\begin{equation}
 Z = \frac{A_f A_i^*}{2m},
\end{equation}
containing the information on the matrix elements of the chosen
operators between the vacuum and the hadron state.

In order to find out at what time $t_0$ the asymptotic range of the hadron
propagator is reached, it is helpful to plot the
effective mass $ m_{\rm eff}(t+1/2)$, 
which is determined from a zero parameter fit of the asymptotic function
$C_\infty(t)$ to the measured propagators $C(t)$
and $C(t+1)$ at two consecutive time slices. When $t$ gets large enough that
the higher lying states 
have disappeared, $m_{\rm eff}$ starts to show a plateau. In general, only
the measured correlators in a time interval $ t\in [t_0,t_1]$ are then used
to fit the parameters. The upper bound $t_1$ can be set to the point
where the signal disappears in the statistical noise, which happens
for all particles but pseudoscalar mesons after a certain temporal range.

\begin{figure}[tb]
\begin{center}
\includegraphics[width=95mm]{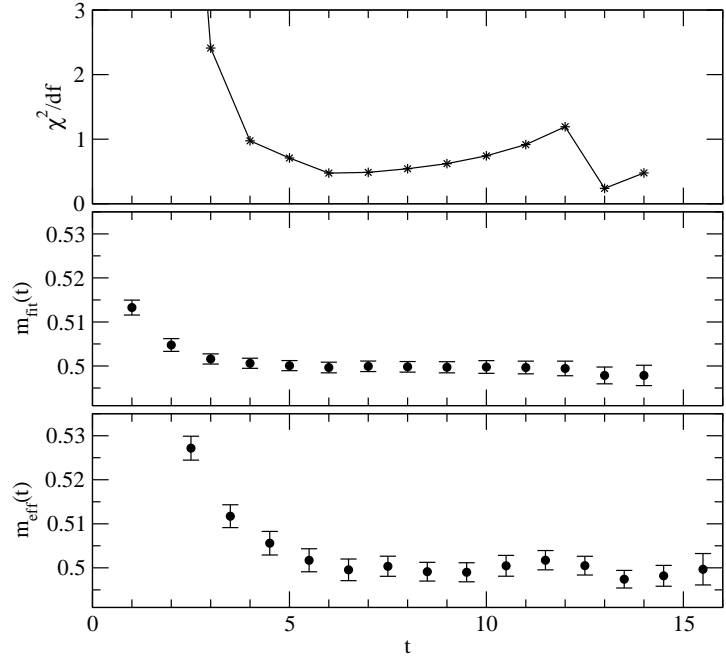}
\end{center}
\vspace{-4mm}
\caption{Effective mass $m_{\rm eff}(t)$, fitted mass $m_{\rm fit}(t_0)$ from a
correlated fit in the range $[t_0,16]$ and the value of $\chi^2/df$
for the fit (from bottom to top) 
for the pseudoscalar meson at $am=0.09$, $\beta=3.0$ and lattice size
$16^3\times 32$. Error bars are from a bootstrap resampling of the
propagators.} 
\label{fig:massdet.PS}
\end{figure}

\begin{figure}[tb]
\begin{center}
\includegraphics[width=95mm]{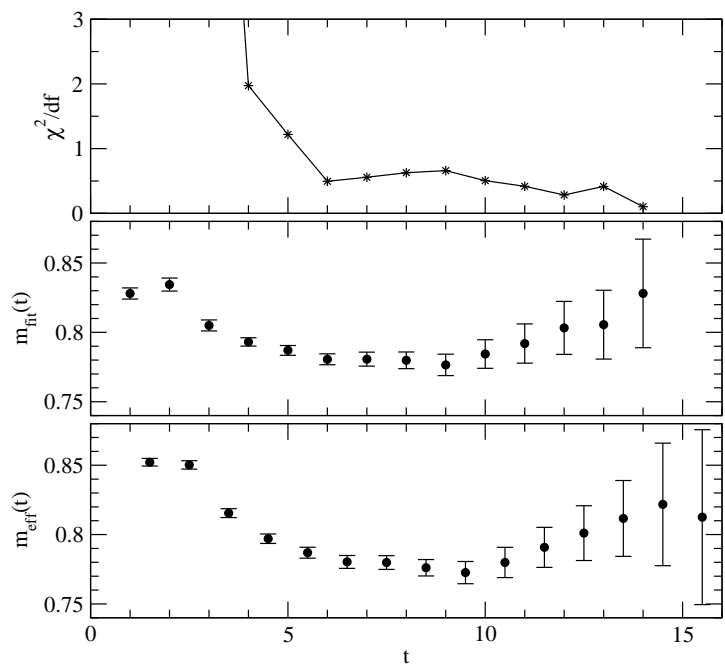}
\end{center}
\vspace{-4mm}
\caption{The same as Fig.~\ref{fig:massdet.PS} for the vector meson.} 
\label{fig:massdet.V}
\end{figure}

\begin{figure}[tb]
\begin{center}
\includegraphics[width=95mm]{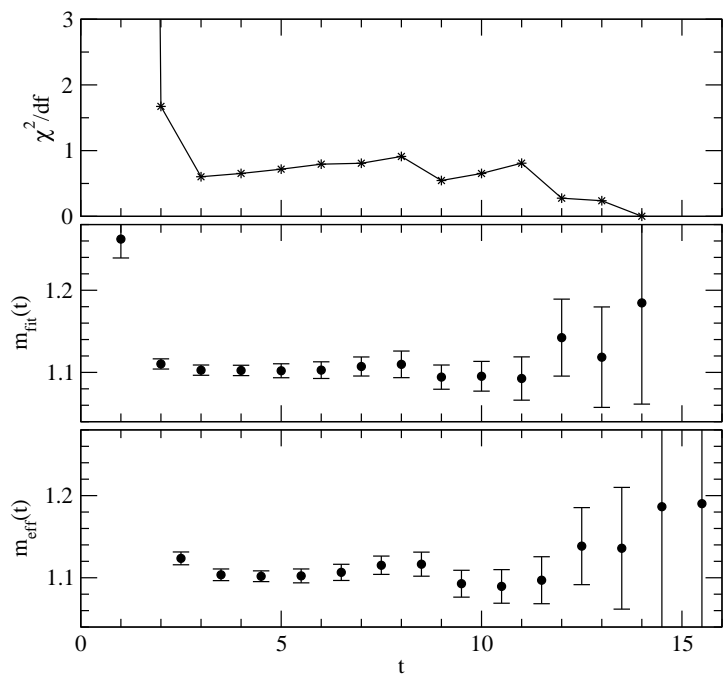}
\end{center}
\vspace{-2mm}
\caption{The same as Fig.~\ref{fig:massdet.PS} for the octet baryon.} 
\label{fig:massdet.N0}
\end{figure}

\begin{figure}[tb]
\begin{center}
\includegraphics[width=95mm]{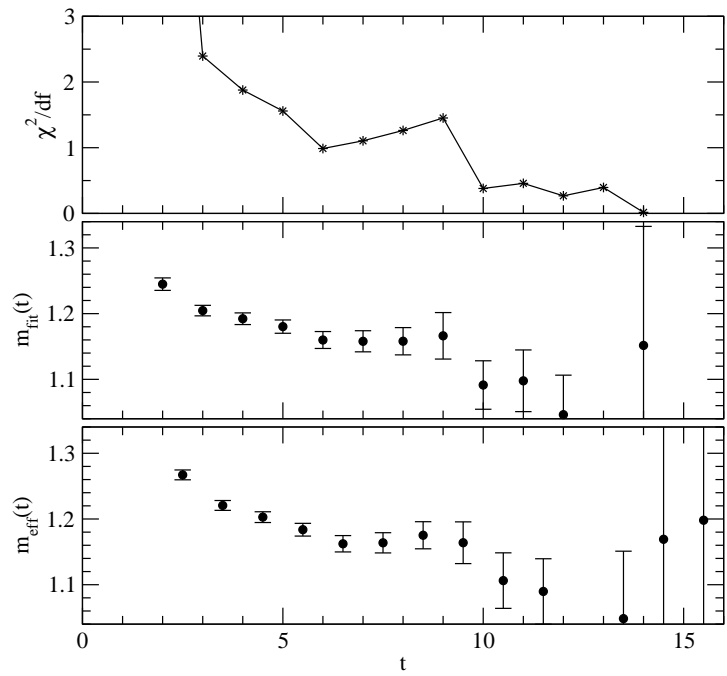}
\end{center}
\vspace{-2mm}
\caption{The same as Fig.~\ref{fig:massdet.PS} for the decuplet baryon.} 
\label{fig:massdet.D}
\end{figure}

In Figs.~\ref{fig:massdet.PS}--\ref{fig:massdet.D}, we show examples 
of effective masses, fitted masses and the quality of the fit given by
the value of $\chi^2/df$ for several particles at intermediate quark
mass. In all cases, $t_1$ was set to $T/2$. For all hadrons,
$\chi^2/df$ quickly decreases and stays then at a value of order
1, and the optimal fit interval starts in the range $3\leq t_0 \leq 6$.

\subsection{Correlated Fits}

\begin{figure}[tb]
\begin{center}
\includegraphics[width=8cm]{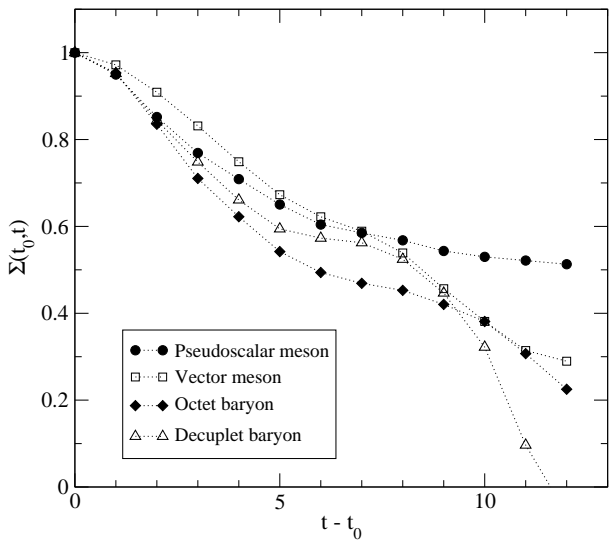}
\end{center}
\vspace{-3mm}
\caption{Temporal correlations of hadron correlators for a typical
time range used in mass fits. The 
correlators are from $N=100$ independent gauge configurations of size
$16^3\times 32$ at
$\beta=3.0$ with smeared sources and point sinks, the quark mass is
$am=0.18$ and the time interval starts at $t_0=4$.} 
\label{fig:covmat}
\end{figure}

Suppose we have measured the hadron propagators\footnote{To avoid confusion
between correlation functions and correlation matrices,
we will denote hadronic correlation functions as hadron
propagators where necessary.} $C^{(i)}(t)$ for $t=0,\dots,T$
on $N$ independent, importance sampled gauge configurations $U^{(i)}$ with
$i=1,\dots,N$.
Trying to fit the gauge average $\overline 
{C}(t)= 1/N \sum_i C^{(i)}(t)$,
we are faced with the problem 
that while the data is uncorrelated in Monte Carlo time~$i$, it is
strongly correlated 
in the temporal direction. The goal is to find the optimal
parameters $m$ and $Z$  in the asymptotic form \eqref{eq:asympt_meson},
taking into account the time 
correlations. This is done by  minimizing the $\chi^2$-function 
\begin{equation}
 \chi^2 = \sum_{t,t^\prime} \left\{\overline{C}(t)-C_\infty(t;m,Z)\right\}
 \, ({\rm Cov})^{-1}(t,t^\prime) \,
 \left\{\overline{C}(t^\prime)-C_\infty(t^\prime;m,Z)\right\} ,
\end{equation}
where the time correlations are encoded in the symmetric, positive definite
covariance matrix
\begin{equation}
 ({\rm Cov}) (t,t^\prime) = \frac{1}{N(N-1)} \sum_{i=1}^{N} \big( C^{(i)}(t)-\overline
 {C}(t) \big)
 \big( C^{(i)}(t^\prime) - \overline{C}(t^\prime) \big) .
\end{equation}
As an illustration of the typical size of time correlations we plot
a row of the normalized correlation matrix 
\begin{equation}
  \Sigma(t,t^\prime) = \frac{({\rm Cov}) (t,t^\prime)} { \sqrt{({\rm Cov})
 (t,t) ({\rm Cov}) (t^\prime,t^\prime)}},
\end{equation}
for different smeared-source hadron propagators in
Fig.~\ref{fig:covmat}. The data is strongly correlated for all hadrons under
consideration. For the pseudoscalar meson, the time correlations do
not die out at 
large $t$, as the signal-to-noise ratio remains constant, whereas for
the other hadrons what mostly remains is uncorrelated statistical
noise. The time correlations of pseudoscalar propagators are even stronger for
point sources \cite{Michael:1995sz}, therefore it is in any case
mandatory to perform correlated fits by including the covariance matrix. 

\subsection{Resampling Methods for Error Estimates}
\label{sect:bootstrap}
The results of the above described procedure are the 
parameters of the fit function, that is the hadron
mass $m$ and the amplitude $Z$. Since these fit parameters are statistical
estimates from a Monte Carlo integration, it is 
necessary to provide an estimate of their statistical errors in order
to judge their reliability. The
measured observables are the hadron correlators, and what is needed is a tool
to estimate the errors of quantities which depend in a complicated
way on these observables. For the sample mean of the correlators, the
standard deviation can be calculated in the usual way, 
\begin{equation} \label{eq:stddev}
 \sigma_{\overline{C}} = \left( \frac{1}{N(N-1)} \sum_{i=1}^{N}
 (C^{(i)} - \overline{C})^2 \right)^{1/2}, 
\end{equation}
but for less trivial
functions of the observables, there is no such function to estimate
the error.
With the advent of powerful computers, robust statistical methods have
been developed which allow to estimate errors of arbitrarily
complicated functions of 
observables with unknown probability distributions in a
straightforward way. These widely used methods are known under the
names jackknife and bootstrap, and they are both based on resampling
of the measured data. We briefly present the jackknife and the
bootstrap resampling techniques in the following. 

Suppose we have a set of $N$ independent and identically distributed
measurements 
\begin{equation}
X_1,\dots,X_N \stackrel{\rm iid}{\sim} F,
\end{equation}
 following an unknown distribution function
$F$. From these data points, an arbitrarily
complicated secondary quantity $\hat\theta(X_1,\dots,X_N)$ is
calculated. What we aim at is an expression for the standard deviation
$\sigma_{\hat\theta} (F)$ 
of the estimator $\hat\theta$. Jackknife resampling requires to calculate the
estimator 
\begin{equation}
\hat\theta_{(i)} = \hat\theta(X_1, \dots, X_{i-1}, X_{i+1},
\dots, X_N),
\end{equation}
on the sample where the data point $X_i$ has been 
dropped, for all $i=1,\dots,N$. The jackknife estimate for the error of
$\hat\theta$ is then defined by
\begin{equation}
 \sigma_{\hat\theta}^{\rm jack} = \left( \frac{N-1}{N} \sum_{i=1}^{N}
 (\hat\theta_{(i)} - 
 \hat\theta_{(\cdot)})^2 \right)^{1/2},
\end{equation}
where $\hat\theta_{(\cdot)} = \sum_{i=1}^{N} \hat\theta_{(i)}/N $ is
the average of the estimator over all jackknife samples.

The jackknife utilizes only $N$ of the $2^N-1$ non-empty subsets of the data
set. The error estimate might thus be improved when more of the
subsets are used. This lead to the development of the bootstrap
\cite{efron:1979}: A bootstrap sample 
\begin{equation}
X^*_1,\dots,X^*_N \stackrel{\rm iid}{\sim} \hat F,
\end{equation}
is a random sample drawn with 
replacement from the observed values $X_1,\dots,X_N$ and follows
the empirical probability distribution $\hat F$ of the data. The bootstrap
error estimate for $\hat\theta$ is defined from the estimator
$\hat\theta^* = \hat\theta(X^*_1,\dots,X^*_N)$ calculated 
on $B$ bootstrap samples by
\begin{equation} 
 \sigma_{\hat\theta}^{\rm boot} = \left( \frac{1}{B-1} \sum_{b=1}^{B}
 (\hat\theta^*_{(b)} - 
 \hat\theta^*_{(\cdot)})^2 \right)^{1/2},
\end{equation}
where  $\hat\theta^*_{(\cdot)} = \sum_{b=1}^{B} \hat\theta^*_{(b)}/B $ 
denotes the average over the bootstrap samples. In the limit
$B\rightarrow \infty$, the bootstrap error is exactly the standard
deviation of the estimator as a function of the empirical probability
distribution $\hat F$, 
\begin{equation}
 \sigma_{\hat\theta}^{\rm boot} \xrightarrow{B\rightarrow\infty}
 \sigma_{\hat\theta}(\hat F) 
 \approx \sigma_{\hat\theta}(F).
\end{equation}
In practice, the
number of bootstrap samples $B$ is finite, and one has to make sure
that it is large enough by varying $B$ and checking
whether the error changes significantly.
But since $\hat F$ is only an estimate for the unknown probability
distribution 
$F$, taking too many bootstrap samples does not help improving the
error estimate for $\sigma_{\hat\theta}(F)$.  For most cases, it is considered
safe to work with values in the range $ 100<B<500$. In our
spectroscopy simulations, the computational cost for the bootstrap is
negligible, and therefore we always calculate $B=1000$ bootstrap samples.

The jackknife and bootstrap procedures also provide an estimate for
the bias $\hat b_{\hat\theta}$ of the estimator $\hat\theta$. The
jackknife estimate of bias is given by 
\begin{equation}
\hat b^{\rm jack}_{\hat\theta} = (N-1) (\hat\theta_{(\cdot)} - \hat\theta),
\end{equation}
while for the bootstrap it is just the difference of the value of
the estimator on the original sample and its mean value on the
bootstrap samples,
\begin{equation} \label{eq:bias_boot}
\hat b^{\rm boot}_{\hat\theta} = \hat\theta^*_{(\cdot)} - \hat\theta.
\end{equation}
The bias-corrected estimator $\tilde\theta = \hat\theta - \hat b$ might
then be used instead of $\hat\theta$ for a better estimate of the true
value $\theta$.

%\section{Meson and Baryon Operators}

\fancyhead[RE]{\nouppercase{\small\it Topological Finite-Volume
Artifacts in Pion Propagators}}
\chapter{Topological Finite-Volume Artifacts in Pion Propagators}
\label{ch:zeromodes}

\begin{figure}[tb]
\begin{center}
\includegraphics[width=10cm]{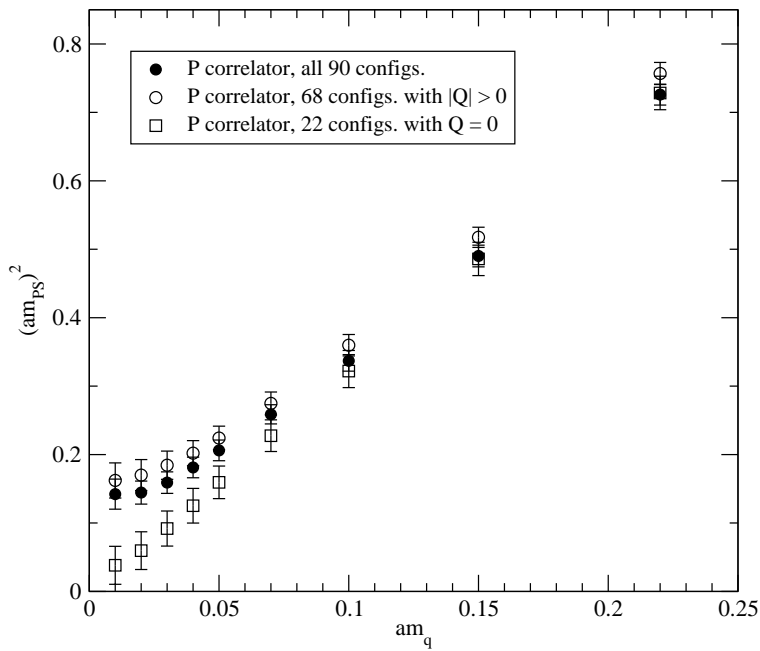}
\end{center}
\vspace{-3mm}
\caption{Chiral limit of the squared pion mass from pseudoscalar (P)
correlators in dependence of gauge field topology. The
overlap-improved FP Dirac operator is used on a set of 90 gauge
configurations of size $6^3\times 16$ at $\beta=3.0$. At small quark
mass, the pion 
evaluated on the full set is dominated by the configurations with
non-trivial topology $|Q|>0$. If 
only the $Q=0$ configurations are considered, no quenched finite-volume
artifacts are seen, and the pion mass goes to zero in the chiral limit.} 
\label{fig:zm_qtop}
\end{figure}

In the chiral limit of quenched QCD, pion\footnote{We denote in this
chapter the
pseudoscalar meson as a pion also for unphysical quark masses.}
propagators suffer from  
unphysical quenching effects which make a thorough examination of
light pions difficult. These quenching artifacts are caused by zero
eigenmodes of the Dirac operator and lead to unphysical divergences
of the pion propagators at $m\rightarrow 0$. In unquenched QCD, gauge
configurations with zero modes are suppressed by the fermion
determinant in the effective action, therefore these divergences are absent.
Because the determinant is set to unity in the quenched theory, the
suppression falls away, and at small quark mass the propagators are
dominated by the zero mode effects. This can be clearly seen in
Fig.~\ref{fig:zm_qtop}, 
where the chiral limit of $m_{\rm PS}^2(m_q)$ is shown
in dependence of the topology of 
the gauge configurations. While the pseudoscalar meson mass goes to
zero in the chiral 
limit on configurations with trivial topology, it deviates as
soon as also configurations with non-zero topological charge are considered
in the Monte Carlo average.

For the study of light pseudoscalar mesons with masses around the physical
mass of the pion, it is therefore  
unavoidable to get rid of these artificial effects. In particular for the
investigation of quenched chiral logarithms in the pseudoscalar mass,
which we undertake  
in Chapter \ref{ch:results_fp}, the effects from zero modes have to be 
properly disentangled from the chiral logarithm, which
produces a measurable signal only at small quark masses.
Due to the explicit breaking of chiral
symmetry, traditional formulations of lattice fermions like Wilson
or Symanzik-improved clover fermions do not allow to identify
topological zero modes unambiguously. 
Only with the development of chiral symmetric lattice Dirac operators, it has
become possible to identify the
zero modes responsible for these unwanted effects.

\begin{figure}[tb]
\begin{center}
\includegraphics[width=80mm]{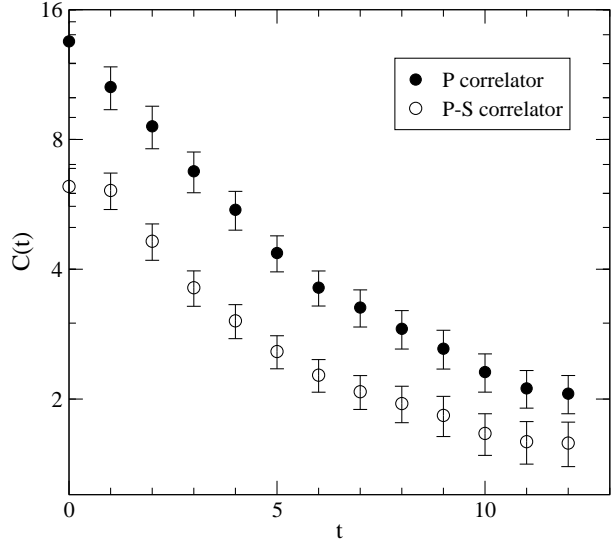}
\end{center}
\vspace{-3mm}
\caption{The kink in the pseudoscalar correlator as reported in
\cite{Dong:2001fm} from our data on the $12^3\times 24$ lattice at
$a \approx 0.16$ fm and quark mass $am_q=0.01$. A closer 
examination shows that even if the fit interval $t \in [6,12] $ is
chosen, which includes only the flatter part, the fitted mass $am_{\rm
PS}=0.186(11)$ is still
considerably larger than for the P-S correlator, where $am_{\rm
PS}=0.136(22)$.} 
\label{fig:zm_m1corr12}
\end{figure}

In a study with the Wilson overlap operator \cite{Dong:2001fm}, the authors
have reported a change in the behavior of the pseudoscalar correlator at large
time, suggesting that the zero modes only contaminate the small $t$
range. As a possible solution, they proposed to fit the mass from the large $t$
tail of the correlator. Fig.~\ref{fig:zm_m1corr12} shows the
reported kink in the correlator
as seen in our data, but even a fit to the flatter region does
not give a pion mass which goes to zero in the chiral
limit. Furthermore, we could not clearly identify such a kink in all
our simulations.

In this chapter we derive and examine two other
solutions of the problem. One solution is based on explicit identification and
subtraction of the zero modes in the quark propagator. The other
solution, originally proposed in \cite{Blum:2000kn}, makes use of the fact that
zero mode effects enter pseudoscalar and scalar meson propagators
equivalently. The zero mode effects can then be subtracted in the
meson propagators. We study these two solutions on a very small
lattice of 
spatial physical extension $L_s\approx 1$ fm, where the zero mode effects are large.

\section{Zero Mode Subtraction of the Quark Propagator}
\label{sect:zm_sub}
The most straightforward way to get rid of  zero mode
contributions is to subtract them directly from the quark
propagator. An exact subtraction however is only possible for an exactly
chiral Dirac operator. We derive in the following the subtraction of
the zero modes for the propagator of the overlap-improved FP Dirac
operator described in 
Chapter \ref{ch:d_ov}, which is not trivial due to the Fixed-Point $R$
appearing in the Ginsparg-Wilson relation.
From the subtracted quark propagators,
meson and baryon correlators can be constructed in the standard way, and
measurements of hadronic quantities derived from these correlators should then
be free of quenched topological finite-volume artifacts. 

\subsection{Spectral Decomposition of the Massless Normal Dirac Operator}
The overlap-improved FP Dirac operator is a solution of the
Ginsparg-Wilson relation 
$\left\{ D, \gamma_5  \right\} = D\gamma_5 2R D $,
where a non-trivial $R$ appears on the right-hand side. Defining
the operator
\begin{equation} \label{eq:def_d_tilde}
\tilde D \equiv \sqrt{2R} D \sqrt{2R},
\end{equation}
the Ginsparg-Wilson relation reduces to the simpler case 
$\{ \tilde D, \gamma_5 \} = \tilde D\gamma_5 \tilde D$.
With the $\gamma_5$-hermiticity of Dirac operators $D^\dag =
\gamma_5 D \gamma_5$, it follows that $\tilde D$ is a normal operator,
\begin{equation}
\tilde D^\dag \tilde D = \tilde D \tilde D^\dag.
\end{equation}
Normal operators can be written as diagonal matrices in an orthonormal
basis of eigenvectors. Thus we can write down the 
spectral decomposition of a matrix element of the massless normal
Dirac operator 
\begin{equation} 
\tilde D_{ij} = \sum_\lambda \lambda \tilde\phi_\lambda(i) \tilde\phi_\lambda^\dag(j),
\end{equation}
where $\lambda$ and $\tilde\phi_\lambda$ are the eigenvalues and the
corresponding 
eigenvectors of $\tilde D$, and the indices $i$ and $j$ contain spin,
color and space-time degrees of freedom.
Since the inversion of a diagonal matrix is trivial, the spectral
decomposition of the quark propagator $\tilde G_{ij}\equiv
(\tilde D^{-1})_{ij}$ is given by
\begin{equation}  \label{eq:spectr_decomp_dtilde}
\tilde G_{kl} = \sum_\lambda \frac{1}{\lambda} \tilde\phi_\lambda(k)\tilde\phi^\dag_\lambda(l).
\end{equation}
The orthonormality condition on the
eigenvectors reads
\begin{equation} 
\sum_l \tilde\phi^\dag_\lambda(l)\tilde\phi_{\lambda^\prime}(l) =
\delta_{\lambda \lambda^\prime}.
\end{equation}
We define the subtracted quark propagator by summing only over the
non-zero eigenmodes of $\tilde D$,
\begin{equation}  \label{eq:qprop_sub}
\tilde G_{kl}^{(\rm sub)} \equiv \sum_{\lambda\neq 0} \frac{1}{\lambda}
\tilde\phi_\lambda(k)\tilde\phi^\dag_\lambda(l).
\end{equation}
In presence of a finite mass, we will invert a matrix of the form $a\tilde
D + b$ with  $a,b\in {\bf R}$, and Eq.~\eqref{eq:qprop_sub} can be written as
\begin{equation}  \label{eq:qprop_sub2}
\tilde G_{kl}^{(\rm sub)} (a,b) = \tilde G_{kl}(a,b) -
\sum_{\lambda=0} \frac{1}{b} \tilde\phi_\lambda(k)\tilde\phi^\dag_\lambda(l).
\end{equation}

\subsection{Basis Transformation}
For practical applications, it is not convenient to work with $\tilde
D$ as defined in Eq.~\eqref{eq:def_d_tilde} due to the appearance of
the square root of $2R$, whose calculation is a non-trivial
numerical problem. A simple basis transformation 
\begin{equation} \label{eq:basis_transform_2r}
\hat \phi \equiv S \tilde \phi,
\end{equation}
 with $S=(2R)^{-1/2}$ helps
 to get rid of this square root. The eigenvalues $\lambda$ remain
unchanged under this transformation.
The application of the basis transformed Dirac operator
\begin{equation} \label{eq:def_hat_d}
\hat D \equiv S \tilde D S^{-1} = D 2R,
\end{equation}
on a vector is then reduced to multiplications with both D and R. From
the definition \eqref{eq:qprop_sub},\eqref{eq:qprop_sub2} we can read
off the subtracted propagator in the new basis,
\begin{equation} \label{eq:qprop_dhat_sub}
\hat G_{ij}^{(\rm sub)}(a,b) \equiv S \tilde G_{ij}^{(\rm sub)}(a,b) S^{-1}  = \hat G_{ij}(a,b) - \sum_{\lambda=0}
\frac{1}{b} \hat\phi_\lambda(i)\hat\phi^\dag_\lambda(k) 2R_{kj},
\end{equation}
where $\hat\phi$ are eigenvectors of $\hat D$ and $\hat G_{kl} \equiv (\hat
D^{-1})_{kl}$. We have to remark that due to the basis transformation,
the Dirac operator $\hat D$ is no longer 
normal, and its eigenvectors are not orthogonal, but
instead 
fulfill the generalized orthonormality condition
\begin{equation} \label{eq:ev_norm_2r} 
\sum_{k,l} \hat\phi^\dag_\lambda(k)2R_{kl}\hat\phi_{\lambda^\prime}(l) =
\delta_{\lambda \lambda^\prime}.
\end{equation}

\subsection{A Cookbook Recipe}
The above discussion is valid for general massless Ginsparg-Wilson
Dirac operators. We consider in the following the overlap-improved FP
Dirac operator \eqref{eq:d_ov_m}.
After basis transformation \eqref{eq:def_hat_d}, the mass dependence
of $\hat D(m)$ is the usual one,
\begin{equation}
\hat D(m) \equiv \left(1-\frac{m}{2}\right) \hat D + m,
\end{equation}
where the $2R$ in the denominator has disappeared.

With these ingredients, we are ready to give a cookbook recipe for
the calculation of zero-mode subtracted quark propagators with the
massive overlap-improved FP Dirac operator. These are the
solutions of the equation
\begin{equation}
   D_{kl}^{(\rm sub)}(m)  g^{(\rm sub)}(k;m) = b(l),
\end{equation}
where the subtracted Dirac operator is defined through
Eqs.~\eqref{eq:def_hat_d} and 
\eqref{eq:qprop_dhat_sub}. The steps to find the solution $g^{(\rm
sub)}(k;m)$ are the following: 

\begin{enumerate}
\item Calculate the few smallest eigenmodes of $\hat D$,
\begin{equation} \label{eq:cookbook1}
\hat D \hat\phi = D2R\hat\phi = \lambda \hat\phi,
\end{equation}
identify the
zero modes, where $\lambda=0$, and normalize them according to \eqref{eq:ev_norm_2r}.
\item Invert $\hat D(m)$ on a source $b(y)$ by solving
the equation 
\begin{equation}
  \hat D_{xy}(m)\hat g(x;m) = b(y),
\end{equation}
for the vector $\hat g(x;m)$.
\item Subtract the zero modes from the solution $\hat g(x;m)$ as
derived in
Eq.~\eqref{eq:qprop_dhat_sub}:
\begin{equation}
\hat g^{(\rm sub)}(x;m) = \hat g(x;m) - \sum_{\lambda=0}
\frac{1}{ m}
\hat\phi_\lambda(x)\hat\phi^\dag_\lambda(z) 2R_{zy} b(y).
\end{equation} \label{step3}
\item To get the inverse of $D$, multiply the result by $2R$:
\begin{equation} \label{eq:cookbook4}
g^{(\rm sub)}(x;m) = 2R \hat g^{(\rm sub)}(x;m).
\end{equation} \label{step4}

\end{enumerate}

We have to remark that in principle it would be more elegant to
perform the basis transformation
\eqref{eq:basis_transform_2r} in the inverse direction with $S=2R$, because
then in the above steps \ref{step3} and \ref{step4} the factor of $2R$
disappears. The reason we 
do not follow this apparently simpler path is that in the
orthonormality condition (and 
in hermitean forms in general, which are used in certain parts of our
code), a multiplication with the inverse of 
$2R$ would show up, which is of course numerically much more demanding.

\section{Zero Mode Contributions in Meson Propagators}
\label{sect:zm_mescorr}
Consider first the case of the normal Dirac operator $\tilde D$,
where a spectral 
decomposition into a complete set of eigenstates is possible.
Inserting the spectrally decomposed quark propagator
\eqref{eq:spectr_decomp_dtilde} 
into the meson propagator \eqref{eq:meson_corr_with_qprop},  we get
\begin{equation} \label{eq:spectral_meson1}
\langle {\mathcal
O}^{\rm M}  (x)  {\mathcal O}^{\rm M \dag} 
(0)\rangle = \sum_{\lambda,\lambda^\prime}
\frac{\Tr \big[ \tilde\phi_\lambda^\dag(x) \Gamma\gamma_5
\tilde\phi_{\lambda^\prime}(x) \big]
\Tr \big[ \tilde\phi_{\lambda^\prime}^\dag(0) \Gamma^*\gamma_5
\tilde\phi_\lambda(0) \big] } {(\lambda+m)(\lambda^\prime+m)} ,   
\end{equation}
where the color and spin indices have been suppressed and only the
spatial index is given explicitly.
Splitting the sum into three contributions and summing over $\vec x$ to
project out zero momentum states leads to
\begin{multline} \label{eq:spectral_meson2}
\sum_{\vec x} \langle {\mathcal
O}^{\rm M}  (x)  {\mathcal O}^{\rm M \dag} 
(0)\rangle = \sum_{\vec x} \left[ \sum_{\lambda=\lambda^\prime=0}
\frac{\Tr \big[ \tilde\phi_\lambda^\dag(x) \Gamma\gamma_5 
\tilde\phi_{\lambda^\prime}(x) \big] \Tr \big[ \tilde\phi_{\lambda^\prime}^\dag(0)
\Gamma^*\gamma_5 
\tilde\phi_\lambda(0) \big] } {m^2} \right.  \\
 +  \left. \sum_{\lambda=0,\,\lambda^\prime\neq 0}
\frac{ \Tr \big[ \tilde\phi_\lambda^\dag(x) \Gamma\gamma_5
\tilde\phi_{\lambda^\prime}(x)\big] \Tr \big[  \tilde\phi_{\lambda^\prime}^\dag(0)
\Gamma^*\gamma_5 
\tilde\phi_\lambda(0) \big] } {m(\lambda^\prime+m)}   \right]
+ \sum_n \frac{|\langle 0|
{\mathcal O}^{\rm M} |n\rangle |^2}{2 E_n}
\ e^{-E_n\tau}.
\end{multline}
The first two summands, which are the pure and mixed zero mode
contributions, are divergent in the chiral limit $m\rightarrow 0$.
The zero mode contributions are finite volume
artifacts expected to diverge with $1/\sqrt{V}$ \cite{Blum:2000kn}, and  the
meson correlator diverges at finite volume  with ${\mathcal
O}(1/m^2)$ and ${\mathcal O}(1/m)$ 
terms in the chiral limit. If the zero modes in the quark propagator
have been subtracted 
as in Eq.~\eqref{eq:qprop_dhat_sub},
these divergences are absent, and also at small quark masses a
good signal for the exponential decay of the meson mass can be
obtained. 

Let us discuss what happens for different meson operators. For the
pseudoscalar, we have $\Gamma=\Gamma^*=\gamma_5$ with our
definition of the Clifford matrices, while for the scalar
meson $\Gamma=\Gamma^*=1$. The zero modes are chiral eigenstates with
$\gamma_5\tilde\phi_{\lambda=0}=\pm\tilde\phi_{\lambda=0}$, hence it follows
that the pseudoscalar and the scalar correlator have the same
contributions from zero modes. By taking the
difference between the two correlators, the
zero mode contributions are therefore exactly cancelled. For the axial
vector current correlator $\Gamma=\gamma_4\gamma_5$, the first summand
in Eq.~\eqref{eq:spectral_meson2} is zero, since $\gamma_4$ couples
different chiral 
sectors and all zero modes on a given gauge configuration have equal
chirality. So for this correlator, the zero modes contribute only in the
mixed term, and therefore it diverges only with $1/m$ in the chiral limit.

If we consider the non-normal Dirac operator $D$, we have 
$G = \sqrt{2R} \tilde G \sqrt{2R}$, and in the numerators of
Eqs.~\eqref{eq:spectral_meson1} and \eqref{eq:spectral_meson2}
additional factors of $\sqrt{2R}$ show up. After subtracting the
zero modes of $\tilde D$ according to 
\eqref{eq:cookbook1}--\eqref{eq:cookbook4}, again the divergence in
the chiral limit is no longer present.

\section{Numerical Results at Small Volume}

\begin{figure}[tb]
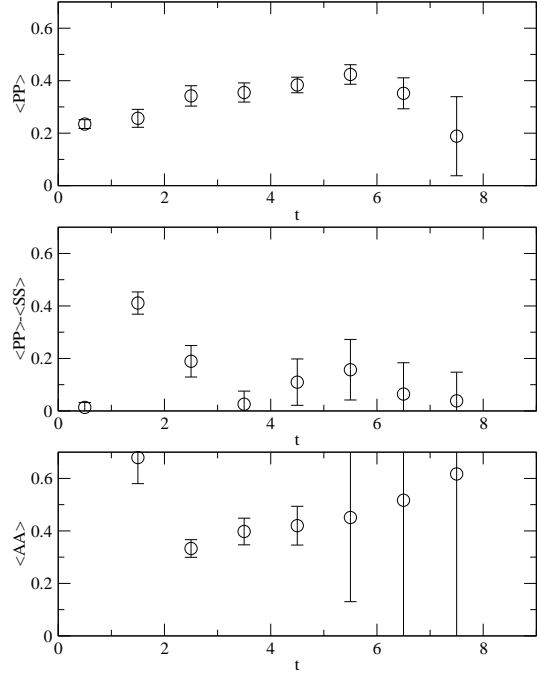

\begin{center}
\includegraphics[width=7cm]{epsf/effmass.6x16_b3.00_o3.P.m1.eps}
\includegraphics[width=7cm]{epsf/effmass.6x16_b3.00_o3.PS.m1.eps}
\includegraphics[width=7cm]{epsf/effmass.6x16_b3.00_o3.A.m1.eps}
\end{center}
\vspace{-3mm}
\caption{ Effective pseudoscalar meson masses at $am=0.01$ from P, P-S and A
correlators at $\beta=3.0$ with the overlap-improved FP Dirac
operator on 90 gauge configurations of size $6^3\times 16$.} 
\label{fig:zm_effmass6}
\end{figure}

\begin{figure}[tb]
\begin{center}
\includegraphics[width=9cm]{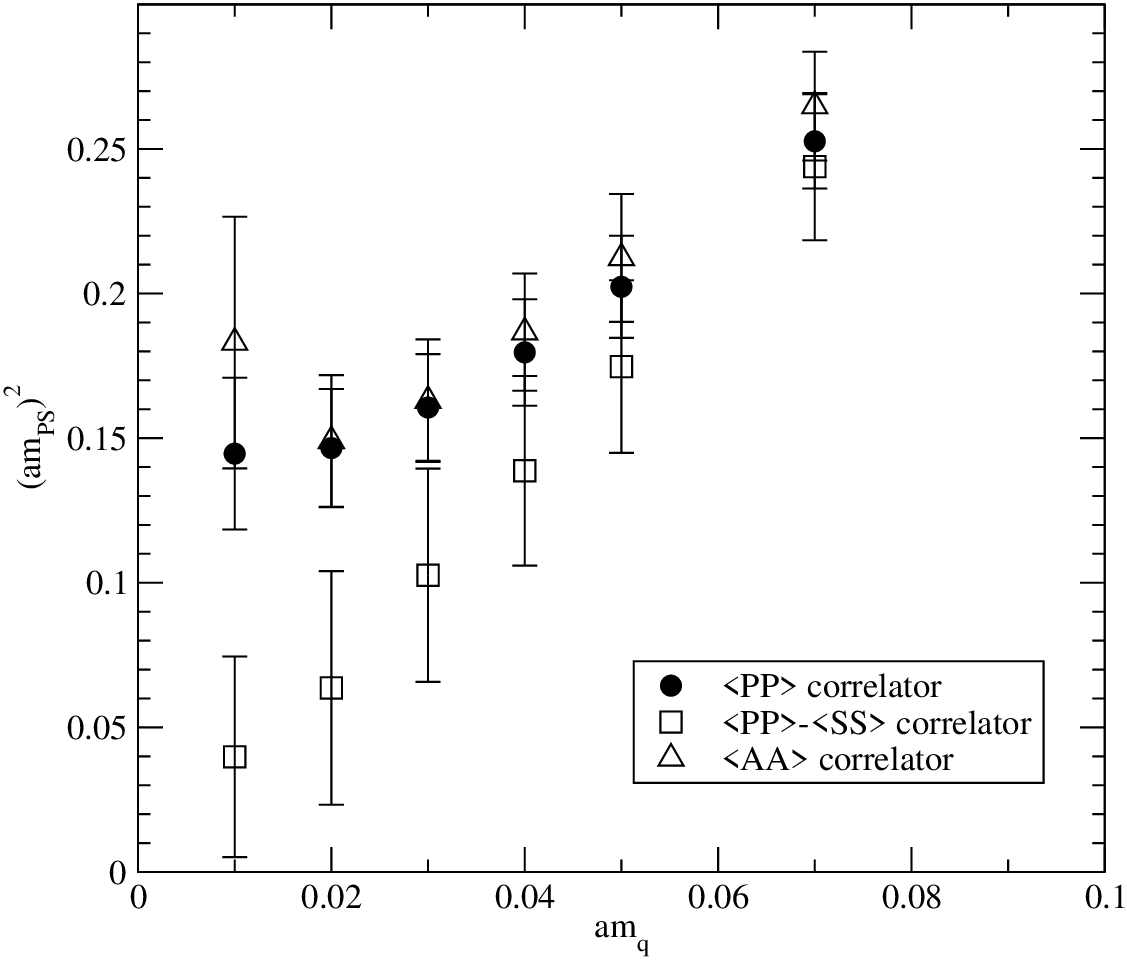}
\end{center}
\vspace{-3mm}
\caption{ Topological quenching artifacts in
squared pseudoscalar meson mass versus bare quark mass at $\beta=3.0$ and
volume $6^3\times 16$. The effective masses at the smallest quark mass
are shown in Fig.~\ref{fig:zm_effmass6}.}
\label{fig:zm_sqrpion6}
\end{figure}

\begin{figure}[tb]
\begin{center}
\includegraphics[width=7cm]{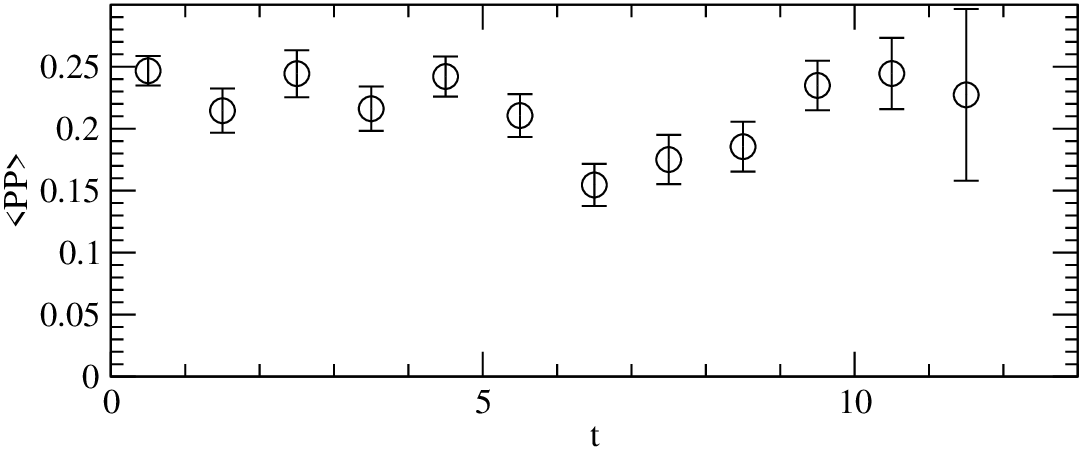}
\includegraphics[width=7cm]{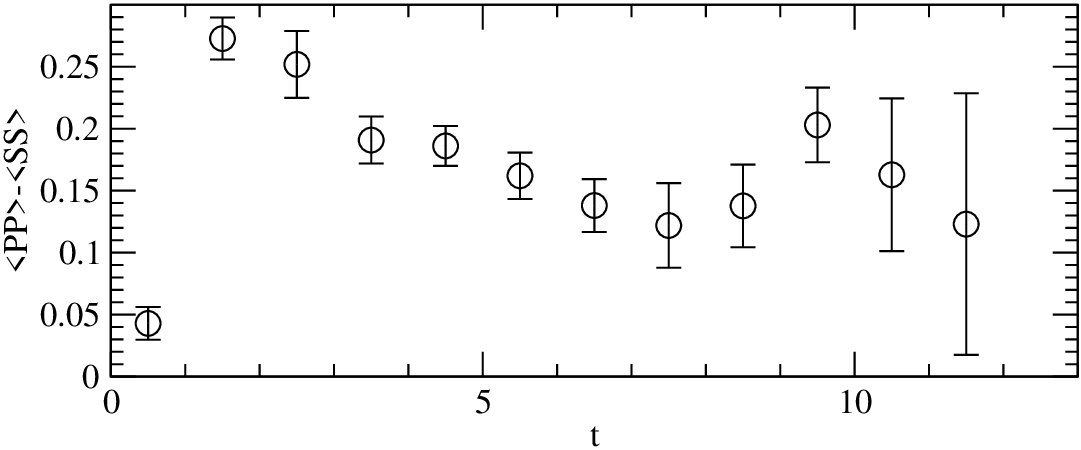}
\includegraphics[width=7cm]{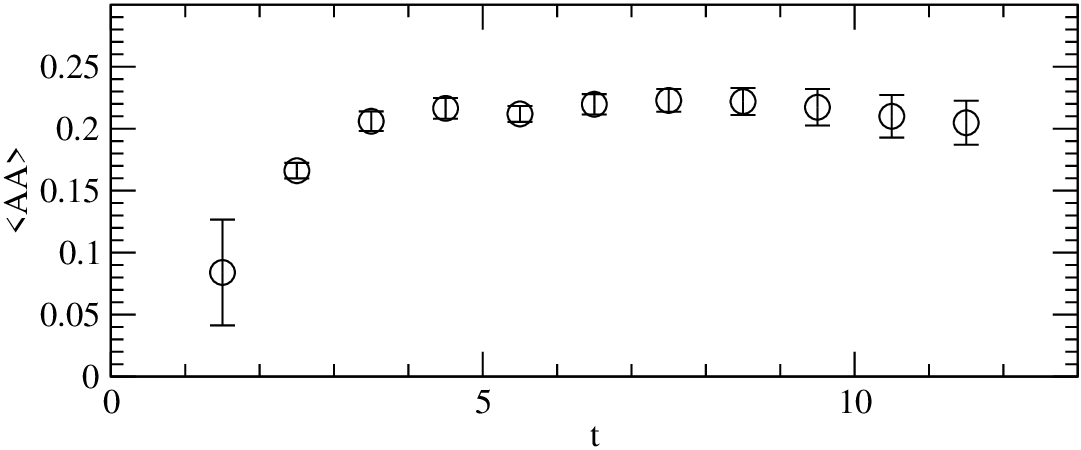}
\end{center}
\vspace{-3mm}
\caption{ Effective pseudoscalar meson masses at $am=0.01$ from P, P-S and A
correlators at $\beta=3.0$ with the overlap-improved FP Dirac
operator on 100 gauge configurations of size $12^3\times 24$. } 
\label{fig:zm_effmass12}
\end{figure}

In order to examine the finite-volume zero mode effects in pion
propagators, 
we perform an exploratory study with the overlap-improved FP Dirac
operator on a very small
lattice of size $6^3\times 16$ at gauge coupling $\beta=3.0$, with a
set of 90 independent gauge configurations. This
lattice volume
amounts to a spatial extension of 1 fm, therefore the signal in the
pseudoscalar channel is
strongly affected by the zero modes.  The Dirac operator is
constructed with a third order Legendre expansion for the 
overlap and with exact treatment of the smallest 10 eigenmodes.
Two different 
strategies are examined to remove the zero mode effects from the pion
propagator:

The first strategy is to remove the zero modes from the pseudoscalar (P)
correlator by building the difference of the pseudoscalar and the scalar
 (P-S) correlator as described in Section \ref{sect:zm_mescorr}. We further 
calculate the zero component axial vector (A) correlator, which also has the
quantum numbers of the pion, and for which the zero mode contributions are
partially cancelled. We show in Fig.~\ref{fig:zm_effmass6} the
effective masses 
at the smallest quark mass $am=0.01$  and in 
Fig.~\ref{fig:zm_sqrpion6} the squared pion mass as a function of the
quark mass. Obviously the different correlators give very different
results at small quark mass. While the chiral limit of the
pseudoscalar correlator
clearly deviates from zero, it is consistent with zero for the P-S
correlator. For the axial correlator, the 
zero mode effects are of the same order as for the pseudoscalar. 
For comparison, we show in Figs.~\ref{fig:zm_effmass12} and
\ref{fig:zm_sqrpion12} the same plots for a larger volume of
$12^3\times 24$, with unchanged gauge coupling $\beta=3.0$, the
same Dirac operator and 100 independent gauge
configurations. There the zero mode effects  
are much smaller due to the larger physical
volume. However, the pion mass from the pseudoscalar correlator still
clearly deviates 
from zero in the chiral limit, while for the P-S correlator it nicely
goes to zero. From 
these results we find that removing the zero mode effects in the
pion at small quark mass by using the P-S correlator works fine.

The second strategy is to explicitly calculate the zero modes
and remove them from the quark propagator as shown in Section
\ref{sect:zm_sub}. This is only possible if the Dirac operator
allows to identify zero modes unambiguously. 
To check whether this requirement is fulfilled for the
overlap-improved FP Dirac operator, we show
in Table \ref{tab:chir} for a
gauge configuration with $Q = 1$ the position of the zero
eigenvalue on the real axis and the first
non-zero eigenvalue as a function of the overlap order. We also list
the chirality of the corresponding 
eigenvector $\phi^{(i)}$, defined by
\begin{equation}
 \chi^{(i)} = \phi^{(i)} \gamma_5 2R \phi^{(i)},
\end{equation}
which is $\pm1$ for zero modes. Obviously, it is
easily possible to separate the zero mode and the first non-zero
mode, even if the eigenvalue $\lambda_0$ is
not very close to zero, because the chirality of the zero mode  approaches
$|\chi_0|\rightarrow 1$ very quickly with increasing order of the
overlap expansion. 

\begin{table}
\begin{center}
\begin{tabular}{c|r|r|r|r} \hline\hline
order & Re $\lambda_0$ &
$1-|\chi_0|$ & Re $\lambda_1$ & Im $\lambda_1$  \\ \hline 
0 & -5.4e-03 &  2.4e-02 & -3.782e-03 & 5.5974e-02 \\ 
1 & -3.1e-03 &  7.0e-04 & -2.487e-03 & 5.5115e-02 \\ 
2 & 5.1e-04  &  1.9e-05 & 2.148e-03 & 5.4854e-02 \\ 
3 & -2.1e-04 &  7.2e-07 & 1.300e-03 & 5.4901e-02 \\ 
4 & -2.5e-06 &  4.5e-08 & 1.508e-03 & 5.4890e-02 \\ 
5 & 1.3e-06  &  3.7e-09 & 1.511e-03 & 5.4889e-02 \\ 
6 & -4.6e-07 &  8.1e-11 & 1.507e-03 & 5.4890e-02 \\
7 & 3.6e-08  &  6.9e-12 & 1.508e-03 & 5.4890e-02 \\  \hline
\end{tabular}
\end{center}
\caption{Flow of the zero mode of a $Q=1$ gauge configuration, its
chirality and the first near-zero mode with increasing overlap order. $(
{\rm Im}\ \lambda_0)$ and $\chi_1$ are zero to machine precision.}  
\label{tab:chir}
\end{table}

\begin{figure}[tb]
\begin{center}
\includegraphics[width=90mm]{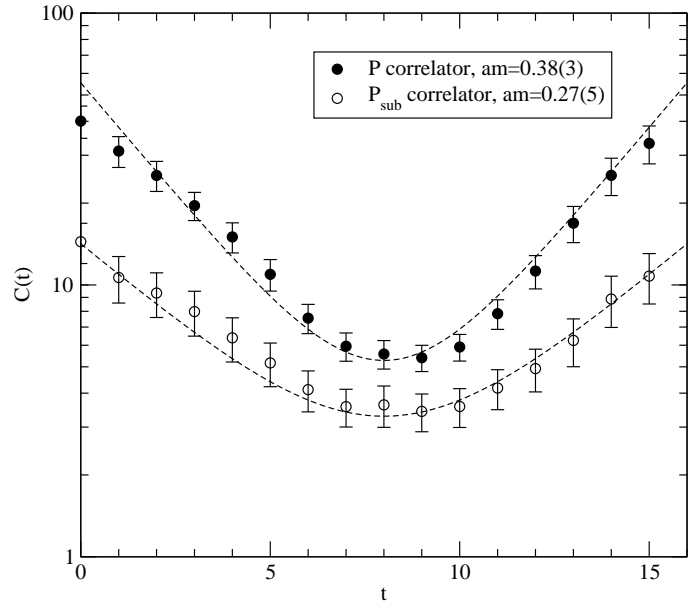}
\end{center}
\vspace{-3mm}
\caption{ Pseudoscalar correlators P and P$_{\rm sub}$  from full
and zero mode subtracted 
quark propagators at the smallest quark mass $ma=0.01$ on the
$6^3\times 16$ lattice at $\beta=3.0$. The dashed lines are
from correlated fits to the range $t\in[4,8]$.}
\label{fig:zmcorr6}
\end{figure}

\begin{figure}[tb]
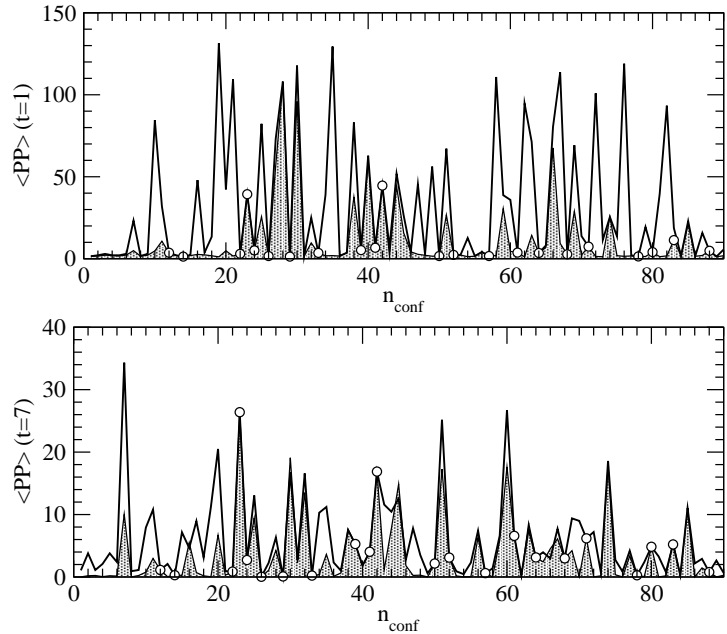

\begin{center}
\includegraphics[width=95mm]{epsf/mctime.6x16.m1t1.both.eps}
\includegraphics[width=95mm]{epsf/mctime.6x16.m1t7.both.eps}
\end{center}
\vspace{-3mm}
\caption{Monte Carlo time evolution of full (thick empty line) and zero mode
subtracted (filled) pseudoscalar correlator at quark mass
$am=0.01$. Shown are the correlators 
at times $t=1$ (top) and $t=7$ (bottom). Empty dots mark
topologically trivial gauge
configurations, where the correlators are equal.}
\label{fig:zm_mctime6}
\end{figure}

We therefore calculate on every gauge configuration the 
eigenvectors $\phi^{(i)}$, $i=1,\dots,10$  of $\hat D$ corresponding to the 10
smallest eigenvalues and computed their chirality, and
the eigenvectors with chirality $|\chi^{(i)}|\geq 0.9$ are considered
to be zero modes. The distribution
of the topological charge for the set of 90 gauge configurations is peaked at
$Q=1$, as can be seen in Table \ref{tab:qtop}.
All zero modes are subtracted from the quark
propagator according to \eqref{eq:qprop_dhat_sub}, and 
with the subtracted quark propagators, pion propagators are
constructed in the usual way. In the pseudoscalar
correlator, the zero mode effects should then disappear, while the
P-S correlator should remain unchanged.\footnote{Indeed our P-S correlators
change only marginally when built from zero mode subtracted quark
propagators.} First we examine the 
correlators at the smallest quark mass $am=0.01$ in Fig.~\ref{fig:zmcorr6}. The
subtraction of the zero modes leads both to a strong decrease in the
amplitude and the mass, as can be seen from the fitted
lines. This effect is illustrated by the Monte Carlo time evolution of
the pseudoscalar correlators in Fig.~\ref{fig:zm_mctime6}. At small time
$t=1$, the correlator fluctuates wildly with Monte Carlo time. The
zero mode subtraction removes most of the peaks at topologically
non-trivial gauge configurations. The average over Monte Carlo time is
therefore 
much smaller for the subtracted correlator. At large time $t=7$, the
picture changes and more peaks remain after zero mode
subtraction. Some of the most prominent peaks (no.~23 and 42) even are
at $Q=0$ gauge 
configurations and therefore do not come from zero modes. The average
correlator decreases  only moderately after
subtraction. Combining the observations at the two times, it follows
that the pion mass gets smaller at $am=0.01$ when determined from the
subtracted propagators.

\begin{figure}[tb]
\begin{center}
\includegraphics[width=90mm]{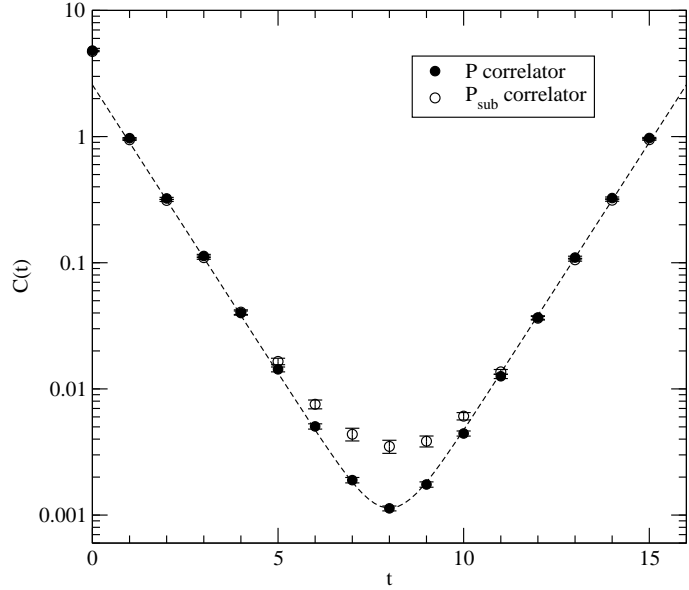}
\end{center}
\vspace{-3mm}
\caption{ Pseudoscalar correlators P and P$_{\rm sub}$ from full
and zero mode subtracted 
quark propagators at the largest quark mass $ma=0.32$ on the
$6^3\times 16$ lattice at $\beta=3.0$. } 
\label{fig:zmcorr12}
\end{figure}

\begin{figure}[tb]
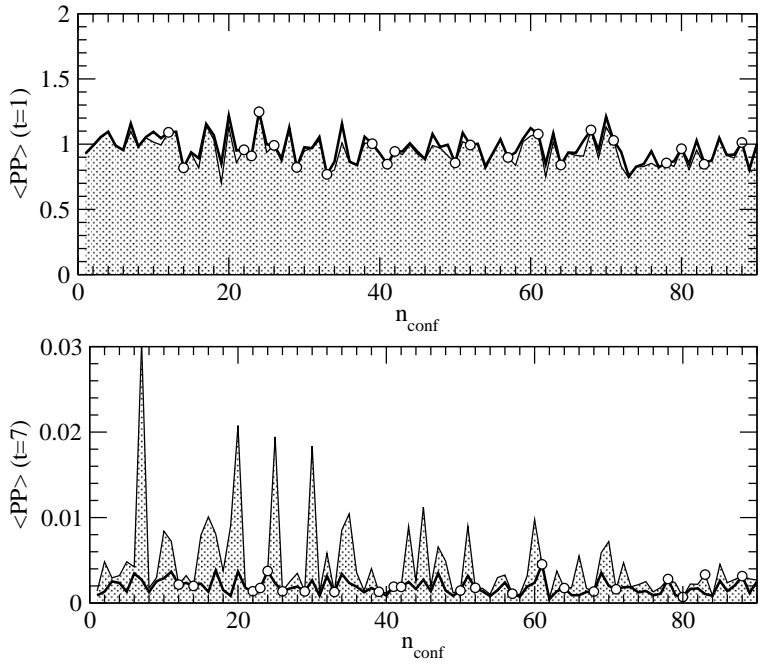

\begin{center}
\includegraphics[width=10cm]{epsf/mctime.6x16.m10t1.both.eps}
\includegraphics[width=10cm]{epsf/mctime.6x16.m10t7.both.eps}
\end{center}
\vspace{-3mm}
\caption{Monte Carlo time evolution of full (thick empty line) and zero mode
subtracted (filled) pseudoscalar correlator at quark mass
$am=0.32$. Shown are the correlators 
at times $t=1$ (top) and $t=7$ (bottom). Empty dots mark
topologically trivial gauge
configurations.}
\label{fig:zm_mctime12}
\end{figure}

The situation is different at very large quark mass
$am=0.32$. The subtracted and unsubtracted pseudoscalar correlators in
Fig.~\ref{fig:zmcorr12} are almost
equivalent at small $t$, but clearly disagree at larger
$t$. Again, the Monte Carlo history in Fig.~\ref{fig:zm_mctime12} helps
illuminating this observation. At $t=1$, there is essentially no
effect from the zero mode subtraction. At $t=7$, after subtraction
there appear some large peaks at topologically non-trivial gauge
configurations which increase the average considerably, while the full
correlator behaves quite smoothly.
There are several explanations for this strange behavior of the meson
correlator at large mass and time. First of all, the third-order
overlap-improved Dirac operator is not 
exactly chiral, therefore the subtraction of the approximate zero
modes leads to small numerical deviations from the exact zero-mode
subtracted quark propagators. These deviations become important at large
mass and large $t$, where the meson
correlator is small, and might thus cause the observed
flattening in Fig.~\ref{fig:zmcorr12}. Second, removing the zero modes
amounts to a modification  
of the quenched theory. The meson correlators then do not necessarily
have to be a sum of exponential functions.
 To rule out the first
possibility, it would be necessary to repeat this examination with a
larger order ${\mathcal O}(10)$ of the overlap expansion in the Dirac operator.
We are however not mainly interested in the large mass behavior of
the meson propagator, where
a reliable pion mass can easily be extracted from the P
correlator. Therefore we do not investigate this effect further.

\begin{table}
\begin{center}
\begin{tabular}{c|c} \hline\hline
 $|Q|$ & $N_{\rm conf}$ \\ \hline
 0 & 22 \\
 1 & 40 \\
 2 & 18 \\
 3 & 9 \\
 4 & 0 \\
 5 & 1 \\ \hline
\end{tabular}
\end{center}
\caption{Distribution of topological charge $Q$ for the 90 gauge
 configurations on the $6^3\times 16$ lattice at $\beta=3.0$.} 
\label{tab:qtop}
\end{table}

\begin{figure}[tb]
\begin{center}
\includegraphics[width=10cm]{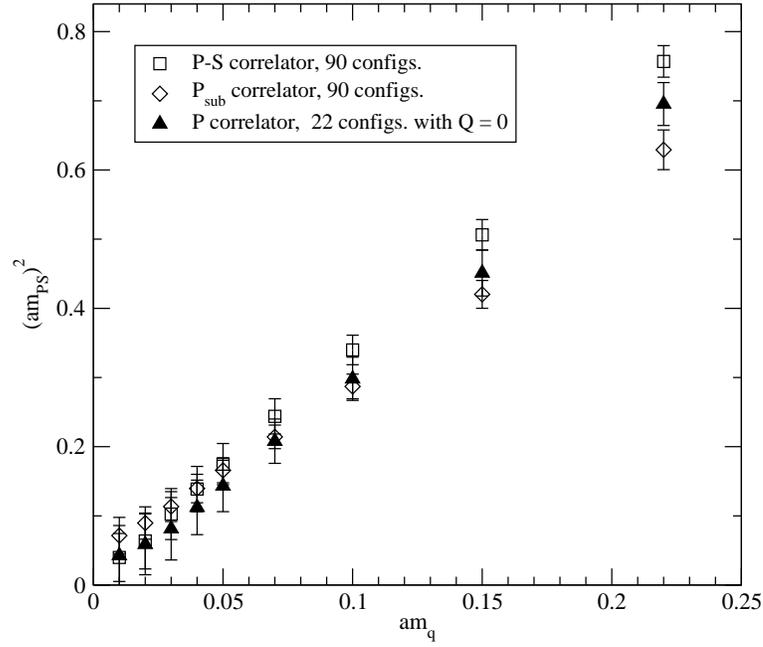}
\end{center}
\caption{Three different methods of removing the quenched topological
finite-volume artifacts in pion correlators. The lattice size is
$6^3\times 16$, and the gauge coupling $\beta=3.0$. Within
errors, all three methods lead to the same results at small quark
mass and go to zero in the chiral limit $m\rightarrow 0$. The zero
mode effects are no longer present. Compared are the P-S correlator,
the $ {\rm P_{sub}}$ correlator from zero mode subtracted quark
propagators and the P correlator evaluated only on gauge
configurations with topological charge $Q=0$. The last method is of
course in general not allowed, as the remaining set of gauge configurations no
longer follows the probability distribution given by the
partition function of the quenched theory.}
\label{fig:zm_remove}
\end{figure}

The results of this small volume study are summarized in
Fig.~\ref{fig:zm_remove}, where the chiral limit of the squared
pseudoscalar meson mass is plotted for the P correlator on $Q=0$
configurations, the zero mode subtracted pseudoscalar correlator
P$_{\rm sub}$ and 
the P-S correlator. All of them agree within errors at small quark
masses and go to zero for $m\rightarrow 0$.

\section{Conclusion}

At small quark masses and fixed volume, the pseudoscalar
meson masses measured from P correlators are distorted by topological
finite volume effects. Forming the difference P-S, the chiral limit of
$(am_{\rm PS})^2$ is consistent with the expectations, confirming
that the observed distortion is due to the topological finite size
effects. Indeed, in the P-S correlators these effects cancel, up to
small chiral symmetry breaking contributions. Furthermore, P-S is a sum
over exponentials with physical meson masses, although both the scalar
and the pseudoscalar mesons enter. For small quark masses however, the pion
dominates.

Unlike P-S, the pseudoscalar correlator P$_{\rm sub}$ built from zero
mode subtracted quark propagators is a strange quantity and does not
need to be a sum of exponential functions. It is therefore better not to use
P$_{\rm sub}$ in actual calculations.

At intermediate quark masses, the P and P-S correlators agree, as
expected. To extract  the pseudoscalar meson mass, we
therefore use the P-S correlator at small quark masses, 
where the P correlator would be contaminated by the zero modes, and the P
correlator at large quark masses, where it would be difficult to disentangle
the contributions of the scalar meson to the P-S correlator. This
provides a reliable determination of the pseudoscalar mass over the
whole range of quark 
masses, and as we will also see in Chapter \ref{ch:results_fp}, an
intersection of 
$(am_{\rm PS})^2$ with the horizontal axis which is consistent with other
determinations.

The best way to avoid any problems with zero mode effects 
is to work at large enough lattice volumes. As will be shown in Section
\ref{sect:fp_zm}, at our 
largest lattice size $L_s\approx 2.5$ fm, the zero modes no longer
contaminate the pion propagator significantly, and it is possible to get
unambiguous answers concerning the chiral limit of pseudoscalar mesons.

\clearpage

\clearpage

\begin{figure}[tb]
\begin{center}
\includegraphics[width=9cm]{epsf/zm.12x24_b3.00_o3.eps}
\end{center}
\vspace{-3mm}
\caption{ Topological quenching artifacts in
squared pseudoscalar meson mass versus bare quark mass at $\beta=3.0$
and volume $12^3\times 24$. The effective masses at the smallest quark mass
are shown in Fig.~\ref{fig:zm_effmass12}.}  
\label{fig:zm_sqrpion12}
\end{figure}

\clearpage

\clearpage

\clearpage

\newpage

\fancyhead[RE]{\nouppercase{\small\it The Light Hadron Spectrum with
Fixed-Point Fermions}}
\chapter{The Light Hadron Spectrum with Fixed-Point Fermions}
\label{ch:results_fp}
Since the first attempts in 1981
\cite{Hamber:1981zn,Weingarten:1982jy}, many lattice studies of
the light hadron
spectrum in quenched QCD have been performed, with quality increasing
with time. The 
first systematic calculation was done in 1993 by the GF11 collaboration
\cite{Butler:1994em}, but today's benchmark is the CP-PACS calculation
\cite{Aoki:1998ri,Yoshie:2001ts} from 1998, which included 
a thorough examination of the chiral and continuum
extrapolations and very high statistics. In their study, the most simple
choice of actions was taken, namely the Wilson plaquette and fermion actions,
and a full year of runs on the dedicated CP-PACS computer with a peak
performance of 614 GFLOPS was
necessary in order to obtain the quenched particle spectrum in a controlled
manner. Because the cut-off effects 
for the Wilson action are known to be large, the
simulation was performed at rather small lattice spacings in the range
$a\approx 0.05$--$0.1$ fm. To avoid finite volume effects, the physical
size was 
chosen to be 3 fm, which required to run on lattices of sizes up to $64^3\times
112$. The Monte Carlo average was taken from up to 800
independent gauge 
configurations. Their result is a physical particle spectrum with very small 
statistical errors, which are on the order of 1--2\% for mesons and 2--3\% for
baryons, and with systematic errors from the extrapolations that are
estimated to be even smaller. The calculated hadron masses agree
qualitatively with the  
experimentally observed spectrum, but the mass values deviate by up to
11\% or $7\sigma$, which is thought to be the error introduced by the
quenched approximation. 

A crucial part in the analysis of the CP-PACS data was the chiral
extrapolation. The quark mass was pushed down to a 
value corresponding to $m_{\rm PS}/m_{\rm V}\approx 0.4$, which is very small
for the Wilson action 
with its inherent chiral symmetry breaking\footnote{Typically,
simulations with Wilson fermions do not go lower than $m_{\rm PS}/m_{\rm V}\approx 0.5$}, and only with the point at
the lowest quark mass it was possible to resolve the non-analytic
contributions predicted by quenched chiral perturbation theory
(Q$\chi$PT) \cite{Bernard:1992mk,Sharpe:1992ft}. It is however important to
investigate whether including such Q$\chi$PT terms leads to
significantly different  
mass values in the chiral limit than when using just low-order
polynomial forms, as usually done in earlier works.

Exact or approximately chiral fermion actions allow to go to much
smaller quark masses than the Wilson action. It is therefore possible to
explore the chiral limit directly, to check the significance of the
Q$\chi$PT terms and thus to increase the reliability of the 
hadron mass extrapolation to the physical quark mass. In this Chapter,
we report the results of a spectroscopy calculation with the parametrized FP
fermion action which probes deep into the
chiral limit and also includes investigations of the scaling properties of
the hadron masses and their finite-volume dependence. With a total
amount of computer resources of about 20 GFLOPS$\times$years theoretical
peak and an effective amount of about 6 GFLOPS$\times$years, this
study in the framework of the BGR collaboration \cite{bgr} by no means
attempts to compete with the above mentioned 
high-statistics calculations. However, in addition to an independent test
for spectroscopy simulations with a new formulation of
the lattice QCD action, we get important 
information from the region of small quark masses, where non-chiral
actions do not allow to perform simulations. Furthermore, this is
one of the first spectroscopy studies with a chiral symmetric fermion
action that examines cut-off effects. 

Before we started this simulation, we carried out some 
tests for spectroscopy with the FP action on a smaller scale, the
results of which are 
published in \cite{Hasenfratz:2001hr,Hasenfratz:2002}. The lattice parameters
of these tests are listed for completeness in Table~\ref{tab:confs_test}. 

\begin{table}
\begin{center}
\begin{tabular}{c|c|c|c|c|c|c|r} \hline\hline
$D$ & $\beta$ & $L_s^3\times L_t$ & $a(r_0)$ & $La(r_0)$ & \# confs. & \# masses & $m_{\rm
PS}/m_{\rm V} $ \\ \hline
FP &  3.0 &  $6^3\times 16 $ & 0.16 fm  & 0.9 fm & 100 & 10 & 0.35--0.8 \\
FP &  3.0 &  $8^3\times 24 $ & 0.16 fm  & 1.3 fm & 100 & 13 & 0.3--0.85 \\
FP &  3.0 &  $9^3\times 24 $ & 0.16 fm  & 1.4 fm & 70 & 12 & 0.3--0.85 \\
ov3 & 3.0 &  $9^3\times 24 $ & 0.16 fm  & 1.4 fm & 28 & 10 & 0.27--0.85 \\
ov3 & 3.2 &  $9^3\times 24 $ & 0.13 fm  & 1.2 fm & 32 & 10 & 0.24--0.89 \\\hline
\end{tabular}
\end{center}
\vspace{-2mm}
\caption{Parameters of exploratory spectroscopy simulations with the FP
and overlap-improved (ov3) FP Dirac operator \cite{Hasenfratz:2001hr,Hasenfratz:2002}.} 
\label{tab:confs_test}
\end{table}

\begin{table}
\begin{center}
\begin{tabular}{c|c|c|c|c|c|r|c} \hline\hline
$D$ & $\beta$ & $L_s^3\times L_t$ & $a(r_0)$ & $La(r_0)$ & \# confs. & $m_{\rm
PS}/m_{\rm V} $  & \# iters.\\ \hline
FP & 3.0 & $8^3\times 24 $ & 0.16 fm  & 1.3 fm & 200 & 0.3--0.88 & 250(39) \\
FP & 3.0 &  $12^3\times 24 $ & 0.16 fm & 1.9 fm & 200 & 0.3--0.88 & 360(55)\\
ov3 & 3.0 & $12^3\times 24 $ & 0.16 fm & 1.9 fm & 100 & 0.21--0.88 & 618(61)\\
FP & 3.0 & $16^3\times 32 $ & 0.16 fm & 2.5 fm & 200 & 0.28--0.88 & 478(74)\\
FP & 3.4 & $12^3\times 24 $ & 0.10 fm & 1.3 fm & 200 & 0.34--0.89 & 306(37)\\
FP & 3.7 & $16^3\times 32 $ & 0.08 fm & 1.3 fm & 100 & 0.34--0.89 & 469(52)\\ \hline
\end{tabular}
\end{center}
\vspace{-2mm}
\caption{Parameters of the spectroscopy simulation described in this
work. The $12^3\times 24 $ gauge configurations at $\beta=3.0$ are the
same for the calculations with the FP and overlap-improved (ov3) FP
Dirac operator. The last column shows the average number of iterations
 required for the inversion of $D$.}
\label{tab:confs}
\end{table}

\section{Simulation Parameters}

For generating the gauge configurations, we use the
parametrized iso\-tro\-pic Fixed-Point gluon action from
\cite{Niedermayer:2000ts}. While this action is relatively
expensive compared to the Wilson action\footnote{A factor of $\sim 60$
in computer time 
is estimated in \cite{urs:diss}.}, generating the gauge
configurations was a  
comparably small effort in terms of computer time in the context of
this work, and therefore we 
could afford using a FP action also for the gluon sector. It was shown
in \cite{Niedermayer:2000ts} that the FP gauge action has small
scaling violations in 
gluonic quantities and that it reproduces topological properties
well. With this gauge action, we produced a set of configurations at
various lattice sizes and gauge couplings as 
shown in Table \ref{tab:confs}, where we also list the lattice spacing
determined from the Sommer scale $r_0\approx 0.5$ fm for the
different values of the gauge coupling\footnote{The detailed analysis
of the scale determination for the FP gluon action is given in
\cite{urs:diss}}. This combination of parameters was 
chosen to allow for a scaling analysis at small physical spatial
lattice size $L_s\approx 1.3$ fm and a 
finite-volume analysis at gauge coupling $\beta=3.0$. The largest
lattice at $\beta=3.0$ has a physical volume large enough to
accommodate hadrons with negligible finite-volume effects. While
this value of the gauge coupling is quite far away from the continuum,
we expect to get on this lattice
precise numbers for hadron masses which can serve as good estimates
for the continuum values, because as we will show, the FP action has
small scaling violations.
We use alternating Metropolis and 
pseudo-overrelaxation sweeps over the lattice, with 2000 sweeps for 
thermalization and 500 sweeps to separate between different
configurations. The number of separation sweeps is a worst-case
estimate based on 
measurements of autocorrelation times for simple gluonic operators
\cite{philipp:diss}. The configurations are then smeared with the RG
inspired two-level hypercubic smearing described in \cite{thomas:diss}
and fixed to Coulomb 
gauge\footnote{The order
of smearing and gauge fixing is a matter of choice. In earlier
studies we first fixed the gauge and then smeared the links, as we
considered the smearing to be part of the definition of the Dirac
operator. For this spectroscopy study, the order was reversed. The
argument was that because after smearing the configurations are
much smoother, the gauge fixing algorithm might have
less problems with Gribov copies.} 
with the algorithm presented in Appendix \ref{app:gaugefix}.
We use periodic boundary conditions for both the gauge and fermion fields.

For the fermion action, we take the parametrized FP Dirac
operator from Chapter \ref{ch:d_fp}, except for one lattice, where the
overlap-improved FP Dirac 
operator from Chapter \ref{ch:d_ov} is used. With the third-order
overlap expansion,  we
decrease the small residual 
chiral symmetry breaking of the parametrization even further and are
able to check what effect the overlap construction has on the mass
spectrum. The quark masses cover a very large range, with the 
smallest value, where $m_{\rm PS}/m_{\rm V}\approx 0.21$, lying close to 
the physical point. This provides us with meaningful data for
chiral fits and allows for a thorough examination of the chiral limit. To
enhance the signal for the hadron correlators, we use Gaussian
smeared sources located at the center of timeslice $t=0$. The
source extension parameter $\gamma$ in Eq.~\eqref{eq:gaussian_source}
is chosen to correspond to a source size of $\sim 0.5$ fm.\footnote{On
the $12^3\times 24$ lattice at $\beta=3.4$, the source size is $\sim
0.3$ fm.} We use point sinks and project to zero momentum by summing
over all spatial sink locations. 
Quark propagators are calculated with the multi-mass BiCGstab inversion
algorithm \cite{Jegerlehner:1996pm} (see also Appendix
\ref{app:multimass}). Source vectors are normalized 
to 1, and as a stopping criterion, we
require the residual to be smaller than some threshold $|r|\leq
\sigma$. 

\begin{figure}[tb]
\begin{center}
\includegraphics[width=65mm]{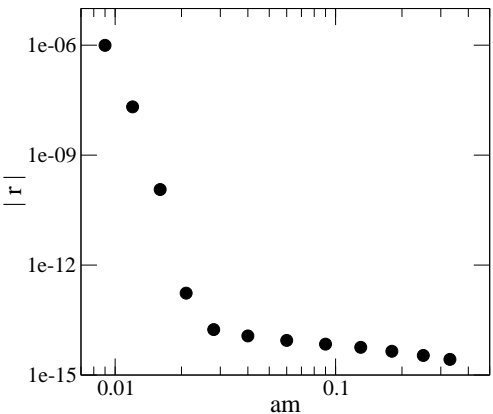}
\end{center}
\vspace{-3mm}
\caption{ Residual for quark propagator inversion at different quark
masses. The tolerance of the multi-mass solver is set to $\sigma=10^{-6}$.}
\label{fig:res_m}
\end{figure}

Due to the fact that a multi-mass solver inverts the Dirac matrix 
at all quark masses simultaneously, the result of the
inversion at larger masses is more accurate than at the smallest one,
where the 
condition number of the Dirac matrix is worst. The value of $\sigma$
therefore determines the accuracy of the quark propagators at the smallest
quark mass, while the propagators at larger masses are calculated to
much higher 
precision, as shown in Fig.~\ref{fig:res_m}. In order to demonstrate
that the error from the truncation of the  
inversion algorithm is much smaller than the statistical error from
the Monte Carlo estimate of the hadron correlators, we show in Table
\ref{tab:resid_cmp} the dependence of pseudoscalar and vector meson
correlators on 
the stopping criterion. Based on this data, we choose a precision
of $\sigma=10^{-6}$ for the inversion of the Dirac operator, which
leads to an error from the truncation of the iterative inversion
algorithm that is 
negligible compared to the statistical error. At our intermediate and
larger quark masses, the   
quark propagator is almost calculated to machine precision. The
average number of iterations needed for the inversion on the various
lattices is also given  
in Table \ref{tab:confs}.\footnote{Interpreting these numbers, one has to keep
in mind that the BiCGStab 
algorithm requires two matrix-vector multiplications with the Dirac
operator per iteration.}

\begin{figure}[tb]
\begin{center}
\includegraphics[width=8cm]{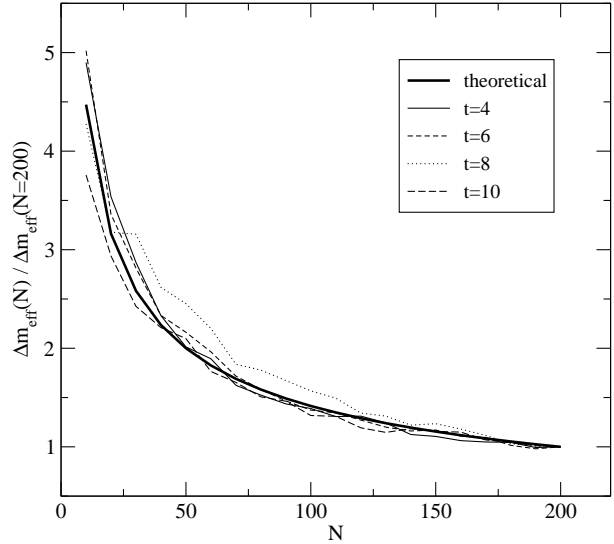}
\end{center}
\vspace{-3mm}
\caption{Dependence of the statistical error in effective pseudoscalar
meson masses
on the number of gauge configurations $N$. The data is taken from the
$\beta=3.40$ configurations on the $12^3\times 24$ lattice, and the
quark mass is $am_q=0.10$. The curves are normalized to 1 at $N=200$. }
\label{fig:effmasserror}
\end{figure}

\begin{figure}[tb]
\begin{center}
\includegraphics[width=8cm]{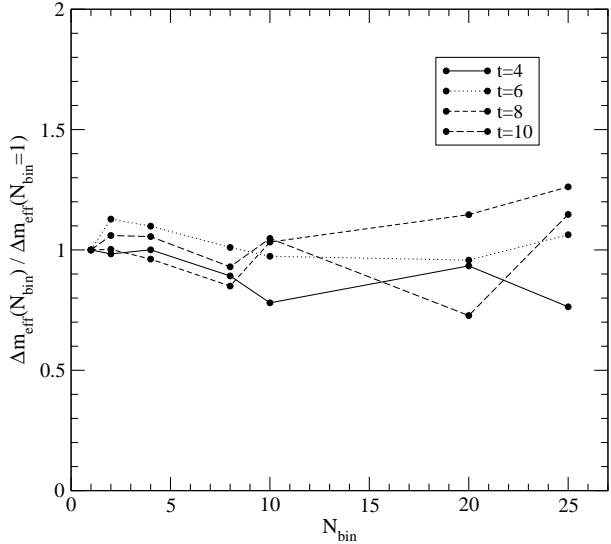}
\end{center}
\vspace{-3mm}
\caption{Dependence of the statistical error in effective pion masses
on the bin size $N_{\rm bin}$. The data is taken from 200
$\beta=3.40$ configurations on the $12^3\times 24$ lattice, and the
quark mass is $am_q=0.10$. The curves are normalized to 1 at $N_{\rm bin}=1$. }
\label{fig:binerror}
\end{figure}

To check whether the computed hadron propagators are statistically
independent, we calculate the statistical error of effective meson
masses in dependence of the number of configurations $N$ used for the
Monte Carlo average. If the gauge configurations are independent, the
bootstrap error is proportional to $1/\sqrt{N}$. In
Fig.~\ref{fig:effmasserror} we plot the error in effective
pseudoscalar meson masses at various values of 
$t$ on the $12^3\times 24$ lattice at $\beta=3.4$. As we can see, the
curves nicely agree with the expected $1/\sqrt{N}$ behavior. For a
second check, we 
collect bins of $N_{\rm bin}$ successive propagators, take the average of each
bin and calculate the statistical error in effective masses from the
$N/N_{\rm bin}$ binned propagators. The resulting bootstrap error, plotted in
Fig~\ref{fig:binerror}, turns 
out to be independent of the bin size $N_{\rm bin}$. These two
observations confirm 
that the number of separation sweeps used in the generation of the gauge
configurations is sufficient to ensure statistical independence.

\begin{table}
\begin{center}
\begin{tabular}{c|c|c|c|c} \hline\hline
correlator  & $am_q$  & $\Delta^{\rm rel}(\sigma=10^{-8})$ & $\Delta^{\rm rel}(\sigma=10^{-6}$) &
 $\Delta^{\rm rel}x(\sigma=10^{-4}$)  
\\ \hline
PS & 0.016 & $3\cdot 10^{-8}$ & $8\cdot 10^{-6}$ & $0.001$  \\
V & 0.016  & $4\cdot 10^{-7}$ & $7\cdot 10^{-5}$ & $0.014$  \\ \hline
PS & 0.04  & $5\cdot 10^{-14}$  & $7\cdot 10^{-11}$ & $4\cdot 10^{-7}$ \\ 
V & 0.04 & $2\cdot 10^{-12}$ & $1\cdot 10^{-9}$ & $4\cdot 10^{-6}$ \\ \hline 
\end{tabular}
\end{center}
\vspace{-2mm}
\caption{ Relative error $\Delta^{\rm rel}(\sigma) = (C(\sigma) - C_{\rm
exact})/C_{\rm exact}$ of meson correlators $C(t=8)$ on one $9^3\times 24$
gauge configuration at $\beta=3.0$. The dependence on the 
tolerance $\sigma$ of the residual in the quark propagator inversion
is shown. The
'exact' result $C_{\rm exact}$ is calculated with tolerance
$\sigma=10^{-12}$. For comparison, the statistical error of PS and V
correlators at $t=8$ is always larger than 1 \% in our simulations.}
\label{tab:resid_cmp}
\end{table}

%\subsection{Autocorrelation}

We list our hadron mass results on the
different lattices for the pseudoscalar and vector mesons and
the octet and decuplet baryons from quarks with degenerate masses, 
together with the temporal fit range and the resulting value of
$\chi^2/df$ for the fit, in Appendix~\ref{app:data}. To account for a possibly
biased determination of the fitted masses we apply 
bootstrap bias correction, as defined in Eq.~\eqref{eq:bias_boot}, to the
fit results.\footnote{ Bias correction has proven relevant only at
the very smallest quark 
masses, where in some cases the fitted hadron masses showed a
noticeable bias. At 
intermediate and large quark masses, the bias is negligible, as can be
seen from the data in Appendix~\ref{app:data}.  }

Fig.~\ref{fig:hadrons} gives an overview of the masses of
the hadrons in dependence of the quark mass 
on the various lattices in our simulation. 
We remind the reader that
due to the use of a multi-mass solver, the results at different quark
masses are highly correlated. This fact has to be taken into account
when interpreting the data presented in this chapter.

The following analysis of our spectroscopy runs has to be considered
preliminary. We try to give in this work some first answers to the
main questions 
that arise from the use of such a new chiral action for the determination of
the hadron spectrum.

\section{Zero Mode Effects}
\label{sect:fp_zm}
In Chapter \ref{ch:zeromodes}, we showed that divergent zero mode
contributions appear in pion 
propagators and examined these quenching artifacts and possible ways
to remove  
them on a small lattice volume. We concluded that it seems sensible to
work with different correlators at different quark masses. We return to this
issue here and
present the zero modes effects seen in the data of our spectroscopy
simulation. 

\begin{figure}[tbp]
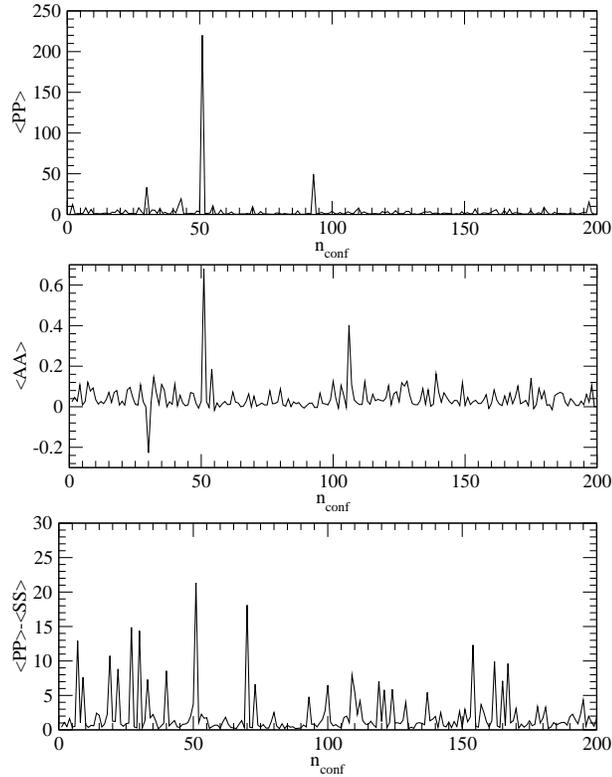

\begin{center}
\includegraphics[width=8cm]{epsf/mctime.08x24.m2t8.P.eps}
\includegraphics[width=8cm]{epsf/mctime.08x24.m2t8.A.eps}
\includegraphics[width=8cm]{epsf/mctime.08x24.m2t8.PS.eps}
\end{center}
\vspace{-4mm}
\caption{Monte Carlo time evolution of pion correlators at bare quark
mass $am_q=0.021$ on lattice size  $8^3\times 24$ with $\beta=3.0$ for
$D^{\rm FP}$. Notice the large difference in the vertical scale of
the P and P-S correlators.}
\label{fig:mctime08}
\end{figure}

\begin{figure}[tbp]
\begin{center}
\includegraphics[width=9cm]{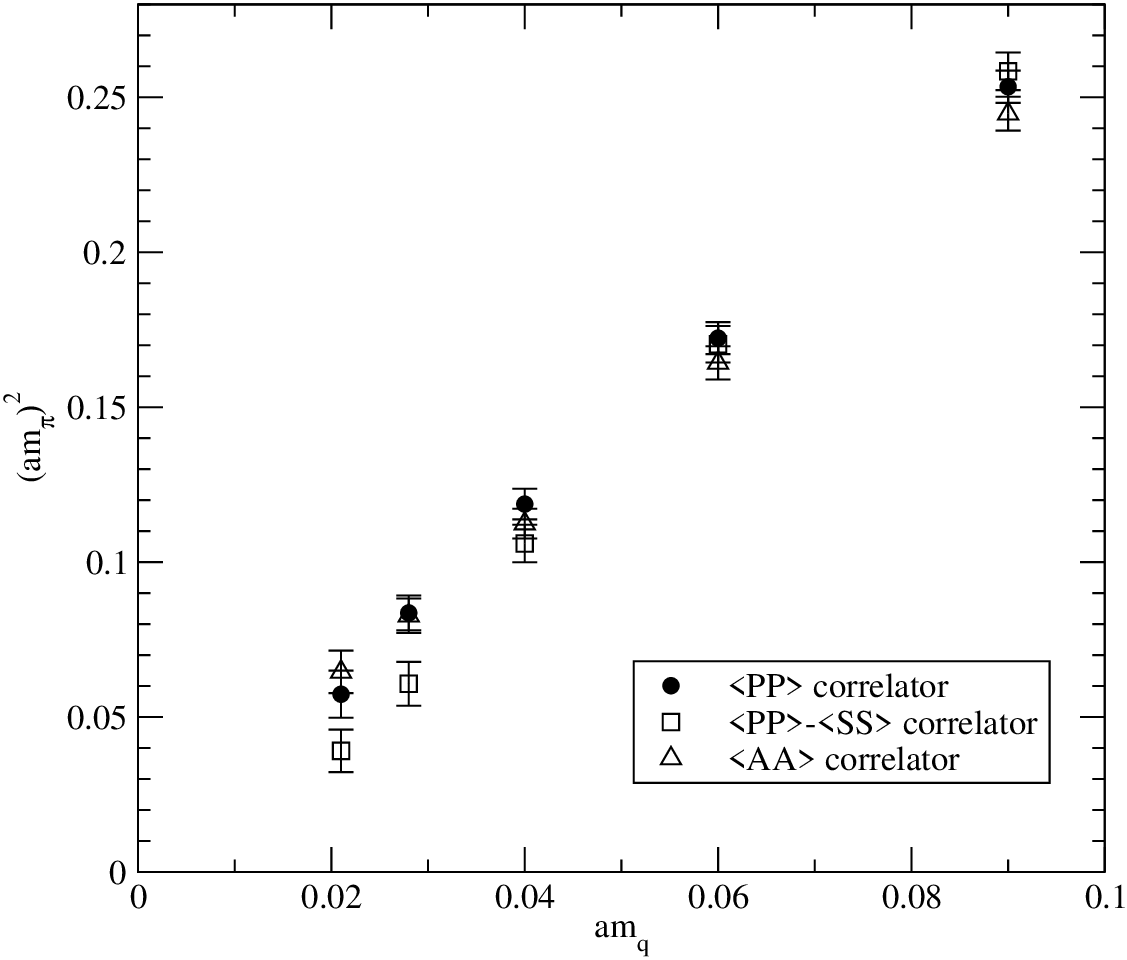}
\end{center}
\vspace{-4mm}
\caption{ Pion mass squared versus bare quark mass for the three pion
correlators on lattice size  $8^3\times 24$ with $\beta=3.0$ for
 $D^{\rm FP}$. }
\label{fig:fp_zm08}
\end{figure}

\begin{figure}[tbp]
\begin{center}
\includegraphics[width=8cm]{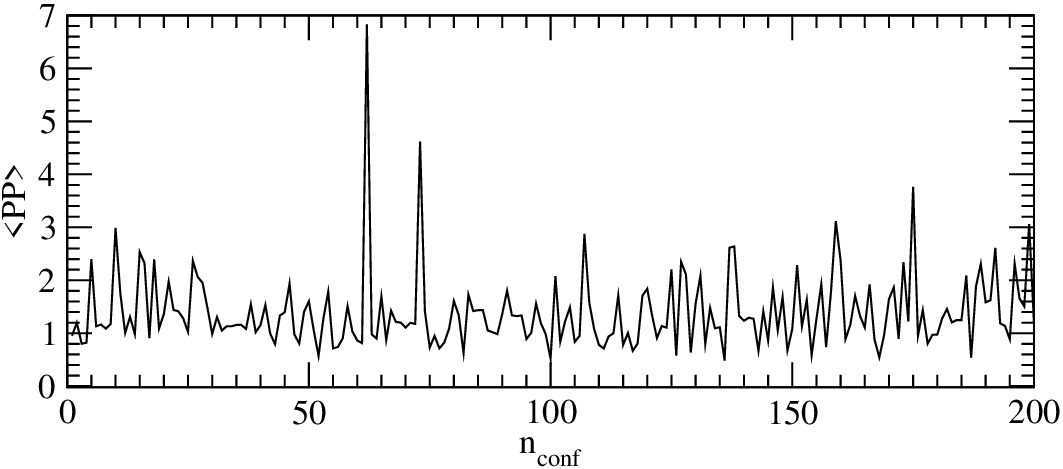}
\includegraphics[width=8cm]{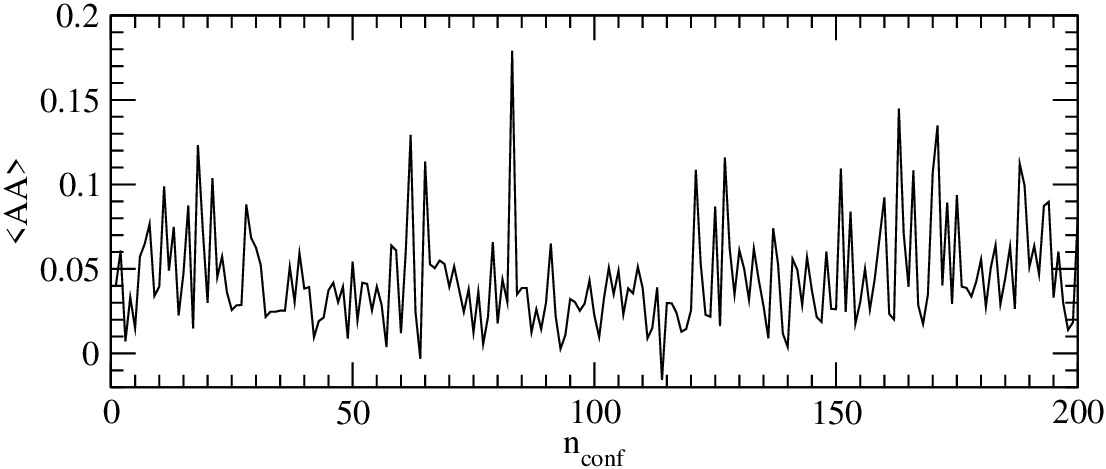}
\includegraphics[width=8cm]{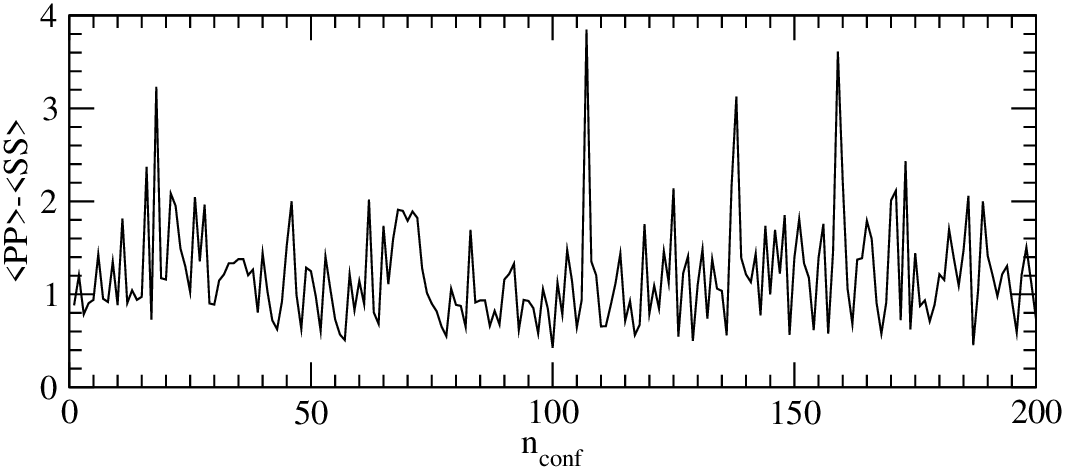}
\end{center}
\vspace{-4mm}
\caption{Monte Carlo time evolution of pion correlators at bare quark
mass $am_q=0.021$ on lattice size  $12^3\times 24$ with $\beta=3.0$ for
$D^{\rm FP}$.}
\label{fig:mctime12}
\end{figure}

\begin{figure}[tbp]
\begin{center}
\includegraphics[width=9cm]{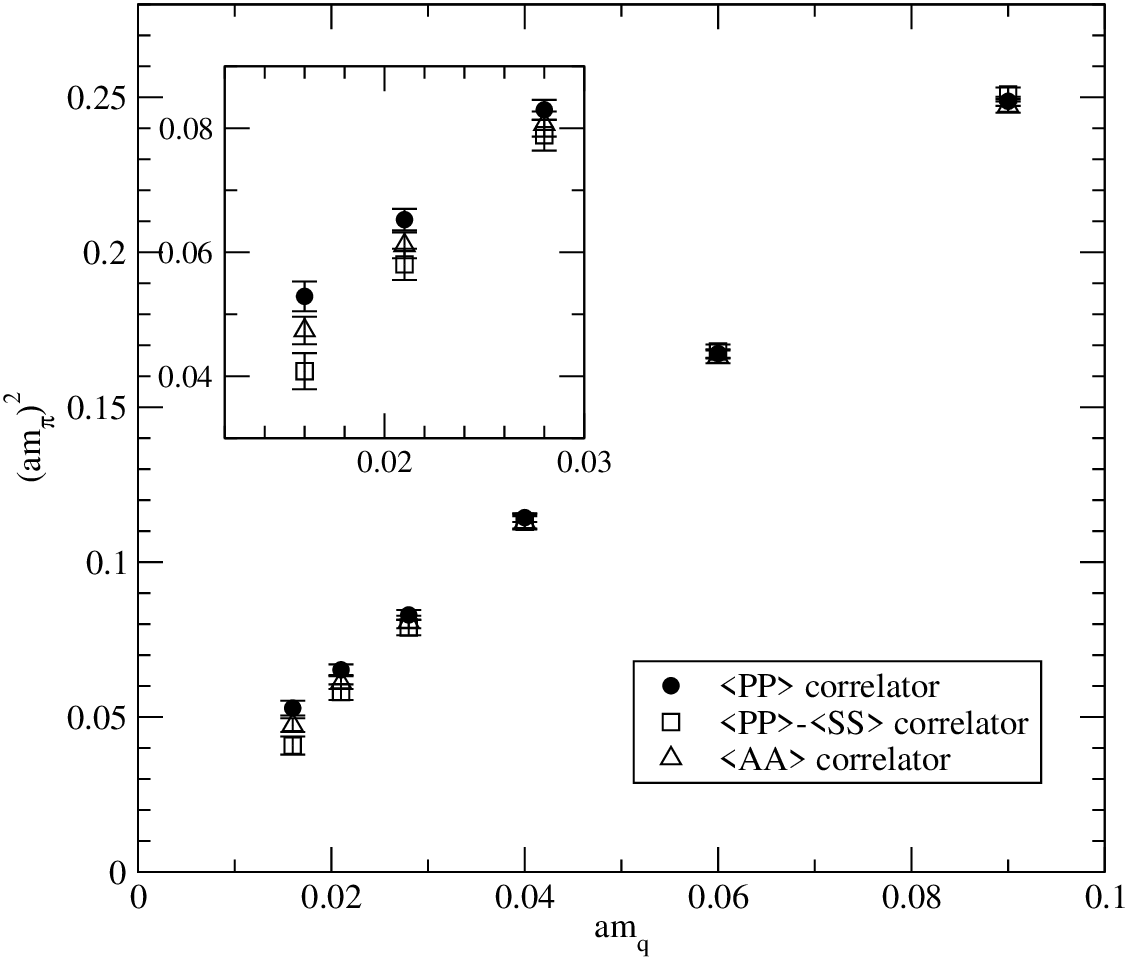}
\end{center}
\vspace{-4mm}
\caption{ Pion mass squared versus bare quark mass for the three pion
correlators on lattice size  $12^3\times 24$ with $\beta=3.0$ for
$D^{\rm FP}$. }  
\label{fig:fp_zm12}
\end{figure}

\begin{figure}[tbp]
\begin{center}
\includegraphics[width=8cm]{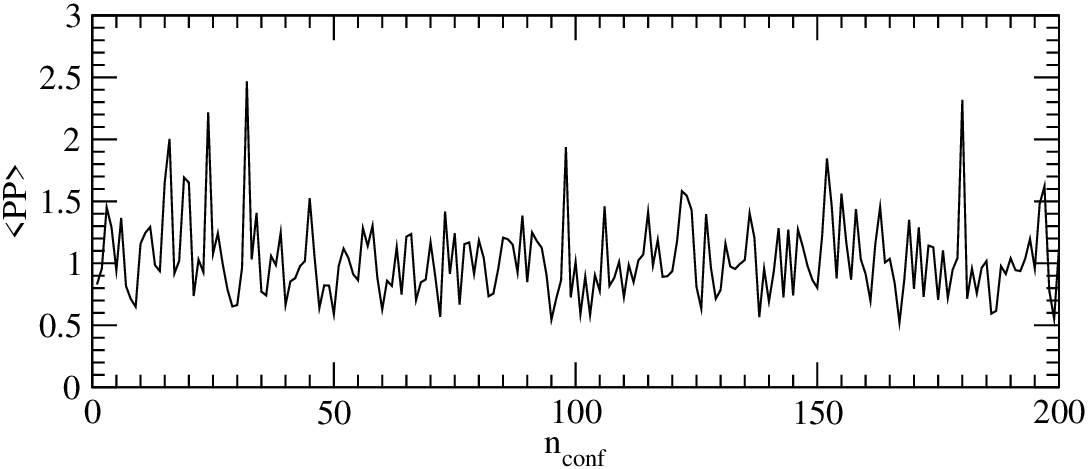}
\includegraphics[width=8cm]{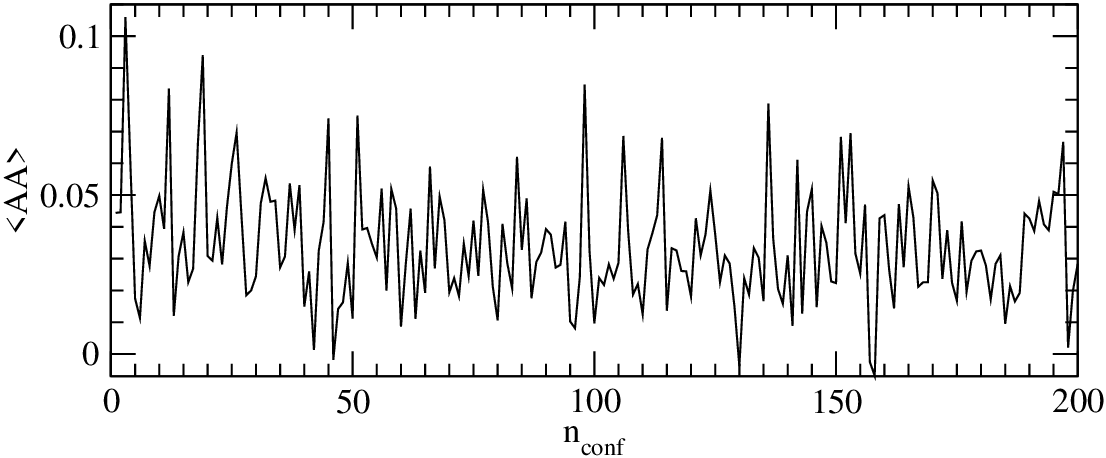}
\includegraphics[width=8cm]{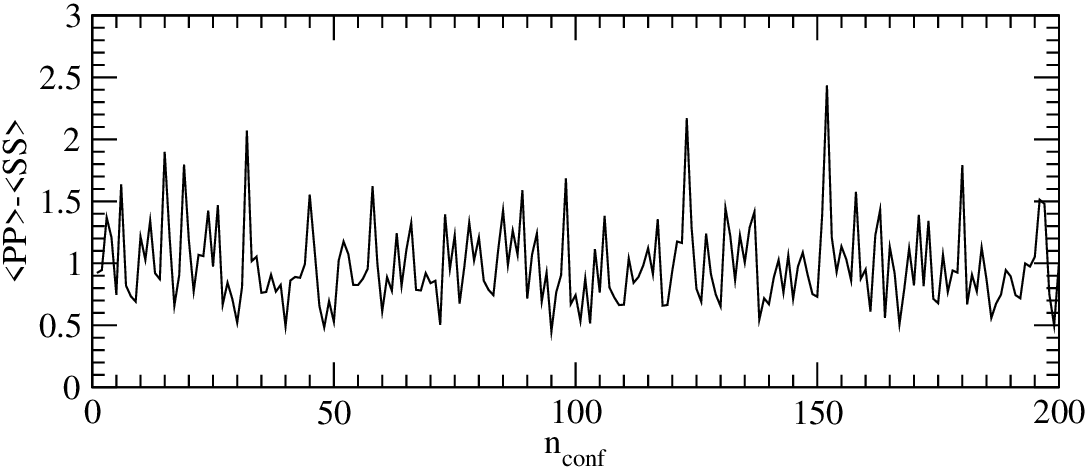}
\end{center}
\vspace{-4mm}
\caption{Monte Carlo time evolution of  pion correlators at bare quark
mass $am_q=0.021$ on lattice size  $16^3\times 32$ with $\beta=3.0$ for
$D^{\rm FP}$.}
\label{fig:mctime16}
\end{figure}

In Figs.~\ref{fig:mctime08}--\ref{fig:fp_zm16} the Monte
Carlo history of 
the three correlators 
P, A, and P-S and the resulting pseudoscalar meson masses are
shown on the three different lattice volumes at gauge coupling
$\beta=3.0$. The values of the
correlators in the Monte Carlo time plots are taken from timeslice $t=8$ and
quark mass $am_q=0.021$. As expected, in
the case of the smallest lattice of size $8^3\times 24$ in
Fig.~\ref{fig:mctime08}, there are a few
very prominent peaks in the P correlator which dominate the gauge average
completely. Also the A correlator is dominated by these peaks, but to
a somewhat smaller extent. In contrast to these heavily contaminated
correlators, many more 
configurations contribute to the average of the P-S
correlator. Considering the absolute  
scale, it is obvious that the zero mode contributions, which are
present in both the
pseudoscalar and scalar correlators, are cancelled to a large extent
in the difference. The resulting pseudoscalar meson 
masses from these three correlators, plotted in Fig.~\ref{fig:fp_zm08}, differ
significantly at the smallest quark masses.

On the lattice with intermediate size $12^3\times 24$, the same
effects in the Monte Carlo history of the correlators can be observed
in Fig~\ref{fig:mctime12},  
but they are much less pronounced. Again, the largest peaks in the P
correlator are cancelled in the P-S correlator. The masses in
Fig.~\ref{fig:fp_zm12} agree much better, but still a systematic
deviation is visible. On the largest lattice
(Fig.~\ref{fig:mctime16}), all three correlators show a fairly smooth
behavior. The mass of the pseudoscalar meson
(Fig.~\ref{fig:fp_zm16}) depends only very little on the choice of
correlator. We summarize this analysis by
listing the relative difference in meson masses from the P and P-S
correlators in dependence of the lattice volume and quark mass in
Table~\ref{tab:zmsize}, 
a quantity which serves as a crude estimate for the size of the
zero mode effects. 

\begin{table}
\begin{center}
%\small
\begin{tabular}{c|c|c|c|c|c} \hline\hline
\multicolumn{2}{c|}{ $8^3\times 24$, $\beta=3.0$} & \multicolumn{2}{c|}{
$12^3\times 24$, $\beta=3.0$} & \multicolumn{2}{c}{ $16^3\times 32$, $\beta=3.0$}\\ 
\multicolumn{2}{c|}{ $V\approx 7$ fm$^4$} & \multicolumn{2}{c|}{
$V\approx 24$ fm$^4$} & \multicolumn{2}{c}{ $V\approx 76$ fm$^4$} \\ \hline
$m_{\rm PS}/m_{\rm V} $ & $\Delta m_{\rm PS}$ & $m_{\rm
PS}/m_{\rm V} $ & $\Delta m_{\rm PS}$ & $m_{\rm PS}/m_{\rm V} $ &
$\Delta m_{\rm PS}$ \\ \hline 
0.28 & 29(13)\% & 0.30  & 14(5)\% & 0.28 & 4(2)\% \\
0.35  & 17(8)\% & 0.35 &  6(3)\% & 0.31 & 3(1)\% \\
0.48 & 6(4)\%  & 0.41  &  3(2)\% & 0.36 & 1(1)\% \\
\hline
\end{tabular}

\end{center}
\vspace{-2mm}
\caption{Estimated size of zero mode effects in pseudoscalar meson masses
from their relative difference when determined from the P, A and
P-S correlators. }  
\label{tab:zmsize}
\end{table}

We conclude from these observations that on our largest lattice
with spatial extension $L_s \approx 2.5$ fm, the zero mode contributions are
sufficiently small to allow for a reliable determination of the mass
in the pseudoscalar channel and the quantitative examination of its
chiral limit. Only at the two smallest quark masses, the results from
the different correlators do not agree within the (already small)
statistical errors. 
 On the smaller lattices, 
the uncertainty grows rapidly with decreasing quark mass, and
therefore it
becomes increasingly hard to keep the
zero mode effects under control. We will in the following
adopt the strategy suggested in Chapter 
\ref{ch:zeromodes} and construct the pseudoscalar meson from the P-S
correlator at small and 
the P correlator at large quark mass.  At the
smaller lattice volumes, this introduces a systematic uncertainty at
the few smallest quark masses, where we can not control how well the
subtraction in the meson correlators works. But as our results will
show, this strategy turns out to be successful in our simulations.

\section{Chiral Extrapolations and Quenched Chiral Logarithms}
With the approximately chiral Fixed-Point fermion action, we are in a
position to 
study the chiral limit of hadron masses in detail. After measuring the
residual additive quark mass renormalization, we examine in this
section the quark mass dependence of the pseudoscalar meson mass
and check for the presence of quenched chiral
logarithms. On our largest lattice, we also calculate the resulting
spectrum of $K$, $K^*$ and $\phi$ mesons and $N$, $\Delta$ and
$\Omega$ baryons after extrapolating the measured data for the
degenerate hadrons to the physical values of the average up and down
quark and the strange quark mass, respectively. 

\subsection{Residual Quark Mass}
\begin{figure}[tbp]
\begin{center}
\includegraphics[width=10cm]{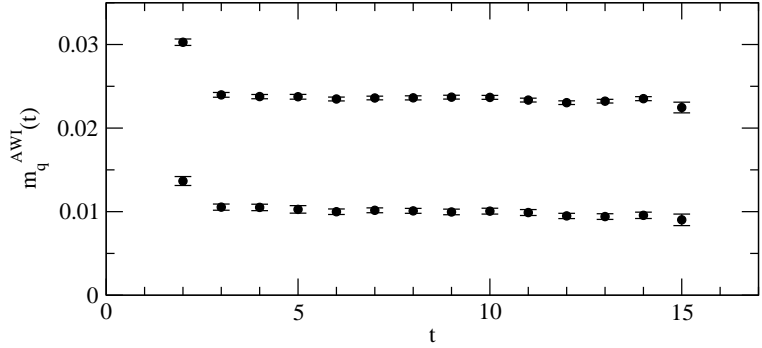}
\end{center}
\vspace{-2mm}
\caption{Unrenormalized quark mass $m_q^{\rm AWI}(t)$ from axial Ward
identity \eqref{eq:awi_mass} for bare quark masses $am_q=0.013$ and
$am_q=0.028$ on $16^3\times 32$ at $\beta=3.0$. }  
\label{fig:awi_effmass}
\end{figure}

We first determine the residual additive quark mass
renormalization introduced by the parametrization of the
Fixed-Point Dirac operator. From the axial Ward identity (AWI), the quark mass 
$m_q^{\rm AWI}$ is given by the large $t$ limit of
\begin{equation} \label{eq:awi_mass}
Z_m m_q^{\rm AWI}(t) = \frac{1}{2}\frac{Z_A\sum_{\vec x}
 \langle\partial_4 A_4(\vec x, t)C(0)\rangle}{Z_P\sum_{\vec 
 x} \langle P(\vec x, t)C(0)\rangle},
\end{equation}
where $C(0)$ is a source operator with the quantum numbers of the pion
and $A_4(\vec x, t)$ and $P(\vec x, t)$ are the local fourth
component axial vector and pseudoscalar currents. As seen in
Fig.~\ref{fig:awi_effmass}, the ratio of the two correlators in
Eq.~\eqref{eq:awi_mass} is flat already at small $t$ and can
easily be fitted to a constant also at very small quark masses.
The unrenormalized AWI quark masses $m_q^{\rm AWI}$ extracted
from our data are listed in Appendix \ref{app:awi}.
Although we do not know the renormalization factors $Z_m$, $Z_S$ and
$Z_A$,\footnote{For exactly chiral actions, $Z_m^{-1}=Z_S$, hence
$Z_A$ can be extracted from Eq.~\eqref{eq:awi_mass}.} the
residual quark mass can be determined from the value of the 
bare quark mass $m_q$ where the 
AWI quark mass vanishes, by linearly extrapolating the measured ratio
of correlators to zero. 

The resulting values of $m_{\rm res}^{\rm (AWI)}
= m_q(m_q^{\rm AWI}=0)$ from a linear fit to the six smallest masses
on each lattice are shown in Table \ref{tab:m_res}.  At $\beta=3.0$,
the residual mass for both the parametrized and the
overlap-improved FP Dirac operator is very close to zero, whereas at
larger $\beta$, 
its value is clearly non-zero. This is not surprising, since we have
optimized the parametrization to $\beta\approx 3.0$ by choosing a
set of gauge configurations with $2.7<\beta <3.4$ to fit the FP
relation. While the fluctuations of the smallest eigenvalues of the Dirac
operator decrease with increasing $\beta$, the central value is
slightly shifted away from zero, leading to a non-vanishing value of $m_{\rm
res}^{\rm (AWI)}$ at larger $\beta$.
Obviously the fluctuation polynomials in the parametrized FP Dirac
operator do not fully absorb this $\beta$-dependence of the eigenvalue
spectrum. 

\begin{table}
\begin{center}
\begin{tabular}{c|c|c} \hline\hline
 $\beta$ & Dirac operator & $m_{\rm res}^{\rm (AWI)}$ \\ \hline
 3.0 & parametrized FP & -0.0006(4) \\
 3.0 & overlap-improved FP & -0.0006(1)\\
 3.4 & parametrized FP & -0.0180(4) \\ 
 3.7 & parametrized FP & -0.0194(2) \\ \hline
\end{tabular}
\end{center}
\vspace{-2mm}
\caption{The residual quark mass determined from the axial Ward identity.}
\label{tab:m_res}
\end{table}

\subsection{The Quenched Chiral Log Parameter $\delta$}
Chiral perturbation theory ($\chi$PT) \cite{Gasser:1982ap} allows to
predict the quark mass dependence of hadrons. In particular
the pion, taking the role of the Goldstone boson 
of spontaneously broken chiral symmetry, should be massless in the
chiral limit, and to lowest order in $\chi$PT, it depends linearly on
the quark mass. Taking into account the quenched approximation,
additional logarithmic terms appear:
Quenched chiral perturbation theory predicts the following dependence
of the pseudoscalar meson mass $m_{\rm PS}$ on the two constituent quark masses
$m_1$ and $m_2$ \cite{Bernard:1992mk,Yoshie:2001ts}:
\begin{multline} \label{eq:chipt_pion}
m_{\rm PS}^2 = A(m_1+m_2) \left[ 1 - \delta
\left( \ln\frac{2Am_1}{\Lambda_\chi^2} + \frac{m_2}{m_2-m_1}
\ln\frac{m_2}{m_1} \right) \right] +
B(m_1+m_2)^2 \\ + {\mathcal O}(m_q^3),
\end{multline}
where $\Lambda_\chi$ is a Q$\chi$PT scale and
of order 1 GeV. The term proportional to $\delta$ is divergent in the
chiral limit and is only present due to quenching.
In the  
case of degenerate quark masses, the divergence can be absorbed into a
redefinition of the quark mass by resummation of the
leading logarithms  \cite{Sharpe:1992ft}. This yields a power form
\begin{equation}
 m_{\rm PS}^2 \propto m_q^{\frac{1}{1+\delta} },
\end{equation}  
for the pseudoscalar mass.
To leading order in
a $1/N_c$-expansion, the value of the parameter $\delta$ is given by
\cite{Yoshie:2001ts} 
\begin{equation} \label{eq:delta}
 \delta = \frac{m_{\eta^\prime}^2 + m_\eta^2 - 2m_K^2}{48\pi^2f_\pi^2}
 \approx 0.18,
\end{equation} 
where the pion decay constant is normalized such that $f_\pi = 93$ MeV.

CP-PACS has estimated a value of $\delta = 0.10(2)$ from their analysis 
of pseudoscalar mesons with non-degenerate quarks and $\delta = 0.09(3)$ from
a fit of Eq.~\eqref{eq:chipt_pion} to pseudoscalar
mesons with degenerate quarks, where the scale was varied in the range $
0.6$ GeV$  < \Lambda_\chi <1.4$ GeV. The drawback of their data 
is the use of the Wilson action with its explicit breaking of chiral
symmetry, which allows to cover only a limited range of quark
masses because of the appearance of exceptional
configurations. Recently a technique  
has been proposed for the Wilson action to shift the would-be zero modes,
that fluctuate far along the positive real axis, 
back to zero \cite{Bardeen:1998gv}. This pole-shifting procedure
amounts to a modification of the quenched theory and prevents
exceptional configurations, thus allowing to go almost down to the physical
pion mass even 
with Wilson fermions. However, one has to assume that the most important
effect of explicit chiral symmetry breaking in the Wilson action is
the resulting 
fluctuation of the zero modes on the real axis. The chiral
properties are not improved for other parts of
the spectrum, and therefore it is not clear how much such a punctual
modification helps. In \cite{Bardeen:2000cz}, the chiral log parameter
$\delta$ has been determined in various ways, amongst others also from
the pseudoscalar meson mass, with the pole-shifted Wilson
action and similar simulation parameters like we use here. The quoted
value averaged over the different determinations is  
$\delta=0.065(13)$, which is a factor of three smaller than the
theoretical estimate, but fairly consistent with the CP-PACS result. 

It would obviously be interesting to determine the value of $\delta$ with a
chiral symmetric action, which is free of problems related
to exceptional configurations and does not only cure the explicit
chiral symmetry breaking 
punctually. In \cite{Dong:2001fm,Dong:2001yf,Dong:2001kv}, 
investigations of the chiral limit with the Wilson overlap
action were performed, and varying values of $\delta$ were found, with
a tendency towards larger values than what was obtained with non-chiral
actions. With our data, we are not only able to quite precisely determine the 
value of $\delta$ from the pseudoscalar meson mass, but we can also
compare the results for the approximately chiral parametrized FP
and the overlap-improved Dirac operator, providing two at least
partially independent results with chiral actions.

First we examine the mass of the pseudoscalar meson with degenerate
quarks, calculated from the P and P-S correlators on various lattices at
$\beta=3.0$. We fit the masses to the polynomial forms   
\begin{eqnarray}
(am_{\rm PS})^2 & = & 2Aa^2(m_q+m_{\rm res}) + 4Ba^3(m_q+m_{\rm res})^2,
\label{eq:fit1}\\ 
(am_{\rm PS})^2 & = & 2Aa^2(m_q+m_{\rm res}) + 4Ba^3(m_q+m_{\rm res})^2
\nonumber \label{eq:fit2} \\
& & + 8Ca^4 (m_q+m_{\rm res})^3,
\end{eqnarray}
and to the forms inspired by quenched chiral perturbation theory
\begin{eqnarray}
(am_{\rm PS})^2 & = & 2Aa^2(m_q+m_{\rm res}) \left[ 1 - \delta
\left( \ln\frac{2Aa^2(m_q+m_{\rm res})}{a^2\Lambda_\chi^2} + 1 \right)
\right] \nonumber \label{eq:fit3}\\ 
 & & + 4Ba^3(m_q+m_{\rm res})^2, \\
(am_{\rm PS})^2 & = & 2Aa^2 (m_q+m_{\rm res})^{\frac{1}{1+\delta} },
\label{eq:fit4}\\ 
(am_{\rm PS})^2 & = & 2Aa^2 (m_q+m_{\rm res})^{\frac{1}{1+\delta} } +
4Ba^3(m_q+m_{\rm res})^2 \label{eq:fit5},
\end{eqnarray}
allowing for a residual additive quark mass renormalization $am_{\rm
res} $. The resulting fit parameters for all three lattices are listed
in Table 
\ref{tab:chirfit}. The errors are determined by bootstrap resampling,
calculating on each of the 500 bootstrap samples the meson masses from a
correlated fit and using these for the chiral fit\footnote{We do
not take into account correlations at different quark masses here.}. 

\begin{figure}[tbp]
\begin{center}
\includegraphics[width=12cm]{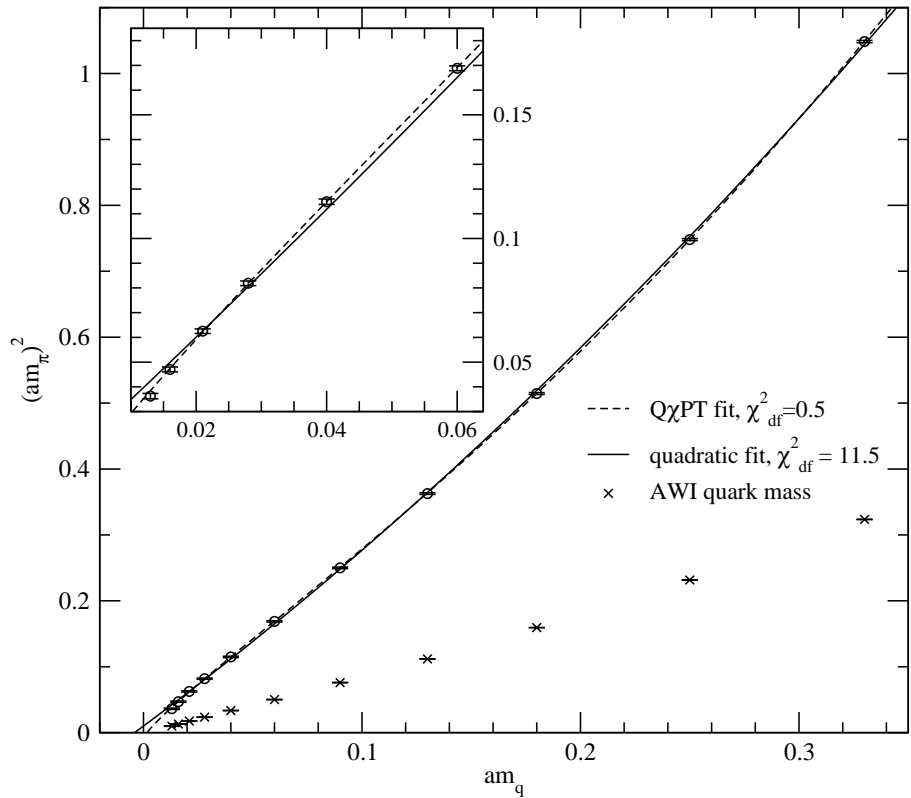}
\end{center}
\vspace{-2mm}
\caption{ Chiral fit to squared pseudoscalar meson mass with a
quadratic function and the logarithmic form predicted by Q$\chi$PT
with $\Lambda_\chi\simeq 1$ GeV, giving $\delta=0.17(1)$. }  
\label{fig:chirlog}
\end{figure}

The most reliable results are obtained on the largest lattice. 
There it is obvious that a quadratic fit with a value of
$\chi^2/df=11.5$ misses the clear negative curvature at small quark
masses in our data. Also a cubic fit with $\chi^2/df=2.9$ can not
account for this curvature well. In contrast, the various Q$\chi$PT fits agree
perfectly with our data, with $\chi^2/df\approx 0.5$. In Fig.~\ref{fig:chirlog}
we compare the logarithmic fit with $\Lambda_\chi=1$ GeV and the
quadratic fit, showing the clear discrepancy when the quenched chiral log is
neglected. 
In Table~\ref{tab:chirfit2} we demonstrate that the results for $\delta$ on
the largest lattice do not differ significantly when 
choosing one single correlator P, A or P-S instead of extracting the
pseudoscalar masses from the P and P-S correlators at different quark mass
according to our proposition.

On the $12\times 24$
lattice, the results for $\delta$ with the
parametrized FP Dirac operator agree very well with those at
$16^3\times 32$, with somewhat 
larger errors. Also 
with the overlap-improved operator, we get consistent values, but the
errors are then even larger due to the smaller statistics.
We compare our values of $\delta$ from the degenerate mesons with
those from \cite{Dong:2001kv} in Table \ref{tab:delta_comp}.
From the results of the fits to the logarithmic form
\eqref{eq:fit3} on the largest lattice, we 
estimate a value of $\delta=0.17(3)$, where the error mostly comes from the
unknown scale $\Lambda_\chi$.

\begin{table}
\begin{center}
\small
\begin{tabular}{l|r|l|r|r|c} \hline\hline
\multicolumn{6}{c}{$16^3\times 32$, $\beta=3.0$, parametrized FP} \\ \hline
 fit form & $ am_{\rm res} $ & $\delta$ & $aA$ & $aB$ & $\chi^2$/df \\ \hline
quadratic \eqref{eq:fit1}& 0.0040(4) &  & 1.22(1) & 0.51(1)   & 11.5(2.0)   \\
cubic \eqref{eq:fit2}, $C=0.55(4)$& 0.0019(5) &  & 1.35(2) & -0.02(5)  & 2.9(8) \\
log \eqref{eq:fit3}, $\Lambda_\chi\simeq 0.6$ GeV& -0.0018(6)& 0.143(9) & 1.46(3) & 0.94(3)  & 0.6(3)  \\
log \eqref{eq:fit3}, $\Lambda_\chi\simeq 0.8$ GeV& -0.0017(5) & 0.157(10) & 1.32(1) & 0.94(3)  & 0.5(2) \\
log \eqref{eq:fit3}, $\Lambda_\chi\simeq 1.0$ GeV & -0.0017(6) & 0.172(12) & 1.21(1) & 0.94(3)  & 0.5(3)\\
log \eqref{eq:fit3}, $\Lambda_\chi\simeq 1.2$ GeV & -0.0017(6) & 0.186(15) & 1.12(1) & 0.94(3)  & 0.5(2)\\
log \eqref{eq:fit3}, $\Lambda_\chi\simeq 1.4$ GeV & -0.0017(6) & 0.200(17) & 1.04(2) & 0.94(3)  & 0.5(2)\\
power \eqref{eq:fit4}, 5 masses& -0.0039(9) & 0.229(40) & 0.86(7)  & & 0.8(5) \\
power \eqref{eq:fit4}, 6 masses & -0.0024(9) & 0.132(30) & 1.05(6)  & & 0.9(4) \\
%power \eqref{eq:fit4}, 7 masses & -0.0008(8) & 0.065(16) & 1.21(3)  & & 1.3(5) \\
power+quadr. \eqref{eq:fit5} & -0.0026(7) & 0.194(16) & 0.86(2) & 0.9(2)  &
0.4(2) \\ \hline
\end{tabular}

\vspace{3ex}

\begin{tabular}{l|r|l|r|r|c} \hline\hline
\multicolumn{6}{c}{$12^3\times 24$, $\beta=3.0$, parametrized FP} \\ \hline
 fit form & $ am_{\rm res} $ & $\delta$ & $aA $& $aB$ & $\chi^2$/df \\ \hline
 quadratic \eqref{eq:fit1} &  0.0043(11) & & 1.21(2) & 0.53(2) & 4.7(1.4)   \\
 cubic \eqref{eq:fit2}, $C=0.53(8)$  & 0.0009(14) & & 1.35(4) & 0.0(1) & 1.0(4) \\
log \eqref{eq:fit3}, $\Lambda_\chi\simeq 0.8$ GeV& -0.0035(16) &
 0.165(18) & 1.34(3) & 0.95(5) & 0.2(1) \\
log \eqref{eq:fit3}, $\Lambda_\chi\simeq 1.0$ GeV & -0.0035(15) &
 0.181(21) & 1.22(2) & 0.95(4) & 0.2(2) \\
log \eqref{eq:fit3}, $\Lambda_\chi\simeq 1.2$ GeV & -0.0035(15) &
 0.196(25) & 1.13(1) & 0.95(4) & 0.2(2) \\
power \eqref{eq:fit4}, 5 masses & -0.0050(15) & 0.164(72) & 1.0(1) &
 & 0.1(1) \\
 power \eqref{eq:fit4}, 6 masses & -0.0029(17) & 0.079(26) & 1.19(5) &
 & 0.3(4) \\
power+quadr. \eqref{eq:fit5} & -0.0045(17) & 0.209(28) & 0.86(3) &
 0.87(4) & 0.1(1) \\ \hline
\end{tabular}

\vspace{3ex}

\begin{tabular}{l|r|l|r|r|c} \hline\hline
\multicolumn{6}{c}{$12^3\times 24$, $\beta=3.0$, overlap-improved} \\ \hline
 fit form & $ am_{\rm res} $ & $\delta$ & $aA$ & $aB$ & $\chi^2$/df \\ \hline
 quadratic \eqref{eq:fit1} &  0.0036(20) & & 1.36(3) & 0.52(3) & 1.3(8)   \\
 cubic \eqref{eq:fit2}, $C=0.55(17)$  & 0.0011(20) & & 1.51(6) & 0.0(2) & 0.4(4) \\
log \eqref{eq:fit3}, $\Lambda_\chi\simeq 0.8$ GeV& -0.0020(22) &
 0.141(35) & 1.50(5) & 0.93(10) & 0.1(2) \\
log \eqref{eq:fit3}, $\Lambda_\chi\simeq 1.0$ GeV & -0.0021(23) &
 0.153(45) & 1.39(3) & 0.94(12) & 0.1(2) \\
log \eqref{eq:fit3}, $\Lambda_\chi\simeq 1.2$ GeV & -0.0020(23) &
 0.163(51) & 1.30(3) & 0.93(12) & 0.1(2) \\
power \eqref{eq:fit4}, 6 masses & -0.0039(24) & 0.21(14) & 1.0(3) &
 & 0.0(1) \\
 power \eqref{eq:fit4}, 7 masses & -0.0037(23) & 0.19(9) & 1.0(2) &
 & 0.0(1) \\
power+quadr. \eqref{eq:fit5} & -0.0020(24) & 0.162(50) & 1.30(3) &
 0.93(11) & 0.1(2) \\ \hline
\end{tabular}
\end{center}
\vspace{-1mm}
\caption{Fit parameters for different forms of the chiral fit to the squared
pseudoscalar meson mass on $\beta=3.0$ lattices. The power form
\eqref{eq:fit4} was fitted 
only to the few smallest masses, where the quadratic dependence is
negligible.}  
\label{tab:chirfit}
\end{table}

\begin{table}
\begin{center}
\small
\begin{tabular}{c|r|l|r|r|c} \hline\hline
\multicolumn{6}{c}{$16^3\times 32$, $\beta=3.0$, parametrized FP} \\ \hline
 correlator & $ am_{\rm res} $ & $\delta$ & $aA$ & $aB$ & $\chi^2$/df \\ \hline
P & -0.0004(6) & 0.158(13) & 1.19(1) & 0.91(3)  & 0.4(3)\\
A & -0.0002(6) & 0.158(13) & 1.18(1) & 0.92(3)  & 0.4(2)\\
P-S & -0.0018(6) & 0.169(15) & 1.22(1) & 0.93(4)  & 0.4(2)\\
P-S at $am_q\leq 0.04$, P else & -0.0017(6) & 0.172(12) & 1.21(1) &
0.94(3)  & 0.5(3)\\ \hline
\end{tabular}
\end{center}
\vspace{-1mm}
\caption{Fit parameters for logarithmic fit with $\Lambda_\chi=1$ GeV
to the squared 
pseudoscalar meson mass on the largest lattice in dependence of the
chosen correlators.} 
\label{tab:chirfit2}
\end{table}

\begin{table}
\begin{center}
\begin{tabular}{c|c|c} \hline\hline
 $\Lambda_\chi$ & $\delta$ (Dong et al. \cite{Dong:2001kv}) & $\delta$
(this work) \\ \hline
 0.6 GeV & 0.23(7)  & 0.143(9) \\
 0.8 GeV & 0.28(11) & 0.157(10) \\
 1.0 GeV & 0.34(17) & 0.172(12) \\ \hline
\end{tabular}
\end{center}
\vspace{-2mm}
\caption{The quenched chiral log parameter $\delta$ from a logarithmic
Q$\chi$PT fit \eqref{eq:fit3} to the pseudoscalar meson mass.}
\label{tab:delta_comp}
\end{table}

\begin{figure}[tbp]
\begin{center}
\includegraphics[width=9cm]{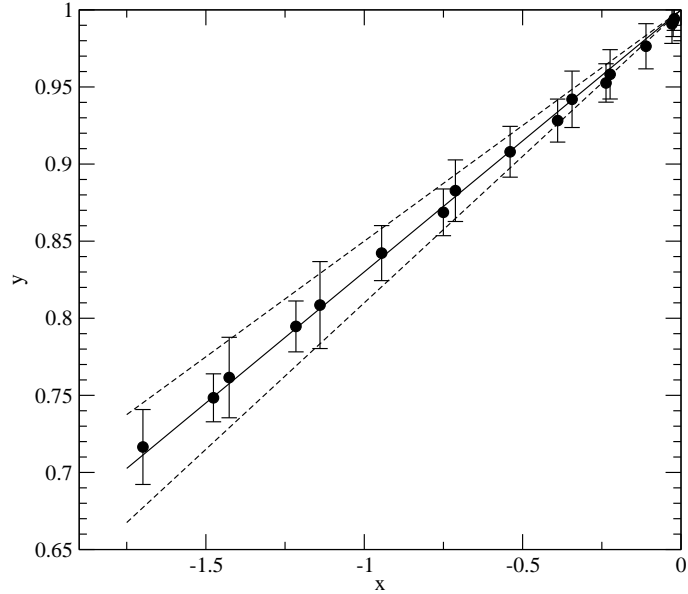}
\end{center}
\vspace{-5mm}
\caption{ The quenched chiral log parameter $\delta$ from cross ratios
\eqref{eq:x_delta}, \eqref{eq:y_delta} of non-degenerate
pseudoscalar meson masses. The  
solid line is a least-squares fit, where the slope gives $\delta=0.17$. The
dashed lines correspond to $\delta=0.15$ and $\delta=0.19$, respectively.}  
\label{fig:delta_nondeg}
\end{figure}

The quenched chiral log parameter $\delta$ can also be determined from
an analysis of the 
non-degenerate pseudoscalar meson masses. The two
quantities
\begin{eqnarray} \label{eq:x_delta}
 x = 2 + \frac{m_1+m_2}{m_1-m_2} \ln\left(\frac{m_2}{m_1}\right), \\
 \label{eq:y_delta} 
 y = \frac{2m_1}{m_1+m_2} \frac{m_{\rm PS,12}^2}{m_{\rm PS,11}^2}
 \cdot \frac{2m_2}{m_1+m_2} \frac{m_{\rm PS,12}^2}{m_{\rm PS,22}^2},
\end{eqnarray}
are related by $y=1 + \delta x + {\mathcal O}(m^2)$, and the terms
proportional to $B(m_1+m_2)$ in Eq.~\eqref{eq:chipt_pion}
cancel \cite{Yoshie:2001ts}. To avoid problems with the residual additive
renormalization, we use the axial
Ward identity quark masses \eqref{eq:awi_mass} as an input for $m_1$
and $m_2$.  
Fig.~\ref{fig:delta_nondeg} shows our data for the non-degenerate
pseudoscalar mesons from the P correlator on the largest
lattice. Only the points from mesons with light enough quarks that the
linear dependence of $y(x)$ remains valid are plotted. For mesons with two
heavy quarks, we encountered a systematic deviation towards smaller values
of $y$ for all $x$, therefore they are omitted here. The slope
of $y(x)$, giving $\delta = 0.170(20)$, is beautifully 
consistent with the above value for the equal quark mass case and
with the theoretical prediction in Eq.~\eqref{eq:delta}.

Taking the intersection of the two determinations of the chiral log
parameter from degenerate and non-degenerate pseudoscalar
mesons, we obtain a value of $\delta=0.17(2)$. In
Table~\ref{tab:delta_comp2}, we compare this final result to
previous determinations from other groups. While our result is
measured at
finite lattice spacing $a\approx 0.16$ fm, the $a$-dependence of
$\delta$ appears to be small \cite{Yoshie:2001ts}.

\begin{table}
\begin{center}
\vspace{2mm}
\begin{tabular}{l|l|c|c} \hline\hline
 Group & year & Fermion action & $\delta$ \\ \hline
 JLQCD \cite{Aoki:1997bs} & 1996 & staggered & 0.05--0.10 \\
 CP-PACS \cite{Yoshie:2001ts} & 1998 & Wilson & 0.10(2) \\
 Bardeen et al. \cite{Bardeen:2000cz} & 2000  & pole-shifted Wilson & 0.065(13) \\
 Dong et al. \cite{Dong:2001kv,Dong:2001yf} & 2001 & Wilson overlap &
 0.23--0.48 \\ 
 This work & 2002 & Fixed-Point & 0.17(2) \\ \hline
\end{tabular}
\end{center}
\vspace{-2mm}
\caption{Compilation of recent results for the quenched chiral log
 para\-me\-ter~$\delta$. Earlier results are reported in \cite{Sharpe:1997ih}.} 
\label{tab:delta_comp2}
\end{table}

As a by-product of this analysis, we estimate a value of $am_{\rm
 res}^{\rm (PS)}=-0.002(2)$ for the residual
 quark mass of the parametrized FP Dirac operator, covering the
 various quenched 
 fits to the pseudoscalar meson mass at $\beta=3.0$. This value
 agrees well with the residual mass 
 determined from the axial Ward identity in Table \ref{tab:m_res},
 which is clearly not the case for $am_{\rm res}$ resulting from quadratic or
 cubic fits to $(am_{\rm PS})^2$. At the
 other values of the gauge coupling $\beta=3.4$
and $3.7$, we get larger values of $am_{\rm res}^{\rm (PS)} =
 -0.016(4)$ and $ -0.012(7)$ respectively, in reasonable
 agreement with the results from the axial Ward identity. This
 coincidence confirms that measuring the pseudoscalar particle
 from the P correlator at large and the P-S correlator at small
 quark mass is a reasonable solution to avoid zero mode effects in the
 pseudoscalar mass.

\subsection{Chiral Extrapolations for Vector Mesons and Baryons}
For vector mesons and baryons, Q$\chi$PT predicts in the continuum limit
\cite{Booth:1997hk,Labrenz:1996jy} 
\begin{equation} \label{eq:chirfit_baryons}
 m(m_{\rm PS}) = m_0 + \delta C_{1/2} m_{\rm PS} + C_1 m_{\rm PS}^2 +
 C_{3/2} m_{\rm PS}^3 + \dots,
\end{equation}
where the $C_i$ are functions of the
coefficients in the quenched chiral Lagrangian. Like in the formula for
the pseudoscalar channel \eqref{eq:chipt_pion},
the term proportional to $\delta$ appears only in the quenched
theory, and according to Q$\chi$PT its coefficient $C_{1/2}$ is negative.

\begin{figure}[tbp]
\begin{center}
\includegraphics[width=9cm]{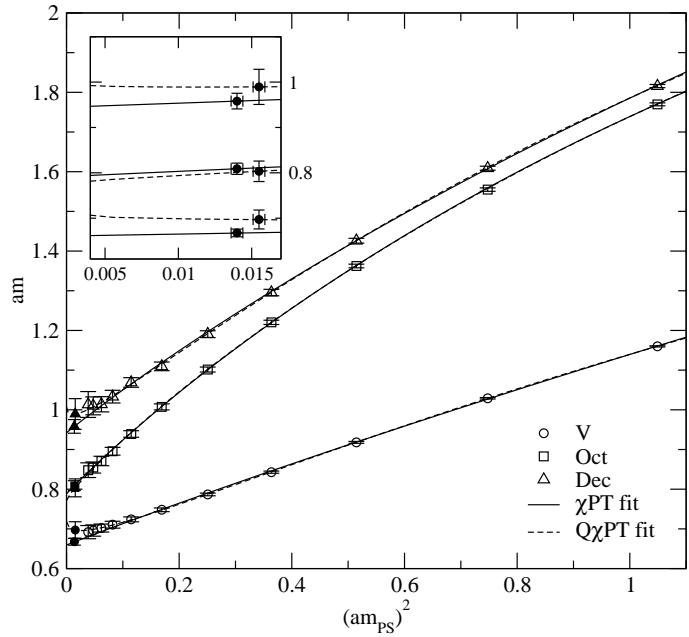}
\end{center}
\vspace{-3mm}
\caption{Chiral extrapolations of vector meson and baryons on the $16\times 32$
lattice at $\beta=3.0$. The solid line is a fit without and the dashed
line a fit including the quenched term in
Eq.~\eqref{eq:chirfit_baryons}. The inset shows the resulting physical
$\rho$, $N$ and $\Delta$ masses at the average up and down quark mass.}  
\label{fig:chir_exp}
\end{figure}

On our largest lattice, we perform chiral extrapolations for the 
vector meson and the octet and decuplet baryons both with and without
the quenched term. The resulting fits are plotted in
Fig~\ref{fig:chir_exp}. For the vector meson, there is some evidence
for the presence of a term linear in $m_{\rm PS}$ with a negative
coefficient, as expected from quenched chiral perturbation theory. The
$\rho$ meson mass from the extrapolation to 
the physical quark mass, defined by $m_{\rm PS}/m_{\rm V}
\approx 0.18$, is used to fix the lattice scale. For the fit including
the quenched term, we get  
$a^{-1} = 1104(32)$~MeV or $a=0.179(4) $~fm, while when setting
$C_{1/2}=0$, we obtain $a^{-1} = 1152(15)$~MeV or 
$a=0.171(2)$~fm. The two values do not
completely coincide, showing that the functional form of the chiral
extrapolation indeed can lead to different results. We use in the following
the value obtained from the quenched fit, which includes in its error
some of the uncertainty in the presence and size of the quenched
term.\footnote{The lattice spacing fixed from the $\rho$ meson mass 
turns out to be somewhat larger than when fixed from $r_0$,
where we obtain $a=0.16(1)$~fm. This uncertainty in the scale
determination is a well-known problem in quenched QCD.}

For the baryons, neither of the two functional forms is clearly favored by our
data. The decuplet mass  
shows some upward curvature at the smallest quark masses, but the
errors are quite large and do not cover the systematic uncertainty in
choosing the fit range, which is increasingly difficult at small quark
mass. What is however evident is the negative curvature in both
baryon masses that gets absorbed by the $\ordo(m_{\rm
PS}^3)$ term, which is also present in ordinary chiral perturbation theory.

\begin{figure}[tbp]
\begin{center}
\includegraphics[width=9cm]{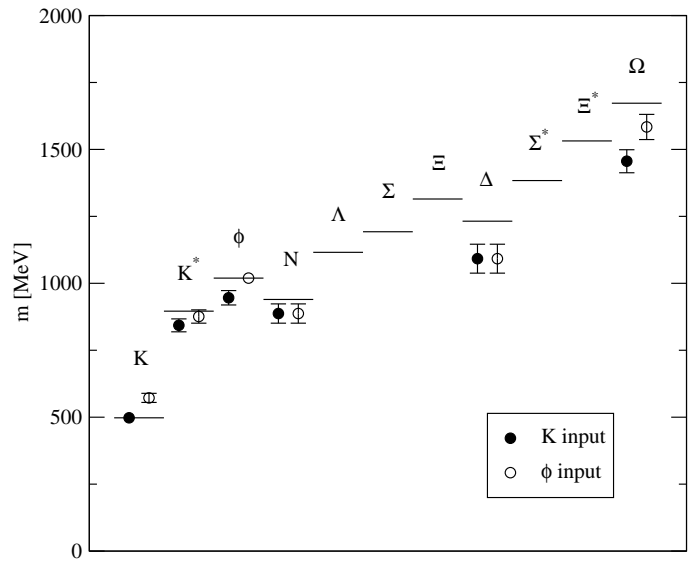}
\end{center}
%\vspace{-3mm}
\caption{Part of the quenched light hadron spectrum
at finite lattice spacing $a\approx 0.16$ fm. The quark mass
$(m_u+m_d)/2$ and the lattice spacing $a$ are determined from the $\pi$
and $\rho$ meson, and the strange quark mass $m_s$ is either
fixed from the $K$ meson (filled symbols) or the $\phi$ meson
(open symbols).}  
\label{fig:hadron_spectrum}
\end{figure}

From a partial analysis of our data, we show in
Fig.~\ref{fig:hadron_spectrum} 
the mass spectrum of hadrons with degenerate light quarks at
$\beta=3.0$ and lattice size $16^3\times 32$. The familiar ambiguity
from fixing the strange 
quark mass either with the $K$ or the $\phi$ meson is evident, and for
both choices the meson hyperfine splitting turns out to be too
small. The chiral 
extrapolation for the baryons, where the quenched term is included, leads to a
nucleon mass which almost agrees with the experimental value, while the
$\Delta$ and $\Omega$ baryon masses come out too small. However, since we are
at finite lattice 
spacing, this discrepancy will not necessarily persist after
a continuum extrapolation.

\section{Physical Finite Size Effects}
When the spatial lattice size is too small to accommodate the wave function
of a hadron, its mass suffers from finite-size corrections. Results
from quenched simulations show that for lattices larger than 2 fm, the
finite volume effects are smaller than 2\% \cite{Bernard:1998qf}. As
only our largest lattice with size $L_s \approx 2.5$ fm
fulfills this requirement, we expect to see significant finite volume
effects in our data on the smaller lattices. Plotting the ratio of
octet baryon to vector 
meson masses in an Edinburgh plot (Fig.~\ref{fig:edinburgh}), we see that
indeed on the 
lattices with spatial size $L_s \approx 1.3$ fm, this ratio stays more or
less constant at $m_{\rm Oct}/m_{\rm V} \approx 1.5$ over the whole range
of quark masses, thus being far too large in the chiral
limit. However, already at lattice size $L_s \approx 1.9$ fm, the results 
do not differ anymore from those on the largest lattice within statistical
errors. 

\begin{figure}[tbp]
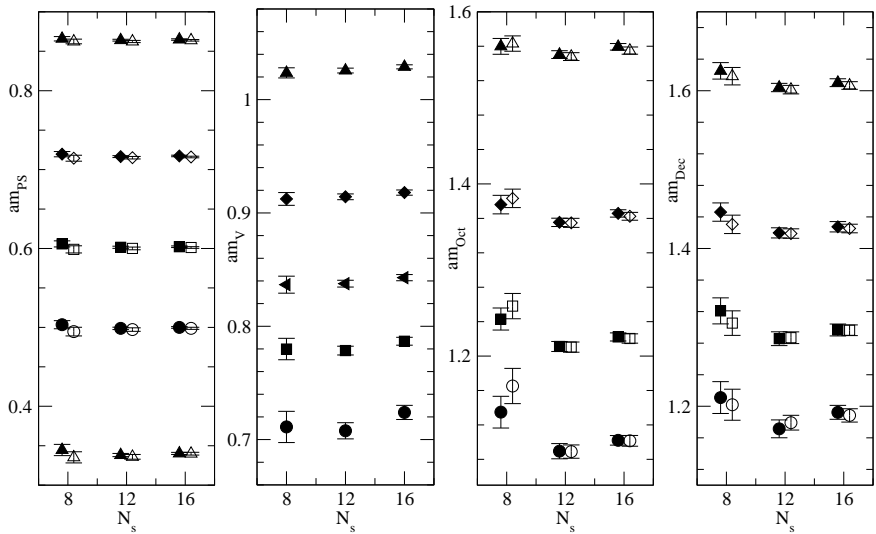

\begin{center}
\includegraphics[width=28mm]{epsf/vol.PS.eps}
\includegraphics[width=28mm]{epsf/vol.V.eps}
\includegraphics[width=28mm]{epsf/vol.N.eps}
\includegraphics[width=28mm]{epsf/vol.D.eps}
\end{center}
\vspace{-5mm}
\caption{Finite volume effects in meson and baryon masses at
$\beta=3.0$. The quark masses are from top to bottom $am_q=0.25$, 0.18,
0.13, 0.09 and for the mesons also $am_q=0.04$. For the
pseudoscalar meson, masses from the A (open symbols) and P (filled)
correlators are shown. For the baryons, full symbols are from the N and
D and open symbols from the N0 and D0 correlators.}
\label{fig:vol}
\end{figure}

We investigate the finite volume effects more closely in
Fig.~\ref{fig:vol} by plotting
the masses of all hadrons against the spatial lattice size. For the
pseudoscalar meson, only the zero mode effects can be seen, leading to
different results for different correlators. Also for the vector
meson, there are no obvious finite-volume effects. The situation is
different for the baryons: On the smallest lattice, the octet mass
increases strongly, independent of the choice of correlator. The same
happens with the decuplet mass, where the effect is slightly smaller. 

We conclude from our data that at $\beta=3.0$, the finite-volume
effects become comparable to our statistical errors already at $N_s=12$,
corresponding to a spatial size of $L_s \approx 1.9$ fm. At
$L_s\approx 1.3$ fm, the baryons are strongly affected by the small
physical volume. The size of the largest lattice $L_s\approx 2.5$ fm
is big enough to provide results which are not affected by
finite-volume effects.

\section{Scaling Properties}
Let us turn to the investigation of the scaling behavior of the
parametrized and overlap-improved FP fermion actions.
A standard test for scaling violations of a given
lattice action is to plot the vector meson mass, which is known to be
particularly sensitive to cut-off effects, against the lattice
spacing. Most groups use in this context the string tension to fix the
scale. Except at $\beta=3.4$, we do not have a direct
calculation of the string tension for our gauge action. Therefore
we use the determination of $r_0/a$ from \cite{urs:diss} instead, where the
interpolating formula\footnote{At the value of $\beta=3.7$, an
extrapolation is needed, because the formula was fitted to the range $2.36
\leq \beta \leq 3.40$. We took this into account by applying increasing
errors of 0.5\%, 0.6\% and 1\% to the value of $r_0/a$ at increasing $\beta$.}
\begin{equation}
 \ln (a/r_0) = -1.1539(18) - 1.0932(68)(\beta - 3) + 0.132(11)(\beta - 3)^2,
\end{equation}
is given and $r_0/\sqrt{\sigma}$ is measured, to set the
scale from the  
string tension. Table~\ref{tab:rho_scaling} shows our 
measurements of the vector meson mass interpolated to\footnote{The
interpolation of the hadron masses is done by fitting 
the Q$\chi$PT formulae \eqref{eq:fit3} and \eqref{eq:chirfit_baryons} to the
measured masses, with bootstrap resampling to determine the errors.} $m_{\rm
PS}/m_{\rm V} = 0.7$ and the resulting value of $m_V/\sqrt\sigma$. 

\begin{figure}[tbp]
\begin{center}
\includegraphics[width=11cm]{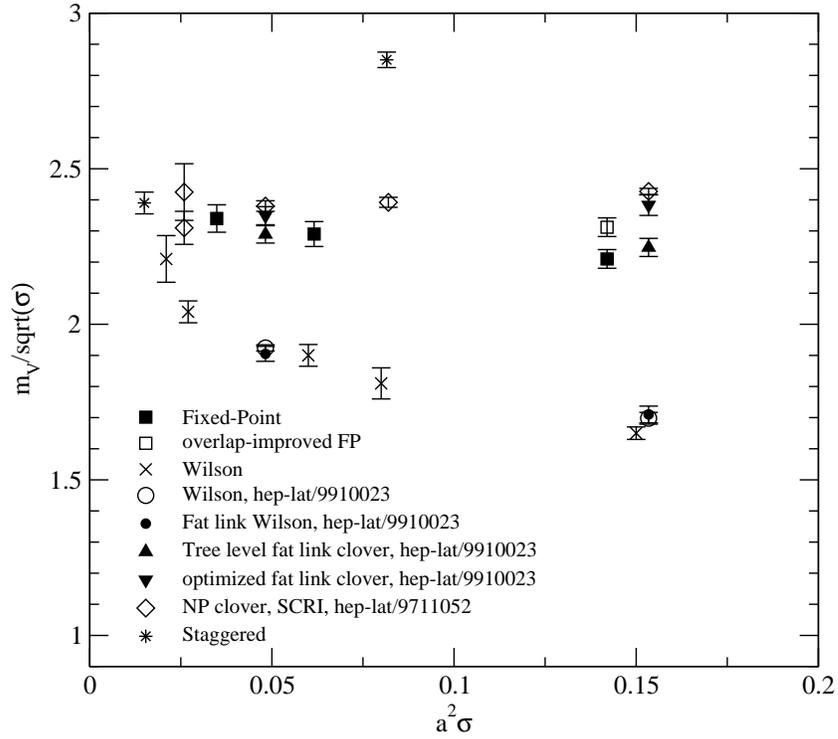}
\end{center}
\caption{Scaling of the vector meson mass. Compared with the FP action is
data from Wilson, staggered and various variants of clover
actions. Filled and open boxes are data from this work. Crosses 
(Wilson) and stars (staggered) are from the compilation of Sharpe
\cite{Sharpe:1998hh}, open diamonds (NP clover) are from
\cite{Edwards:1998nh} and the fat link data and the open circles
(Wilson) are from \cite{Stephenson:1999ns}. The rightmost point for the
FP action is from a larger physical lattice size (with $N_s=16$) than
the others, as there is no significant volume dependence seen in the data in
Table~\ref{tab:rho_scaling}.}  
\label{fig:rho_scaling}
\end{figure}

Our data is plotted in comparison with results from other 
fermion actions in Fig.~\ref{fig:rho_scaling}. Wilson fermions have
large $\mathcal{O}(a)$ and 
 unimproved staggered fermions large $\mathcal{O}(a^2)$ effects, which
are clearly seen in the scaling of the vector meson mass. Fat
 links do not help to improve the situation. For
 the various types of clover actions shown, only small
$\mathcal{O}(a^2)$ effects remain, 
 as it also seems to be the case for the FP action. While the
 parametrization of the FP action introduces cut-off effects to all
 orders, we do not see evidence for $\mathcal{O}(a)$ effects here. 

\begin{figure}[tbp]
\begin{center}
\includegraphics[width=9cm]{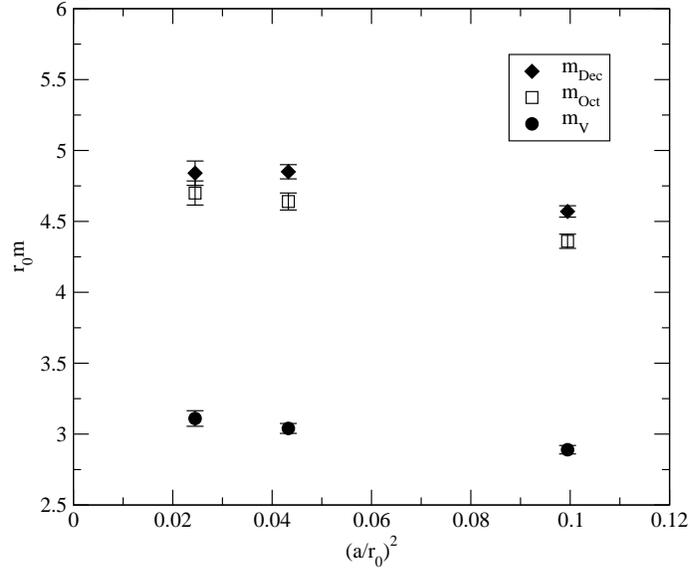}
\end{center}
\vspace{-2mm}
\caption{Scaling of hadrons on spatial lattice size $L_s\approx 1.3$~fm at $m_{\rm
PS}/m_{\rm V} =0.78$ for the parametrized FP Dirac operator.} 
\label{fig:scaling2}
\end{figure}

Because the
 conversion to the string tension 
 introduces an additional error, it is not clear from this plot whether the
 scaling violations of the FP action are significant. We therefore
 determine the hadron masses in units of $r_0$ and plot them in
 Fig.~\ref{fig:scaling2}. For all hadrons under consideration, the
point at $\beta=3.0$ 
 does not coincide within errors with the data at smaller lattice
 spacings. One has to take into account that the statistical error in
$a/r_0$ does not fully cover the systematic uncertainty in
determining the lattice scale at this $\beta$. Furthermore a scaling
study in such a small physical volume encompasses the danger of substantial
contributions from small differences in the lattice volume at the
various values of the gauge coupling, because the volume dependence
of the hadron masses at $L_s\approx 1.3$ is on the onset of getting
strong \cite{Aoki:2000kp}.
Whether our data indicates the presence of small
$\mathcal{O}(a^2)$  cut-off effects in our parametrization of
the FP fermion action or results from the determination of the Sommer
scale $r_0$ and the related uncertainty in the lattice volume
therefore needs further investigation. 

\begin{figure}[tbp]
\begin{center}
\includegraphics[width=10cm]{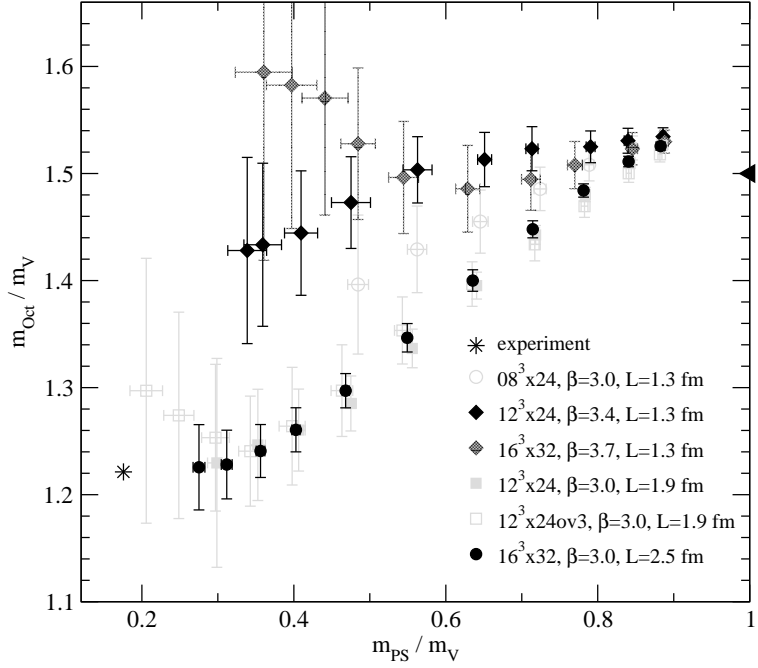}
\end{center}
\vspace{-5mm}
\caption{Our data compiled into an Edinburgh plot. Shown are the
experimental point (star) and the theoretical prediction $m_{\rm
Oct}/m_{\rm V}=1.5$ for the large mass limit (triangle). The three
lattices with smallest physical volume show large finite-volume
effects due to the increased mass of the octet baryon.}
\label{fig:edinburgh}
\end{figure}

\begin{figure}[tbp]
\begin{center}
\includegraphics[width=85mm]{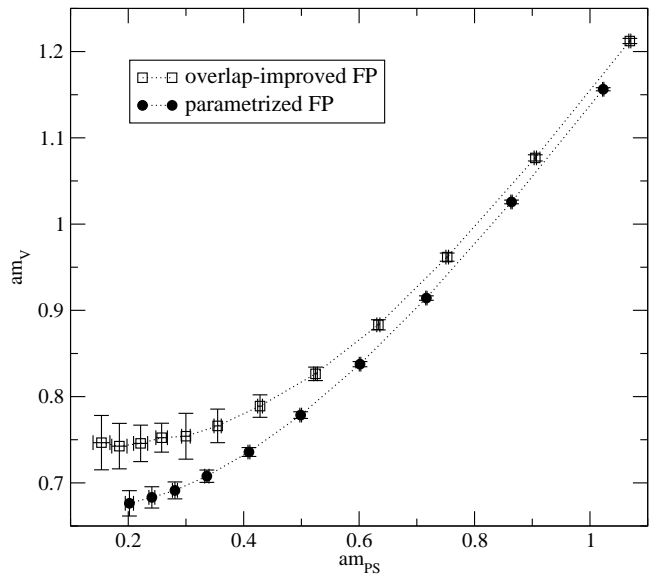}
\end{center}
\vspace{-2mm}
\caption{Comparison of vector meson mass for parametrized FP and
overlap-improved FP Dirac operators.} 
\label{fig:PvsV}
\end{figure}

A striking observation is that the vector meson mass turns out to be
significantly larger for the overlap-improved Dirac operator than for the FP
operator. We illustrate this by 
plotting $am_{\rm V}$ against $am_{\rm PS}$ in Fig.~\ref{fig:PvsV}. Over the
whole range of quark masses covered, a clear discrepancy is seen. This
is to some extent surprising, since the 
overlap expansion is only carried out to third order and therefore one
might expect that the properties of the input operator are only slightly
changed. The discrepancy vanishes in the Edinburgh plot
Fig.~\ref{fig:edinburgh}, where the results on the $12^3\times 24$
lattice agree for the FP and overlap-improved operators. The
difference in the cut-off effects thus cancels in these mass ratios.

From our results we therefore find that the
overlap-improvement changes the scaling behavior of the fermion
action. In Fig.~\ref{fig:rho_scaling}, the point for the
overlap-improved operator at 
$\beta=3.0$ seems to agree with the continuum value obtained
by extrapolating all the data to a single point at $a^2\sigma=0$, but this
coincidence can be misleading due to a possible overall shift from the
scale determination. To conclude whether the scaling violations are
decreased or increased by the overlap, additional measurements with
the overlap-improved FP Dirac operator at different 
values of $\beta$ are needed.

\begin{table}
\begin{center}
\small
\begin{tabular}{c|c|c|c|c|c|c|c} \hline\hline
$\beta$ & $N_s$ & $D$ & $a/r_0$ & $a\sqrt{\sigma}$  &
$am_q$ & $am_{\rm V}$ & $m_{\rm V}/\sqrt{\sigma}$ \\ \hline
3.7 & 16 & FP & 0.1565(16) & 0.187(2) & 0.076(3) & 0.438(7) & 2.34(4.4) \\ 
3.4 & 12 & FP & 0.2080(13) & 0.248(3) & 0.095(2) & 0.568(6) & 2.29(4) \\
3.0 & 8 & FP & 0.3154(16) & 0.376(4) & 0.123(2) & 0.824(10) & 2.19(4) \\ \hline
3.0 & 12 & FP & 0.3154(16) & 0.376(4) & 0.123(2) & 0.827(4) & 2.20(3) \\  
3.0 & 16 & FP & 0.3154(16) & 0.376(4) & 0.123(2) & 0.833(4) & 2.21(3) \\ \hline
3.0 & 12 & ov & 0.3154(16) & 0.376(4) & 0.120(3) & 0.871(9) & 2.31(3) \\ 
\hline
% $am_{\rm Oct}$ &  $m_{\rmOctV}/\sqrt{\sigma}$
% 0.665(10) & 3.56(11)\\
% 0.861(10)  & 3.47(10)\\
% 1.220(16) & 3.24(9) \\
\end{tabular}
\end{center}
\vspace{-2mm}
\caption{Data for the determination of the scaling violations in the
vector meson mass at $m_{\rm PS}/m_{\rm V} =0.7 $. To convert the scale to the string tension, we take
the value of $r_0\sqrt{\sigma}=1.193(10) $ from \cite{urs:diss}, which
was shown not to depend on $\beta$ significantly for the FP gauge action.} 
\label{tab:rho_scaling}
\end{table}

\begin{table}
\begin{center}
\small
\vspace{2mm}
\begin{tabular}{c|c|c|c|c|c|c|c|c} \hline\hline
$\beta$ & $N_s$ & $am_q$ & $am_{\rm V}$ & $r_0 m_{\rm V} $ & $am_{\rm
Oct}$ & $r_0 m_{\rm Oct} $ & $am_{\rm Dec}$ & $ r_0 m_{\rm Dec}$ \\ \hline
3.7 & 16 & 0.105(2) & 0.487(7) & 3.11(5) & 0.736(11) & 4.70(9) &
0.757(11) & 4.84(9) \\ 
3.4 & 12 & 0.133(2) & 0.632(6) & 3.04(3) & 0.965(10) & 4.64(6) &
1.009(9) & 4.85(5)\\ 
3.0 & 8 & 0.179(3) & 0.910(8) & 2.89(3) & 1.374(14) & 4.36(5) &
1.440(11) & 4.57(4) \\ 
\hline
\end{tabular}
\end{center}
\vspace{-2mm}
\caption{Scaling of hadron masses in units of $r_0$ at $m_{\rm
PS}/m_{\rm V} =0.78$ and lattice size $L_s\approx 1.3$ fm.} 
\label{tab:scaling2}
\end{table}

\section{Hadron Dispersion Relations}
\begin{figure}[tbp]
\begin{center}
\includegraphics[width=8cm]{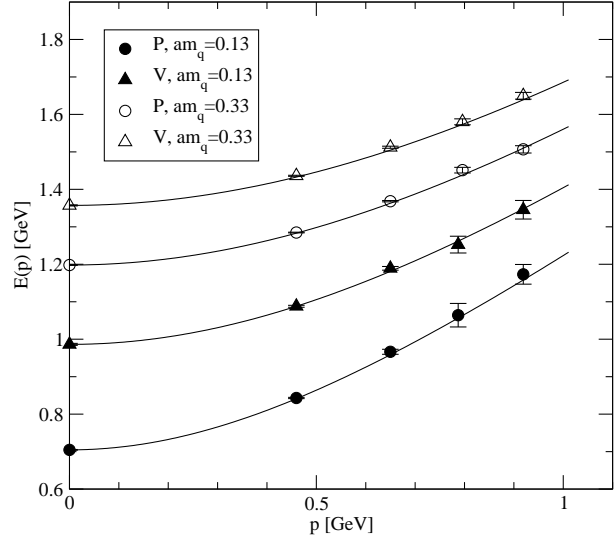}
\end{center} 
\vspace{-2mm}
\caption{Energy-momentum dispersion relation $E(\vec p)$ for pseudoscalar and
vector mesons on the $16^3\times 32$ lattice with $\beta=3.0$. The
solid lines show the continuum value with $E(0)$ given by the measured mass.}
\label{fig:disprel}
\end{figure}

Another quantity that measures the magnitude of
discretization errors of a given action is the energy-momentum
dispersion relation for hadrons $E(\vec p) = m^2 + \vec p\, ^2$, or
equivalently the squared speed of light 
\begin{equation}
 c^2(\vec p\,) = \frac{E^2(\vec p\,) - m^2}{\vec p\,^2},
\end{equation}
which should be $c^2=1$ for all momenta according to the continuum
dispersion relation. At large lattice 
spacings, unimproved lattice actions are known to suffer from substantial
deviations from the continuum relation. Even for
$\mathcal{O}(a)$-improved clover fermions, $c^2$
deviates by 20\%--30\% at $a=0.25$ fm and $p\approx 0.6$ GeV for
pseudoscalar and vector mesons \cite{Alford:1998yy}.
Our coarsest lattice spacing is $a=0.16$ fm, therefore we can not
compare our results directly to this data. However, the
energy-momentum dispersion relation for our 
pseudoscalar and vector mesons calculated with the parametrized FP
Dirac operator on the lattice of size $16^3\times 32$, shown in
Fig.~\ref{fig:disprel}, is consistent
with $c^2=1$ over the whole range of momenta considered. The
parametrization of the Fixed-Point Dirac operator therefore seems to
conserve the classically perfect properties very well also for the
dispersion relation. 

\begin{figure}[tbp]
\begin{center}
\includegraphics[width=9cm]{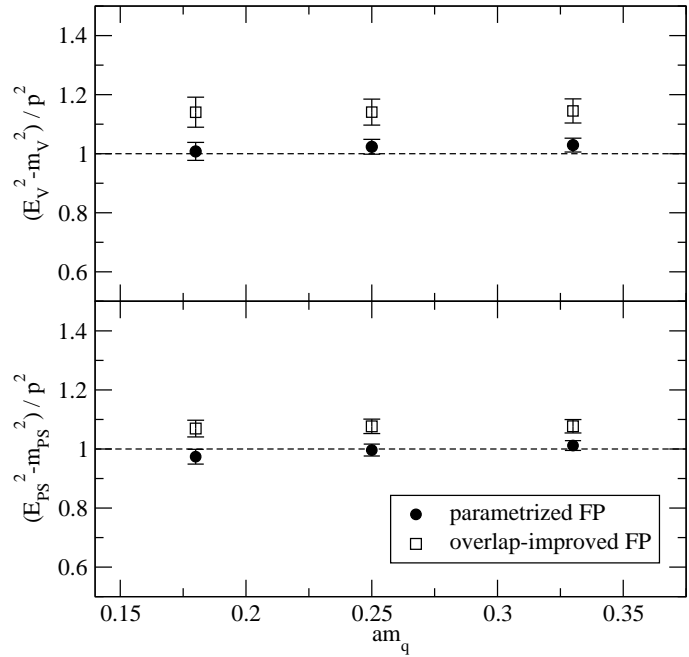}
\end{center}
\vspace{-2mm}
\caption{Comparison of squared speed of light $c^2$ from meson dispersion
relations at the smallest non-zero momentum
for parametrized FP and overlap-improved Dirac operators on the
$12^3\times 24$ lattice with $\beta=3.0$.} 
\label{fig:c_fpov}
\end{figure}

It is interesting to check how
the overlap-improved FP Dirac operator performs. We plot the squared speed of
light at the smallest non-zero momentum on the $12^3\times 24$ 
lattice with $\beta=3.0$ in Fig.~\ref{fig:c_fpov}, and compare the
results from the parametrized FP and the overlap-improved FP Dirac
operator at the 
three largest quark masses. While the data from the FP operator again agrees
with $c^2=1$ within errors for both 
mesons, the results from the overlap-improved operator are too large by
7(3)\% for the pseudoscalar and 14(5)\% for the vector meson. The
overlap construction therefore seems to drive the hadronic dispersion
relation of the FP operator away from the continuum form. The
situation appears to be analogous to the case of the free Wilson
operator, where the dispersion
relation is also deteriorated by the overlap \cite{Niedermayer:1998bi}.

\begin{figure}[tb]
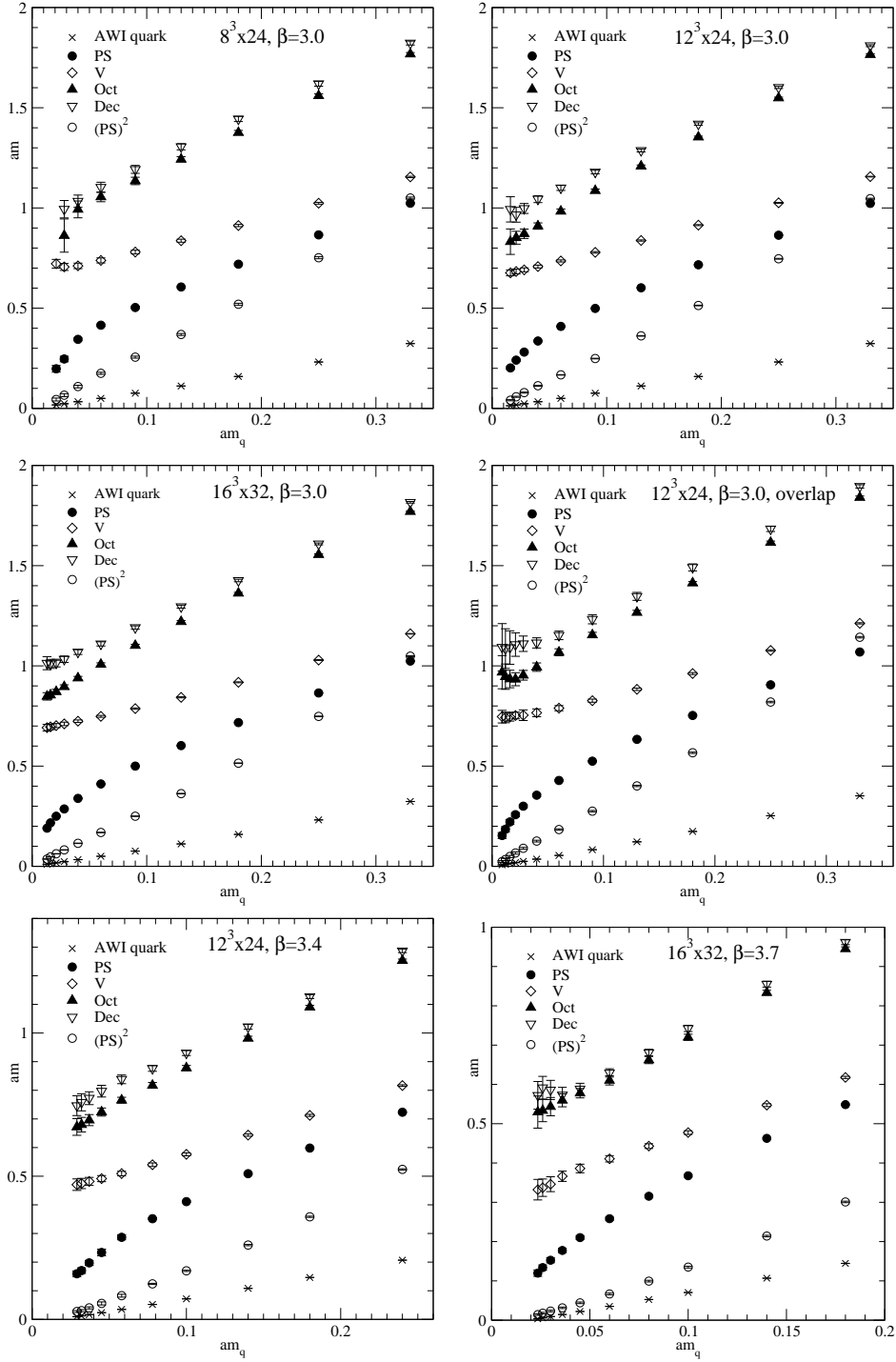

%\begin{center}
\begin{tabular}{cc}
\hspace{-5mm}
\includegraphics[width=60mm]{epsf/hadrons.08x24_b3.00_fp.200.bw.eps} &
\includegraphics[width=60mm]{epsf/hadrons.12x24_b3.00_fp.200.bw.eps} \\
\hspace{-5mm}
\includegraphics[width=60mm]{epsf/hadrons.16x32_b3.00_fp.200.bw.eps} &
\includegraphics[width=60mm]{epsf/hadrons.12x24_b3.00_o3.100.bw.eps} \\
\hspace{-5mm}
\includegraphics[width=60mm]{epsf/hadrons.12x24_b3.40_fp.200.bw.eps} &
\includegraphics[width=60mm]{epsf/hadrons.16x32_b3.70_fp.100.bw.eps} \\
\end{tabular}
%\end{center}
\caption{Compilation of hadron masses versus quark mass from the various
lattices. }
\label{fig:hadrons}
\end{figure}

\begin{figure}[tbp]
\begin{center}
\includegraphics[width=9cm]{epsf/zm.16x32_b3.00_fp.eps}
\end{center}
\vspace{-4mm}
\caption{ Pion mass squared versus bare quark mass for the three pion
correlators on lattice size  $16^3\times 32$ with $\beta=3.0$ for
$D^{\rm FP}$. }
\label{fig:fp_zm16}
\end{figure}

\clearpage

\fancyhead[RE]{\nouppercase{\small\it Conclusions and Prospects}}
\chapter{Conclusions and Prospects}

After the beautiful properties of classically perfect Fixed-Point
actions have been 
verified in various models, and recently a sophisticated FP $SU(3)$
gauge action has been parametrized and successfully tested, we have constructed
in this work and in a parallel thesis a FP fermion action for lattice
QCD. In this thesis it is applied to
calculations of the quenched light hadron spectrum.

\subsubsection{Properties of Fixed-Point Fermions}
We find that the parametrization of FP fermions in QCD is a feasible
task. The resulting Dirac operator is rather complex, including the
full Clifford structure and hypercubic couplings, and uses two-level
hypercubic RG smeared links as a gauge input. We have presented a way
to efficiently build the gauge paths that are stored in the 
matrix representing the Dirac operator. Technically the multiplication
of such a Dirac operator with a vector is 36 times more costly than for
the Wilson 
operator, but the overall factor
might be smaller in actual simulations due to the faster convergence of
the iterative inversion. The computational cost also depends strongly
on the architectural details of the computer. 

What one gets at this moderately higher price is a lattice fermion
action which preserves 
chiral symmetry to a high level and has largely reduced cut-off
effects. The presence of approximate chiral symmetry manifests itself in the
eigenvalue spectrum, which is very close to the exact Ginsparg-Wilson
case, in the eigenvalue flow of discretized exact instantons, 
and in the small residual additive quark mass renormalization
determined in spectroscopy calculations. Due to the good chiral properties, we
are able to perform lattice simulations at very small quark masses,
corresponding to $m_{PS}/m_V<0.3$.

Like chiral symmetry, reduced cut-off effects are a consequence of the close
approximation to the perfect action that is achieved by our
parametrization. A high level of improvement compared to standard
actions is observed in the scaling of hadron masses, where only
small---if any---$\ordo(a^2)$ effects seem to remain. Moreover, the
hadron dispersion 
relation shows an impressive agreement with the continuum form.

For applications where very good, but not exact chiral symmetry is needed, the
parametrized FP Dirac operator therefore provides a highly competitive
alternative to domain wall or Wilson overlap fermions, with the additional
advantage of reduced cut-off effects. Furthermore, the parametrized FP
fermion action is ultralocal, thus avoiding possible problems with
suboptimal localization properties observed for the Wilson overlap.

\subsubsection{Overlap with Fixed-Point Fermions}
If exact chiral symmetry is needed, the overlap construction is the
only known solution. We examined the consequences of taking the FP
instead of the Wilson Dirac operator as a starting point for the
overlap. First of all, since the FP operator is almost chiral, only
an expansion to low order is needed for the overlap, illuminated also
by the far better behavior of the FP kernel in the exact
treatment of the smallest eigenmodes for the calculation of the
inverse square root. The need for only a low-order overlap expansion partially
compensates the higher computational cost of the FP kernel in the
overall cost for the FP overlap operator. 
In this work, we considered a overlap expansion to third order and
examined the effect on hadron spectroscopy. The localization properties
of this operator are much better both for the
couplings and in reproducing the zero mode of a discretized
exact instanton. 

The cut-off effects of the overlap-improved FP Dirac operator are
another issue. While the overlap construction removes
effects to $\ordo(a)$, the higher-order effects might even increase.
In our measurements, the hadron dispersion relation for instance seems
to get worse compared to 
the FP operator. The scaling violations for the vector meson
mass at $a=0.16$ fm can not be adequately judged due to uncertainties
in the scale determination. We do not have enough data here
to make a definite statement about the cut-off effects of the overlap
with FP kernel, but the
first results indicate that the improvement from the overlap needs not
be large and might even be negative for some quantities. This however
has to be taken under the premise that the starting operator is
already highly improved. 
 
Overall, we find that the overlap with FP kernel yields an operator
which has better locality properties than the standard Wilson overlap. Whether
in applications where exact chiral symmetry is needed it is better
suited than the Wilson overlap remains to be examined. We refer to the
thesis of Thomas J\"org for a further discussion of applications and
results with the FP overlap. 

\subsubsection{Physical Results}
Exploiting the chiral symmetry of FP fermions, we have
studied topological finite-volume effects in quenched pion propagators,
which are induced by zero modes of the Dirac operator. A recently
suggested solution, amounting to the subtraction of the effects in the
meson propagators, turns out to be the most practical and efficient way
to remove these effects. The explicit calculation and subtraction of
the zero modes in the quark propagators does not appear to be
competitive. Applying the former method, we find that the 
intercept of the squared pseudoscalar meson mass with zero is
consistent with the 
determination from the axial Ward identity quark mass. 

Having
clarified the complication from zero modes, we confirm the presence of
the quenched 
chiral logarithm in the squared pseudoscalar meson mass and measure
its coefficient. The resulting value is significantly larger than for
previous measurements with non-chiral actions, but consistent with
the theoretical expectation.

In a preliminary analysis of our data, we  find some hints, but no clear
evidence, for quenched terms in the chiral extrapolations of vector mesons and
baryons. If however such terms have to be taken into account, the
value of the hadrons at the physical mass of the light quarks might
be substantially affected at least at fixed lattice spacing.

\subsubsection{Prospects}

The results for hadron spectroscopy with FP fermions are
encouraging. Further work leads into various
directions. First, our experiences from the parametrization and the
simulations give insight into the strengths and weaknesses of
the current parametrization, which might help in finding a set of
parameters that
describes the perfect action even better. Second, the application of FP
fermions to physical problems like pion scattering is promising. 
Third, the construction of chiral currents and FP operators might
complement the FP QCD action in the future.

In the context of the BGR collaboration, a comparison of these results
to hadron spectroscopy simulations with a different chiral formulation
of lattice fermions is under way.

\begin{appendix}
\fancyhead[RE]{\nouppercase{\small\it Non-Perturbative Gauge Fixing}}
\chapter{Non-Perturbative Gauge Fixing} 
\label{app:gaugefix}

In lattice QCD, the functional integrals used to determine physical
observables are well-defined due to the finite number of
space-time points and the
gauge fields being elements of a compact group. In general it is therefore not
necessary to fix the gauge. For certain applications however, amongst
which are the computation of gauge-dependent quantities
like gluon propagators or matrix elements used in
non-perturbative renormalization techniques \cite{Martinelli:1995ty},
it is unavoidable to 
work in a fixed gauge background. Fixing the gauge is also a
way to make life easy when using extended sources for
calculating quark propagators, as it is 
then not necessary to ensure gauge invariance by hand.
Thus, it is nice to have a fast and reliable algorithm to
numerically fix a lattice gauge configuration to a certain
gauge. Although lattice gauge fixing is mainly a technical aid for
doing calculations, a fair amount of work has been done on this
subject by the lattice community, as a recent review of the current
status shows \cite{Giusti:2001xf}.

\section{Gauge Fixing and the Lattice}
On the lattice, the fundamental variables for the gauge degrees of freedom are
not the continuum fields $\mathcal{A}_\mu(x)$ themselves, but the matrices
$U_\mu(x)$, which are group elements of $SU(3)$ in
the fundamental representation and are formally defined as parallel
transporters of the color interaction between lattice sites,
\begin{equation} \label{eq:ugauge}
  U_\mu(x) \equiv e^{iag\mathcal{A}_\mu(x)},
\end{equation}
so that the $U_\mu(x)$ fields live on lattice links. We define the
lattice gauge potential
\begin{equation} \label{eq:lattice_a}
  A_\mu(x) \equiv \frac{1}{2iag} \Bigl. \left[ U_\mu(x) - U_\mu^\dag(x) \right]
  \Bigr|_{\rm traceless}, 
\end{equation}
which is suggested by the formal relation
\eqref{eq:ugauge} between lattice and 
continuum gauge fields $\mathcal{A}_\mu(x)$. Note that 
$A_\mu(x)$ is equivalent to the continuum gauge field $\mathcal{A}_\mu(x)$
only in the continuum limit $a\rightarrow 0$. While
Eq. \eqref{eq:lattice_a} is 
a common way to define the lattice gauge potential, it is not
unique, and other definitions which differ only by irrelevant terms are
perfectly allowed. The choice of one particular definition then
also leads to a particular solution of a given gauge fixing
condition. However, by comparing Green's functions, it has been
checked that in the continuum limit  
the continuum gluon field described by different definitions of
$A_\mu(x)$ on the lattice is unique \cite{Giusti:1998ur}.

Under a local gauge transformation, the lattice gauge field $U_\mu(x)$
transforms like
\begin{equation}
  U_\mu(x) \longrightarrow U_\mu^G(x) = G(x)U_\mu(x)G(x+\hat\mu),
\end{equation}
where $G(x)$ are elements of the gauge group SU(3) living on lattice sites.
To fix the gauge, a condition $f(U_\mu^G(x))=0$ is introduced, which
should pick out one configuration per gauge orbit. In
general however, there are multiple solutions to this equation for a given
gauge configuration. These solutions belonging to the same
gauge orbit are called Gribov copies.
The question to what extent this Gribov ambiguity introduces systematic
uncertainties in lattice results is not definitely answered.
While for certain problems like the calculation of the
photon propagator in compact U(1) \cite{Nakamura:1991ww,Bornyakov:1993yy} or studies of center vortices
\cite{Bornyakov:2000ig,Kovacs:1999st}, Gribov copies are known to distort
measurements heavily, the effect seems to be barely distinguishable from
the statistical 
noise for measurements of the axial vector renormalization factor
$Z_A$ \cite{Martinelli:1993dq,Paciello:1994gs,Conti:1996dw} and $B$
meson correlators with smeared sources \cite{Paciello:1992gy}. In the
latter study, which was performed on $10^3\times20$ lattices with the
Wilson gauge action at $\beta=6.0$, the Gribov noise could actually be
identified, and 
it was argued that for larger lattices, its size could become
significant compared to the statistical noise. Hence one has to keep in
mind that the noise from Gribov copies is a possible source of errors
in measurements of hadronic correlators, if we use gauge fixing in
combination with extended, gauge-dependent operators. 

The gauge
fixing condition can be freely chosen. The most common choices in
lattice QCD belong to the general class of $\lambda$-gauges, which are
characterized by a continuum gauge fixing condition
\begin{equation} \label{eq:continuum_lambda_gauges}
\lambda \partial_0 {\mathcal A}_0 + \partial_i {\mathcal A}_i = 0.
\end{equation}
For $\lambda=1$ one gets Landau and for $\lambda=0$ Coulomb gauge.
These two conditions are
equivalent to finding the extremal value of the lattice functional
\begin{equation} \label{eq:gf_functional}
  F_U[G] = - \mbox{Re} \Tr \sum_x\sum_{\mu=1}^l U_\mu^{G(x)}(x),
\end{equation}
where the second sum runs over the spacial indices only ($l=3$) for Coulomb
gauge and over all space-time indices ($l=4$) for Landau gauge.
Again, exact equivalence between the lattice and continuum gauge fixing
conditions holds only in the continuum limit, and one might consider
to reduce discretization error with improved gauge fixing conditions
\cite{Bonnet:1999bw}. 
Since it
is numerically impossible to find the global minimum of the
functional \eqref{eq:gf_functional}, which would be a unique solution
up to a global gauge transformation, we specify to 
take any local minimum as our gauge-fixed configuration.
Local minima of  \eqref{eq:gf_functional} are numerically found in an
iterative procedure. There are several gauge fixing algorithms on the
market, we present here the Los Alamos method with improved
convergence by stochastic overrelaxation.  

\section{The Los Alamos Algorithm with Stochastic Overrelaxation}
In this method, introduced by De Forcrand and Gupta \cite{deforcrand:1989aa}, the minimizing functional
is rewritten using the auxiliary variable
\begin{equation}
  w(x) = \sum_{\mu=1}^l \left( U_\mu(x) + U_\mu^\dag(x-\hat\mu) \right),
\end{equation}
so that the sum over the lattice sites in \eqref{eq:gf_functional} is replaced by a
sum over only half the lattice sites. If we assign the colors red and
black to the lattice sites in a checkerboard manner, the functional reads
\begin{equation} \label{eq:gaugefix_functional_redblack}
  F_U[G] = - \frac{1}{2}\mbox{Re} \Tr \sum_{x \in {\rm red\ or\ black}} w^G(x).
\end{equation}
The basic idea is now to subsequently transform the gauge fields on the red
and black lattice sites separately in a way that the minimizing
functional monotonically decreases in every iteration step. The
gauge transformation $G(x)$ is therefore chosen to be unity on the red(black)
lattice sites at even(odd) iteration steps. Under a local gauge
transformation, the field $U_\mu(x)$ then transforms like
\begin{eqnarray}
  U_\mu(x) &\longrightarrow& U_\mu^G(x) = G(x)U_\mu(x), \nonumber \\
  U_\mu^\dag(x) &\longrightarrow& U_\mu^{\dag G}(x) = U_\mu^\dag(x) G^\dag(x).
\end{eqnarray}
This gauge transformation amounts to one step in the
iterative process. For the variable $w(x)$ introduced above, the
gauge transformation reads
\begin{equation}
  w(x) \longrightarrow w^G(x) = G(x)w(x).
\end{equation}
The transformation $G(x)$ is now chosen independently on every
other lattice site such that
\begin{equation}
  {\rm Re} \Tr\ G(x)w(x) \geq {\rm Re} \Tr\ w(x).
\end{equation}
We choose $G(x)$ to be the projection of $w(x)$ onto the $SU(3)$ group
manifold which maximizes the left hand side of the equation. This is done by 
iterative maximization of $SU(2)$ subgroups \cite{Cabibbo:1982zn}.
After each step the roles of the red and black sites are interchanged.

To overcome problems with critical slowing down when fixing the
gauge on large lattices,
several acceleration methods are discussed in the literature, amongst
which are Fourier
preconditioning \cite{Davies:1988vs}, overrelaxation
\cite{Mandula:1990vs} and
multigrid schemes \cite{Hulsebos:1992na,Cucchieri:1998ew}. A very
simple method is 
stochastic overrelaxation \cite{deforcrand:1989aa}, 
which is based on the idea to overdo the local maximization once in a
while. More precisely, with a probability $0\leq p_{\rm or}\leq 1$ one
applies a local 
gauge transformation $G^2(x)$ instead of $G(x)$.
 For $p_{\rm or}=0$ there is no
change to the algorithm, while for $p_{\rm or}=1$ the algorithm does
not converge. For
intermediate values of $p_{\rm or}$, a dramatic speedup in the convergence
can be reached \cite{Cucchieri:1998ta,Suman:1993mg}. However, the
optimal value for the 
parameter $p_{\rm or}$ depends quite strongly on the
lattice volume and the smoothness of the gauge configuration. It is therefore
necessary to optimize $p_{\rm or}$ for every lattice size and lattice
spacing separately to get fast convergence.

\subsection{Convergence Criterion}
It remains to define a criterion which tells us when 
the desired accuracy is reached and the algorithm can be stopped. Besides
monitoring the
value of the functional $F_U[G]$ during the
process, we determine the gauge fixing accuracy by measuring its first
derivative 
\begin{equation}
\sigma = \sum_x {\rm tr} \left[ \Delta^G(x) {\Delta^G}^\dag (x) \right],
\end{equation}
where
\begin{equation}
\Delta^G(x) = \sum_\mu \left[ A_\mu^G(x) - A_\mu^G(x-\hat\mu) \right],
\end{equation}
is a discretized version of the
derivative in the continuum gauge fixing
condition \eqref{eq:continuum_lambda_gauges},
and $A_\mu^G(x)$ is the gauge transformed lattice gauge potential
\eqref{eq:lattice_a}. At a local minimum of the gauge fixing
functional, $\sigma$ 
vanishes. For our purposes, we stop the algorithm if $\sigma < 10^{-8}$. 

\begin{figure}[tb]
\begin{center}
\includegraphics[width=9cm]{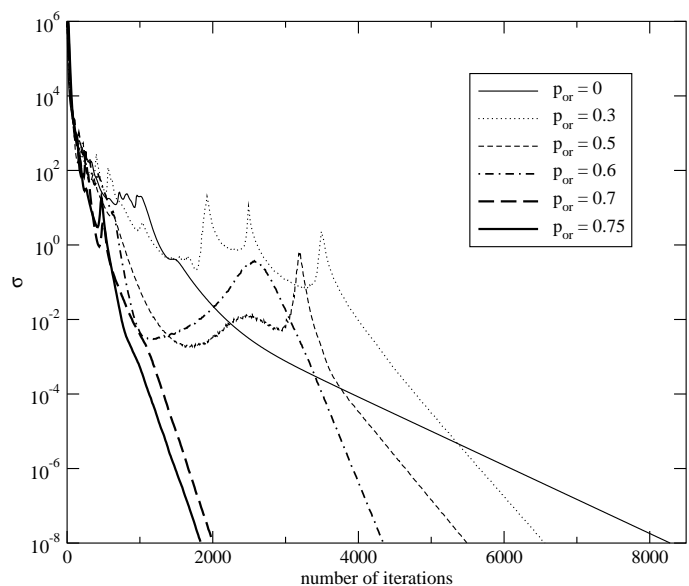}
\end{center}
\caption{Convergence of the Los Alamos gauge fixing algorithm for different
values of the stochastic overrelaxation parameter $p_{\rm or}$ on a
$9^3\times24$ gauge configuration with $\beta=3.0$ fixed to Landau gauge. For
$p_{\rm or}= 0.8$ the algorithm did not converge, but stayed at
$\sigma\approx 10^{-2}$ after $\sim 1000$ iterations.} 
\label{fig:gaugefix_convergence_9x24}
\end{figure}

\subsection{Tuning of the Overrelaxation Parameter $p_{\rm or}$}
\begin{figure}[tb]
\begin{center}
\includegraphics[width=9cm]{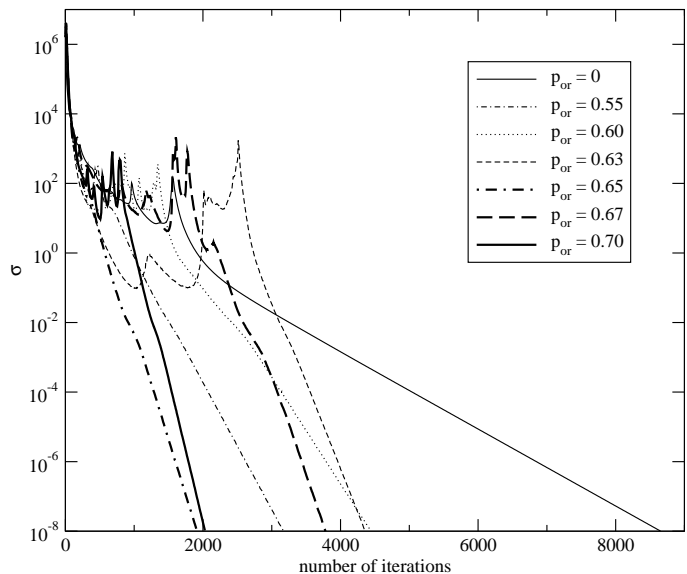}
\end{center}
\caption{Convergence of the Landau gauge fixing for different
values of the stochastic overrelaxation parameter $p_{\rm or}$ on a
$12^3\times24$ gauge configuration with $\beta=3.4$. For $0.55\leq
p_{\rm or} \leq 0.7$, the total number of iterations does 
not depend smoothly on the stochastic overrelaxation parameter. For
$p_{\rm or}\geq 0.75$, the gauge fixing did not converge.} 
\label{fig:gaugefix_convergence_12x24}
\end{figure}

\begin{figure}[tb]
\begin{center}
\includegraphics[width=9cm]{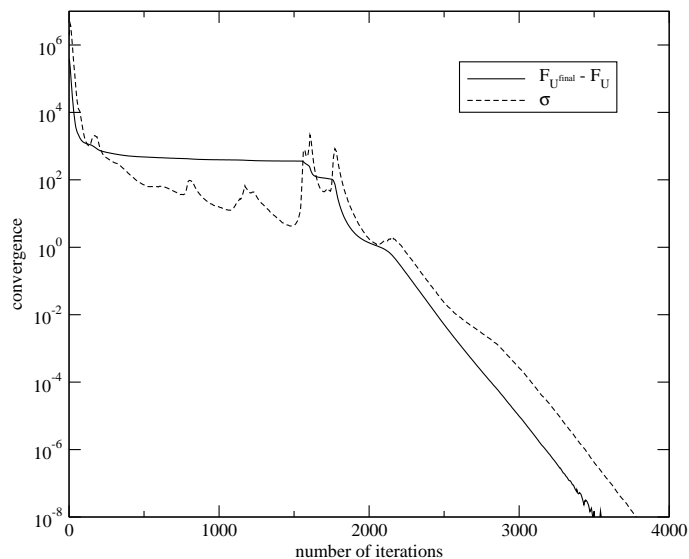}
\end{center}
\caption{The difference of the gauge fixing functional from its final
value and the convergence criterion $\sigma$ for a 
$12^3\times24$ gauge configuration with $\beta=3.4$ fixed to Landau
gauge for $p_{\rm or}=0.67$. The plot shows that the algorithm tends
to fall into several other
local minima before finally converging.} 
\label{fig:gaugefix_convergence_diff}
\end{figure}

%The tuning of the stochastic parameter $p_{\rm or}$, is a task which in
%principle has to
%be repeated for every lattice volume and every value of the gauge
%coupling.
It is easily possible to gain a factor of 4 in
the number of iterations needed to reach a given gauge fixing
accuracy when accelerating the algorithm by stochastic
overrelaxation, as  
the gauge fixing procedure is very sensitive to the
value of $p_{\rm or}$. 
Fig.~\ref{fig:gaugefix_convergence_9x24} shows a plot of the
convergence history on one 
configuration in dependence of the stochastic
parameter. The number of iterations needed to fix the gauge is reduced
from over 8000 to below 2000 when setting
$p_{\rm or}=0.75$. Unfortunately, the optimization of $p_{\rm or}$ is not a
very 
stable procedure, and it can happen that for $p_{\rm or}> 0$, the
gauge fixing takes much longer on certain gauge
configurations than on the others. As seen in
Fig.~\ref{fig:gaugefix_convergence_12x24}, 
this is related to the fact that shortly after the beginning, $\sigma$
starts to fluctuate wildly for some time, and only then a monotonic
decrease is observed. A possible explanation of this behavior is that
for a certain time it is not clear which Gribov copy the algorithm is
going to choose. In Fig.~\ref{fig:gaugefix_convergence_diff} we plot
the difference of the gauge fixing functional
\eqref{eq:gaugefix_functional_redblack} from its final 
value together with $\sigma$. Several plateaus in the value of $F_U$
show up before the algorithm decides which local minimum to take.

As soon as the region of monotonic decrease of $\sigma$ is reached, the
convergence is much faster with stochastic overrelaxation than
without. It is also obvious from
Fig~\ref{fig:gaugefix_convergence_12x24} that small changes in
$p_{\rm or}$ can lead to 
a very different behavior in this fluctuating region, resulting in
factors of 2 in the total number of iterations. Still,
compared to the $p_{\rm or}=0$ case, a significant reduction is achieved. 

In our tests, it appeared that
a value of $p_{\rm or}$ slightly below the point where convergence
is lost worked well in most cases. As the computer time needed for the gauge
fixing was considerably smaller than for the generation of the gauge
configurations, a further optimization of the procedure was not necessary.

\section{Coulomb vs.~Landau Gauge}
When one decides to work in a fixed gauge background in order to measure the hadron
spectrum with extended operators, one has to choose a specific
gauge.
There is a theoretical argument why Landau gauge is not well-suited for
the measurement of time-dependent correlation
functions $C(t) = \langle \mathcal{O}_f(t)\mathcal{O}_i(0) \rangle$ which are
used to extract 
masses: 
Consider a gauge configuration $U$, which is fixed to Landau gauge:
\begin{equation}
U \longrightarrow U_{\rm gf}.
\end{equation}
 Suppose we measure a spatially extended, 
gauge-dependent operator $\mathcal{O}(U_{\rm gf},t=0)$ on the time slice
$t=0$ of the gauge-fixed configuration. Such an operator might for
example be the smeared source of a hadron
correlator. Now change one single gauge link of the configuration $U$
on a time slice $t \neq 0$, for example by rotating it into the
opposite direction:
\begin{equation}
 U^\prime_\mu(x)  =
\left\{ \begin{array}{ll} 
- U_\mu(x) & \mbox{if $\mu=1$ and $x=(x_0,t_0\neq0)$ }, \\
\ \ U_\mu(x) & \mbox{otherwise}.
\end{array}
\right.
\end{equation}
Fixing the resulting gauge configuration $U^\prime$ to Landau gauge,
\begin{equation}
U^\prime \longrightarrow U^\prime_{\rm gf} ,
\end{equation}
the gauge-fixed configuration $U^\prime_{\rm gf}$ differs
globally from $U_{\rm gf}$, since in the Landau gauge fixing process,
spatial and 
temporal links enter in the minimizing functional \eqref{eq:gf_functional}.
It follows that although the operator $\mathcal{O}$ is
thought to be defined only on the time slice $t=0$, 
its value depends on the gauge fields on $t\neq
0$ time slices, and it is different on the two
gauge fixed configurations:
\begin{equation}
\mathcal{O}(U_{\rm gf}^\prime,t=0) \neq \mathcal{O}(U_{\rm gf},t=0).
\end{equation}
Landau gauge fixing therefore introduces
a non-local interaction in the time direction which
might spoil the signal of hadron correlators.

The situation is different with Coulomb gauge: In the minimizing
functional the links in the time direction do not enter, hence the gauge
fixing is performed independently on each time slice. If a link on a
time slice $t\neq 0$ is changed, the gauge fixed configuration on
time slice $t=0$ is not affected, and the measured operator will not
change. 

In a study where hadron correlators with wall sources were
compared on configurations fixed to both Coulomb and Landau gauge, a
significant difference of order $3\sigma$ has been found for the
mass of the $\Delta$ meson \cite{Chen:1995cp}, which might be an
effect of this 
non-local interaction in the Landau gauge fixing. For our
hadron spectroscopy study, we
therefore fix the configurations to Coulomb gauge.

\fancyhead[RE]{\nouppercase{\small\it QCD on Large Computers}}
\chapter{QCD on Large Computers}

The non-perturbative approach to Quantum Chromodynamics as defined by
the lattice regularization yields a theoretical formulation of the
strong nuclear force which is
ideal for treatment on large computers. The simulation of a realistic
problem, like the scattering of two pions or the decay $K \rightarrow
\pi\pi$, turns out to be a priori
numerically very demanding due to the large number of degrees of
freedom involved. The following arguments make it clear why lattice
QCD calculations are hard:
First, the spatial lattice volume has to be large enough,
so that the wave functions of all involved particles fit into the box
of side length
$L_s=N_sa$ without getting squeezed. The temporal size $L_t=N_ta$ also
has to be large, as one 
is mainly interested in the asymptotic behavior of correlation functions
at large Euclidean time (see Chapter \ref{chapter:had_spect}). Second,
to get rid of 
discretization errors, the continuum 
limit $a\rightarrow 0$ has to be 
taken by performing simulations at several lattice spacings
$a$ and extrapolating the measured observables to the continuum. This
implies that it is necessary to work with lattice spacings $a$ small
enough to ensure a controlled extrapolation, and therefore $N_s$ and
$N_t$ get large quickly. Third, as the
computational effort grows with inverse powers of the quark mass,
simulations are mostly performed at values larger than the physical
mass. Hence, another extrapolation from results
calculated at several higher quark masses is necessary to obtain
physical values.

As shown in Section \ref{sect:qprop}, the numerically demanding
part in quenched lattice QCD calculations is the inversion of the
Dirac operator, which is a large sparse 
complex matrix of rank $4 N_c V$ on a lattice with volume $V = N_s^3\times
N_t$. This matrix inversion can be done very efficiently on
massively parallel computers, because it boils down to many
complex multiplication and addition operations, and only 
minimal communication is necessary between different processors.
Large LQCD calculations are done either on commercial machines
or on custom-built computers dedicated to lattice QCD.
Commercial supercomputers are multi-purpose machines and have to
perform well on a very diverse range of problems. Their
architecture is therefore highly sophisticated in order to process all
kinds of complex
code efficiently. Due to the simple computational structure of lattice
QCD calculations, building a dedicated
computer is in general the much more cost-effective alternative.  
 The lattice groups which have been running the largest simulations
in the last few years all use custom-built machines: The Japanese
CP-PACS computer in Tsukuba \cite{Iwasaki:1998qr}, several machines of
the European APE 
project \cite{Alfieri:2001xm} and the Anglo-American
QCDSP \cite{Chen:1998gj} computers  and its successor QCDOC
\cite{Boyle:2001dc}. 
For our work, we had in contrary access to various multi-purpose
machines spanning a large range
from workstations to supercomputers. Tables
\ref{tab:computers_1} and  \ref{tab:computers_2} 
show the different platforms that were used for our simulations and
their most relevant properties.\footnote{Unfortunately, we could not
run simulations on the IBM Power4 in Manno yet, as it took up user
operations later than expected.}  The runs on the Hitachi were
performed in the framework of the BGR collaboration \cite{bgr}.

\begin{table}[tb]
\begin{center}
\begin{tabular}{l|l|l|r|r|r} \hline\hline
Computer & Location  & Type & $N_{\rm node}$ & $N_{\rm CPU}$  &
 $\nu_{\rm CPU}$ \\ 
 & & & & & [MHz]  \\ \hline  
AMD Athlon & ITP Bern &  scalar & 1 & 1 & 1200 \\
AlphaServer DS20  & ITP Bern & scalar & 2 & 2 & 500 \\
NEC SX-5/16 & CSCS Manno& vector & 10 & 10 & 250 \\
IBM Power4  &  CSCS Manno&  scalar & 8 &256 & 1300 \\
Hitachi SR8000-F1 & LRZ M\"unchen & scalar & 168 & 1344 & 375 \\ \hline
\end{tabular}
\caption{Computers used for the simulations, with the number of nodes
 $N_{\rm node}$, the number of processors $N_{\rm CPU}$ and the CPU frequency
 $\nu_{\rm CPU}$. A node is defined here as the largest unit for which explicitly
 parallel code is not necessary. This might either be a single processor or a
 set of processors on which automatic parallelization by the compiler
 is provided.}
\label{tab:computers_1}
\end{center}
\end{table}

\begin{table}[tb]
\begin{center}
\begin{tabular}{l|r|r|r|r} \hline\hline
Computer & $M_{\rm tot}$ & $M_{\rm node}$ & 
$P_{\rm tot}$ & $P_{\rm node}$ \\
 & [GB]& [GB] & [GFLOPS] & [GFLOPS] \\ \hline  
AMD Athlon  & 1 & 1 &  2 & 2\\
AlphaServer DS20   & 4 & 4 & 2 & 1\\
NEC SX-5/16 & 64 & 64 & 80 & 8\\
IBM Power4   & 768 & 96 & 1330 & 166 \\
Hitachi SR8000-F1  & 928 & 8  & 2016 & 8 \\ \hline
\end{tabular}
\caption{Main memory and performance of the different computers, each
listed for the 
total machine and per node. The theoretical peak
performance $P$ is given in units of $10^9$
floating point operations per second (GFLOPS).} 
\label{tab:computers_2}
\end{center}
\end{table}

In Appendix \ref{sect:technical_data}, we summarize some technical details
of the two computers we mostly worked on. Then we present in Appendix 
\ref{app:perf} benchmark measurements from the 
spectroscopy simulations on the Hitachi and in particular the performance
of our code under MPI parallelization. Finally we give in Appendix
\ref{app:multimass} a brief introduction to modern matrix inversion
algorithms and their extension to shifted linear systems, which is
very useful for performing QCD simulations at many quark masses.

\section{Specifications of Utilized Supercomputers}
\label{sect:technical_data}
It is well-known that what is called a
supercomputer at a given time might become
inferior to a desktop PC only a few years later, and therefore technical
details outdate quickly and are of marginal interest. But as
for our simulations the architecture of the involved
computers differed quite a lot, and the process of making the simulations
run efficiently on the available platforms depends strongly on
the technical details of the machines, we present the main
properties of the two mostly used computers in some more detail.

\subsection{The NEC SX-5/16}
The Swiss Center for Scientific Computing in Manno installed in early 2000 a
NEC SX-5 parallel-vector computer with eight processors and 64 GB of shared
main memory. In  2001, two more processors were added. At
the time it went operational, it was ranked at position 242
in the list of the top 500 supercomputers \cite{dongarra:2000aa}.
This computer differs from the other machines considered here by its vector
architecture, which allows to obtain a very high single-processor
performance: The basic building block is a 250 MHz CPU 
with 16 vector pipelines capable of processing two floating-point
instructions each per 
clock cycle, leading to a theoretical peak performance of 8
GFLOPS per CPU. Although it is possible to run parallel processes, we
only worked with scalar code due to the small number of installed
processors and the long waiting time of the parallel queues. 

The SX-5 is a good choice
if one needs a lot of memory and does not want to write parallel code,
as long as the problem is 
well-suited for vectorization. The  compiler supports automatic
vectorization, so it is possible to easily migrate existing scalar code
from a workstation to the SX-5. However, to reach good performance, it
is essential to optimize the time-critical parts by hand or with
compiler directives. Moreover, code which does not
vectorize well, like the algorithms used for the Monte Carlo update
of the gauge configurations, ends up running extremely slow. The
reason is that vector instructions need some 
time to be initialized, therefore the vectors have to
be as long as possible in order to make the overhead
irrelevant. In the program code, this manifests itself in the length
of loops in which vector instructions appear. If these loops are
not long enough, as it happens when doing manipulations on $SU(3)$
matrices, it is then even
faster to run the program in scalar mode, which means that the machine
is in this case inferior to every desktop computer due to its
comparably low clock frequency.

%\subsection{The IBM Power4}
%This machine, which started user operations in early 2002, is one of
%the first computers using the Power4 architecture from IBM. It is
%built from eight nodes, each with 32 Power4 processors and 96 GB of main
%memory. The Power4 chip actually consists of two identical processors
%running at 1.3 GHz with separate Level 1 and shared Level 2
%caches. Four of these chips are packed onto a module, and four modules
%make up a node. Each of the 32 processors of one node provides two
%floating point units capable of 
%performing each a fused multiply-add instruction simultaneously,
%yielding four 
%floating point operations per clock cycle per processor. The theoretical
%peak performance of one processor accordingly is 5.2 GFLOP/s, giving a
%total of 1.3 TFLOP/s for the whole machine.

\subsection{The Hitachi SR8000-F1}
A very different kind of architecture is provided by the Hitachi
SR8000-F1 at Leibniz Rechenzentrum in M\"unchen. This 
massively parallel scalar computer was also installed in early 2000
and was at
that time the world's fastest computer dedicated to academic research,
ranked at 5th position of all supercomputers
\cite{dongarra:2000aa}. At installation time, it 
comprised of $112\times 8$ modified Power3 processors running at a clock
frequency of 375 MHz. Each unit is able to process two multiply/add
instructions simultaneously. Eight processors are grouped in
a node, which 
can be treated by the programmer like a single CPU, and each
node has access to 8 GB of main memory\footnote{A few special nodes
have 16 GB of main memory}. The
theoretical peak performance per node is $8\times 1.5 = 12$
GFLOPS. In January 2002, the machine was upgraded from 112 to 168
nodes, reaching more than 2 TFLOPS peak performance. The inter-node
communication is realized with a multi-dimensional crossbar
delivering a bi-directional peak bandwidth of 1 GB/s.

Programs running on the SR8000 have to be parallelized in order make
use of multiple nodes. The parallelization 
within the eight processors of one node is automatically done by the
compiler, which also provides a hardware-based
'pseudo-vectorization' facility that imitates a 
vector processor. In our applications, it seemed that compared to a
real vector computer, the SR8000 was more tolerant when running code
that does not vectorize well, resulting in significantly better
performance for such programs. However, to make full use of
the computer's capabilities, it is also 
necessary to tune the programs by optimizing the arrangement of array
elements and loops and by placing appropriate compiler directives for
parallelization and pseudo-vectorization at the time-critical code segments.

\section{Measurements of Parallel Performance}
\label{app:perf}

For larger simulations on the Hitachi SR8000, explicitly
parallel code is required. The common 
standard for programming the communication between different nodes of
a computer or even different computers is the Message Passing
Interface (MPI). In lattice QCD, explicit parallelization of the
quark propagator code for $n$ nodes is in general a very
simple task, since with an iterative solver
the inversion of the Dirac operator reduces to matrix-vector
multiplications. 
The Dirac operator can be split into $n$ parts containing $12V/n$ rows, 
and it only remains to 
write a distributed matrix-vector multiplication. However, for our
implementation of the FP Dirac operator, parallelization is not
completely trivial, because in the process of constructing the
matrix it can occur that a gauge path calculated on a certain 
node needs to be stored on a different node. This is merely a consequence
of the way our low-level routines for building the gauge paths in
the Dirac operator are
designed, and not a problem of the FP Dirac operator itself. 
Therefore, to make the construction of the Dirac operator work in
parallel, some inter-node communication is
needed. The Fixed-Point R operator is free of this complication, so
parallelizing its construction is trivial. In order to have a
completely parallel 
code, also the vectors of size $12V$ appearing in the different 
algorithms like the matrix inverter or the eigenvalue solver have to be 
distributed.

Not only in terms of computation time, but also in terms of storage
the parallelization of the FP Dirac operator is crucial. The memory
needed for storing the Dirac operator on a lattice of volume $V$ is
$12\times 81\times 12\times V\times 16$ bytes\footnote{We use double
precision complex numbers, requiring 16 bytes, in all our code.}, 
which exceeds the shared main memory available on 
most machines even at moderate lattice sizes. 
Table \ref{tab:memory_req} lists the storage 
requirements for various elements of our code at the lattice volumes
used in the simulations.

\begin{table}[tb]
\begin{center}
\begin{tabular}{c|r|r|r} \hline\hline
 array & $8^3\times 24$ & $12^3\times 24$ & $16^3\times 24$ \\ \hline
$D^{\rm FP}$ & 2.1 GB & 7.2 GB & 22.8 GB \\
$R^{\rm FP}$ & 0.15 GB & 0.5 GB & 1.5 GB\\ 
$U$ & 6.8 MB & 23 MB & 72 MB \\
$b$  &   2.3 MB   &  7.6 MB    &    24 MB \\\hline
\end{tabular}
\end{center}
\caption{Memory requirements for storing the parametrized FP Dirac and R
operators, a gauge configuration $U$ and a vector $b$ of size
$12V$ at different lattice volumes $V$. }
\label{tab:memory_req}
\end{table}

\begin{table}[tb]
\begin{center}
\begin{tabular}{c*{3}{|cc}} \hline\hline
 & \multicolumn{2}{c|}{$8^3\times 24$} & \multicolumn{2}{c|}{$12^3\times
24$} & \multicolumn{2}{c}{$16^3\times 24$} \\ 
\raisebox{1.5ex}[1.5ex]{\# nodes} & $t_{\rm dot}$ [$ms$] & $\omega$ &
$t_{\rm dot}$ [$ms$] & $\omega$ & $t_{\rm dot}$ [$ms$] & $\omega$ \\ \hline
1 & 0.33 & 1.0 & 0.89 & 1.0 & 2.60 & 1.00 \\
2 & 0.31 & 1.9 & 0.59 & 1.3 & 1.43 & 1.10 \\
4 & 0.27 & 3.3 & 0.41 & 1.8 & 0.82 & 1.26 \\
8 & 0.25 & 6.1 & 0.34 & 3.1 & 0.54 & 1.66 \\
16 & 0.39 & 19 & 0.44 & 7.9 & 0.44 & 2.71 \\ \hline
\end{tabular}
\end{center}
\caption{Time im milliseconds and overhead factor for dot product of two
complex vectors of size $12V$ at different levels of MPI
parallelization on the Hitachi 
SR8000. On the smallest
volume, there is essentially no gain from parallelization.}
\label{tab:t_dot}
\end{table}

\begin{table}[tb]
\begin{center}
\begin{tabular}{c*{3}{|cc}} \hline\hline
 & \multicolumn{2}{c|}{$8^3\times 24$} & \multicolumn{2}{c|}{$12^3\times
24$} & \multicolumn{2}{c}{$16^3\times 32$} \\ 
\raisebox{1.5ex}[1.5ex]{\# nodes} & $t_D$ [$s$] & $\omega$ & $t_D$
[$s$] & $\omega$ &$t_D$ & $\omega$ \\ \hline 
1    &  347 & 1.00   &   &  & & \\ 
2    &  192 & 1.11   & 610 & 1.00  & &  \\ 
4    &  98  & 1.13   & 323 & 1.06  & &  \\ 
8    &  57  & 1.31   & 188 & 1.16  & 930 & 1.00 \\
16   &      &        &     &       & 570 & 1.23 \\ \hline
\end{tabular}
\end{center}
\caption{Construction time in seconds and overhead factor for building up the Fixed-Point
Dirac operator as 
a function of lattice size and parallelization level. There is a small
parallelization overhead due to communication occurring when gauge
paths cross node boundaries.}
\label{tab:t_d}
\end{table}

\begin{table}[tb]
\begin{center}
\begin{tabular}{c*{3}{|cc}} \hline\hline
 & \multicolumn{2}{c|}{$8^3\times 24$} & \multicolumn{2}{c|}{$12^3\times
24$} & \multicolumn{2}{c}{$16^3\times 32$} \\
\raisebox{1.5ex}[1.5ex]{\# nodes} & $t_R$ [$s$] & $\omega$ & $t_R$
[$s$] & $\omega$ & $t_R$ &  $\omega$  \\ \hline 
1    & 221 & 1.0  & &  & &\\ 
2    & 111 & 1.0  & 371 & 1.0 &  &  \\ 
4    &  54 & 1.0  & 187 & 1.0 &  & \\ 
8    &  28 & 1.0  & 94  & 1.0 & 296 & 1.0 \\
16   &     &       &     &      & 148 & 1.0 \\ \hline
\end{tabular}
\end{center}
\caption{Construction time in seconds and overhead factor for building
up the Fixed-Point R operator as 
a function of lattice size and parallelization level. No overhead for
parallelization is seen, as there is no communication over node
boundaries necessary.} 
\label{tab:t_r}
\end{table}

\begin{table}[tb]
\begin{center}
\begin{tabular}{c*{4}{|cc}} \hline\hline
 & \multicolumn{2}{c|}{$8^3\times 24$} & \multicolumn{2}{c|}{$12^3\times
24$} & \multicolumn{2}{c|}{$12^3\times 24$} &
\multicolumn{2}{c}{$16^3\times 32$} \\ 
 \# nodes & \multicolumn{2}{c|}{(FP)} & \multicolumn{2}{c|}{(FP)} &
 \multicolumn{2}{c|}{(overlap)} & \multicolumn{2}{c}{(FP)} \\  
  & $t_{\rm iter}$ [$s$] & $\omega$ &
$t_{\rm iter}$ [$s$] & $\omega$ & $t_{\rm iter}$ &  $\omega$ & $t_{\rm
iter}$ &  $\omega$  \\ \hline  
1   &  0.564 & 1.00 &  &  & & & &\\ 
2   &  0.313 & 1.11 &  1.045 & 1.00 & & & & \\ 
4   &  0.185 & 1.31 &  0.619 & 1.18 &  5.36 & 1.00 & & \\ 
8   &  0.122 & 1.73 &  0.405 & 1.55 &  3.57 & 1.33 & 1.30 & 1.00 \\
16  &        &      &        &      &       &      & 1.05 & 1.62 \\ \hline
\end{tabular}
\end{center}
\caption{Time im seconds and overhead factor for one iteration of the
matrix inversion 
algorithm as a function of lattice size and parallelization level for
the parametrized FP and the overlap Dirac operator. One
iteration requires two matrix-vector products of both the Dirac and R
operator and some additional dot products of two vectors. For the overlap,
a third order Legendre expansion of the inverse square root is used with the
100 lowest eigenmodes treated exactly.}
\label{tab:t_iter}
\end{table}

\begin{table}[tb]
\begin{center}
\begin{tabular}{c*{2}{|ccc}} \hline\hline
 & \multicolumn{3}{c|}{$8^3\times 24$} & \multicolumn{3}{c}{$12^3\times
24$} \\  
 \# nodes & \multicolumn{3}{c|}{(FP)} & \multicolumn{3}{c}{(FP)} \\  
  & $t_D$ [\%] & $t_R$ [\%]  & MFLOPS &  $t_D$ [\%] & $t_R$ [\%]  & MFLOPS  \\
 \hline 
1 &   56 &   16  &  4200 & & & \\ 
2 &   55 &   18  &  3800 &   60  & 19 &  4160 \\ 
4 &   53 &   22  &  3240 &   59 &  23 &  3570 \\ 
8 &   51 &   27  &  2560 &   57  & 29 &  2850 \\ \hline
\end{tabular}

\vspace{2ex}

\begin{tabular}{c*{2}{|ccc}} \hline\hline
 & \multicolumn{3}{c|}{$12^3\times 24$} &
\multicolumn{3}{c}{$16^3\times 32$} \\ 
 \# nodes & \multicolumn{3}{c|}{(overlap)} &
\multicolumn{3}{c}{(FP)} \\ 
  & $t_D$ [\%] & $t_R$ [\%]  & MFLOPS &  $t_D$ [\%] & $t_R$ [\%]  &
  MFLOPS \\ \hline
4 &    54  & 22  & 3660 & & & \\ 
8 &    50  & 26 &  2750 & 56 & 29 & 2770 \\ 
16 &     &     &       &       51 & 34 & 1700 \\ \hline
\end{tabular}

\end{center}
\caption{Overall performance measurements for typical quark propagator
 runs on the Hitachi SR8000. Shown are the percentage of the overall
 time spent for the $D$ and $R$ multiplication routines including
 communication of the input vector and the overall speed in MFLOPS
 per node with 
 theoretical peak speed of 12 GFLOPS. The overall run time is measured
 from program start to finish and thus includes I/O time, MPI
 initialization and finalization and constructing of $D$ and $R$ operators. }
\label{tab:overall_perf}
\end{table}

It is clear that the communication between different
nodes introduces an overhead, and that this parallelization overhead
increases with 
the number of nodes. Therefore it is important to find a reasonable
balance between the gain in wall-clock time and the loss in CPU time due to
parallelization. To quantify the parallelization  overhead, we list in
Tables \ref{tab:t_dot}--\ref{tab:t_iter} for several computational
tasks the wall-clock 
time and the overhead factor $\omega$, which is defined as
\begin{equation}
 \omega = n \frac{t_n}{t_{\rm ref}},
\end{equation}
where $n$ is the number of nodes, $t_n$ is the wall clock time for the
task running on $n$ nodes and $t_{\rm ref}$ is the wall-clock
time for the smallest $n$ on which it was possible to run the task due
to memory 
limitations. As shown in Table \ref{tab:t_dot}, the dot product of two
vectors does not really profit from parallelization on the smallest
lattice. The situation changes drastically on the larger lattices, where
the distribution of the vectors is crucial if one does not want to have
severe slowing down in algorithms which perform many vector operations. For
the construction of the FP Dirac operator (Table \ref{tab:t_d}),
parallelization works well. The overhead introduced by the above
mentioned inter-node communication is small for
reasonable ratios of the lattice volume $V$ and the number of nodes $n$.
That the construction of the $R$ operator parallelizes trivially can be
seen in Table \ref{tab:t_r}.

 The crucial quantity for quenched QCD simulations is 
the time needed for one iteration of the matrix inversion algorithm,
which is listed in Table
\ref{tab:t_iter}. This is by far the most time-consuming task, and as
one sees the overhead is still considerable. On the
$12^3\times 24$ lattice, for example, there is only $50\%$ gain in
time when going from 4 to 8 nodes, so it is advisable to run the
simulations on the smallest $n$ possible for a given $V$. Comparing
the results on the different lattice 
sizes running on 8 nodes shows that the time increases proportionally
to the volume. Hence it seems that the code does not profit anymore
from longer loops as they appear for larger volumes.

Typical benchmark measurements of the overall performance 
for the quark propagator inversion in hadron spectroscopy
are given in Table \ref{tab:overall_perf}. 
These values depend on the number of iterations needed for
the inversion of the Dirac operator, thus they only 
give an estimate for the performance of simulations in which the smallest
quark mass is given by those used in our runs. 
The bottom line is that
our code is reasonably efficient, running at an overall rate of around 30\%
of peak performance in the production runs. This number noticeably increases
when considering only the matrix inversions, since the construction of the
$D$ and $R$ operators and the time for I/O and MPI setup decrease overall
performance: The matrix-vector multiplications without communication
run at 6.3 GFLOPS per node for the Dirac operator $D$ and at
8.6 GFLOPS per node for the $R$ operator, which is remarkably fast.

\section{Matrix Inversion Techniques}
\label{app:multimass}

\begin{figure}[htb]
\begin{center}
\includegraphics[width=10cm]{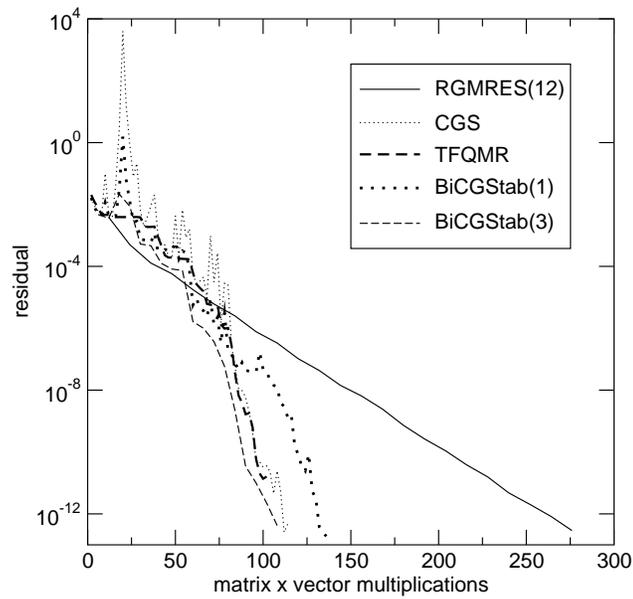}
\end{center}
\caption{Convergence history of quark propagator inversion for various
algorithms. All but the RGMRES algorithm converge after a
comparable number of matrix-vector multiplications. While CGS shows
large fluctuations of the 
residual, the convergence of the transpose-free QMR (TFQMR) and
BiCGStab is smoother.} 
\label{fig:inversion_convergence}
\end{figure}

The key element for efficient simulations of quenched QCD is a fast
matrix inversion algorithm. As the rank of the Dirac operator matrix is far
too large to perform an exact inversion, the methods of choice are
iterative procedures, and the most widely used algorithms for QCD are
variants of Krylov subspace methods. Consider the linear system of equations
\begin{equation} \label{eq:linsys}
Mx=b, 
\end{equation}
where $M\in {\bf C}^{R\times R}$ is in general a non-hermitean matrix and
$b\in {\bf C}^R$ is the source vector on which the inversion is carried
out. Choosing an initial guess $x_0$, the initial residual $r_0 =
b-Mx_0$ is defined. The Krylov subspace $K^n$ is then given by
\begin{equation}
 K^n = {\rm span} \{M^mr_0;  m=0,\dots,n\}.
\end{equation}
The common feature of all variants of Krylov space solvers is that the
solution of \eqref{eq:linsys} is iteratively approximated using an orthogonal
basis of 
the Krylov subspace $K^n$. The iterative solution $x_n$ is of the form
\begin{equation}
 x_n = x_0 + q_{n-1}(M) r_0,
\end{equation}
where $q_{n-1}$ is a polynomial of maximum degree $n-1$. The iterative
residual $r_n$ is therefore
\begin{equation}
 r_n = \big(1-M q_{n-1}(M)\big) r_0.
\end{equation}
The best-known Krylov space solver is the Conjugate Gradient (CG)
algorithm, which however only works for hermitean matrices. Refined
types of algorithms are the  Conjugate 
Gradient Squared (CGS)  \cite{sonneveld:89}, Quasi-Minimal Residual (QMR)
\cite{freund_nachtigal:91}, Generalized Minimal Residual (GMRES)
\cite{saad:86} or 
Bi-Conjugate Gradient (BiCG) \cite{vandervorst:92}
algorithms. These more sophisticated methods show either
faster convergence or increased stability and are also applicable
to non-hermitean matrices. We compared the convergence 
of the matrix inversion for several algorithms on a toy lattice of size
$4^4$, with $M$ given by the parametrized FP Dirac operator. The results in
Fig.~\ref{fig:inversion_convergence} show that the 
stabilized Bi-Conjugate Gradient (BiCGStab), transpose-free QMR and
CGS algorithm 
perform comparably well in this test. The order $l$ in the
BiCGStab($l$) variants denotes the order of the subspaces which are
intermediately orthogonalized in the iteration process. Closer
investigations of the properties of 
the different methods have shown that BiCGStab is in general a good
and reliable 
choice \cite{Frommer:1994vn,Gupta:1996zv,Cella:1996nn}.

As QCD simulations are in generally done at several quark masses, a
significant computational gain 
can be obtained using multi-mass solvers \cite{Frommer:1995ik}, which
exploit the fact  
that it is possible to get the solution of the shifted linear system
\begin{equation}
 (M + \sigma ) x = b,
\end{equation}
for a whole set of values $\sigma\in {\bf C}$ at the cost of only one
inversion. In QCD this implies that the cost of a multi-mass inversion is
equivalent to the cost of a single inversion at the smallest quark mass. 
For our spectroscopy calculations, we worked with the multi-mass BiCGStab
algorithm in \cite{Jegerlehner:1996pm}\footnote{A typo in the
algorithm had to be corrected.}. The drawback of this method is that
it is no longer possible to improve the condition number 
of $M$ by using preconditioning techniques, because the starting guess
$x_0$ for the multi-mass inversion is required to be zero
\cite{Jegerlehner:1996pm}. If one however wants to invert at a large
number of quark masses, this disadvantage is more than compensated by
the gain from needing only one inversion.

\fancyhead[RE]{\nouppercase{\small\it Conditions on the Dirac Operator
from Discrete Symmetries}}
\chapter{Conditions on the Dirac Operator from Discrete Symmetries}

While a lattice transcription of the continuum quark action allows
many possible discretizations of the Dirac operator, it is essential that
any lattice Dirac operator has the same properties under discrete
symmetry transformations as in the continuum. We derive here the 
transformation properties under reflection of a coordinate axis and
charge conjugation from the basic properties of the Dirac spinors and
the gauge fields. The representation of the Clifford algebra which
is used is given in Appendix \ref{app:conventions}.

\section{Reflection of an Axis} \label{app:axis_refl}
A reflection in direction of a coordinate axis $\eta$ with
$\eta=1,\dots,4$ can be described by a unitary operator
$\mathcal{P}_\eta = \mathcal{P}_\eta^\Psi \mathcal{P}_\eta^U$. The
operator $\mathcal{P}_\eta^\Psi$ acts on the fermion fields, 
\begin{align}
  \mathcal{P}_\eta^\Psi\Psi(n){\mathcal{P}_\eta^\Psi}^{-1} &=
  P_\eta\Psi(\tilde n), \label{par1}\\ 
\mathcal{P}_\eta^\Psi\bar\Psi(n){\mathcal{P}_\eta^\Psi}^{-1} &=
\bar\Psi(\tilde n)P_\eta^{-1}, \label{par2} 
\end{align}
where $\tilde n$ is the reflected lattice space-time variable,
\begin{equation}
\label{eq:reflect_n}
\tilde n_\mu \equiv \begin{cases} -n_\mu \quad  &\textrm{for } \mu=\eta, \\
n_\mu  &\textrm{for } \mu\neq\eta,
\end{cases}
\end{equation}
and $P_\eta$ is a matrix in Dirac space (in our representation
$P_\eta=\gamma_\eta\gamma_5$). On the other hand, $\mathcal{P}_\eta^U$
acts on the gauge fields,
\begin{equation}\label{par3}
\mathcal{P}_\eta^U U_\mu(n){\mathcal{P}_\eta^U}^{-1} =
U_\mu^{\mathcal{P}_\eta}(n), 
\end{equation}
where the reflected gauge field $U_\mu^{\mathcal{P}_\eta}$ is defined as
\begin{equation}
U_\mu^{\mathcal{P}_\eta}(n) \equiv \begin{cases} 
U_\eta^\dagger(\tilde n - \hat\eta) &\textrm{for } \mu=\eta, \\
U_\mu(\tilde n) \quad&\textrm{for } \mu\neq\eta.
\end{cases}
\end{equation}
The lattice fermion action
\begin{equation} \label{act}
  S[\Psi,\bar\Psi,U_\mu] = \sum_{n,n^\prime} \bar\Psi(n)
  D(n,n^\prime,U_\mu(n)) \Psi(n), 
\end{equation}
has to be symmetric under reflections of the $\eta$-axis:
\begin{equation} \label{eq:axsym}
 S[\Psi,\bar\Psi,U_\mu]  =
\mathcal{P}_\eta S[\Psi,\bar\Psi,U_\mu] \mathcal{P}_\eta^{-1}.
\end{equation}
Inserting the action \eqref{act} into \eqref{eq:axsym} and using
$\mathcal{P}_\eta \mathcal{P}_\eta^{-1}= {\bf 1} $ twice, we get
\begin{equation}
  S[\Psi,\bar\Psi,U_\mu] = \sum_{n,n^\prime}
  \mathcal{P}_\eta\bar\Psi(n) \mathcal{P}_\eta^{-1}\mathcal{P}_\eta
  D(n,n^\prime,U_\mu(n))\mathcal{P}_\eta^{-1}\mathcal{P}_\eta\Psi(n) 
  \mathcal{P}_\eta^{-1}.
\end{equation}
The fermion and gauge fields transform as specified in
Eqs.~\eqref{par1}--\eqref{par3}:
\begin{equation} \label{imd}
  S[\Psi,\bar\Psi,U_\mu] = \sum_{n,n^\prime} \bar\Psi(\tilde n)P_\eta^{-1}
  D(n,n^\prime,U_\mu^{\mathcal{P}_\eta}(n)) P_\eta\Psi(\tilde n).
\end{equation}
It remains to reorder the summation over the $\eta$-component of the
lattice variables $n$ and $n^\prime$. As it does not matter whether
the sum over a variable is performed from above or from below,
\begin{equation}
  \sum_{n_\eta=-\infty}^\infty F(n) = \sum_{n_\eta=\infty}^{-\infty}
  F(n) = \sum_{n_\eta=-\infty}^\infty F(\tilde n), 
\end{equation}
the argument of the summand can be reflected
without changing the value of the sum. Applying this to
Eq.~\eqref{imd}, we get  
\begin{equation}
  S[\Psi,\bar\Psi,U_\mu] = \sum_{n,n^\prime} \bar\Psi(n)P_\eta^{-1}
  D(\tilde n,\tilde n^\prime,U_\mu^{\mathcal{P}_\eta}(\tilde n))
  P_\eta\Psi(n),  
\end{equation}
which provides us the condition for the Dirac operator
\begin{equation}
  D(n,n^\prime,U_\mu(n)) = P_\eta^{-1} D(\tilde n,\tilde
  n^\prime,U_\mu^{\mathcal{P}_\eta}(\tilde n)) P_\eta \, ,
\end{equation}
 when comparing with the original action \eqref{act}.

\section{Charge Conjugation} \label{app:charge_conjg}
The unitary charge conjugation operator $\mathcal{C}$
acts on the fermion and gauge fields like \cite{Peskin:1995ev}
\begin{align}
  \mathcal{C}\Psi\mathcal{C}^{-1} &= C \bar\Psi^T,  \\
  \mathcal{C}\bar\Psi\mathcal{C}^{-1} &= \Psi^T C^{-1}, \\
  \mathcal{C} U_\mu\mathcal{C}^{-1} &= U_\mu^*,
\end{align}
where the charge conjugation matrix fulfills $C\gamma_\mu^TC^{-1} =
-\gamma_\mu$ and can be expressed in our representation by
$C=\gamma_2\gamma_4$. Invariance of the action under
$\mathcal{C}$-transformations means  
\begin{equation}
  \bar\Psi D(U_\mu)\Psi = \mathcal{C} \bar\Psi D(U_\mu) \Psi\mathcal{C}^{-1}.
\end{equation}
Inserting $\mathcal{C}^{-1}\mathcal{C}$ twice and using $C^T=C^{-1}$, we get
\begin{align}
  \bar\Psi D(U_\mu)\Psi &= \Psi^T C^{-1}
 D(U_\mu^*) C \bar\Psi^T  \nonumber \\
 &= \bar\Psi C^{-1} D(U_\mu^*)^T C\Psi,
\end{align}
where the transposition extends over all index spaces. The
transformation property of the Dirac operator under charge conjugation
is therefore
\begin{equation}
  D(U_\mu) = C^{-1} D(U_\mu^*)^T C,
\end{equation}
completing the set of conditions from the $\mathcal{C}$,
$\mathcal{P}$ and (Euclidean) $\mathcal{T}$ symmetries. 

\fancyhead[RE]{\nouppercase{\small\it Collection of Data}}
\chapter{Collection of Data}

\section{Hadron Masses}
\label{app:data}

For each lattice, we list the bare input quark masses, the bias-corrected
masses from correlated 
fits to the hadron propagators, the value of 
$\chi^2/df$ for the fit and the fit range. The numbers in
brackets and the superscripts denote bootstrap
errors and bias \eqref{eq:bias_boot}, respectively.

\subsection{Pseudoscalar Mesons}

Pseudoscalar mesons are determined from the pseudoscalar (P), fourth
component axial vector (A), and pseudoscalar minus scalar (P-S) correlators.

\vspace{3mm}

\begin{table}[htb]
\hspace{-8mm}
\small
\begin{tabular}{l*{3}{|l|c|c}} \hline\hline
 \multicolumn{10}{c}{$8^3\times 24$, $\beta=3.0$, parametrized FP } \\ \hline
$am_q$ & $am_{\rm PS}$(P) & $\chi^2_{df}$ & $t$ & $am_{\rm PS}$(A) &
$\chi^2_{df}$ & $t$ & $am_{\rm PS}$(P-S) & $\chi^2_{df}$ & $t$  \\
\hline
0.021 & $0.240(15)^{+8}$ & 1.9 & [4,12] & $0.254(13)^{-1}$ & 0.7 & [5,12] & $0.198(16)^{+8}$ & 2.3 & [4,12] \\ 
0.028 & $0.289(9)^{+2}$ & 2.4 & [4,12] & $0.288(9)^{+0}$ & 0.9 & [5,12] & $0.246(13)^{+6}$ & 2.3 & [5,12] \\ 
0.04 & $0.345(7)^{+2}$ & 2.6 & [4,12] & $0.335(7)^{+1}$ & 1.4 & [5,12] & $0.326(9)^{+2}$ & 3.0 & [5,12] \\ 
0.06 & $0.415(6)^{+2}$ & 2.4 & [4,12] & $0.405(6)^{+1}$ & 1.8 & [5,12] & $0.413(7)^{+1}$ & 3.2 & [5,12] \\ 
0.09 & $0.503(5)^{+1}$ & 2.2 & [4,12] & $0.495(5)^{+1}$ & 1.9 & [5,12] & $0.508(5)^{+1}$ & 3.6 & [6,12] \\ 
0.13 & $0.606(3)^{+1}$ & 2.0 & [4,12] & $0.599(4)^{+1}$ & 1.8 & [5,12] & $0.614(4)^{+1}$ & 3.3 & [6,12] \\ 
0.18 & $0.720(3)^{+1}$ & 1.7 & [4,12] & $0.714(3)^{+1}$ & 1.7 & [5,12] & $0.728(3)^{+0}$ & 2.8 & [6,12] \\ 
0.25 & $0.866(2)^{+0}$ & 1.6 & [4,12] & $0.862(2)^{+1}$ & 1.8 & [5,12] & $0.875(3)^{+0}$ & 2.2 & [6,12] \\ 
0.33 & $1.024(2)^{+0}$ & 1.8 & [4,12] & $1.022(2)^{+0}$ & 2.1 & [5,12] & $1.031(3)^{+0}$ & 1.9 & [7,12] \\ 
 \hline
\end{tabular}
\caption{Pseudoscalar meson masses on $8^3\times 24$ lattice at
$\beta=3.0$ with $D^{\rm FP}$.}
\end{table}

\begin{table}[htb]
\hspace{-11mm}
\small
\begin{tabular}{l*{3}{|l|c|c}} \hline\hline
 \multicolumn{10}{c}{$12^3\times 24$, $\beta=3.0$, parametrized FP }\\ \hline
$am_q$ & $am_{\rm PS}$(P) & $\chi^2_{df}$ & $t$ & $am_{\rm PS}$(A) &
$\chi^2_{df}$ & $t$ & $am_{\rm PS}$(P-S) & $\chi^2_{df}$ & $t$  \\
\hline
0.016 & $0.2300(51)^{-8}$ & 0.9 & [4,12] & $0.2177(50)^{-2}$ & 2.2 & [4,12] & $0.2020(70)^{+22}$ & 0.6 & [6,12] \\ 
0.021 & $0.2555(33)^{-2}$ & 0.6 & [4,12] & $0.2472(41)^{+4}$ & 1.5 & [4,12] & $0.2409(51)^{+10}$ & 0.6 & [6,12] \\ 
0.028 & $0.2881(27)^{+1}$ & 0.6 & [4,12] & $0.2840(35)^{+2}$ & 0.9 & [4,12] & $0.2809(43)^{+9}$ & 0.5 & [6,12] \\ 
0.04 & $0.3382(20)^{+2}$ & 0.5 & [4,12] & $0.3359(29)^{+2}$ & 0.5 & [4,12] & $0.3362(37)^{+5}$ & 0.4 & [6,12] \\ 
0.06 & $0.4090(16)^{+2}$ & 0.4 & [4,12] & $0.4077(25)^{+1}$ & 0.5 & [4,12] & $0.4099(27)^{+2}$ & 0.6 & [7,12] \\ 
0.09 & $0.4987(14)^{+1}$ & 0.7 & [4,12] & $0.4972(22)^{+0}$ & 1.1 & [4,12] & $0.5009(22)^{+1}$ & 0.9 & [7,12] \\ 
0.13 & $0.6016(13)^{+1}$ & 1.3 & [4,12] & $0.6001(17)^{+2}$ & 2.2 & [4,12] & $0.6052(17)^{+0}$ & 0.8 & [7,12] \\ 
0.18 & $0.7165(11)^{+1}$ & 1.8 & [4,12] & $0.7151(15)^{+1}$ & 3.3 & [4,12] & $0.7213(15)^{+0}$ & 0.7 & [7,12] \\ 
0.25 & $0.8640(11)^{+0}$ & 2.1 & [4,12] & $0.8624(13)^{+1}$ & 4.5 & [4,12] & $0.8703(14)^{+0}$ & 0.7 & [7,12] \\ 
0.33 & $1.0229(11)^{+1}$ & 2.2 & [4,12] & $1.0212(12)^{+1}$ & 5.5 & [4,12] & $1.0304(13)^{+1}$ & 1.1 & [7,12] \\ 
\hline
\end{tabular}
\caption{Pseudoscalar meson masses on $12^3\times 24$ lattice at
$\beta=3.0$ with $D^{\rm FP}$.}
\end{table}

\begin{table}[htb]
\hspace{-9mm}
\small
\begin{tabular}{l*{3}{|l|c|c}} \hline\hline
 \multicolumn{10}{c}{$12^3\times 24$, $\beta=3.0$, overlap-improved } \\ \hline
$am_q$ & $am_{\rm PS}$(P) & $\chi^2_{df}$ & $t$ & $am_{\rm PS}$(A) &
$\chi^2_{df}$ & $t$ & $am_{\rm PS}$(P-S) & $\chi^2_{df}$ & $t$  \\ \hline
0.009 & $0.196(11)^{+3}$ & 2.1 & [4,12] & $0.218(6)^{+0}$ & 1.3 & [4,12] & $0.154(14)^{+7}$ & 1.4 & [4,12] \\ 
0.012 & $0.217(9)^{+2}$ & 2.0 & [4,12] & $0.235(6)^{+0}$ & 1.1 & [4,12] & $0.185(13)^{+6}$ & 1.2 & [4,12] \\ 
0.016 & $0.245(6)^{+2}$ & 2.0 & [4,12] & $0.256(5)^{+1}$ & 1.0 & [4,12] & $0.221(12)^{+5}$ & 1.3 & [4,12] \\ 
0.021 & $0.274(5)^{+2}$ & 1.9 & [4,12] & $0.282(5)^{+1}$ & 1.2 & [4,12] & $0.258(10)^{+4}$ & 1.7 & [5,12] \\ 
0.028 & $0.309(4)^{+1}$ & 1.9 & [4,12] & $0.315(4)^{+1}$ & 1.6 & [4,12] & $0.300(7)^{+3}$ & 2.0 & [5,12] \\ 
0.04 & $0.360(3)^{+1}$ & 1.8 & [4,12] & $0.363(4)^{+0}$ & 1.9 & [4,12] & $0.355(6)^{+2}$ & 2.2 & [5,12] \\ 
0.06 & $0.433(3)^{+1}$ & 1.9 & [4,12] & $0.432(3)^{+1}$ & 2.3 & [4,12] & $0.428(4)^{+1}$ & 2.3 & [6,12] \\ 
0.09 & $0.526(2)^{+1}$ & 2.3 & [4,12] & $0.524(3)^{+1}$ & 2.6 & [4,12] & $0.525(3)^{+1}$ & 2.0 & [6,12] \\ 
0.13 & $0.633(2)^{+1}$ & 2.5 & [4,12] & $0.631(2)^{+1}$ & 2.9 & [4,12] & $0.634(2)^{+1}$ & 2.3 & [7,12] \\ 
0.18 & $0.753(2)^{+1}$ & 2.5 & [4,12] & $0.752(2)^{+0}$ & 3.1 & [4,12] & $0.756(2)^{+0}$ & 1.9 & [7,12] \\ 
0.25 & $0.905(2)^{+0}$ & 2.1 & [4,12] & $0.905(2)^{+0}$ & 3.1 & [4,12] & $0.911(2)^{+0}$ & 1.7 & [7,12] \\ 
0.33 & $1.069(1)^{+0}$ & 1.7 & [4,12] & $1.069(2)^{+0}$ & 3.2 & [4,12] & $1.076(2)^{+0}$ & 1.7 & [7,12] \\ 
\hline
\end{tabular}
\caption{Pseudoscalar meson masses on $12^3\times 24$ lattice at
$\beta=3.0$ with the overlap-improved $D_{\rm ov}^{\rm FP}$.}
\end{table}

\begin{table}[htb]
\hspace{-12mm}
\small
\begin{tabular}{l*{3}{|l|c|c}} \hline\hline
 \multicolumn{10}{c}{$16^3\times 32$, $\beta=3.0$, parametrized FP} \\ \hline
$am_q$ & $am_{\rm PS}$(P) & $\chi^2_{df}$ & $t$ & $am_{\rm PS}$(A) &
$\chi^2_{df}$ & $t$ & $am_{\rm PS}$(P-S) & $\chi^2_{df}$ & $t$  \\
\hline
0.013 & $0.1984(31)^{+7}$ & 1.2 & [4,16] & $0.1984(33)^{+6}$ & 0.8 & [4,16] & $0.1904(28)^{+17}$ & 0.9 & [4,16] \\ 
0.016 & $0.2224(19)^{+3}$ & 1.2 & [4,16] & $0.2227(26)^{+3}$ & 0.7 & [4,16] & $0.2168(22)^{+10}$ & 0.9 & [4,16] \\ 
0.021 & $0.2534(16)^{+2}$ & 1.0 & [4,16] & $0.2538(24)^{+3}$ & 0.5 & [4,16] & $0.2500(18)^{+7}$ & 0.9 & [4,16] \\ 
0.028 & $0.2892(13)^{+2}$ & 0.8 & [4,16] & $0.2893(20)^{+3}$ & 0.4 & [4,16] & $0.2862(16)^{+5}$ & 0.7 & [6,16] \\ 
0.04 & $0.3402(13)^{+0}$ & 0.8 & [5,16] & $0.3399(17)^{+2}$ & 0.3 & [4,16] & $0.3389(15)^{+4}$ & 0.5 & [6,16] \\ 
0.06 & $0.4109(12)^{+1}$ & 0.7 & [5,16] & $0.4100(14)^{+3}$ & 0.1 & [4,16] & $0.4116(14)^{+2}$ & 0.4 & [6,16] \\ 
0.09 & $0.5000(10)^{+2}$ & 0.7 & [5,16] & $0.4988(13)^{+2}$ & 0.1 & [4,16] & $0.5017(15)^{+0}$ & 0.5 & [8,16] \\ 
0.13 & $0.6022(10)^{+1}$ & 0.6 & [6,16] & $0.6012(11)^{+2}$ & 0.2 & [4,16] & $0.6045(12)^{+1}$ & 1.0 & [9,16] \\ 
0.18 & $0.7173(9)^{+1}$ & 0.8 & [6,16] & $0.7160(10)^{+1}$ & 0.3 & [4,16] & $0.7196(11)^{+1}$ & 1.5 & [9,16] \\ 
0.25 & $0.8649(9)^{+1}$ & 0.8 & [6,16] & $0.8639(9)^{+1}$ & 0.2 & [5,16] & $0.8665(11)^{+0}$ & 1.7 & [11,16] \\ 
0.33 & $1.0239(8)^{+1}$ & 0.6 & [6,16] & $1.0230(8)^{+1}$ & 0.2 & [5,16] & $1.0256(11)^{+1}$ & 1.1 & [11,16] \\ 
 \hline
\end{tabular}
\caption{Pseudoscalar meson masses on $16^3\times 32$ lattice at
$\beta=3.0$ with $D^{\rm FP}$.} 
\end{table}

\begin{table}[htb]
\hspace{-7mm}
\small
\begin{tabular}{l*{3}{|l|c|c}} \hline\hline
 \multicolumn{10}{c}{$12^3\times 24$, $\beta=3.4$, parametrized FP } \\ \hline
$am_q$ & $am_{\rm PS}$(P) & $\chi^2_{df}$ & $t$ & $am_{\rm PS}$(A) &
$\chi^2_{df}$ & $t$ & $am_{\rm PS}$(P-S) & $\chi^2_{df}$ & $t$  \\
\hline
0.029 & $0.191(6)^{+0}$ & 1.3 & [4,12] & $0.183(5)^{-1}$ & 0.3 & [3,12] & $0.159(9)^{+3}$ & 0.5 & [7,12] \\ 
0.032 & $0.203(4)^{+0}$ & 1.1 & [4,12] & $0.197(5)^{-1}$ & 0.4 & [3,12] & $0.170(9)^{+3}$ & 0.3 & [8,12] \\ 
0.037 & $0.224(4)^{+0}$ & 1.1 & [4,12] & $0.218(4)^{+0}$ & 0.6 & [3,12] & $0.197(8)^{+2}$ & 0.2 & [8,12] \\ 
0.045 & $0.251(5)^{+0}$ & 2.3 & [7,12] & $0.248(4)^{+0}$ & 0.9 & [3,12] & $0.234(11)^{+1}$ & 0.0 & [10,12] \\ 
0.058 & $0.294(3)^{+0}$ & 2.8 & [7,12] & $0.291(3)^{+0}$ & 1.4 & [3,12] & $0.287(8)^{+0}$ & 0.0 & [10,12] \\ 
0.078 & $0.355(3)^{+0}$ & 3.3 & [7,12] & $0.352(2)^{+0}$ & 1.8 & [3,12] & $0.355(6)^{+0}$ & 0.4 & [10,12] \\ 
0.1 & $0.415(2)^{+0}$ & 3.9 & [7,12] & $0.411(2)^{+0}$ & 1.8 & [3,12] & $0.417(4)^{+0}$ & 1.4 & [10,12] \\ 
0.14 & $0.514(2)^{+0}$ & 4.3 & [7,12] & $0.509(2)^{+0}$ & 1.5 & [3,12] & $0.518(3)^{+0}$ & 4.0 & [10,12] \\ 
0.18 & $0.603(2)^{+0}$ & 4.3 & [7,12] & $0.598(1)^{+0}$ & 1.2 & [3,12] & $0.609(2)^{+0}$ & 6.8 & [10,12] \\ 
0.24 & $0.729(1)^{+0}$ & 3.8 & [7,12] & $0.723(1)^{+0}$ & 1.0 & [3,12] & $0.736(2)^{+0}$ & 9.1 & [10,12] \\ 
\hline
\end{tabular}
\caption{Pseudoscalar meson masses on $12^3\times 24$ lattice at
$\beta=3.4$ with $D^{\rm FP}$.}
\end{table}

\begin{table}[htb]
\hspace{-7mm}
\small
\begin{tabular}{l*{3}{|l|c|c}} \hline\hline
 \multicolumn{10}{c}{$16^3\times 32$, $\beta=3.7$, parametrized FP } \\ \hline
$am_q$ & $am_{\rm PS}$(P) & $\chi^2_{df}$ & $t$ & $am_{\rm PS}$(A) &
$\chi^2_{df}$ & $t$ & $am_{\rm PS}$(P-S) & $\chi^2_{df}$ & $t$  \\
\hline
0.0235 & $0.133(11)^{-4}$ & 0.6 & [6,16] & $0.120(8)^{+2}$ & 0.7 & [7,16] & $0.120(12)^{+5}$ & 2.2 & [7,14] \\ 
0.026 & $0.143(8)^{-2}$ & 0.8 & [6,16] & $0.134(6)^{+2}$ & 0.8 & [7,16] & $0.135(10)^{+3}$ & 2.0 & [7,15] \\ 
0.03 & $0.159(6)^{-1}$ & 1.0 & [6,16] & $0.153(6)^{+1}$ & 0.7 & [7,16] & $0.155(7)^{+2}$ & 2.0 & [7,15] \\ 
0.036 & $0.183(4)^{+0}$ & 1.2 & [6,16] & $0.177(5)^{+1}$ & 0.4 & [7,16] & $0.186(6)^{+2}$ & 2.1 & [7,15] \\ 
0.045 & $0.216(3)^{+1}$ & 1.4 & [6,16] & $0.210(4)^{+2}$ & 0.4 & [7,16] & $0.225(5)^{+2}$ & 2.2 & [8,16] \\ 
0.06 & $0.264(3)^{+1}$ & 1.4 & [6,16] & $0.258(4)^{+1}$ & 0.8 & [7,16] & $0.277(4)^{+1}$ & 1.5 & [8,16] \\ 
0.08 & $0.321(2)^{+0}$ & 1.1 & [6,16] & $0.315(3)^{+1}$ & 0.9 & [7,16] & $0.335(3)^{+0}$ & 1.1 & [9,16] \\ 
0.1 & $0.372(2)^{+1}$ & 0.9 & [6,16] & $0.368(2)^{+1}$ & 1.0 & [7,16] & $0.386(3)^{+0}$ & 1.1 & [9,16] \\ 
0.14 & $0.465(1)^{+0}$ & 0.9 & [6,16] & $0.463(2)^{+0}$ & 0.6 & [8,16] & $0.478(2)^{+0}$ & 1.6 & [9,16] \\ 
0.18 & $0.550(1)^{+0}$ & 1.4 & [6,16] & $0.549(1)^{+0}$ & 0.8 & [8,16] & $0.562(2)^{+0}$ & 1.7 & [9,16] \\ 
\hline
\end{tabular}
\caption{Pseudoscalar meson masses on $16^3\times 32$ lattice at
$\beta=3.7$ with $D^{\rm FP}$.}
\end{table}

\clearpage
\subsection{Vector Mesons, $m_{\rm PS}/m_{\rm V}$ and $m_{\rm Oct}/m_{\rm V}$}
Together with the bias-corrected fitted masses for the vector 
meson, we list the mass ratios $m_{\rm PS}/m_{\rm
V}$ and $m_{\rm Oct}/m_{\rm V}$, where the pseudoscalar mass is
taken from fits to the P-S correlator at small and the P correlator at
large quark mass.
% and for the octet baryon in each case the more
%reliable of the two correlators N and N0 is used.

\nopagebreak

\begin{table}[htb]
\small
\begin{center}
\begin{tabular}{l|l|c|c|l|l} \hline\hline
 \multicolumn{6}{c}{$8^3\times 24$, $\beta=3.0$, parametrized FP } \\ \hline
$am_q$ & $am_{\rm V}$ & $\chi^2_{df}$ & $t$ & $m_{\rm PS}/m_{\rm V}$ & $m_{\rm Oct}/m_{\rm V}$  \\ \hline
0.021 & $0.722(21)^{+2}$ & 1.5 & [4,12] & 0.274(24) & N/A  \\ 
0.028 & $0.706(17)^{+6}$ & 1.3 & [4,12] & 0.349(21) & 1.221(21) \\ 
0.04 & $0.711(13)^{+5}$ & 1.2 & [4,12] & 0.485(13) & 1.396(64) \\ 
0.06 & $0.738(12)^{+4}$ & 1.2 & [4,12] & 0.562(12) & 1.429(40) \\ 
0.09 & $0.780(9)^{+2}$ & 1.4 & [4,12] & 0.645(10) & 1.455(29) \\ 
0.13 & $0.837(7)^{+3}$ & 1.6 & [4,12] & 0.724(7) & 1.486(20) \\ 
0.18 & $0.912(5)^{+2}$ & 1.4 & [4,12] & 0.789(6) & 1.508(15) \\ 
0.25 & $1.024(4)^{+2}$ & 1.1 & [4,12] & 0.846(4) & 1.524(11) \\ 
0.33 & $1.155(3)^{+1}$ & 1.1 & [4,12] & 0.887(3) & 1.531(8) \\ \hline
\end{tabular}

\vspace{2mm}

\begin{tabular}{l|l|c|c|l|l} \hline\hline
 \multicolumn{6}{c}{$12^3\times 24$, $\beta=3.0$, parametrized FP } \\ \hline
$am_q$ & $am_{\rm V}$ & $\chi^2_{df}$ & $t$ & $m_{\rm PS}/m_{\rm V}$ & $m_{\rm Oct}/m_{\rm V}$  \\ \hline
0.016 & $0.676(14)^{+2}$ & 1.7 & [4,12] & 0.299(12) & 1.230(97) \\ 
0.021 & $0.683(12)^{+1}$ & 1.7 & [4,12] & 0.353(9) & 1.247(51) \\ 
0.028 & $0.691(9)^{+2}$ & 1.6 & [4,12] & 0.406(8) & 1.260(38) \\ 
0.04 & $0.708(7)^{+1}$ & 1.7 & [4,12] & 0.475(7) & 1.285(25) \\ 
0.06 & $0.736(5)^{+0}$ & 1.8 & [4,12] & 0.556(4) & 1.337(18) \\ 
0.09 & $0.779(3)^{+0}$ & 1.7 & [4,12] & 0.641(3) & 1.395(12) \\ 
0.13 & $0.838(3)^{+0}$ & 1.6 & [4,12] & 0.718(3) & 1.443(8) \\ 
0.18 & $0.914(2)^{+0}$ & 1.5 & [4,12] & 0.784(2) & 1.480(6) \\ 
0.25 & $1.026(2)^{+0}$ & 1.2 & [4,12] & 0.842(2) & 1.510(5) \\ 
0.33 & $1.156(1)^{+0}$ & 0.8 & [4,12] & 0.885(1) & 1.526(4) \\ \hline
\end{tabular}

\vspace{2mm}

\begin{tabular}{l|l|c|c|l|l} \hline\hline
 \multicolumn{6}{c}{$16^3\times 32$, $\beta=3.0$, parametrized FP } \\ \hline
$am_q$ & $am_{\rm V}$ & $\chi^2_{df}$ & $t$ & $m_{\rm PS}/m_{\rm V}$ & $m_{\rm Oct}/m_{\rm V}$  \\ \hline
0.013 & $0.692(16)^{+2}$ & 1.7 & [5,16] & 0.275(7) & 1.226(39) \\ 
0.016 & $0.696(14)^{+0}$ & 1.9 & [5,16] & 0.312(7) & 1.228(32) \\ 
0.021 & $0.702(10)^{+0}$ & 2.0 & [5,16] & 0.356(5) & 1.241(24) \\ 
0.028 & $0.711(8)^{+0}$ & 2.0 & [5,16] & 0.403(5) & 1.261(20) \\ 
0.04 & $0.724(6)^{+0}$ & 1.8 & [5,16] & 0.468(4) & 1.297(16) \\ 
0.06 & $0.748(4)^{+0}$ & 1.3 & [5,16] & 0.549(3) & 1.347(13) \\ 
0.09 & $0.787(3)^{+0}$ & 1.2 & [5,16] & 0.635(3) & 1.400(10) \\ 
0.13 & $0.843(2)^{+0}$ & 1.5 & [5,16] & 0.714(2) & 1.448(7) \\ 
0.18 & $0.918(2)^{+0}$ & 1.6 & [5,16] & 0.781(2) & 1.484(6) \\ 
0.25 & $1.029(1)^{+0}$ & 1.4 & [5,16] & 0.841(1) & 1.511(4) \\ 
0.33 & $1.160(1)^{+0}$ & 1.1 & [5,16] & 0.883(1) & 1.526(3) \\ \hline
\end{tabular}

\end{center}
\vspace{-2mm}
\caption{Vector meson masses and mass ratios $m_{\rm PS}/m_{\rm V}$
and $m_{\rm Oct}/m_{\rm V}$ on the three lattice of size $8^3\times
24$, $12^3\times 24$ and $16^3\times 32$ at
$\beta=3.0$ with $D^{\rm FP}$.}
\label{tab:vmass}
\end{table}

\begin{table}[htb]
\small
\begin{center}
\begin{tabular}{l|l|c|c|l|l} \hline\hline
 \multicolumn{6}{c}{$12^3\times 24$, $\beta=3.0$, overlap-improved} \\ \hline
$am_q$ & $am_{\rm V}$ & $\chi^2_{df}$ & $t$ & $m_{\rm PS}/m_{\rm V}$ & $m_{\rm Oct}/m_{\rm V}$  \\ \hline
0.009 & $0.747(31)^{+6}$ & 2.7 & [3,11] & 0.206(21) & 1.297(123) \\ 
0.012 & $0.743(26)^{+5}$ & 2.4 & [3,11] & 0.249(19) & 1.274(96) \\ 
0.016 & $0.746(21)^{+6}$ & 1.9 & [3,11] & 0.297(18) & 1.253(68) \\ 
0.021 & $0.752(16)^{+5}$ & 1.5 & [3,11] & 0.343(15) & 1.241(51) \\ 
0.028 & $0.754(26)^{+7}$ & 1.6 & [4,11] & 0.398(17) & 1.264(54) \\ 
0.04 & $0.766(19)^{+4}$ & 1.4 & [4,11] & 0.463(14) & 1.297(42) \\ 
0.06 & $0.789(13)^{+2}$ & 1.1 & [4,11] & 0.543(10) & 1.353(31) \\ 
0.09 & $0.826(7)^{+1}$ & 0.9 & [4,11] & 0.635(7) & 1.397(20) \\ 
0.13 & $0.883(5)^{+1}$ & 0.9 & [4,11] & 0.717(5) & 1.434(15) \\ 
0.18 & $0.962(4)^{+1}$ & 1.0 & [4,11] & 0.783(4) & 1.469(10) \\ 
0.25 & $1.077(3)^{+0}$ & 0.8 & [4,11] & 0.841(3) & 1.500(8) \\ 
0.33 & $1.212(2)^{+0}$ & 0.5 & [4,11] & 0.882(2) & 1.518(7) \\  \hline
\end{tabular}
\end{center}
\vspace{-2mm}
\caption{Vector meson masses and mass ratios on the $12^3\times 24$ lattice at
$\beta=3.0$ with overlap-improved $D_{\rm ov}^{\rm FP}$.}
\end{table}

\begin{table}[htb]
\small
\begin{center}
\begin{tabular}{l|l|c|c|l|l} \hline\hline
 \multicolumn{6}{c}{$12^3\times 24$, $\beta=3.4$, parametrized FP } \\ \hline
$am_q$ & $am_{\rm V}$ & $\chi^2_{df}$ & $t$ & $m_{\rm PS}/m_{\rm V}$ & $m_{\rm Oct}/m_{\rm V}$  \\ \hline
0.029 & $0.470(20)^{+2}$ & 0.5 & [7,12] & 0.339(25) & 1.428(87) \\ 
0.032 & $0.474(17)^{+1}$ & 0.6 & [7,12] & 0.359(24) & 1.433(76) \\ 
0.037 & $0.482(13)^{+1}$ & 0.7 & [7,12] & 0.410(21) & 1.444(58) \\ 
0.045 & $0.492(10)^{+1}$ & 0.7 & [7,12] & 0.475(25) & 1.473(43) \\ 
0.058 & $0.509(8)^{+1}$ & 0.7 & [7,12] & 0.563(19) & 1.503(31) \\ 
0.078 & $0.540(6)^{+1}$ & 1.1 & [7,12] & 0.651(9) & 1.513(25) \\ 
0.1 & $0.576(5)^{+1}$ & 1.6 & [7,12] & 0.713(7) & 1.523(21) \\ 
0.14 & $0.644(3)^{+0}$ & 2.1 & [7,12] & 0.791(5) & 1.525(15) \\ 
0.18 & $0.712(3)^{+0}$ & 2.2 & [7,12] & 0.840(4) & 1.531(12) \\ 
0.24 & $0.817(2)^{+0}$ & 1.8 & [7,12] & 0.886(3) & 1.534(8) \\ \hline
\end{tabular}
\end{center}
\vspace{-2mm}
\caption{Vector meson masses and mass ratios $m_{\rm PS}/m_{\rm V}$
and $m_{\rm Oct}/m_{\rm V}$ on $12^3\times 24$ lattice at
$\beta=3.4$ with $D^{\rm FP}$.}
\end{table}

\begin{table}[htb]
\small
\begin{center}
\begin{tabular}{l|l|c|c|l|l} \hline\hline
 \multicolumn{6}{c}{$16^3\times 32$, $\beta=3.7$, parametrized FP } \\ \hline
$am_q$ & $am_{\rm V}$ & $\chi^2_{df}$ & $t$ & $m_{\rm PS}/m_{\rm V}$ & $m_{\rm Oct}/m_{\rm V}$  \\ \hline
0.0235 & $0.332(26)^{+11}$ & 1.4 & [8,16] & 0.360(37) & 1.595(75) \\ 
0.026 & $0.337(22)^{+9}$ & 1.0 & [8,16] & 0.397(33) & 1.582(33) \\ 
0.03 & $0.346(18)^{+7}$ & 0.9 & [8,16] & 0.441(30) & 1.570(9) \\ 
0.036 & $0.366(13)^{+5}$ & 1.1 & [8,16] & 0.485(22) & 1.528(70) \\ 
0.045 & $0.386(10)^{+3}$ & 1.3 & [8,16] & 0.544(19) & 1.496(52) \\ 
0.06 & $0.411(8)^{+2}$ & 1.4 & [8,16] & 0.629(15) & 1.486(40) \\ 
0.08 & $0.443(6)^{+2}$ & 1.2 & [8,16] & 0.712(12) & 1.495(28) \\ 
0.1 & $0.477(4)^{+1}$ & 1.1 & [8,16] & 0.770(9) & 1.508(22) \\ 
0.14 & $0.547(3)^{+1}$ & 1.1 & [8,16] & 0.846(6) & 1.523(15) \\ 
0.18 & $0.618(2)^{+1}$ & 1.3 & [8,16] & 0.888(4) & 1.530(10) \\ \hline
\end{tabular}
\end{center}
%\vspace{-3mm}
\caption{Vector meson masses and mass ratios $m_{\rm PS}/m_{\rm V}$
and $m_{\rm Oct}/m_{\rm V}$ on $16^3\times 32$ lattice at
$\beta=3.7$ with $D^{\rm FP}$.}
\end{table}

\clearpage
\subsection{Octet Baryons}

For each lattice, the bias-corrected fitted masses for the N and N0
correlators, the value of $\chi^2/df$ for the fit and the 
fit range are given. The correlator which was used for the ratio
$m_{\rm Oct}/m_{\rm V}$ is marked with a star in each table.

\vspace{3ex}

\begin{table}[htb]
\small
\begin{center}
\begin{tabular}{l*{2}{|l|c|c}} \hline\hline
 \multicolumn{7}{c}{$8^3\times 24$, $\beta=3.0$, parametrized FP } \\ \hline
$am_q$ & $am_{\rm Oct}$(N)$^*$ & $\chi^2_{df}$ & $t$ & $am_{\rm Oct}$(N0) &
$\chi^2_{df}$ & $t$  \\ \hline
0.028 & $0.862(82)^{+29}$ & 1.0 & [4,12] & $1.024(54)^{+15}$ & 0.7 & [4,12] \\ 
0.04 & $0.993(41)^{+0}$ & 1.8 & [4,12] & $1.043(39)^{+11}$ & 0.7 & [4,12] \\ 
0.06 & $1.055(23)^{+2}$ & 2.0 & [4,12] & $1.090(28)^{+11}$ & 1.2 & [4,12] \\ 
0.09 & $1.135(18)^{+4}$ & 1.6 & [4,12] & $1.165(20)^{+9}$ & 1.8 & [4,12] \\ 
0.13 & $1.243(12)^{+4}$ & 1.3 & [4,12] & $1.258(14)^{+7}$ & 2.1 & [4,12] \\ 
0.18 & $1.376(10)^{+4}$ & 1.2 & [4,12] & $1.383(10)^{+6}$ & 2.0 & [4,12] \\ 
0.25 & $1.560(9)^{+4}$ & 1.2 & [4,12] & $1.563(8)^{+4}$ & 2.1 & [4,12] \\ 
0.33 & $1.768(7)^{+3}$ & 1.7 & [4,12] & $1.771(7)^{+3}$ & 2.4 & [4,12] \\ \hline
\end{tabular}

\vspace{3ex}

\begin{tabular}{l*{2}{|l|c|c}} \hline\hline
 \multicolumn{7}{c}{$12^3\times 24$, $\beta=3.0$, parametrized FP } \\ \hline
$am_q$ & $am_{\rm Oct}$(N)$^*$ & $\chi^2_{df}$ & $t$ & $am_{\rm Oct}$(N0) &
$\chi^2_{df}$ & $t$  \\ \hline
0.016 & $0.832(63)^{+0}$ & 1.3 & [4,12] & $0.755(34)^{+20}$ & 0.8 & [3,12] \\ 
0.021 & $0.852(32)^{+1}$ & 1.2 & [4,12] & $0.829(23)^{+7}$ & 1.5 & [3,12] \\ 
0.028 & $0.871(23)^{+3}$ & 1.1 & [4,12] & $0.868(17)^{+3}$ & 1.9 & [3,12] \\ 
0.04 & $0.910(15)^{+2}$ & 0.9 & [4,12] & $0.918(13)^{+2}$ & 1.9 & [3,12] \\ 
0.06 & $0.983(11)^{+2}$ & 1.2 & [4,12] & $0.991(9)^{+2}$ & 1.4 & [3,12] \\ 
0.09 & $1.086(8)^{+2}$ & 1.9 & [4,12] & $1.089(7)^{+2}$ & 1.2 & [3,12] \\ 
0.13 & $1.208(6)^{+1}$ & 1.8 & [4,12] & $1.210(5)^{+1}$ & 1.2 & [3,12] \\ 
0.18 & $1.353(4)^{+1}$ & 1.4 & [4,12] & $1.355(5)^{+1}$ & 1.3 & [3,12] \\ 
0.25 & $1.549(4)^{+1}$ & 1.2 & [4,12] & $1.548(4)^{+1}$ & 1.6 & [3,12] \\ 
0.33 & $1.765(4)^{+1}$ & 1.4 & [4,12] & $1.763(4)^{+1}$ & 1.9 & [3,12] \\  \hline
\end{tabular}

\vspace{3ex}

\begin{tabular}{l*{2}{|l|c|c}} \hline\hline
 \multicolumn{7}{c}{$16^3\times 32$, $\beta=3.0$, parametrized FP } \\ \hline
$am_q$ & $am_{\rm Oct}$(N) & $\chi^2_{df}$ & $t$ & $am_{\rm Oct}$(N0)$^*$ &
$\chi^2_{df}$ & $t$  \\ \hline
0.013 & $0.766(31)^{+19}$ & 1.1 & [4,16] & $0.848(18)^{+1}$ & 0.9 & [3,16] \\ 
0.016 & $0.799(19)^{+6}$ & 1.2 & [4,16] & $0.855(13)^{+1}$ & 1.2 & [3,16] \\ 
0.021 & $0.833(12)^{+3}$ & 1.5 & [4,16] & $0.871(11)^{+1}$ & 1.5 & [3,16] \\ 
0.028 & $0.872(10)^{+2}$ & 1.6 & [4,16] & $0.896(9)^{+1}$ & 1.8 & [3,16] \\ 
0.04 & $0.927(8)^{+1}$ & 1.4 & [4,16] & $0.939(8)^{+1}$ & 1.9 & [3,16] \\ 
0.06 & $1.004(6)^{+1}$ & 1.0 & [4,16] & $1.007(7)^{+2}$ & 1.3 & [3,16] \\ 
0.09 & $1.101(5)^{+1}$ & 0.7 & [4,16] & $1.102(6)^{+2}$ & 0.6 & [3,16] \\ 
0.13 & $1.221(4)^{+1}$ & 0.9 & [4,16] & $1.220(5)^{+2}$ & 0.8 & [3,16] \\ 
0.18 & $1.364(4)^{+1}$ & 1.2 & [4,16] & $1.362(4)^{+2}$ & 1.1 & [3,16] \\ 
0.25 & $1.558(3)^{+2}$ & 1.3 & [4,16] & $1.555(4)^{+2}$ & 1.0 & [3,16] \\ 
0.33 & $1.773(3)^{+1}$ & 1.3 & [4,16] & $1.770(3)^{+1}$ & 0.8 & [3,16] \\  \hline
\end{tabular}

\end{center}
%\vspace{-3mm}
\caption{Octet baryon masses on three lattice of size $8^3\times
24$, $12^3\times 24$ and $16^3\times 32$ at
$\beta=3.0$ with $D^{\rm FP}$.}
\end{table}

\begin{table}[htb]
\small
\begin{center}
\begin{tabular}{l*{2}{|l|c|c}} \hline\hline
 \multicolumn{7}{c}{$12^3\times 24$, $\beta=3.0$, overlap-improved  } \\ \hline
$am_q$ & $am_{\rm Oct}$(N)$^*$ & $\chi^2_{df}$ & $t$ & $am_{\rm Oct}$(N0) &
$\chi^2_{df}$ & $t$  \\ \hline
0.009 & $0.968(82)^{-59}$ & 1.7 & [3,10] & $0.792(122)^{+29}$ & 1.6 & [3,12] \\ 
0.012 & $0.946(63)^{-35}$ & 1.4 & [3,10] & $0.877(90)^{+15}$ & 1.5 & [3,12] \\ 
0.016 & $0.935(43)^{-14}$ & 1.4 & [3,10] & $0.940(63)^{-3}$ & 1.4 & [3,12] \\ 
0.021 & $0.933(32)^{+0}$ & 1.3 & [3,10] & $0.966(44)^{+0}$ & 1.3 & [3,12] \\ 
0.028 & $0.953(24)^{+2}$ & 1.2 & [3,10] & $0.972(26)^{+0}$ & 1.2 & [3,12] \\ 
0.04 & $0.994(20)^{+2}$ & 1.0 & [3,10] & $0.992(18)^{+2}$ & 1.2 & [3,12] \\ 
0.06 & $1.068(17)^{+0}$ & 1.1 & [3,10] & $1.047(15)^{+4}$ & 1.8 & [3,12] \\ 
0.09 & $1.154(13)^{+1}$ & 1.5 & [3,10] & $1.143(10)^{+3}$ & 1.6 & [3,12] \\ 
0.13 & $1.266(10)^{+2}$ & 1.3 & [3,10] & $1.262(8)^{+2}$ & 0.6 & [3,12] \\ 
0.18 & $1.413(7)^{+3}$ & 1.0 & [3,10] & $1.409(8)^{+2}$ & 0.4 & [3,12] \\ 
0.25 & $1.615(7)^{+1}$ & 1.1 & [3,10] & $1.611(8)^{+2}$ & 1.1 & [3,12] \\ 
0.33 & $1.840(7)^{+1}$ & 1.1 & [3,10] & $1.835(6)^{+2}$ & 1.2 & [3,12] \\  \hline
\end{tabular}
\end{center}
%\vspace{-3mm}
\caption{Octet baryon masses on $12^3\times 24$ lattice at
$\beta=3.0$ with overlap-improved $D_{\rm ov}^{\rm FP}$.}
\end{table}

\begin{table}[htb]
\small
\begin{center}
\begin{tabular}{l*{2}{|l|c|c}} \hline\hline
 \multicolumn{7}{c}{$12^3\times 24$, $\beta=3.4$, parametrized FP } \\ \hline
$am_q$ & $am_{\rm Oct}$(N) & $\chi^2_{df}$ & $t$ & $am_{\rm Oct}$(N0)$^*$ &
$\chi^2_{df}$ & $t$  \\ \hline
0.029 & $0.564(47)^{+19}$ & 0.8 & [4,12] & $0.672(29)^{+2}$ & 1.3 & [4,12] \\ 
0.032 & $0.585(40)^{+15}$ & 0.6 & [4,12] & $0.680(25)^{+2}$ & 1.1 & [4,12] \\ 
0.037 & $0.637(24)^{+8}$ & 0.5 & [4,12] & $0.696(19)^{+2}$ & 0.8 & [4,12] \\ 
0.045 & $0.696(16)^{+4}$ & 0.5 & [4,12] & $0.724(14)^{+2}$ & 0.5 & [4,12] \\ 
0.058 & $0.756(11)^{+3}$ & 0.6 & [4,12] & $0.765(10)^{+1}$ & 0.3 & [4,12] \\ 
0.078 & $0.811(8)^{+1}$ & 0.8 & [5,12] & $0.817(9)^{+1}$ & 0.4 & [5,12] \\ 
0.1 & $0.874(8)^{+1}$ & 1.4 & [6,12] & $0.878(8)^{+1}$ & 0.6 & [6,12] \\ 
0.14 & $0.977(7)^{+1}$ & 1.1 & [7,12] & $0.981(7)^{+1}$ & 0.4 & [7,12] \\ 
0.18 & $1.088(5)^{+1}$ & 1.0 & [7,12] & $1.090(6)^{+1}$ & 0.4 & [7,12] \\ 
0.24 & $1.253(5)^{+1}$ & 1.0 & [7,12] & $1.253(5)^{+0}$ & 0.6 & [7,12] \\  \hline
\end{tabular}
\end{center}
%\vspace{-3mm}
\caption{Octet baryon masses on $12^3\times 24$ lattice at
$\beta=3.4$ with $D^{\rm FP}$.}
\end{table}

\begin{table}[htb]
\small
\begin{center}
\begin{tabular}{l*{2}{|l|c|c}} \hline\hline
 \multicolumn{7}{c}{$16^3\times 32$, $\beta=3.7$, parametrized FP } \\ \hline
$am_q$ & $am_{\rm Oct}$(N) & $\chi^2_{df}$ & $t$ & $am_{\rm Oct}$(N0)$^*$ &
$\chi^2_{df}$ & $t$  \\ \hline
0.0235 & $0.537(63)^{-22}$ & 0.8 & [5,16] & $0.530(40)^{+2}$ & 1.4 & [5,14] \\ 
0.026 & $0.528(42)^{-8}$ & 0.8 & [5,16] & $0.533(28)^{+1}$ & 1.4 & [5,14] \\ 
0.03 & $0.532(23)^{+2}$ & 0.8 & [6,16] & $0.543(23)^{+3}$ & 1.5 & [6,14] \\ 
0.036 & $0.540(17)^{+2}$ & 0.6 & [6,16] & $0.560(16)^{+2}$ & 1.4 & [6,14] \\ 
0.045 & $0.556(11)^{+3}$ & 0.4 & [6,16] & $0.578(11)^{+3}$ & 0.9 & [6,16] \\ 
0.06 & $0.599(12)^{+6}$ & 0.6 & [7,16] & $0.610(11)^{+6}$ & 0.9 & [8,16] \\ 
0.08 & $0.659(11)^{+5}$ & 0.8 & [8,16] & $0.662(9)^{+5}$ & 0.7 & [8,16] \\ 
0.1 & $0.722(8)^{+3}$ & 0.8 & [8,16] & $0.720(7)^{+4}$ & 0.5 & [8,16] \\ 
0.14 & $0.835(6)^{+3}$ & 1.0 & [8,16] & $0.833(6)^{+3}$ & 0.6 & [8,16] \\ 
0.18 & $0.944(5)^{+2}$ & 1.0 & [8,16] & $0.945(5)^{+2}$ & 0.8 & [8,16] \\ \hline
\end{tabular}
\end{center}
%\vspace{-3mm}
\caption{Octet baryon masses on $16^3\times 32$ lattice at
$\beta=3.7$ with $D^{\rm FP}$.}
\end{table}

\clearpage
\subsection{Decuplet Baryons}

For each lattice, the bias-corrected fitted masses for the D and D0
correlators, the value of $\chi^2/df$ for the fit and the 
fit range are given.

\vspace{5mm}

\begin{table}[htb]
\small
\begin{center}
\begin{tabular}{l*{2}{|l|c|c}} \hline\hline
 \multicolumn{7}{c}{$8^3\times 24$, $\beta=3.0$, parametrized FP } \\ \hline
$am_q$ & $am_{\rm Dec}$(D) & $\chi^2_{df}$ & $t$ & $am_{\rm Dec}$(D0) &
$\chi^2_{df}$ & $t$  \\ \hline
0.028 & $0.994(43)^{+15}$ & 0.7 & [3,12] & $0.952(57)^{+24}$ & 0.7 & [3,12] \\ 
0.04 & $1.034(30)^{+13}$ & 0.7 & [3,12] & $1.038(35)^{+12}$ & 0.6 & [3,12] \\ 
0.06 & $1.103(25)^{+12}$ & 0.6 & [3,12] & $1.115(25)^{+9}$ & 0.6 & [3,12] \\ 
0.09 & $1.192(19)^{+9}$ & 0.9 & [3,12] & $1.202(19)^{+9}$ & 0.9 & [3,12] \\ 
0.13 & $1.305(15)^{+8}$ & 1.3 & [3,12] & $1.305(15)^{+8}$ & 1.3 & [3,12] \\ 
0.18 & $1.444(12)^{+6}$ & 0.9 & [4,12] & $1.431(11)^{+6}$ & 1.7 & [3,12] \\ 
0.25 & $1.620(12)^{+5}$ & 0.5 & [5,12] & $1.618(11)^{+5}$ & 0.5 & [4,12] \\ 
0.33 & $1.823(9)^{+3}$ & 0.9 & [5,12] & $1.821(9)^{+4}$ & 0.5 & [5,12] \\  \hline
\end{tabular}

\vspace{3ex}

\begin{tabular}{l*{2}{|l|c|c}} \hline\hline
 \multicolumn{7}{c}{$12^3\times 24$, $\beta=3.0$, parametrized FP } \\ \hline
$am_q$ & $am_{\rm Dec}$(D) & $\chi^2_{df}$ & $t$ & $am_{\rm Dec}$(D0) &
$\chi^2_{df}$ & $t$  \\ \hline
0.016 & $0.973(38)^{-5}$ & 0.8 & [3,12] & $0.993(62)^{-5}$ & 1.8 & [3,12] \\ 
0.021 & $1.005(31)^{-5}$ & 1.2 & [3,12] & $0.966(37)^{+1}$ & 1.3 & [3,12] \\ 
0.028 & $1.023(24)^{-3}$ & 1.6 & [3,12] & $0.997(24)^{+3}$ & 0.9 & [3,12] \\ 
0.04 & $1.012(25)^{+3}$ & 2.0 & [4,12] & $1.044(17)^{+1}$ & 0.9 & [3,12] \\ 
0.06 & $1.075(15)^{+1}$ & 2.2 & [4,12] & $1.100(12)^{+1}$ & 1.8 & [3,12] \\ 
0.09 & $1.169(10)^{+1}$ & 2.5 & [4,12] & $1.179(9)^{+1}$ & 2.2 & [3,12] \\ 
0.13 & $1.283(8)^{+1}$ & 2.8 & [4,12] & $1.287(7)^{+1}$ & 1.6 & [3,12] \\ 
0.18 & $1.418(6)^{+1}$ & 2.5 & [4,12] & $1.419(6)^{+1}$ & 1.0 & [3,12] \\ 
0.25 & $1.602(5)^{+1}$ & 1.6 & [4,12] & $1.601(5)^{+1}$ & 0.7 & [3,12] \\ 
0.33 & $1.811(4)^{+1}$ & 1.0 & [4,12] & $1.810(4)^{+1}$ & 0.6 & [3,12] \\  \hline
\end{tabular}

\vspace{3ex}

\begin{tabular}{l*{2}{|l|c|c}} \hline\hline
 \multicolumn{7}{c}{$16^3\times 32$, $\beta=3.0$, parametrized FP } \\ \hline
$am_q$ & $am_{\rm Dec}$(D) & $\chi^2_{df}$ & $t$ & $am_{\rm Dec}$(D0) &
$\chi^2_{df}$ & $t$  \\ \hline
0.013 & $1.013(33)^{-1}$ & 1.7 & [3,16] & $0.933(40)^{+4}$ & 1.1 & [3,16] \\ 
0.016 & $1.010(24)^{-3}$ & 1.7 & [3,16] & $0.948(34)^{+4}$ & 1.7 & [3,16] \\ 
0.021 & $1.014(19)^{+0}$ & 1.7 & [3,16] & $0.962(26)^{+4}$ & 2.1 & [3,16] \\ 
0.028 & $1.033(16)^{+0}$ & 1.7 & [3,16] & $0.989(20)^{+3}$ & 1.9 & [3,16] \\ 
0.04 & $1.069(13)^{+1}$ & 1.7 & [3,16] & $1.033(14)^{+3}$ & 2.0 & [3,16] \\ 
0.06 & $1.110(10)^{+1}$ & 1.6 & [4,16] & $1.099(11)^{+1}$ & 2.2 & [3,16] \\ 
0.09 & $1.191(8)^{+1}$ & 1.9 & [4,16] & $1.188(8)^{+1}$ & 2.4 & [3,16] \\ 
0.13 & $1.296(8)^{+1}$ & 2.3 & [4,16] & $1.296(6)^{+1}$ & 2.5 & [3,16] \\ 
0.18 & $1.426(6)^{+2}$ & 2.4 & [4,16] & $1.425(5)^{+1}$ & 2.5 & [3,16] \\ 
0.25 & $1.609(5)^{+2}$ & 2.0 & [4,16] & $1.607(4)^{+2}$ & 2.2 & [3,16] \\ 
0.33 & $1.816(4)^{+1}$ & 1.8 & [4,16] & $1.814(4)^{+1}$ & 1.7 & [3,16] \\  \hline
\end{tabular}

\end{center}
%\vspace{-3mm}
\caption{Decuplet baryon masses on three lattice of size $8^3\times
24$, $12^3\times 24$ and $16^3\times 32$ at
$\beta=3.0$ with $D^{\rm FP}$.}
\end{table}

\begin{table}[htb]
\small
\begin{center}
\begin{tabular}{l*{2}{|l|c|c}} \hline\hline
 \multicolumn{7}{c}{$12^3\times 24$, $\beta=3.0$, overlap-improved } \\ \hline
$am_q$ & $am_{\rm Dec}$(D) & $\chi^2_{df}$ & $t$ & $am_{\rm Dec}$(D0) &
$\chi^2_{df}$ & $t$  \\ \hline
0.009 & $1.093(117)^{-20}$ & 0.5 & [2,10] & $0.940(105)^{+55}$ & 0.9 & [2,12] \\ 
0.012 & $1.086(98)^{-20}$ & 0.6 & [2,10] & $1.032(79)^{+29}$ & 1.1 & [2,12] \\ 
0.016 & $1.091(83)^{-24}$ & 0.9 & [2,10] & $1.069(74)^{+24}$ & 1.5 & [2,12] \\ 
0.021 & $1.106(58)^{-11}$ & 1.1 & [2,10] & $1.105(64)^{+16}$ & 1.8 & [2,12] \\ 
0.028 & $1.110(38)^{-2}$ & 1.2 & [2,10] & $1.135(54)^{-2}$ & 1.9 & [2,12] \\ 
0.04 & $1.115(25)^{+6}$ & 1.2 & [2,10] & $1.132(32)^{+10}$ & 1.4 & [2,12] \\ 
0.06 & $1.152(21)^{+12}$ & 1.5 & [2,10] & $1.142(28)^{+15}$ & 0.7 & [2,12] \\ 
0.09 & $1.232(22)^{+9}$ & 1.7 & [2,10] & $1.225(19)^{+11}$ & 0.4 & [2,12] \\ 
0.13 & $1.347(19)^{+8}$ & 1.7 & [2,10] & $1.344(13)^{+7}$ & 0.3 & [2,12] \\ 
0.18 & $1.491(17)^{+5}$ & 1.8 & [2,10] & $1.483(12)^{+4}$ & 0.4 & [2,12] \\ 
0.25 & $1.683(12)^{+5}$ & 1.8 & [2,10] & $1.673(10)^{+3}$ & 0.9 & [2,12] \\ 
0.33 & $1.895(8)^{+2}$ & 1.9 & [2,10] & $1.887(9)^{+2}$ & 1.9 & [2,12] \\  \hline
\end{tabular}
\end{center}
%\vspace{-3mm}
\caption{Decuplet baryon masses on $12^3\times 24$ lattice at
$\beta=3.0$ with $D_{\rm ov}^{\rm FP}$.}
\end{table}

\begin{table}[htb]
\small
\begin{center}
\begin{tabular}{l*{2}{|l|c|c}} \hline\hline
 \multicolumn{7}{c}{$12^3\times 24$, $\beta=3.4$, parametrized FP } \\ \hline
$am_q$ & $am_{\rm Dec}$(D) & $\chi^2_{df}$ & $t$ & $am_{\rm Dec}$(D0) &
$\chi^2_{df}$ & $t$  \\ \hline
0.029 & $0.746(34)^{+4}$ & 1.1 & [4,12] & $0.743(38)^{-2}$ & 0.9 & [4,12] \\ 
0.032 & $0.758(29)^{+4}$ & 0.9 & [4,12] & $0.754(33)^{+2}$ & 0.8 & [4,12] \\ 
0.037 & $0.772(22)^{+1}$ & 0.7 & [4,12] & $0.775(25)^{+0}$ & 0.7 & [4,12] \\ 
0.045 & $0.797(20)^{+0}$ & 0.6 & [4,12] & $0.800(17)^{+0}$ & 0.6 & [4,12] \\ 
0.058 & $0.838(15)^{+1}$ & 0.8 & [4,12] & $0.833(18)^{+4}$ & 0.6 & [5,12] \\ 
0.078 & $0.876(9)^{+1}$ & 0.3 & [5,12] & $0.877(10)^{+2}$ & 0.7 & [5,12] \\ 
0.1 & $0.930(8)^{+1}$ & 0.6 & [6,12] & $0.931(9)^{+1}$ & 1.1 & [6,12] \\ 
0.14 & $1.022(8)^{+2}$ & 1.0 & [7,12] & $1.022(9)^{+2}$ & 1.1 & [7,12] \\ 
0.18 & $1.127(6)^{+1}$ & 2.0 & [7,12] & $1.125(7)^{+1}$ & 1.5 & [7,12] \\ 
0.24 & $1.287(5)^{+1}$ & 3.2 & [7,12] & $1.284(6)^{+1}$ & 2.0 & [7,12] \\  \hline
\end{tabular}
\end{center}
%\vspace{-3mm}
\caption{Decuplet baryon masses on $12^3\times 24$ lattice at
$\beta=3.4$ with $D^{\rm FP}$.}
\end{table}

\begin{table}[htb]
\small
\begin{center}
\begin{tabular}{l*{2}{|l|c|c}} \hline\hline
 \multicolumn{7}{c}{$16^3\times 32$, $\beta=3.7$, parametrized FP } \\ \hline
$am_q$ & $am_{\rm Dec}$(D) & $\chi^2_{df}$ & $t$ & $am_{\rm Dec}$(D0) &
$\chi^2_{df}$ & $t$  \\ \hline
0.0235 & $0.572(35)^{+13}$ & 0.9 & [3,12] & $0.506(93)^{+26}$ & 1.0 & [5,16] \\ 
0.026 & $0.592(28)^{+0}$ & 1.0 & [3,12] & $0.560(93)^{+34}$ & 0.6 & [6,16] \\ 
0.03 & $0.586(24)^{+1}$ & 0.7 & [4,12] & $0.589(44)^{+8}$ & 0.8 & [6,16] \\ 
0.036 & $0.573(19)^{+8}$ & 1.0 & [4,14] & $0.587(23)^{+5}$ & 0.9 & [6,16] \\ 
0.045 & $0.588(14)^{+8}$ & 1.2 & [4,14] & $0.590(16)^{+7}$ & 0.6 & [6,16] \\ 
0.06 & $0.629(10)^{+6}$ & 1.6 & [4,14] & $0.643(15)^{+7}$ & 0.7 & [7,16] \\ 
0.08 & $0.681(8)^{+5}$ & 1.7 & [4,14] & $0.696(12)^{+6}$ & 1.2 & [7,16] \\ 
0.1 & $0.742(7)^{+4}$ & 1.5 & [6,14] & $0.748(10)^{+5}$ & 1.4 & [7,16] \\ 
0.14 & $0.855(7)^{+2}$ & 1.5 & [6,16] & $0.852(8)^{+4}$ & 1.3 & [7,16] \\ 
0.18 & $0.962(6)^{+2}$ & 1.6 & [6,16] & $0.960(6)^{+3}$ & 1.3 & [7,16]
\\ \hline

\end{tabular}
\end{center}
%\vspace{-3mm}
\caption{Decuplet baryon masses on $16^3\times 32$ lattice at
$\beta=3.7$ with $D^{\rm FP}$.}
\end{table}

\clearpage

\section{Unrenormalized AWI Quark Masses}
\label{app:awi}

The listed values and bootstrap errors of the unrenormalized quark masses
from the axial Ward identity are determined by
averaging the measurements of $am_q^{\rm AWI}(t)$ in
Eq.~\eqref{eq:awi_mass} over the time range $t\in[t_1,t_2]$. 

\vspace{2ex}

\begin{table}[htb]
\small
\begin{center}
\begin{tabular}{l|c|c|l|c|c} \hline\hline
 \multicolumn{3}{c}{$16^3\times 32$, $\beta=3.0$} & \multicolumn{3}{|c}{$12^3\times 24$, $\beta=3.0$} \\ 
 \multicolumn{3}{c}{ parametrized FP } & \multicolumn{3}{|c}{ overlap-improved FP } \\ \hline
$am_q$ & $am_q^{\rm AWI}$ & $t$ & $am_q$ & $am_q^{\rm AWI}$ & $t$  \\ \hline
 &   &                     & 0.009 & 0.00767(4) & [3,10]  \\
0.013 & 0.0100(3) & [4,14] & 0.012 & 0.01037(4) & [3,10] \\
0.016 & 0.0131(3) & [4,14] & 0.016 & 0.01398(5) & [3,10] \\
0.021 & 0.0176(2) & [4,14] & 0.021 & 0.01850(6) & [3,10] \\
0.028 & 0.0235(2) & [4,14] & 0.028 & 0.02486(6) & [3,10] \\
0.04 & 0.0335(2) & [4,14] & 0.04 & 0.03583(8) & [3,10] \\
0.06 & 0.0502(2) & [4,14] & 0.06 & 0.0544(1) & [4,10] \\
0.09 & 0.0759(2) & [4,14] & 0.09 & 0.0829(1) & [4,10] \\
0.13 & 0.1118(2) & [6,14] & 0.13 & 0.1222(2) & [4,10] \\
0.18 & 0.1594(2) & [6,14] & 0.18 & 0.1741(2) & [4,10] \\
0.25 & 0.2316(2) & [7,14] & 0.25 & 0.2523(3) & [4,10] \\
0.33 & 0.3235(2) & [7,14] & 0.33 & 0.3516(4) & [4,10] \\ \hline
\end{tabular}
\end{center}
%\vspace{-3mm}
\caption{AWI quark masses on $\beta=3.0$ lattices for parametrized and
 overlap-improved FP Dirac operators.}
\end{table}

\begin{table}[htb]
\small
\begin{center}
\begin{tabular}{l|c|c|l|c|c} \hline\hline
 \multicolumn{3}{c}{$12^3\times 24$, $\beta=3.4$} &
 \multicolumn{3}{|c}{$16^3\times 32$, $\beta=3.7$} \\  
 \multicolumn{3}{c}{parametrized FP} &
 \multicolumn{3}{|c}{parametrized FP} \\ \hline 
$am_q$ & $am_q^{\rm AWI}$ & $t$ & $am_q$ & $am_q^{\rm AWI}$ & $t$  \\ \hline
0.029 & 0.0094(3) & [6,10] & 0.0235 & 0.0034(2) & [4,14] \\
0.032 & 0.0122(2) & [6,10] & 0.026 & 0.0058(2) & [4,14] \\
0.037 & 0.0166(2) & [6,10] & 0.030 & 0.0093(2) & [4,14] \\
0.045 & 0.0235(2) & [6,10] & 0.036 & 0.0145(1) & [4,14] \\
0.058 & 0.0347(2) & [6,10] & 0.045 & 0.0222(1) & [4,14] \\
0.078 & 0.0521(2) & [6,10] & 0.06 & 0.0352(1) & [4,14] \\
0.10 & 0.0717(1) & [6,10] & 0.08 & 0.0527(1) & [4,14] \\
0.14 & 0.1083(1) & [6,10] & 0.10 & 0.0704(1) & [6,14] \\
0.18 & 0.1464(1) & [7,10] & 0.14 & 0.1069(1) & [8,14] \\
0.24 & 0.2068(1) & [7,10] & 0.18 & 0.1446(1) & [8,14] \\\hline
\end{tabular}
\end{center}
%\vspace{-3mm}
\caption{AWI quark masses on $\beta=3.4$ and $\beta=3.7$ lattices.}
\end{table}

\fancyhead[RE]{\nouppercase{\small\it Conventions}}
\chapter{Conventions} \label{app:conventions}

\section{Dirac Algebra in Minkowski Space}
%\subsection{Definitions}
In Minkowski space, the Dirac algebra is defined by the
anticommutation relation 
\begin{equation}
  \{\gamma_M^\mu,\gamma_M^\nu\}=2g^{\mu\nu} \cdot {\mathbf 1}.
\end{equation}
From the elements $\gamma^\mu_M$ of the Dirac algebra, we construct the tensor
\begin{equation}
  \sigma_M^{\mu\nu} \equiv \frac{1}{2i}[\gamma_M^\mu,\gamma_M^\nu],
\end{equation}
and the pseudoscalar
\begin{equation}
\gamma_M^5 \equiv i\gamma_M^0\gamma_M^1\gamma_M^2\gamma_M^3,
\end{equation}
which satisfies $\gamma_M^5  \gamma_M^5 = {\mathbf 1}$ and ${\gamma_M^5}^\dagger = \gamma_M^5$. The set of 16 elements
\begin{equation}
  \Gamma_M \equiv  \left\{ {\mathbf 1}, \gamma_M^\mu, \sigma_M^{\mu\nu},
  \gamma_M^\mu\gamma_M^5, i\gamma_M^5  \right\},
\end{equation}
with $\mu\le\nu$ forms a basis of the Dirac algebra and satisfies $
\gamma^0\Gamma_M^\dagger\gamma^0=\Gamma_M$.

%\subsection{Weyl Representation}
The Weyl representation of the Dirac algebra is given by the four-dimensional matrices
\begin{equation}
  \gamma_M^0 = \begin{pmatrix} 0 & {\mathbf 1} \\ {\mathbf 1} & 0 \end{pmatrix}, \quad\gamma_M^i = \begin{pmatrix} 0 & \sigma^i \\ -\sigma^i & 0 \end{pmatrix},
\end{equation}
where $\sigma^i$ are the Pauli matrices
\begin{equation}
  \sigma_1 = \begin{pmatrix} 0 & 1 \\ 1 & 0 \end{pmatrix}, \quad
  \sigma_2 = \begin{pmatrix} 0 & -i \\ i & 0 \end{pmatrix}, \quad
  \sigma_3 = \begin{pmatrix} 1 & 0 \\ 0 & -1 \end{pmatrix}.
\end{equation}
With this definition, the basis element $\gamma_M^5$ is diagonal:
\begin{equation}
  \gamma_M^5 = \begin{pmatrix} -{\mathbf 1} & 0 \\ 0 & {\mathbf 1} \end{pmatrix}.
\end{equation}
%\subsection{Properties of the Dirac Matrices}
We conclude this section by listing some transformation properties of
the Dirac matrices in our convention. Under hermitean conjugation, we have
\begin{equation}
 {\gamma_M^0}^\dagger=\gamma_M^0; \quad {\gamma_M^i}^\dagger=-\gamma_M^i.
\end{equation}
Under complex conjugation the Dirac matrices transform as
\begin{equation}
  {\gamma_M^0}^*=\gamma_M^0, \quad {\gamma_M^1}^*=\gamma_M^1, \quad {\gamma_M^2}^*=-\gamma_M^2, \quad {\gamma_M^3}^*=\gamma_M^3,
\end{equation}
and finally, the transposition properties are
\begin{equation}
{\gamma_M^0}^T=\gamma_M^0; \quad {\gamma_M^1}^T=-\gamma_M^1; \quad {\gamma_M^2}^T=\gamma_M^2; \quad {\gamma_M^3}^T=-\gamma_M^3.
\end{equation}

\section{Analytic Continuation to Euclidean Space}
\label{app:euclidean_space}
Euclidean space-time is reached from Minkowski space-time by
analytic continuation, rotating the time direction onto the imaginary axis:
\begin{equation}
  x^0 \rightarrow -ix^4; \quad x^i \rightarrow x^i.
\end{equation}
In Euclidean space-time, the Dirac (or Clifford) algebra satisfies the
anticommutation rule 
\begin{equation} \label{eq:clifford_rule}
  \{\gamma_\mu,\gamma_\nu\}=2\delta_{\mu\nu} \cdot {\mathbf 1}.
\end{equation}
Due to the trivial Euclidean metric, upper and lower
indices are the same. We find that the Euclidean matrices 
\begin{equation}
  \gamma_4 \equiv \gamma_M^0, \quad \gamma_i \equiv -i\gamma_M^i,
\end{equation}
satisfy Eq.~\eqref{eq:clifford_rule}. 

From the properties of the Dirac matrices in Minkowski
space, it then follows that all the $\gamma_\mu$ are hermitean. In the
Euclidean version of the Weyl representation, $\gamma_2$ and
$\gamma_4$ are real and symmetric, while $\gamma_1$ and $\gamma_3$ are
purely imaginary and antisymmetric. Like in Minkowski space, we define
the tensor element
\begin{equation}
  \sigma_{\mu\nu} \equiv \frac{1}{2i}[\gamma_\mu,\gamma_\nu].
\end{equation}
The pseudoscalar $\gamma_5$ is taken to be the same as in Minkowski space:
\begin{equation}
  \gamma_5 \equiv  \gamma^5_M = -\gamma_1\gamma_2\gamma_3\gamma_4.
\end{equation}
The set
\begin{equation}
  \Gamma \equiv \left\{ {\mathbf 1}, \gamma_\mu, \sigma_{\mu\nu},
  i\gamma_\mu \gamma_5, 
  \gamma_5  \right\} ,
\end{equation}
with $\mu\le\nu$ then forms a hermitean basis $\Gamma^\dagger=\Gamma$
with elements 
\begin{equation}
  \Gamma = \left\{ S, V, T, A, P  \right\} ,
\end{equation}
transforming like scalars (S), vectors (V), tensors (T), axial vectors
(A) and pseudoscalars (P).

%\section{Transformation Properties of the Dirac Operator under
%Discrete Symmetries} 

%From the properties of the Dirac spinor $\Psi$ under discrete symmetry
%transformations (see e.g.~Peskin, Schr\"oder), one finds the following
%requirements for the Dirac operator 
%assuming that the action $S=\int \bar\Psi D \Psi$ remains invariant:

%\begin{center}
%\begin{tabular}{|r|c|c|} \hline
% & $D_M(t,\vec x)$ & $D(\vec x, x^4)$ \\ \hline
% h.c. & $\gamma^0_M D_M^\dagger \gamma^0_M$ & $\gamma^5 D^\dagger
% \gamma^5$\\
% $\mathcal{P}$ & $i\gamma^0_m D_M(t,-\vec x) (-i\gamma^0_M)$ & $\gamma^4
% D(-\vec x, x^4) \gamma^4$ \\
% $\mathcal{C}$ & $-i\gamma^0_m\gamma^2_M D_M^T i \gamma^0_M\gamma^2_M$ &
% $\gamma^4\gamma^2 D^T \gamma^2\gamma^4$ \\
% $\mathcal{P}_0$, $\mathcal{P}_4$ & $\gamma^0_M\gamma^5_M D_M(-t,\vec x)
% \gamma^5_M\gamma^0_M$ & $\gamma^4\gamma^5 D(\vec x,-x^4) \gamma^5\gamma^4$ \\
%\hline
%\end{tabular} 
%\end{center}

\end{appendix}

\fancyhead[RE]{\nouppercase{\small\it Acknowledgements}}
\chapter*{Acknowledgements}
\addcontentsline{toc}{chapter}{Acknowledgements}

Many people have played an important role in making this work
possible. Peter Hasenfratz introduced me to the subject of
perfect actions and gave me the opportunity
to work in the fascinating field of lattice QCD. Ferenc
Niedermayer provided a wealth 
of ideas and advice from which I could profit.  

During my thesis, I had the chance to collaborate with a 
number of people: Primarily I have to mention my fellow PhD
student Thomas J\"org, with whom I worked closely on the construction
of the Fixed-Point fermions. Kieran Holland participated during an
essential phase in the 
project, parametrized the Fixed-Point $R$ operator and implemented the
multi-mass inverter used in this work. In the first period of my
thesis, I could rely on Urs Wenger and  
Philipp R\"ufenacht, who were so kind to share their experience with
perfect gauge actions and to answer my naive
questions. Thanks to these people, I could also enjoy not only the
scientific part of the lattice conferences and workshops.

I would like to thank Tom DeGrand for providing me the possibility of
a memorable six-weeks stay in Boulder, for giving me insight into a
range of technical 
problems and for sharing his scaling data. Thanks to Fernando Perez,
with whom I shared the office there, I could
experience some spectacular rock climbing in the Boulder area. Christof
Gattringer initiated the BGR collaboration, consisting of the groups 
in Regensburg, Graz and Bern, due to which we could run simulations on
the Hitachi 
computer in M\"unchen, and he contributed gauge configurations
and results from the chirally improved operator. I further thank
Christian Lang 
for discussions on the gauge fixing algorithm. 
I thank the Swiss Center for Scientific Computing in Manno and the Leibniz
Rechenzentrum in M\"unchen for granting access to their supercomputers
and offering technical support.
Thanks go to Ottilia H\"anni for dealing with the administrative
hassles. At last, I thank the present and former members of the
institute in Bern, 
who created an enjoyable atmosphere.

Most importantly, I was
so fortunate to enjoy the social environment, support and love of my
friends and 
family in all 
these years. I am deeply indebted to them. 

This work has been supported by the Schweizerischer Nationalfonds
and by the European Community's Human Potential Programme
under HPRN-CT-2000-00145 Hadrons/Lattice QCD, BBW Nr. 99.0143.

\fancyhead[RE]{\nouppercase{\small\it Bibliography}}
\addcontentsline{toc}{chapter}{Bibliography}
\bibliography{lattice}

%%%%% Curriculum %%%%%
%\newpage
%\cleardoublepage
%\newpage
%\pagestyle{empty}
%\cleardoublepage
%\newpage
%\pagestyle{empty}
%\include{cv}

\end{document}